\DocumentMetadata 
{
	lang		= en-US, % default language
	pdfversion  = 1.7,
	pdfstandard = a-2b,
}

\documentclass[twoside]{mitthesis}
\usepackage{listings}%   documentation is here https://ctan.org/pkg/listings

\usepackage[version=4]{mhchem}%   documentation at https://ctan.org/pkg/mhchem

\usepackage{lipsum}
\IfPackageAtLeastTF{lipsum}{2021/09/20}{\setlipsum{auto-lang=false}}{} % to avoid a hyphenation warning

\usepackage{float}

\usepackage{booktabs}% publication quality tables (https://ctan.org/pkg/booktabs)

\usepackage{array}% Additional options for column formats (https://ctan.org/pkg/array)

\usepackage{dcolumn}% For alignment of numbers on the decimal place (https://ctan.org/pkg/dcolumn) 
  \newcolumntype{d}[1]{D{.}{.}{#1}}% use with dcolumn package. Note: dcolumns are set in math mode.
\usepackage{longtable}% typeset multi-page tables (https://ctan.org/pkg/longtable)

\usepackage[nopatch=footnote]{microtype}% typographic fine-tuning (https://ctan.org/pkg/microtype)
\usepackage{multicol}

\usepackage[style=ext-numeric-comp,giveninits=true,maxbibnames=10,sorting=none,language=american]{biblatex}

    \AtEveryBibitem{%
      \ifentrytype{article}{%
        \renewbibmacro{in:}{}% Removes "In:" for articles
		\renewbibmacro*{issue+date}{% Print date without parentheses around it
		\iffieldundef{issue}%
        	{}%
        	{\printfield{issue}%
         	\setunit*{\addcomma\space}
        	}%
			\printdate
		}
      }{}
    }
	\renewbibmacro*{volume+number+eid}{%
        \printtext[bold]{\printfield{volume}}%    bold volume (issue number) , eid
        \iffieldundef{number}
          {} % do nothing
          {\printtext[parens]{\printfield{number}}} % print number in parentheses
        \setunit{\bibeidpunct}%
        \printfield{eid}
    }

\usepackage{setspace}%  documentation at https://ctan.org/pkg/setspace
\hypersetup{%
	pdfsubject={Applications of Flux Compactifications in F-Theory},
	pdfkeywords={Massachusetts Institute of Technology, MIT, Shing Yan Li, string theory, string phenomenology, F-theory, Standard Model, GUT, compactification, flux},
	pdfurl={},
	pdfcontactemail={},
	pdfauthortitle={},
}

\usepackage{amssymb}% \mathbb, \mathfrak, and other AMS symbols

\newcommand{\SU}{\mathrm{SU}}
\newcommand{\SO}{\mathrm{SO}}
\newcommand{\U}{\mathrm{U}}

\newcommand{\gsm}{G_{\rm SM}}

\def\Z{\mathbb{Z}}
\def\C{\mathbb{C}}
\def\F{\mathbb{F}}

\def\P{\mathbb{P}}

\newcommand*{\cy}{CY~}
\def\mred{M_{\rm red}}

\usepackage{verbatim}

\DeclareMathOperator{\Erf}{erf}
\DeclareMathOperator{\Erfc}{erfc}
\DeclarePairedDelimiterXPP\erf[1]{\Erf\mkern1mu}(){}{#1}% increase to 2mu with stix2 font
\DeclarePairedDelimiterXPP\erfc[1]{\Erfc\mkern1mu}(){}{#1}

\begin{document}
%%% edit the following commands to match your thesis %%%%%%%%%%

\title{Applications of Flux Compactifications in F-Theory}

% \Author{Author full name}{Author department}[Author's first PREVIOUS degree][Author's second PREVIOUS degree][...
% Note that third, fourth, fifth, and sixth arguments are optional [] and may be omitted
%
 \Author{Shing Yan Li}{Department of Physics}
% \Author{Luisa Hernández}{Department of Research}[B.S. Mechanical Engineering, UCLA, 2018][M.S. Stellar Interiors, Vulcan Science Academy, 2020]
% \Author{Thurston Howell III}{Department of Economics}[MBA, Ferengi School of Management, 2022]
%
% note on these names: most of the following names are made up, but Silas Holman was a physics professor at MIT in the 19th century.

% Use once for each degree fulfilled by thesis
% For two degrees from one department, leave the department argument blank for the second degree {}.
% For one degree from two departments, leave the degree argument blank for the second department {}.
%
\Degree{Doctor of Philosophy in Physics}{Department of Physics}
% \Degree{Master of Science in Physics}{}
% \Degree{Bachelor of Science in Mechanical Engineering}{Department of Mechanical Engineering}

% If there is more than one supervisor, use the \Supervisor command for each.
% NB: department field added with v1.21 in late 2025. It is used ONLY on the Thesis Committee page.
%
 \Supervisor{Washington Taylor}{Professor of Physics}[Department of Physics]
% \Supervisor{Secunda Castor}{Professor of Research \\ & Professor of Knowledge \\ & Professor of Reason}[Department of Research ]
% \Supervisor{Quintus Castor}{Professor of Log Dams}[Department of Civil \& Environmental Engineering \\ & Department of All Faculty Who Are Not Members of a Department] 
% Separate multiple titles or departments with "\\ &".  If not preceded by "\\", an "&" will be converted to the string "&" (an "\&" by itself will be converted to "and"). 

% Professor who formally accepts theses for your department (e.g., the Graduate Officer, Professor Sméagol,...)
% If more than one department gives the degree, use once for each department
%
\Acceptor{Scott Hughes}{Professor of Physics}{Associate Department Head} % \\ & Third title}
% \Acceptor{Quarta Castor}{Professor of Lodge Building}{Graduate Officer, Department of Mechanical Engineering}
%%%  If you need to reduce vertical space, put the acceptor title in the second argument and leave the third blank, {}.
% \Acceptor{Primus Castor}{Professor and Undergraduate Officer, Department of Physics}{}

% Date degree will be issued: \DegreeDate{Month}{year}
% Valid degree months are February, May, June, or September
\DegreeDate{September}{2026}

% Date that final thesis is submitted to department
\ThesisDate{August 14, 2026}

% If at least one thesis reader is included, a committee members page will be automatically generated.
% The MIT libraries do not set a format for the committee members page, and if you prefer to use your own format,
% omit ALL readers here and insert your page just before the abstract.
%
\Reader{Jesse Thaler}{Professor of Physics}{Department of Physics}
\Reader{Netta Engelhardt}{Associate Professor of Physics}{Department of Physics}

% Are your readers running off the page?  You can move the page heading upward with this command:
% \CMPShortenTop

%%%%%%  Choose whether to have a CREATIVE COMMONS License  %%%%%%%%%%%%%%%%%%%%%%%%%%%%%%%%%%%%%%
%
% If you are using a cc license, uncomment the following line and insert details of your cc license here.
%
\CClicense{CC BY-NC-ND 4.0}{https://creativecommons.org/licenses/by-nc-nd/4.0/}
%

%%%%%%%  Solutions for overflowing titlepage  %%%%%%%%%%%%%%%%%%%%%%%%%%%%%%%%%%%%%%%%%%%%%%%%%%%

% If your title page is overflowing (from too many names, degrees, etc.):
%
% (a) you can reduce the 12pt and 18pt skips between various blocks to 6pt with this command:
%
% \Tighten
%
% (b)  you can scale down the Signature block at the bottom with this command:
%
% \SignatureBlockSize{\small}  %or this one \SignatureBlockSize{\footnotesize}
%
% (c) you can put the acceptor name and title onto two lines, rather than three like this:
%
% \Acceptor{Tertius Castor}{Professor and Graduate Officer, Department of Research}{}
%
% (d) you can change the font size of the author name[s] with
%
%	\AuthorNameSize{\normalsize}
%
% (e) and you can omit any previous degrees from the title page, instead mentioning them in the biographical sketch

% Also, if you prefer to keep the text toward the top of the page with most white space at the bottom, you
% can use this command to squash all of the vertical glue (stretchy space) with this command:
%
% \Squash 
%
% This command is useful when the text has not already reach the bottom of the page, since the glue gets squashed automatically
% when the page is too full.

%%%%%%%%%%%%%%%%%%%%%%%%%%%%%%%%%%%%%%%%%%%%%%%%%%%%%%%%%%%%%%%%%%%%%%%%%%%%%%%%%%%%%%%%%%%%%%%%%

%%% Make titlepage
\maketitle

%%%%%%%%% Content that you need to write follows! %%%%%%%%%%%%%%%%%%%%%%%%%%%%%%%%%%%%%%%%%%%%%%

% \includeonly{acknowledgments,biography,chapter1,chapter2,...,appendixa,...} 
%   for usage of includeonly, see https://latexref.xyz/_005cinclude-_0026-_005cincludeonly.html

%%% Frontmatter (write this material in the mentioned files)  %%%%%%%%%%%%%%%%%%%%%%%%%%%%%%%%%%%

% If you are not using the automatic committee members page (by completing at least one \Reader macro), 
% you can insert a page here.
% \include{committee_members}% This page is optional. Edit the file committee_members.tex 

% The abstract environment creates all the required headings and footers. 
% You only need to the text of the abstract in the file abstract.tex
\begin{abstract}
	The Standard Model of particle physics can describe most of observed 
physics, with three fundamental forces in terms of gauge symmetries, three generations of
quarks and leptons, and the Higgs boson. Despite its phenomenological success, the 
origin of these specific choices of gauge
groups and matter content remains one of the most fundamental open questions
in physics. String theory provides a top-down approach to this problem, in which all the
observed physics is in principle determined by the geometry of the six compactified extra
dimensions. It is notoriously hard, however, to construct explicit solutions in string
theory that exactly resemble the realistic 4D low-energy physics.

F-theory, a strongly coupled version of type IIB string theory, offers a tractable
framework for exploring the set of possible vacua in string theory, known as the
string landscape. F-theory gives rise to the currently known largest set of string vacua,
since the low-energy physics, namely the gauge groups, matter content, and Yukawa couplings, 
are fully encoded in the singular geometry of elliptically fibered Calabi-Yau manifolds,
which can be analyzed using tools from algebraic geometry. To study the connection to 
particle physics, this thesis focuses on 4D F-theory compactifications 
with fluxes. By applying fluxes on rigid exceptional gauge groups, which are ubiquitous in the F-theory landscape, we can reproduce 
the Standard Model gauge groups and chiral matter 
content without fine-tuning many moduli. Therefore, it is natural to look for realistic 
low-energy physics in these compactifications.

First, we develop a general formalism for gauge symmetry breaking with fluxes. Vertical
and remainder fluxes can be used to break a rigid gauge group, with the former 
simultaneously inducing chiral matter, even when the unbroken gauge group has no complex
representations. The number of generations of chiral matter, known as the chiral index,
is controlled by the fluxes and the intersection-theoretic data of the compactification.
We show that it is fairly likely to have three generations of chiral matter due to the
linear Diophantine structure in the chiral index. We focus on the realization of the 
Standard Model gauge groups and chiral matter by breaking rigid $E_7,E_6$ gauge groups 
with an intermediate $\SU(5)$ gauge group, and provide examples of explicit constructions.

Then, the phenomenological aspects of rigid $E_7$ models are studied in detail. We find 
that many other Standard-Model like features are naturally compatible with the $E_7$ 
models. For example, dimension-4 and dimension-5 proton decay are ubiquitously 
suppressed.  Many of these features are due to the group-theoretic structure of $E_7$
and its F-theory geometry. In particular, a set of approximate global symmetries descends
from the $E_7$ group, leading to exponential suppression of undesired couplings. These
features suggest a new set of grand unified theories based on the $E_7$ group and its 
string theory construction, which are
qualitatively different from the conventional grand unified theories
using gauge groups such as $\SU(5),\SO(10)$, and $E_6$.

Finally, we study a separate application of flux breaking in F-theory, namely 
constructions of abelian gauge symmetries with exotic charges. By breaking a rigid nonabelian group to
a $\U(1)$ gauge group using vertical flux, very large charges can arise in the massless or 
light spectrum. We give an explicit construction in 4D F-theory in which the vector-like
matter carries charges as large as 657, which is much larger than the previously known
bounds in F-theory. We heuristically argue that this result may provide an upper bound 
on charges for light fields under decoupled $\U(1)$ gauge groups in the F-theory landscape. % use \input rather than \include because we're inside an environment
\end{abstract}

%% acknowledgments.tex

% From mitthesis package
% Version: 1.02, 2024/06/19
% Documentation: https://ctan.org/pkg/mitthesis

\chapter*{Acknowledgments}
\pdfbookmark[0]{Acknowledgments}{acknowledgments}

These six years of my PhD journey have truly been a roller-coaster, full of surprises
and turning points. Walking from the string landscape to the fitness landscape, my past 
self would have had no idea, and probably would not have believed, 
what kind of researcher I would become by the end of my PhD. A significant part of this 
life-changing journey comes from the support and guidance of the people around me.

First, I would like to thank my research supervisor, Washington Taylor. He is always kind
and encouraging, making me feel comfortable pursuing my research ideas, asking questions, 
and expressing my opinions. I have learned from him not only his deep insights in physics, but
also his attitude as a researcher: pay full attention to scientific details while 
keeping the big research goal in mind, and be curious and open-minded on a broad range
of topics. He has also provided me valuable advice on academic writings and 
presentations. Most importantly, he has always been very supportive of all the major career decisions I 
made, even though I chose to leave string theory at the end. We even collaborated on a
project on theoretical ecology, which was a very unique and fun experience.

Next, I would like to thank my collaborators in string theory, Patrick Jefferson, James
Gray, and Lara Anderson, whose collaboration made many of my research projects possible. I also enjoyed
talking to other postdocs in our research group, Manki Kim, Richard Nally, Thomas 
Harvey, and our academic alumni, Andrew Turner, Nikhil Raghuram, and Yi-Nan Wang. They
have taught me a wide variety of knowledge in string theory, as well as general lessons on research. I would also like to thank my former supervisors, Yi Wang, Henry
Tye, Kam Tuen Law, Cliff Burgess, and David Chernoff during my undergraduate studies, Hong Liu during the first year
of my PhD, and Kevin Li during my internship in quantitative finance. In particular,
I would like to thank Mehran Kardar, Pankaj Mehta, and Akshit Goyal for their great
help and guidance when I was switching my research from string theory to statistical 
physics and biophysics during the late PhD years. They have all played an important role in shaping my research trajectory.

The Center for Theoretical Physics (now CTP--LI) has provided me with an excellent 
environment that supports my PhD studies. I would like to thank the administrative 
staff, especially Scott Morley, for keeping things running smoothly in the background. I 
would like to thank my academic advisor, Tracy Slatyer, and my thesis readers, 
Jesse Thaler and Netta Engelhardt, for their support throughout the completion of my PhD. I would also 
like to thank Daniel Harlow, Lina Necib, Maxim Metlitski, Julien Taileur, and 
Kevin Burdge for the opportunity to serve as a teaching assistant in their courses, which was a wonderful experience.
I would like to thank my fellow classmates, Nico, Rikab, Josh, Ryan, Elba, and Zhiquan, for studying the qualification exam together and other conversations on physics.

Outside of academia, I would like to thank my family, especially my parents, who have always supported my decision to
leave Hong Kong and pursue the path as I wished. I would like to thank my online riddle
community, which provided me with a refreshing source of entertainment when I got tired of research.

Finally, I would like to thank my partner Sarah, who recently became my 
collaborator. We have shared happiness and faced challenges together both in academia and 
in life, since we met during our undergraduate studies. We complement each other intellectually as well as personally. With her unwavering support, I have been able to persist in academia, especially during my transition to biophysics. I am very fortunate to have her love in my life.% acknowledgments.tex (.tex extension is presumed by \include) 

% \include{biography}% biography.tex (optional, see https://libraries.mit.edu/distinctive-collections/thesis-specs/#format)

%%% Table of contents and lists of stuff (delete unused lists, i.e., if no tables or figures) %%%%%

\tableofcontents
\listoffigures
\listoftables

%%% Chapters of thesis  %%%%%%%%%%%%%%%%%%%%%%%%%%%%%%%%%%%%%%%%%%%%%%%%%%%%%%%%%%%%%%%%%%%%%%%%%%%

%% If you want to use "double spacing", you should start here...

\onehalfspacing
% \doublespacing

\chapter{Motivation}
\label{chap:motivation}

\section*{The Standard Model and grand unified theories}

It is well known that the Standard Model (SM) of particle physics gives fully quantum
descriptions of three of the fundamental forces, namely the strong, weak, and 
electromagnetic forces, with the gauge group $\SU(3)\times\SU(2)\times\U(1)$, possibly with a $\mathbb Z_6$ quotient. 
Together with three generations of quarks and leptons, and the Higgs boson, the SM has 
been successful in capturing almost all of physics with energies as
high as the TeV scale at the Large Hadron Collider. To make these predictions, the SM 
requires the values of 19 free parameters that can only be obtained from experimental 
measurements, namely the gauge coupling constants, the fermion masses and mixing 
angles, the CP violation phases, and the Higgs mass and vacuum expectation value.

The structure of the SM also suggests that there may be even more 
fundamental theories at higher energies that unify the three fundamental forces. If we
extrapolate the three gauge coupling constants to high energies by running their 
renormalization group flows reversely, we see that the couplings converge around a single
value at an energy around $10^{16}$ GeV. This convergence, known as gauge coupling 
unification, is even more accurate in the
minimally supersymmetric Standard Model (MSSM), which extends the SM by a minimal amount
of supersymmetry, introducing superpartner particles into the matter spectrum. The 
gauge coupling unification and other observations such as charge quantization suggest 
the potential existence of grand unified theories (GUTs) at high energies.

In such theories,
the three fundamental forces descend from a single gauge group, and the three generations of matter are grouped into representations of the 
higher gauge group. Canonical candidates for the GUT group include $\SU(5),\SO(10)$, and
$E_6$. In particular, $\SO(10)$ is the minimal GUT group in which a single representation
contains all the SM chiral matter plus a right-handed neutrino of one generation. 
Note that the resulting SM gauge group from breaking the GUT group has the $\mathbb Z_6$ quotient. On the 
other hand, GUTs face phenomenological challenges especially in proton decay. For 
example, the current
absence of proton decay in experiments already rules out the simplest $\SU(5)$ GUT, which predicts a proton lifetime of around $10^{30}$ years \cite{Langacker:1980js}.

\section*{Open problems in fundamental physics}

While the SM has been successful in describing most of observed physics, it is clear that the theory is 
incomplete for describing the whole Universe. There are important open questions in cosmology, and even in the formulation of the SM itself.

On the cosmological side, the SM does not incorporate gravity, the remaining fundamental 
force, since a quantum theory of gravity would require going beyond the field-theoretic
framework used by the SM. Moreover, the SM does not explain the identity of dark matter,
which has so far been observed to interact with ordinary matter only through gravity but accounts for around 30\% 
of the total content of the Universe, nor dark energy, which is responsible for 
accelerated expansion of the Universe and accounts for around 70\% of the total 
content. At the intersection with particle physics, the amount of matter-antimatter 
asymmetry from the SM is not sufficient to explain the observed surplus of matter 
resulting from baryogenesis in the early Universe.

Within particle physics, the neutrino masses and mixing have been observed in 
experiments but they are not explained within the SM. Explaining fine-tuning problems in the SM
such as the Higgs hierarchy problem, i.e., why the Higgs mass is much smaller than 
radiative corrections from new physics at high energies such as the Planck scale, and the
strong CP problem, i.e., why the QCD $\theta$-angle is very close to zero rather than
finite, require physics beyond the SM that has not been observed.

More fundamentally, the origin of the SM gauge groups, matter content, and the values of
the 19 free parameters remains unexplained within the SM itself. Why is 
the SM gauge group as simple as $\SU(3)\times\SU(2)\times\U(1)$, which can also be 
unified into $\SU(5)$? Why are there exactly three generations of matter, but not other
numbers? What determines the values of the 19 free parameters? Why does our Universe have
three large dimensions? To answer these questions, we need to go beyond the framework of
SM and even quantum field theories.

\section*{String theory as a top-down approach}

The answers to the above questions may lie in a full theory at high energies that unifies all
the fundamental forces and describes all observed phenomena.
So far, string theory is the only consistent framework for unifying gravity with other
fundamental forces within a quantum description. In string theory, the fundamental objects
are no longer point particles as in the SM, but one-dimensional strings, which can be
closed or open. Closed strings propagate in the bulk of spacetime and describe gravity
with graviton states in a way that is compatible with general relativity. Meanwhile,
gauge dynamics are localized on higher dimensional objects known as branes, and the gauge
fields and charged matter are described by open strings stretched between branes. 
Therefore, strings naturally unify gravity and gauge dynamics into the same framework.

On the other hand, there are complications when applying string theory to our Universe.
The version of string theory we understand the most, known as the critical superstring
theory, is only consistent at the quantum level with ten spacetime dimensions. To 
describe our Universe, six of the spatial dimensions must therefore be compactified into a
small manifold, and in many cases additional objects such as branes, fluxes, and 
orientifolds must be incorporated along with the compactification. To preserve the
minimal amount of supersymmetry ($\mathcal N=1$) in four dimensions, we compactify the
six spatial dimensions into Calabi-Yau (CY) manifolds, which are complex K\"ahler
manifolds with vanishing first Chern class.

String compactifications give rise to an interesting perspective on how the physical laws
in four dimensions are determined. In string theory, there are no free parameters except
the string scale. Instead, the low-energy physics including gauge groups, matter content,
and parameters such as particle masses and couplings, are all in principle determined by
the geometry of the compactification. The set of possible compactifications, known as
the string landscape, is widely believed to be finite but enormously large (at least
$10^{272000}$ string vacua \cite{TaylorWangVacua}). As a result, string theory heavily constrains the
possible low-energy effective theories, which have infinite possibilities from the 
quantum field theory perspective. The idea that only few effective field theories 
coupled to gravity can be UV-completed in string theory has been pushed forward by the
Swampland program \cite{VafaSwamp,OoguriVafaSwamp}. By studying the structure of the 
string landscape, in particular by searching for SM-like features,
we hope to understand the constraints on physics beyond the SM from a top-down perspective.

Despite decades of research in string phenomenology, it is currently still too difficult to construct string compactifications that
realize all the exact details of observed phenomenology in the SM; 
see \cite{Cvetic:2022fnv,Marchesano:2024gul} for recent reviews of the progress so far. Given the large size
of the string landscape, rather than the existence of exact SM solutions, it would be
more useful to think about distributions across the landscape, and study whether SM-like
features can arise as statistically natural solutions in the landscape. In other words,
we would like to understand whether SM-like features are ubiquitous in the landscape or 
require extensive fine-tuning. To answer these questions, we need a systematic framework
to explore large parts of the string landscape.

\section*{F-theory and string phenomenology}

In this thesis, we focus on F-theory \cite{VafaF-theory,MorrisonVafaI,MorrisonVafaII}, 
which is a strongly coupled version of type IIB string theory; see \cite{WeigandTASI} 
for an extensive review. In type IIB string theory,
gauge dynamics can be described by the strongly coupled physics of 7-branes. It was noticed
that one can replace the strong coupling (axio-dilaton) profile by inserting an auxiliary torus at each point
of the compact space, giving a space known as an elliptic fibration. The brane physics is then
translated into singularities of the fibration, which is purely geometrical and can be analyzed using powerful tools from
algebraic geometry. In particular, 4D F-theory models can be obtained by compactifying on
elliptic CY fourfolds. The gauge groups, matter representations, and interactions in the
models can then be read off from singularities of different codimensions in the geometry. There
is also flux background, which characterizes other aspects of the models such as gauge
symmetry breaking and chiral matter spectrum. This approach allows us to explore the strongly
coupled regime in the string landscape, leading to the largest known set of solutions in the
landscape. 

This approach of constructing string theory solutions has very fruitful applications.
It can be used to explore different corners
in the string landscape, and understand how they are related to other string theory constructions
by string dualities. F-theory has also shed light on problems related to algebraic
geometry \cite{Grassi:2000we,Grassi:2011hq}, nonperturbative quantum field theories \cite{HeckmanMorrisonVafa,Heckman:2015bfa,HeckmanRudeliusSCFT}, and many other subjects. Most importantly, it gives rise to various types of constructions that realize large parts (but not
necessarily all) of the SM-like features, making these solutions an ideal playground for
string phenomenology.

As a result, there have been many attempts to build models with the SM gauge group
$G_{\rm SM}=\SU(3)\times \SU(2)\times \U(1)/\mathbb{Z}_6$ in
F-theory (while building the SM gauge group and matter without the $\mathbb Z_6$ quotient appears to be much harder \cite{TaylorTurnerGeneric}).  Starting from
\cite{Donagi:2008ca,BeasleyHeckmanVafaI,BeasleyHeckmanVafaII,DonagiWijnholtGUTs},
F-theory GUTs have been constructed, using
gauge groups of $\SU(5)$
\cite{Blumenhagen:2009yv,Marsano:2009wr,Grimm:2009yu,KRAUSE20121,Braun:2013nqa},
$\SO(10)$ \cite{Chen:2010ts}, etc. (See \cite{HeckmanReview} for
review) Recently, $10^{15}$ explicit solutions of directly tuned
$\gsm$ 
 were found in the string landscape \cite{CveticEtAlQuadrillion}. It
has also been argued that the SM matter representations generically
appear when $\gsm$ is directly tuned
\cite{TaylorTurnerGeneric,Raghuram:2019efb}. These results signal that
a considerable portion of the landscape may contain SM-like models.

These models cannot be the most generic or natural SMs in
the landscape, however. All the preceding gauge groups arise from
fine-tuning many moduli. In contrast, most F-theory
compactification bases have strong
curvature that enforces rigid (a.k.a\ {\it geometrically
  non-Higgsable} \cite{MorrisonTaylor4DClusters}) gauge symmetries,
which are present throughout the whole branch of moduli space
\cite{TaylorWangMC,HalversonLongSungAlg,TaylorWangLandscape}.
Furthermore, on many bases these rigid gauge factors forbid tuning additional factors like  $\gsm$.

A generic F-theory model with desired gauge groups and matter content can arise more 
naturally from the
geometric rigid gauge symmetries without tuning moduli. 
Turning on fluxes inside a larger rigid group can break the rigid gauge group
into smaller gauge groups, while inducing chiral and vector-like matter in the broken
gauge groups. This mechanism, known as \emph{flux breaking} in 4D F-theory flux
compactifications, generates an incredibly rich variety of string theory models with generic 
compactification geometries, hence will be the central focus of this thesis.

\section*{Thesis outline}

This thesis is based on the author's published papers 
\cite{Li:2021eyn,Li:2022aek,Li:2023rqf,Li:2024uwf}.

The remaining chapters of this thesis are organized as follows. To set up the 
stage, we first review various aspects of F-theory in Chapter~\ref{chap:review}. We 
formulate the framework of F-theory starting from the basics of string theory. Then,
we review the techniques in F-theory that are used throughout this thesis. We introduce
the Weierstrass model to describe elliptic fibrations in CY manifolds. From the
singularities in this model, we read off the gauge groups, matter content, and Yukawa 
couplings. We conclude the review with the formalism and techniques of fluxes in 
F-theory, which are the central focus of this thesis.

The next three chapters describe the author's original research in detail. In 
Chapter~\ref{chap:fluxbreaking}, we develop a general formalism for gauge symmetry 
breaking with fluxes, and apply the formalism to SM-like constructions in F-theory, 
based on \cite{Li:2021eyn,Li:2022aek}. Vertical
and remainder fluxes can be used to break a rigid gauge group, with the former 
simultaneously inducing chiral matter, even when the unbroken gauge group has no complex
representations. The number of generations of chiral matter, known as the chiral index,
is controlled by the fluxes and the intersection-theoretic data of the compactification.
We show that it is fairly likely to have three generations of chiral matter due to the
linear Diophantine structure in the chiral index. We focus on the realization of the 
Standard Model gauge groups and chiral matter by breaking rigid $E_7,E_6$ gauge groups 
with an intermediate $\SU(5)$ gauge group, and provide an example of explicit construction using the $E_6$ group.

In Chapter~\ref{chap:e7}, we study in detail the phenomenological aspects of rigid $E_7$ models, based on \cite{Li:2024uwf}. We find 
that many other Standard-Model like features are naturally compatible with the $E_7$ 
models. For example, dimension-4 and dimension-5 proton decay are ubiquitously 
suppressed.  Many of these features are due to the group-theoretic structure of $E_7$
and its F-theory geometry. In particular, a set of approximate global symmetries descends
from the $E_7$ group, leading to exponential suppression of undesired couplings. We 
also analyze other phenomenological aspects including Higgs, Yukawa, and neutrino 
sectors, as well as gauge coupling unification. We complement this analysis with 
explicit constructions of the $E_7$ models, and comparison with existing F-theory SM-like
constructions in the literature.

In Chapter~\ref{chap:u1}, we study a separate application of flux breaking in F-theory, namely 
constructions of abelian gauge symmetries with exotic charges, based on \cite{Li:2023rqf}. By breaking a rigid nonabelian group to
a $\U(1)$ gauge group using vertical flux, very large charges can arise in the massless or 
light spectrum. We give an explicit construction in 4D F-theory in which the vector-like
matter carries charges as large as 657, which is much larger than the previously known
bounds in F-theory. We heuristically argue that this result may provide an upper bound 
on charges for light fields under decoupled $\U(1)$ gauge groups in the F-theory 
landscape. We also construct even larger charges by coupling the $\U(1)$ factor to other
gauge groups.

Finally, in Chapter~\ref{chap:conclusion}, we summarize the main original results from
the previous chapters. We discuss the limitations of these results and suggest future directions.

\section*{Use of generative AI}

In preparing the thesis, the author used Claude (by Anthropic) as an editorial tool. The
usages include:
\begin{itemize}
    \item Refining a thesis plan written by the author in the initial stage of writing the thesis;
    \item Summarizing standard literature by extracting parts relevant to this thesis,
    which are incorporated with the author's original writing;
    \item Merging the author's original papers into this thesis, rearranging review
    materials and maintaining consistent references;
    \item Formatting equations and tables into \LaTeX commands;
    \item Suggesting language improvements for the written thesis text.
\end{itemize}
No generative AI tools were used in the author's original research including 
calculations, analysis, and writeups. Moreover, no generative AI tools were used in 
writing the original text of this thesis. The author takes full responsibility for the
accuracy and integrity of the material presented in this thesis.% .tex extension is presumed
\chapter{Review of F-theory}
\label{chap:review}

In this chapter, we review the basics of string theory and F-theory that are used
throughout the thesis, to prepare for the original research presented in the next
chapters. In \S\ref{sec:review:strings-to-ftheory}, we briefly introduce basic ingredients
of string theory, building towards the definition of F-theory. After constructing 
fundamental objects such as superstrings and D-branes, we describe CY 
compactifications for 4D physics at low energies. From this viewpoint, we introduce 
M-theory and string dualities, which are important in formulating F-theory. Finally,
we discuss type IIB string theory and 7-branes in detail, and establish F-theory as
a nonperturbative version of type IIB string theory, by representing its axio-dilaton
background with geometry.

In \S\ref{sec:review:basics}, we cover the common techniques used for studying
F-theory compactifications. We first introduce the Weierstrass model for describing 
compactifications on elliptically fibered CY manifolds. Then, we turn to gauge groups
in F-theory arising from codimension-1 singularities of the elliptic fibrations. 
Different gauge groups correspond to different types of singularities according to the
Kodaira classification, and can be systematically constructed using the Tate models.
To study these singularities in detail, we resolve the singularities by blowing up the
elliptic fibers. We also discuss the important distinction between tuned and rigid gauge
groups, which forms the basis of this thesis.
After that, we study matter content and Yukawa couplings in F-theory,
which correspond to codimension-2 and codimension-3 singularities of the elliptic 
fibrations. To construct a large class of elliptically fibered CY manifolds, we also 
briefly introduce toric geometry, a computational framework from algebraic geometry.
Finally, the $G_4$ flux formalism in F-theory is described in detail. We focus on its
construction, classification and usages, consistency conditions the fluxes need to 
satisfy, and tools from intersection theory for studying the fluxes. These are the main
tools for the research in this thesis.

\section{From string theory to F-theory}
\label{sec:review:strings-to-ftheory}

To understand the concepts of the string landscape and string phenomenology, as well as
to prepare for studying F-theory, it is useful to first review the relevant elements 
in string theory more broadly. The review in this section is largely based on \cite{Becker:2006dvp,WeigandTASI}.

\subsection{Closed bosonic strings and superstrings}
\label{sec:review:strings-to-ftheory:basics}

We start with the most fundamental object in string theory, namely one-dimensional 
strings propagating in spacetime. Recall that a point particle propagating in spacetime
traces a one-dimensional worldline. Similarly, a string sweeps out a two-dimensional 
worldsheet $\Sigma$ in a $D$-dimensional target spacetime. A natural candidate of the string dynamics is
to minimize the area of its worldsheet, leading to the Nambu-Goto action
\begin{equation}
\label{eq:review:NG}
    S_{\rm NG} = -\frac{1}{2\pi\alpha'} \int_\Sigma d^2\sigma \sqrt{-\det \gamma_{ab}}\,,
\end{equation}
where $\gamma_{ab} = \partial_a X^\mu \partial_b X_\mu$ is the induced metric on the 
worldsheet, $X^\mu$ are bosonic fields on the worldsheet representing target space coordinates,
$\sigma^a = (\tau,\sigma)$ are the worldsheet coordinates, and $\alpha'$ is the 
Regge slope related to the string tension $T=1/2\pi\alpha'$. The corresponding string 
length scale is $l_s=\sqrt{\alpha'}$, which should have much higher energy than the 
collider energy scale. Due to the
presence of the square root in $S_{\rm NG}$, it is more tractable to introduce an 
auxiliary and independent worldsheet metric $h_{ab}$ and study the Polyakov action instead
\begin{equation}
\label{eq:review:Polyakov}
    S_{\rm P} = -\frac{1}{4\pi\alpha'} \int_\Sigma d^2\sigma \sqrt{-h}\, h^{ab} \partial_a X^\mu \partial_b X_\mu\,.
\end{equation}
The Polyakov action is classically equivalent to the Nambu-Goto action after integrating
out $h_{ab}$. Apart from Poincar\'e invariance in 2D, $S_{\rm P}$ also
has local reparametrization and Weyl invariances. Therefore, $S_{\rm P}$
defines a 2D conformal field theory (CFT) on the worldsheet.

To quantize the worldsheet CFT consistently, the Weyl invariance must be preserved at 
the quantum level. Due to the Weyl anomaly, however, this can be true only when
the central charge of the worldsheet CFT vanishes, which fixes the spacetime dimension 
$D$. For the above theory, known as the critical bosonic string theory, anomaly cancellation requires
$D=26$. After quantizing the left- and right-moving modes of closed strings consistently,
the theory results in an infinite tower of string states with masses separated by the
string scale. Different string states correspond to different types of particles at low
energies. Throughout this thesis, it suffices to consider the massless sector of the 
tower, which consists of the graviton or the spacetime metric $g_{\mu\nu}$, the 
Kalb-Ramond two-form $B_2$, and a scalar field $\phi$ called the dilaton that controls the 
coupling strength of string interactions $g_s=e^\phi$. It is worth noting that in string
theory, all couplings are promoted to vacuum expectation values of some scalar fields
instead of external parameters. In this manner, we see that gravity naturally
emerges from the dynamics of closed strings.

On the other hand, bosonic string theory suffers from two issues. First, the tower of 
string states also contains a tachyon with negative mass squared, meaning that the vacuum
in this theory is unstable. Second, the tower only consists of bosonic states and there
is no fermion in this theory. To resolve these issues, we extend $S_{\rm P}$ with 
worldsheet fermions $\psi^\mu$, resulting in a new action
\begin{equation}
\label{eq:review:RNS}
    S_{\rm RNS} = -\frac{1}{4\pi\alpha'} \int_\Sigma d^2\sigma \sqrt{-h}\, \left(\partial_a X^\mu \partial^a X_\mu+\bar\psi^\mu\rho^a\nabla_a\psi_\mu\right)\,,
\end{equation}
where $\rho^a$ are the 2D Dirac matrices. It turns out that $S_{\rm RNS}$ is 
supersymmetric apart from being a CFT. This is known as the Ramond-Neveu-Schwarz (RNS)
formalism of type II superstrings based on worldsheet supersymmetry. This theory can be
consistently quantized only when $D=10$, again due to the Weyl anomaly; this is the number of spacetime dimensions 
shared by all critical superstring theories.

The matter spectrum in superstring theories is much richer since we can impose either
periodic (Ramond or R) or antiperiodic (Neveu-Schwarz or NS) boundary 
conditions for the worldsheet fermions. For closed strings, there are two independent
choices for the left- and right-moving modes. In addition, to maintain a consistency 
condition known as modular invariance of the one-loop partition function, we have to
impose the so called GSO projection after quantizing the left- and right-moving string modes. The projection removes states in the NS sector with an odd number
of fermionic excitations. As a result, the tachyonic state is removed from the matter
spectrum, and at the same time the 10D spacetime gains $\mathcal N = 2$ supersymmetry. After the projection, the massless closed string spectrum is organized as
follows. In the NS-NS sector, we have the graviton, the two-form, and the dilaton as for
bosonic strings. The R-R sector gives rise to additional form fields, but the exact field
content depends on the chiralities of the left- and right-moving modes of the fermionic ground states in the R sector. 
More specifically,
\begin{itemize}
    \item In type IIA string theory, the two chiralities are opposite, resulting in a
    1-form potential $C_1$ and a 3-form potential $C_3$ in the R-R sector;
    \item In type IIB string theory, the two chiralities are the same, resulting in a
    $C_0$ axion, a 2-form potential $C_2$ similar to $B_2$, and a self-dual 4-form potential $C_4$ with the field strength $F_5=dC_4$ satisfying $F_5=*F_5$.
\end{itemize}
Finally, the R-NS and NS-R sectors contain fermionic superpartners of the 
above bosonic states.

There are five consistent superstring theories in 10D. Apart from type II string 
theories, there are also type I and two heterotic string theories. We focus on type IIB
string theory, which will be further reviewed in \S\ref{sec:review:strings-to-ftheory:IIB}.

\subsection{Open strings and D-branes}
\label{sec:review:strings-to-ftheory:Dbranes}

So far, we have focused on closed strings, which have no ends and give rise to gravity.
On the other hand, there are also open strings with two ends that give rise to gauge 
dynamics. This is why string theory can unify gravity and gauge theories into a single
framework. We now study open strings in detail.

From the equations of motion of the worldsheet theory, the ends of an open string satisfy
either Dirichlet or Neumann boundary conditions. As a result, the ends are constrained
to move on some hypersurface with $p\leq D-1$ spatial dimensions.
As noticed by Polchinski \cite{Polchinski:1995mt}, these hypersurfaces are in fact not 
just boundary conditions but themselves dynamical objects, known as D$p$-branes (where D 
stands for Dirichlet). Unlike fundamental strings, one can show that the brane tension $T_p$ is
inversely proportional to the string coupling $g_s$, hence D-branes are inherently 
nonperturbative and are missing in the quantum description of perturbative string 
theory.

Quantizing open strings that end on D-branes results in vector bosons that 
behave as gauge fields localized on the D-brane worldvolume, hence open
strings give rise to gauge dynamics. The corresponding gauge group comes from the 
configuration of multiple D-branes. Namely, each end of an open string carries a label,
known as a Chan-Paton factor, that identifies which brane the end lies on, resulting in
additional degrees of freedom. As an example, each end of an open string has $N$ choices for a stack of 
$N$ D-branes; the two ends give rise to $N^2$ states in total, corresponding to the
adjoint representation of $\U(N)$. Therefore, a $\U(N)$ gauge group is realized on the
D-brane stack worldvolume. To realize other types of gauge groups, different types of
branes and orientifolds, as well as nonlocal brane configurations, may be involved.

In superstring theories, D$p$-branes are charged under certain R-R form fields to
ensure their stability. As a result, D-branes are BPS states and preserve half of
the spacetime supersymmetry. The R-R form fields present in the massless spectrum in 
turn constrain the possible numbers of spatial dimensions $p$ for D-branes. Given a
$(p+1)$-form potential $C_{p+1}$, its electric charges $\mu_p$ are carried by D$p$-branes,
while its magnetic charges $\mu_{6-p}$ are carried by D$(6-p)$-branes. As a result, type
IIA and IIB string theories support D-branes with even and odd dimensions respectively.
Note that 
D3-branes can carry both electric and magnetic charges of $C_4$, consistent with the 
self-duality of $C_4$. The usual Dirac quantization 
condition now implies $\mu_p \mu_{6-p}\in 2\pi \mathbb Z$.

Similar to the worldsheet theory of strings, we can also write down the worldvolume 
action for D-branes, which contains both the induced metric $G$ and the worldvolume
gauge field strength $F$. Effectively, we are treating the gauge dynamics from open
strings ending on the D-branes as fluctuations of the brane worldvolume itself. The
resulting action is the Dirac-Born-Infeld (DBI) action plus Chern-Simons terms
\begin{equation}
\label{eq:review:DBI}
    S_{Dp} = -T_p \int_{\Sigma} d^{p+1}\sigma\, \sqrt{-\det(G + 2\pi\alpha' F)} + \mu_p \int_{\Sigma} \sum_n C_n \wedge e^{2\pi\alpha' F}\,.
\end{equation}
We recall that $T_p$ and $\mu_p$ are the D-brane tension and R-R charge. With this action,
we can study the gauge dynamics in terms of the worldvolume Yang-Mills theory.

\subsection{Compactification and Calabi-Yau manifolds}
\label{sec:review:strings-to-ftheory:CY}

Since the critical superstring theories require 10D spacetime, they cannot be directly 
applied to describe our 4D Universe. The most common approach to resolve this issue is to
compactify the six extra spatial dimensions into a compact internal manifold $M_6$. 
In the simplest setting, we assume that the full spacetime $M_{1,9}$ can be factorized into a product form
$M_{1,9}=\mathbb R^{1,3}\times M_6$, where $\mathbb R^{1,3}$ is the 4D Minkowski 
spacetime. In the presence of flux background, one may also consider warped 
compactifications in which the metric of the noncompact dimensions is transformed by a 
conformal factor depending on the internal compact coordinates. At energy scales below
the Kaluza-Klein (KK) scale $m_{\rm KK}\sim {\rm Vol}(M_6)^{-1/6}$, the momentum modes
in the compact dimensions are heavy and only the zero modes contribute in the 4D 
effective theory. In this way, string compactifications may potentially describe our
Universe if $m_{\rm KK}$ is much higher than the collider energy scale.

In the past literature, there was an alternative approach, namely the brane-world 
scenario, which does not assume small extra dimensions; instead, a 4D universe is 
localized on a brane (or a stack of branes) in noncompact 10D spacetime. Nevertheless, it was then realized
that this scenario is phenomenologically not viable, especially because gravity tends
to leak into the large extra dimensions and violate Newton's law in 4D; see examples in
\cite{Li:2020mgc}. We focus on compactifications in this thesis.

To gain better control on the 4D effective theory, we typically require the 
compactifications to preserve a minimal amount of supersymmetry in 4D. This requires the
internal manifold $M_6$ to have an $\SU(3)$ holonomy group, which is satisfied by complex
K\"ahler manifolds with vanishing first Chern class $c_1$, known as Calabi-Yau 
threefolds (CY3). By Yau's theorem, these manifolds admit Ricci-flat metrics and solve the
vacuum Einstein's equation without a cosmological constant. The algebraic conditions
satisfied by CY manifolds allow us to study many aspects of string compactifications
without the explicit metric, although there has been recent advance in numerical 
computation of CY metrics using modern neural networks \cite{Ashmore:2019wzb,Anderson:2020hux,Jejjala:2020wcc,Douglas:2020hpv,Larfors:2021pbb,Larfors:2022nep,Gerdes:2022nzr}. In type II string theories, CY
compactifications lead to $\mathcal N=2$ supersymmetry in 4D, which can be further
broken to $\mathcal N=1$ by additional objects including branes, fluxes, and 
orientifolds.

In string compactifications, the low-energy physics in 4D is determined by the geometry
of the internal manifold and the associated objects. Recall that gauge dynamics are 
localized on brane worldvolumes. Therefore, gauge dynamics can be realized in 4D using
branes that fully fill the noncompact dimensions, and possibly wrap around some cycles
in the compact dimensions. In addition, there are multiple scalar fields from the moduli
of the internal manifold. Roughly speaking, there are $h^{1,1}$ K\"ahler moduli that control the 
sizes of cycles in the manifold, and $h^{2,1}$ complex structure moduli that control the shape of
the manifold. The numbers $h^{1,1},h^{2,1}$ are topological data of the manifold called
Hodge numbers. On the other hand, since we do not observe any of these scalar fields in 
our Universe, the moduli fields need to be lifted to high masses and stabilized to certain
vacuum expectation values through some dynamical mechanism, typically by turning on 
fluxes. This procedure, known as moduli stabilization, is one of the main challenges in
building phenomenologically viable string compactifications.

It would be wonderful if there were a unique choice of CY3 that happens to describe our
Universe. While CY onefolds and twofolds are indeed unique (tori and K3 surfaces, respectively),
there is an enormously large number of possibilities for CY3. 
It is widely believed, however, that
the number of CY3 is still finite. The set of possible CY3 and the associated flux data
is called the string landscape, representing a finite set of possible ``universes'', or 
vacua to be more precise, that
can be realized with string compactifications. Even if one of the string vacua does describe
our Universe, so far there is no clear principle to explain why this particular string
vacuum would be realized. There are multiple possible scenarios:
\begin{itemize}
    \item There may be some
    missing consistency conditions or anthropic reasons that select a particular string 
    vacuum against the others;
    \item All the possible string vacua may actually be realized as a ``multiverse'',
    and our Universe is only one of the existing universes.
    \item There may only be a single universe in reality, which realizes one of the string vacua randomly drawn from the string landscape.
\end{itemize}
Nevertheless, regardless of the interpretation of the string landscape, it is useful to
explore the landscape and study its structure, so that we can understand better how our 
Universe may be realized in the string landscape.

In the next chapters on F-theory, we will in fact focus more on CY fourfolds (CY4), that
are manifolds with $\SU(4)$ holonomy and satisfying the same K\"ahler and Chern class 
constraints. There are three independent Hodge numbers $h^{1,1},h^{2,1},h^{3,1}$, with
the first and last ones counting the K\"ahler and complex structure moduli respectively.

\subsection{String dualities and M-theory}
\label{sec:review:strings-to-ftheory:Mtheory}

Although there are five apparently independent superstring theories in 10D, they are in fact
closely related through string dualities, especially under compactifications. String
dualities are one of the few tools that allow us to study nonperturbative aspects of
string theories. As we will see, these dualities also play an essential role in 
the construction of F-theory.

There are mainly two types of string dualities. Under the context of string 
compactifications, ``T-duality'' relates two string theories with different radii of a
circle in the internal manifold. If a circle on one side has radius $R$, the 
corresponding circle on the other side will have radius $\alpha'/R$. An immediate 
consequence of the radius map is that the winding and momentum modes of closed strings,
which have energies proportional to $R/\alpha'$ and $1/R$ respectively, are exchanged
under T-duality. For open strings, one can also show that the Dirichlet and Neumann 
boundary conditions are exchanged under T-duality, hence T-duality maps between D-branes
of odd and even dimensions. Although there are multiple T-dualities between different
string theories, below we will use the fact that type IIA and IIB string theories are
T-dual to each other.

``S-duality'' relates two string theories with string couplings $g_s$ and $1/g_s$ 
respectively. Although we motivate S-duality in the context of string theories, it
can also be understood as the strong-weak duality in quantum field theories. The 
simplest example is the exchange of electric and magnetic fields in electromagnetism,
which also exchanges electric and magnetic charges proportional to $e$ and $1/e$ 
respectively. A more nontrivial example can be found in the 4D $\mathcal N=4$ 
super-Yang-Mills (SYM) theory, in which the theory is invariant under ${\rm SL}(2,\mathbb Z)$
modular transformations of the complexified coupling
\begin{equation}
\label{eq:review:SYM}
    \tau=\frac{\theta}{2\pi}+\frac{4\pi i}{g^2}\,,
\end{equation}
where $\theta$ and $g$ are the vacuum angle and the coupling respectively. When 
$\theta=0$, the modular transformation $\tau\rightarrow -1/\tau$ indeed corresponds to
$g\rightarrow 4\pi/g$. This S-duality also manifests if we treat the SYM as the 
dimensional reduction of 6D $(2,0)$ SCFT compactified on a torus, in which $\tau$
becomes the complex structure modulus of the torus and clearly obeys modular invariance.
We will see a very similar S-duality in type IIB string theory and F-theory.

The five superstring theories in 10D can be related to each other by a series of string
dualities. In fact, they can also be understood as different limits of a single 11D
theory called M-theory. Although a complete formulation of M-theory remains unknown, its
low-energy limit should be descirbed by the 11D supergravity \cite{Cremmer:1978km}, which is the unique 
supergravity in 11D. The bosonic part of the supergravity action is
\begin{equation}
\label{eq:review:11dsugra}
    S_{11} = \frac{1}{2\kappa_{11}^2}\left[\int d^{11}x\,\sqrt{-g}\left(R - \frac{1}{2}|G_4|^2\right) - \frac{1}{6}\int C_3 \wedge G_4 \wedge G_4\right]\,,
\end{equation}
where $\kappa_{11}$ is the 11D gravitational coupling, $g$ is the 11D metric, and $C_3$
is a 3-form potential with $G_4=dC_3$. M-theory should also support M2- and M5-branes,
which carry electric and magnetic charges of $C_3$ respectively.

M-theory is related to the five superstring theories by compactifying on one dimension,
possibly followed by a series of S- and T- dualities \cite{HullTownsend,Witten:1995ex}. The simplest example is the duality
between type IIA string theory and M-theory compactified on $S^1$, with the radius of 
$S^1$ corresponding to the string coupling $g_s$ in type IIA. Therefore, the 
uncompactified M-theory is a strong coupling limit of type IIA. More specifically, the 11D
metric $g$ reduces to the 10D metric, the $C_1$ potential, and the dilaton, while the 
11D 3-form potential $C_3$ reduces to the 10D 3-form potential and the $B_2$ potential.
M2- and M5-branes correspond to D2- and NS5-branes if not wrapping the $S^1$, and 
correspond to the fundamental F1-string and D4-branes if wrapping the $S^1$. The KK
momentum modes and monopoles of the $S^1$ correspond to D0- and D6-branes. Indeed, these
are the full spectrum of the type IIA string theory at low energies.

\subsection{Type IIB string theory and 7-branes}
\label{sec:review:strings-to-ftheory:IIB}

Before introducing F-theory, we shall study type IIB string theory and its string
dualities in detail. In particular, we discuss the properties of 7-branes in type IIB,
which act as sources of axio-dilaton background and make the theory inherently strongly coupled.

We first recall the low-energy spectrum of type IIB. In the NS-NS sector, there are the
metric $g_{\mu\nu}$, the Kalb-Ramond 2-form $B_2$, and the dilaton $\phi$. In the R-R
sector, there are the $C_0$ axion, a 2-form $C_2$, and a self-dual 4-form $C_4$. There
are F1-strings and NS5-branes charged under $B_2$, $D(-1)$-instantons and D7-branes 
charged under $C_0$, D1-strings and D5-branes charged under $C_2$, and D3-branes charged
under $C_4$. It is useful to define the axio-dilaton $\tau$ as the combination
\begin{equation}
\label{eq:review:axio-dilaton}
    \tau = C_0 + ie^{-\phi} = C_0 + \frac{i}{g_s}\,.
\end{equation}
At low energies, the theory is described by type IIB supergravity with the Einstein frame action
\begin{equation}
\label{eq:review:typeIIBaction}
    S_{\rm IIB}=\frac{1}{2\kappa_{10}^2} \int d^{10}x \sqrt{-g} \left[ R - \frac{\partial_\mu \tau \partial^\mu \bar{\tau}}{2(\text{Im}\,\tau)^2} - \frac{|G_3|^2}{2\,\text{Im}\,\tau} - \frac{|\tilde{F}_5|^2}{4} \right] + \frac{1}{8i\kappa_{10}^2} \int \frac{C_4 \wedge G_3 \wedge \bar{G}_3}{\text{Im}\,\tau}\,,
\end{equation}
where $\kappa_{10}$ is the 10D gravitational coupling, and
\begin{align}
\label{eq:review:IIBactionDef}
    G_3&=dC_2-\tau dB_2\,,\\
    \tilde F_5&=dC_4-\frac{1}{2}C_2\wedge dB_2+\frac{1}{2}B_2\wedge dC_2\,.
\end{align}
This action must be supplemented by the self-duality constraint $F_5=*F_5$. From the 
above form of the action, we can see that it is invariant under ${\rm SL}(2,\mathbb R)$ modular transformations
\begin{equation}
\label{eq:review:SL2Z}
    \tau \;\mapsto\; \frac{a\tau + b}{c\tau + d}\,, \qquad \begin{pmatrix} C_2 \\ B_2 \end{pmatrix} \;\mapsto\; \begin{pmatrix} a & b \\ c & d \end{pmatrix}\begin{pmatrix} C_2 \\ B_2 \end{pmatrix}\,,\quad  \begin{pmatrix} a & b \\ c & d \end{pmatrix}\in \mathrm{SL}(2,\mathbb{R})\,,
\end{equation}
with the other fields remaining invariant. Due to instanton effects, however, this symmetry
is broken to ${\rm SL}(2,\mathbb Z)$ at the quantum level. This is the S-duality of type
IIB string theory to itself, similar to the one in $\mathcal N=4$ SYM.

Since the 2-forms $B_2,C_2$ transform as a doublet under the S-duality, the same is also
true for F1- and D1-strings, and more types of strings can be generated by the S-duality 
transformations. In general, $(p,q)$-strings where ${\rm gcd}(p,q)=1$ carry charges 
$p,q$ under $B_2,C_2$ respectively. In particular, F1- and D1-strings correspond to 
$(1,0)$- and $(0,1)$-strings respectively. Similarly, we can define $[p,q]$ 7-branes as
the 7-branes that $(p,q)$-strings can end on; D7-branes correspond to $[1,0]$ 7-branes.

Type IIB compactifications with 7-branes are interesting not only because the gauge dynamics
can be localized on the 7-branes, but also because 7-branes are charged under 
axio-dilaton and can create a complex spatial profile of the string coupling $g_s$. If
there are nonlocal configurations of 7-branes with different $[p,q]$ charges, there is no 
S-duality frame such that the string coupling is uniformly weak, hence the theory becomes
inherently strongly coupled.

Here is an example with a D7-brane. Recall that D7-branes have
charge 1 under $C_0$ but are uncharged under $\phi$. Let us describe the two transverse
dimensions as a complex plane with coordinate $z$, and let a D7-brane locate at $z=z_0$.
One can show by solving Poisson-like equations that the axio-dilaton profile near the D7-brane is \cite{Greene:1989ya}
\begin{equation}
\label{eq:review:tauNearD7}
    \tau(z)\sim \frac{1}{2\pi i}\log(z-z_0)\,.
\end{equation}
Due to the logarithmic branch cut, we see that the D7-brane induces a monodromy 
$\tau\rightarrow\tau+1$ around the brane, which is indeed an ${\rm SL}(2,\mathbb Z)$ 
transformation. The axio-dilaton profiles and monodromies of
other $[p,q]$ 7-branes can be obtained by transforming the above with S-dualities.

Different gauge groups can be constructed by different combinations of $[p,q]$ 7-branes.
A stack of $N$ D7-branes gives rise to an $\SU(N)$ gauge group (instead of $\U(N)$, since
the $\U(1)$ factor is broken through the St\"uckelberg mechanism). An interesting example
is the stack of 4 D7-branes and an O7-plane, which supports an $\SO(8)$ gauge group. Since the
O7-plane has $C_0$ charge $-4$, the charges cancel and the brane stack does not induce
any nontrivial axio-dilaton background. Importantly, nonlocal brane configurations are 
required for engineering exceptional gauge groups $E_6,E_7,E_8$, hence these gauge
groups cannot be described with perturbative type IIB compactifications.

If we can systematically study the 7-brane dynamics in type IIB compactifications, they will allow
us to explore the strongly coupled regime of the string landscape and give rise to a
large set of string vacua. 
As we will introduce below, such a task can be accomplished 
with the framework of F-theory.

\subsection{The definition of F-theory}
\label{sec:review:strings-to-ftheory:ftheory}

From the above discussion, we see that a nonperturbative formulation is required to 
describe strongly coupled type IIB compactifications with 7-branes. F-theory 
\cite{VafaF-theory,MorrisonVafaI,MorrisonVafaII} achieves this
by utilizing the ${\rm SL}(2,\mathbb Z)$ S-duality. The key observation is that, similar
to $\mathcal N=4$ SYM, since the S-duality group is the same as modular invariance of
a torus, we can represent the spatially varying axio-dilaton profile by inserting a
torus or \emph{elliptic curve} at each point in space, and geometrizing the axio-dilaton
as the complex structure modulus of the elliptic curve. Formally, we consider an 
\emph{elliptic fibration} $Y$, which is a complex $(n+1)$-dimensional compact manifold, over a complex 
$n$-dimensional base $B$, which is the internal manifold of a nonperturbative type IIB
compactification. Since 7-branes source the axio-dilaton with singularities at the brane 
positions, the 7-brane configurations in the compactification are now represented by
singularities of the elliptic fibration. Therefore, both the axio-dilaton background and 
7-brane physics now become parts of the compactification geometry.

Since there is now an axio-dilaton background in the Einstein equation, $B$ is no longer
CY in general. Instead, the equations of motion now imply that the total space $Y$ should
be CY to preserve minimal supersymmetry. Hence, we require that $B$ is still a compact
K\"ahler manifold, with an effective anticanonical class $-K_B$. These conditions allow
us to study the elliptic fibration with powerful tools from algebraic geometry. Note that
although we consider compactifications of a 12D spacetime on $Y$, the two fiber dimensions are only 
bookkeeping devices for the nonperturbative type IIB physics instead of physical 
dimensions. In particular, there is no supergravity in 12D and the 12D spacetime is not
physical.

Despite the origin from type IIB string theory, when studying F-theory it is usually more
useful to consider its dual M-theory picture. More precisely, F-theory compactified on
$Y\times S^1$ is dual to M-theory compactified on $Y$, and the limit of infinite radius 
of the $S^1$, i.e., F-theory compactified on $Y$, corresponds to the zero fiber volume
limit on the M-theory side. The duality can be seen as follows. Consider the elliptic 
fiber over $B$ as $T^2=S^1_A\times S^1_B$ in M-theory. As discussed in \S
\ref{sec:review:strings-to-ftheory:Mtheory}, shrinking the size of $S^1_A$ to zero 
corresponds to the type IIA limit compactified on the $S^1_B$ fibration over $B$. 
Through T-duality, this is also dual to type IIB compactified on the dual circle 
$\tilde S^1_B$ fibration over $B$. The zero size limit of $S^1_B$ corresponds to the infinite
size limit of $\tilde S^1_B$, in which we recover type IIB compactified on $B$, i.e., 
F-theory compactified on $Y$. Indeed, we have shrunk the elliptic fiber, i.e., both
$S^1_A$ and $S^1_B$, to zero volume on the M-theory side.

As discussed previously, the gauge dynamics of F-theory compactifications are encoded
in the structure in singularities of the elliptic fibration. Complex codimension-1
singularities on the base encode the positions of 7-branes, hence the gauge groups. In this 
framework, nonperturbative brane configurations including exceptional gauge groups can
also be realized in the same way as perturbative brane configurations. Codimension-2
singularities correspond to intersections of branes and encode the matter content apart
from the vector multiplets. Finally codimension-3 singularities correspond to Yukawa 
couplings between the matter. We see that 4D F-theory compactifications contain 
singularities of all three codimensions, hence provide the necessary ingredients for
phenomenology. 4D F-theory also supports fluxes, which induce chiral matter as we will 
see. On the other hand, 8D F-theory only contains vector multiplets, and 6D F-theory
contains vector and hypermultiplets instead of chiral multiplets. In the next section,
we will develop tools to study the elliptic fibrations and their singularities in 4D F-theory models.

\section{Basics of F-theory compactifications}
\label{sec:review:basics}

In this section, we review the tools used throughout this thesis for studying F-theory
compactifications, particularly in 4D. We discuss how singularities of the elliptic 
fibrations give rise to gauge groups, matter content, and Yukawa couplings, and provide
various methods for constructing these models explicitly. We also review $G_4$ fluxes in
detail, which play an essential role in the next research chapters. The review in this 
section is largely based on \cite{WeigandTASI,Li:2022aek}.

\subsection{Weierstrass model for elliptic fibrations}
\label{sec:review:basics:weierstrass}

Recall that the internal manifold of an F-theory compactification is an elliptically 
fibered CY manifold $Y$ over a base $B$. The elliptic fibration in a general F-theory model can be
described by treating the elliptic curve as a (1D) CY
hypersurface in the ambient projective space $\mathbb P^{2,3,1}$ (i.e., the hypersurface
with homology class the same as the anticanonical class of $\mathbb P^{2,3,1}$) with homogeneous 
coordinates $[x:y:z]$. The projective space is defined as the space 
$\mathbb C^3\setminus{(0,0,0)}$ modulo the equivalence relation
\begin{equation}
\label{eq:review:P231}
    (x,y,z)\simeq (\lambda^2x,\lambda^3y,\lambda z)\,.
\end{equation}
The internal manifold $Y$ is then given by the locus of
\begin{equation}
\label{eq:review:Weierstrass}
    y^2=x^3+fxz^4+gz^6\,,
\end{equation}
where $f,g$ are sections of line bundles $\mathcal O(-4K_B),\mathcal O(-6K_B)$ respectively such that $Y$ is CY. This is known as a
Weierstrass model. While there are many ways to parametrize elliptic fibrations as
hypersurfaces of complete intersections in ambient spaces, it is known that every 
elliptic fibration is birationally equivalent to a Weierstrass model.

The Weierstrass model always has a section $[x:y:z]=[1:1:0]$ called the zero section,
denoted by $D_0$. Outside of the zero section, it is more convenient to use the 
inhomogeneous coordinates with $z=1$ and write
\begin{equation}
\label{eq:review:WeierstrassInhomo}
    y^2=x^3+fx+g\,.
\end{equation}
As an example, for a 4D F-theory model with the base $B=\mathbb P^3$, the anticanonical
class is $-K_B=4H$ where $H$ is the hyperplane class of $B$. Then, $f \in \mathcal{O}(16H)$ 
and $g \in \mathcal{O}(24H)$ are degree-16 and degree-24 polynomials in the homogeneous coordinates of $B$.

The loci where the elliptic fibration becomes singular can be found by the 
conditions $P=dP=0$, where $P=0$ is the hypersurface equation of $Y$. For the 
Weierstrass model, it can be shown that the singularities are along the loci where the discriminant
\begin{equation}
\label{eq:review:discriminant}
    \Delta=4f^3+27g^2\,,
\end{equation}
vanishes, where $\Delta\in\mathcal O(-12K_B)$. Meanwhile, the axio-dilaton $\tau$ is given by
\begin{equation}
\label{eq:review:jinvariant}
    j(\tau) = \frac{4 \cdot (24 f)^3}{\Delta} = \frac{1728 \cdot 4f^3}{4f^3 + 27g^2}\,,
\end{equation}
where
\begin{equation}
\label{eq:review:jinvariantExpansion}
    j(\tau)=e^{-2\pi i\tau}+744+196884 e^{2\pi i\tau}+\cdots\,,
\end{equation}
is the $j$-invariant. Indeed, we see that $\tau$ diverges on $\{\Delta=0\}$. Therefore,
$\{\Delta=0\}$ determines the 7-brane loci in type IIB language. As we will discuss 
next, the vanishing behavior of $f,g,\Delta$ determines the gauge dynamics localized on
these 7-branes.

\subsection{Gauge groups from codimension-1 singularities}
\label{sec:review:basics:gaugegroups}

Consider a base divisor (algebraic subspace at codimension one in the base)
given by an irreducible codimension-one locus $\Sigma=\{s=0\}$ contained within
the vanishing locus of $\Delta$. The degree of the fiber singularity
at generic points on the divisor $\Sigma$ is determined by the orders
of vanishing of $f,g,\Delta$. When the orders are sufficiently high,
the internal manifold $Y$ itself becomes singular, and a non-abelian gauge
group $G$ is supported on the divisor. We call such a divisor a gauge
divisor. In general we abuse notation and use $\Sigma$ to denote both
the divisor and its homology class. The ``geometric'' gauge group, up to monodromies, can be
determined by the vanishing orders according to the classification by
Kodaira and N\'eron \cite{Kodaira,Neron,BershadskyEtAlSingularities}
(see Table \ref{tab:Kodaira}). This geometry, however, does not fully
determine the physical gauge group since, as described in the next chapters, it may
be broken by a flux background. While the above only considers a single geometric
non-abelian gauge factor, the same kind of
analysis directly generalizes to the case of multiple geometric non-abelian
gauge factors, as the gauge divisors are just local features in the
geometry of $B$, although there can be further complications when geometric
non-abelian gauge factors intersect.

\begin{table}[t]
    \centering
    \begin{tabular}{|c|c|c|c|c|c|}
    \hline
    Type & ord($f$) & ord($g$) & ord($\Delta$) & Singularity & Symmetry algebra \\
    \hline\hline
    $I_0$ & $\geq 0$ & $\geq 0$ & 0 & / & / \\
    $I_1$ & 0 & 0 & 1 & / & / \\
    $II$ & $\geq 1$ & 1 & 2 & / & / \\
    $III$ & 1 & $\geq 2$ & 3 & $A_1$ & $\mathfrak{su}(2)$ \\
    $IV$ & $\geq 2$ & 2 & 4 & $A_2$ & $\mathfrak{sp}(1)$ or $\mathfrak{su}(3)$ \\
    $I_n$ & 0 & 0 & $n\geq 2$ & $A_{n-1}$ & $\mathfrak{sp}([n/2])$ or $\mathfrak{su}(n)$ \\
    $I^*_0$ & $\geq 2$ & $\geq 3$ & 6 & $D_4$ & $\mathfrak{g}_2$ or $\mathfrak{so}(7)$ or $\mathfrak{so}(8)$ \\
    $I^*_n$ & 2 & 3 & $n\geq 7$ & $D_{n-2}$ & $\mathfrak{so}(2n-5)$ or $\mathfrak{so}(2n-4)$ \\
    $IV^*$ & $\geq 3$ & 4 & 8 & $E_6$ & $\mathfrak{f}_4$ or $\mathfrak{e}_6$ \\
    $III^*$ & 3 & $\geq 5$ & 9 & $E_7$ & $\mathfrak{e}_7$ \\
    $II^*$ & $\geq 4$ & 5 & 10 & $E_8$ & $\mathfrak{e}_8$ \\
    \hline
    non-min & $\geq 4$ & $\geq 6$ & $\geq 12$ & \multicolumn{2}{|c|}{incompatible with CY condition} \\
    \hline
    \end{tabular}
    \caption{Kodaira classification of singular elliptic fibers, mapping vanishing orders to non-abelian gauge groups up to monodromies.}
    \label{tab:Kodaira}
\end{table}

On the other hand, geometric abelian $\U(1)$ gauge groups arise in a more complicated 
way. Namely, they are localized on additional rational sections of the elliptic 
fibrations other than the zero section. More formally, there are $\U(1)$ factors if the
Mordell-Weil group of rational sections, in which the zero section serves as the 
identity element, have a nonzero rank. In this thesis, we focus on F-theory models 
without nontrivial Mordell-Weil group, although we still construct $\U(1)$ factors with
other mechanisms such as fluxes.

As the geometry of $Y$ becomes singular in the presence of a
non-abelian gauge divisor $\Sigma$, to have well-defined geometric
quantities such as intersection numbers for the geometry, 
the usual procedure is to follow the M-theory approach and to
blow up the singular locus by $\mathbb P^1$'s, resulting in a smooth
resolved CY manifold $\hat Y$. The resolution introduces a set of
exceptional divisors $D_i$ ($i=1,2,...,\mathrm{rank}(G)$) in the
$\hat Y$, which are the $\mathbb P^1$-fibers over $\Sigma$. These new
divisors correspond to the Dynkin nodes of the group supported on
$\Sigma$ (associated with simple roots $\alpha_i$), and their intersections match with 
the structure of the Dynkin diagram; this geometry will be demonstrated in \S
\ref{sec:review:basics:tate} and discussed more in \S\ref{subsec:review:intersections}.
This precise correspondence between fiber geometry and gauge dynamics is the key idea
that motivates F-theory.
In accord with the Shioda-Tate-Wazir theorem
\cite{shioda1972,Wazir}, the divisors $D_I$ on $\hat Y$ are
spanned by the zero section $D_0$, the pullbacks
of base divisors $\pi^*D_\alpha$ (which we also call $D_\alpha$
depending on context), and exceptional divisors $D_i$. There will be additional divisors
associated with more rational sections if the Mordell-Weil group is nontrivial.
Notice that
there is no unique choice of the resolution and $D_i$'s, although 
consistency of the theory requires that the physics is independent of such a
choice. 
The resolution independence of the physics and of certain relevant
aspects of
the intersection form on CY fourfolds
(as found in \cite{Jefferson:2021bid} and reviewed in 
\S\ref{subsec:review:intersections}) suggests that these
quantities should have a natural geometric interpretation directly in
the context of the singular geometry; although this is not yet well
understood from a pure mathematics perspective.

The origin of the gauge fields associated with $G$ can be understood in the dual
M-theory picture. M-theory compactified on the resolved manifold $\hat Y$ encodes the
Coulomb branch of the gauge theory, in which the non-abelian group $G$ is broken down
to $\U(1)^{{\rm rank}(G)}$. Denoting the fiber curves in $D_i$ as $\mathbb P^1_i$,
M2-branes wrapping a curve $\sum_i n_i\mathbb P^1_i$ correspond to massive W-bosons
associated with the root $-\sum_i n_i\alpha_i$, with mass given by the volume of the 
curve. In the F-theory limit where the fiber volume shrinks to zero, these W-bosons 
become massless and the gauge group is enhanced back to $G$. Meanwhile, the Cartan 
gauge fields $A_i$ come from the reduction of $C_3$ potential on exceptional divisors
\begin{equation}
\label{eq:review:CartanFromC3}
    C_3\supset \sum_i A_i\wedge [D_i]\,,
\end{equation}
where $[D_i]$ are the 2-forms Poincar\'e-dual to $D_i$. Combining the W-bosons and the
Cartan gauge bosons gives the adjoint representation of the nonabelian gauge field of 
$G$. Apart from the vector multiplet of $G$, there may also be additional adjoint matter
depending on the geometry of $\Sigma$.

\subsection{Tate models and their resolutions}
\label{sec:review:basics:tate}

While the Weierstrass model can parametrize the most general elliptic fibrations, the
expressions of the coefficients $f,g$ are usually very complicated to work with. In many cases, it is
useful to focus on a particularly simple subclass of models called the Tate models,
given by the hypersurface in $\mathbb P^{2,3,1}$
\begin{equation}
\label{eq:review:Tate}
    y^2 + a_1 xyz + a_3 y z^3 = x^3 + a_2 x^2 z^2 + a_4 x z^4 + a_6 z^6 \,,
\end{equation}
where the sections $a_n \in \mathcal{O}(-n K_B)$ are related to the Weierstrass 
coefficients $f$, $g$ through birational transformations by
\begin{equation}
\label{eq:review:Tate2Weierstrass}
    f=-\frac{1}{48}\left(b_2^2-24b_4\right)\,,\quad g=\frac{1}{864}\left(b_2^3-36b_2b_4+216b_6\right)\,,
\end{equation}
where 
\begin{equation}
\label{eq:review:Tate2WeierstrassDef}
    b_2=4a_2+a_1^2\,,\quad b_4=2a_4+a_1a_3\,,\quad b_6=4a_6+a_3^2\,.
\end{equation}
To construct a gauge group $G$ over a gauge divisor $\Sigma=\{s=0\}$, we simply impose 
the corresponding vanishing orders of $a_n$ on $s$ 
\cite{BershadskyEtAlSingularities,KatzEtAlTate,HuangTaylorLargeHodge}, listed in Table \ref{tab:tatealg}.

The form of the Tate models allows us to resolve the singularities of the elliptic
fibration in simple ways. To illustrate this, we work through a resolution of the $\SU(3)$
Tate model in detail. From Table \ref{tab:tatealg}, we choose the vanishing orders to be 
$(0,1,1,2,3)$. Therefore, the hypersurface equation of $Y$ is (setting $z=1$ for simplicity)
\begin{equation}
\label{eq:review:SU3Tate}
    y^2 + a_1xy + a_{3,1}sy = x^3 + a_{2,1}sx^2 + a_{4,2}s^2x + a_{6,3}s^3\,,
\end{equation}
where $a_{n,m}\in \mathcal O(-nK_B-m\Sigma)$.

\begin{table}[H]
\begin{center}
\begin{tabular}{|c|c|c|c|c|c|c|c|} \hline
type & group & $ a_1$ &
$a_2$ & $a_3$ &$ a_4 $& $ a_6$ &$\Delta$ \\ \hline $I_0 $ & --- &$ 0 $ &$ 0
$ &$ 0 $ &$ 0 $ &$ 0$ &$0$ \\ \hline $I_1 $ & --- &$0 $ &$ 0 $ &$ 1 $ &$ 1
$ &$ 1 $ &$1$ \\ \hline $I_2 $ &$SU(2)$ &$ 0 $ &$ 0 $ &$ 1 $ &$ 1 $ &$2$ &$
2 $ \\ \hline $I_{3}^{ns} $ & $Sp(1)$ &$0$ &$0$ &$2$ &$2$ &$3$ &$3$ \\ \hline
$I_{3}^{s}$ & $SU(3)$ &$0$ &$1$ &$1$ &$2$ &$3$ &$3$ \\ \hline
$I_{2n}^{ns}$ &$ Sp(n)$ &$0$ &$0$ &$n$ &$n$ &$2n$ &$2n$ \\ \hline
$I_{2n}^{s}$ &$SU(2n)$ &$0$ &$1$ &$n$ &$n$ &$2n$ &$2n$ \\ \hline
$I_{2n}^{s}$ (exotic) &$SU(2n)^\circ$ &$0$ &$2$ &$n-1$ &$n+1$ &$2n$ &$2n$ \\ \hline
$I_{2n+1}^{ns}$ &$Sp(n)$ & $0$ &$0$ &$n+1$ &$n+1$ &$2n+1$ &$2n+1$
\\ \hline $I_{2n+1}^s$ &$SU(2n+1)$ &$0$ &$1$ &$n$ &$n+1$ &$2n+1$ &$2n+1$
\\ \hline $II$ & --- &$1$ &$1$ &$1$ &$1$ &$1$ &$2$ \\ \hline $III$ &$SU(2)$ &$1$
&$1$ &$1$ &$1$ &$2$ &$3$ \\
\hline $IV^{ns} $ &$Sp(1)$ &$1$ &$1$ &$1$
&$2$ &$2$ &$4$ \\ 
\hline $IV^{s}$ &$SU(3)$ &$1$ &$1$ &$1$ &$2$ &$3 \; \; \; (2)^\star$ &$4$
\\ \hline $I_0^{*\,ns} $ &$G_2$ &$1$ &$1$ &$2$ &$2$ &$3$ &$6$ \\ \hline
$I_0^{*\,ss}$ &$SO(7)$ &$1$ &$1$ &$2$ &$2$ &$4$ &$6$ \\ \hline $I_0^{*\,s}
$ &$SO(8)$ &$1$ &$1$ &$2$ &$2$ &$(4, 3)^\star$ & $6$ \\ 
%\hline $I_{1}^{*\,ns}$
%&$SO(9)$ &$1$ &$1$ &$2$ &$3$ &$4$ &$7$ \\ \hline $I_{1}^{*\,s}$ &$SO(10) $
%&$1$ &$1$ &$2$ &$3$ &$5$ &$7$ \\ \hline $I_{2}^{*\,ns}$ &$SO(11)$ &$1$ &$1$
%&$3$ &$3$ &$5$ &$8$ \\ \hline $I_{2}^{*\,s}$ &$SO(12)$ &$1$ &$1$ &$3$
%&$3$ &$6$&$8$\\ 
\hline 
$I_{2n-3}^{*\,ns}$ &$SO(4n+1)$ &$1$ &$1$ &$n$ &$n+1$
&$2n$ &$2n+3$ \\ \hline $I_{2n-3}^{*\,s}$ &$SO(4n+2)$ &$1$ &$1$ &$n$ &$n+1$
&$2n+1 \; \; \; (2n)^{\star}$ &$2n+3$ \\ \hline $I_{2n-2}^{*\,ns}$ &$SO(4n+3)$ &$1$ &$1$ &$n+1$
&$n+1$ &$2n+1$ &$2n+4$ \\ \hline $I_{2n-2}^{*\,s}$ &$SO(4n+4)$ &$1$ &$1$
&$n+1$ &$n+1$ &$2n+2 \; \; \; (2n + 1)^{\star}$ 
&$2n+4$ \\ \hline $IV^{*\,ns}$ &$F_4 $ &$1$ &$2$ &$2$ &$3$ &$4$
&$8$\\ 
\hline $IV^{*\,s} $ &$E_6$ &$1$ &$2$ &$2$ &$3$ &$5 \; \; \;(4)^{\star}$ & $8$\\ \hline
$III^{*} $ &$E_7$ &$1$ &$2$ &$3$ &$3$ &$5$ & $9$\\ \hline $II^{*} $
&$E_8\,$ &$1$ &$2$ &$3$ &$4$ &$5$ & $10$ \\ \hline
 non-min & --- &$ 1$ &$2$ &$3$ &$4$ &$6$ &$12$ \\ \hline
\end{tabular}
\end{center}
\caption{The gauge groups and the corresponding vanishing orders of coefficients $a_n$ in the Tate models. In particular,
the alternate tuning of $\SU(2N)$ gives exotic matter content. Groups and
tunings marked with asterisks require additional monodromy conditions.}
 \label{tab:tatealg}
\end{table}

While $f,g$ does not vanish on $\Sigma$,
the discriminant $\Delta$ has the form
\begin{equation}
\label{eq:review:SU3discriminant}
    \Delta=\frac{1}{16}a_1^3\left(a_1a_{2,1}a_{3,1}^2 - a_{3,1}^3 - a_1^2 a_{3,1} a_{4,2} + a_1^3 a_{6,3}\right)s^3+O(s^4)\,.
\end{equation}
Therefore, by Table \ref{tab:Kodaira} there is indeed an $\SU(3)$ gauge group supported
on $\Sigma$.
One can check that $Y$ is singular along the locus
\begin{equation}
\label{eq:review:SU3singularLocus1}
    x=y=s=0\,.
\end{equation}
To resolve this 
singularity, we blow up $Y$ to $Y_1$ by redefining
\begin{equation}
\label{eq:review:SU3blowup1}
    (x,y,s) \;\longmapsto\; (x e_1,y e_1,s e_1)\,,
\end{equation}
where $e_1$ is a new projective coordinate. Then, the hypersurface equation becomes
\begin{equation}
\label{eq:review:SU3TateResolved1}
    e_1^2\left(y^2+a_1xy+a_{3,1}sy-e_1x^3-a_{2,1}e_1sx^2-a_{4,2}e_1s^2x-a_{6,3}e_1s^3\right)=0\,.
\end{equation}
Note that we have to rescale $x,y,s$ simultaneously such that there are common factors of $e_1$.
The locus $\{e_1=0\}$ is an exceptional divisor in the ambient space, and the remaining factor is
the hypersurface equation of $Y_1$. Although we have introduced a new coordinate $e_1$,
it is also associated with a new scaling relation
\begin{equation}
\label{eq:review:newScaling}
    (x,y,s,e_1)\simeq (\lambda x,\lambda y,\lambda s,\lambda^{-1} e_1)\,.
\end{equation}
Therefore, the dimension is unchanged. Due to the new scaling relation, now $x,y,s$ 
cannot vanish simultaneously on $Y_1$, hence the above singularity has been resolved. 

Nevertheless, $Y_1$ is still singular along the locus
\begin{equation}
\label{eq:review:SU3singularLocus2}
    y=e_1=a_1x+a_{3,1}s=a_1a_{2,1}a_{3,1}^2 - a_{3,1}^3 - a_1^2 a_{3,1} a_{4,2} + a_1^3 a_{6,3}=0\,.
\end{equation}
As we can see in \eqref{eq:review:SU3discriminant} and will discuss more in 
\S\ref{subsec:review:matter}, the locus $\{s=a_1a_{2,1}a_{3,1}^2 - a_{3,1}^3 - a_1^2 a_{3,1} a_{4,2} + a_1^3 a_{6,3}=0\}$ is a codimension-2 locus where
the $\SU(3)$ singularity on $\Sigma$ is enhanced to $\SU(4)$. To resolve this 
singularity, we blow up $Y_1$ to $\hat Y$ by redefining
\begin{equation}
\label{eq:review:SU3blowup2}
    (y,e_1) \;\longmapsto\; (y e_2,e_1 e_2)\,,
\end{equation}
where $e_2$ is another new coordinate. Then, the hypersurface equation becomes
\begin{equation}
\label{eq:review:SU3TateResolved2}
    e_2\left(e_2y^2+a_1xy+a_{3,1}sy-e_1x^3-a_{2,1}e_1sx^2-a_{4,2}e_1s^2x-a_{6,3}e_1s^3\right)=0\,.
\end{equation}
The locus $\{e_2=0\}$ is another exceptional divisor in the ambient space, and the 
remaining is the hypersurface equation of $\hat Y$. One can check that $\hat Y$ is smooth
and we have resolved all the singularities. We summarize the above resolution as
\begin{equation}
\label{eq:review:SU3resolution}
    \hat Y\stackrel{(y,e_1|e_2)}{\longrightarrow}Y_1\stackrel{(x,y,s|e_1)}{\longrightarrow}Y\,.
\end{equation}
Note again that this is only one of the possible routes to resolve an $\SU(3)$ singularity in an elliptic fibration.

The intersections of $\{e_1=0\},\{e_2=0\}$ with $\hat Y$ give the exceptional divisors
in $\hat Y$
\begin{align}
\label{eq:review:SU3exceptionalDivisors}
    D_1&:\quad \{e_1=e_2y^2+a_1xy+a_{3,1}sy=0\}\,,\\
    D_2&:\quad \{e_2=a_1xy+a_{3,1}sy-e_1x^3-a_{2,1}e_1sx^2-a_{4,2}e_1s^2x-a_{6,3}e_1s^3=0\}\,.
\end{align}
We see that $D_1$ and $D_2$ intersect along the locus
\begin{equation}
\label{eq:review:SU3exceptionalDivInt}
    e_1=e_2=a_1x+a_{3,1}s=0\,.
\end{equation}
Since all the terms are linear, $D_1$ and $D_2$ intersect along one irreducible 
component. Equivalently, the $\mathbb P^1$ fiber curves in $D_1,D_2$ intersect at one 
point. Therefore, the fiber geometry indeed matches the Dynkin diagram of $\SU(3)$.

\subsection{Tuned and rigid gauge groups}
\label{sec:review:basics:rigid}

While the associated (non-abelian) gauge group factor can be easily
determined when given a singular gauge divisor, it is interesting to
consider the possible different origins of these singularities and
associated groups.  In particular, there are two main classes of gauge
group factors, namely tuned and rigid groups, which have qualitatively
different origins.

Tuned gauge groups are easily understood using the general
description of a Weierstrass model given in the previous section. 
Such gauge groups are obtained on a divisor in any base by fine-tuning (many) complex
structure moduli.
Roughly speaking,
we can do a local expansion of the
Weierstrass model around the divisor $\Sigma$:
\begin{align}
    f &= f_0+f_1 s+f_2 s^2+...\,, \nonumber \\
    g &= g_0+g_1 s+g_2 s^2+... \,,
\label{eq:review:local-expansion}
\end{align}
where the coefficient functions live in various line bundles.  By
fine-tuning these Weierstrass coefficients $f_i,g_i$, over a divisor
whose normal bundle is not strongly negative, we can get various
orders of vanishing up to $(4,6)$.  In this way, any gauge group
factor in Table \ref{tab:Kodaira} can be tuned over many divisors, such as
the plane $H$ in the simple base $\P^3$.

On the other
hand, many F-theory bases contain rigid gauge groups, which do not
require any fine-tuning like that described above, and are therefore
present throughout the whole set of moduli space branches associated
with elliptic fibrations over that base
\cite{MorrisonTaylorClusters,MorrisonTaylor4DClusters}.  Such rigid
gauge groups arise when a divisor $\Sigma$ has a sufficiently negative
normal bundle $N_\Sigma$; the associated strong curvature forces
sufficiently high degrees of singularity on the Weierstrass model that
a non-abelian gauge factor automatically arises on $\Sigma$.  Since
the gauge group does not depend on any moduli, there is no geometric
deformation that can break the gauge group. From the low-energy
perspective such a deformation corresponds to Higgsing, so these
groups are also called (geometrically) non-Higgsable gauge
groups. They can, however, be broken by certain types of flux
background, which we demonstrate in the next chapters.  
And, when supersymmetry is broken, these groups can also be broken by
the standard Higgs mechanism by a massive charged scalar Higgs field
in the usual way.
Therefore, to avoid confusion
we refer to these gauge factors that are forced by geometry
as ``rigid'' gauge groups in this thesis.  Exploration of the landscape of
allowed bases for elliptic CY threefolds and fourfolds, giving
6D and 4D F-theory models respectively, has given strong evidence that
the vast majority of F-theory bases support multiple disjoint clusters
of rigid gauge factors
\cite{TaylorWangMC,HalversonLongSungAlg,TaylorWangLandscape}.
Indeed, the only bases that do not support rigid gauge factors are
essentially the weak Fano bases, which form a tiny subset of
the full set of allowed bases (for example, for surfaces for 6D
F-theory models the only bases without rigid gauge factors are the
generalized del Pezzo surfaces, which contain no curves of
self-intersection below $-2$; among toric bases these represent only a
handful of the roughly 60,000 possible base surfaces, and for
threefold bases the weak Fano bases are an even smaller fraction of
the full set of possibilities).

The possible rigid gauge groups  in 4D F-theory models have been
completely classified \cite{MorrisonTaylor4DClusters}, in terms of
single factors and intersecting pairs of gauge factors that may arise. Unlike
% the above
gauge groups that can be realized through tuned
Weierstrass models, not all gauge groups
in Table \ref{tab:Kodaira} can be rigid. For a single gauge factor,
the possible rigid gauge algebras are
\begin{equation}
\label{eq:review:rigid-single}
    \mathfrak{su}(2), \mathfrak{su}(3), \mathfrak{g}_2, 
    \mathfrak{so}(7), \mathfrak{so}(8), \mathfrak{f}_4,
    \mathfrak{e}_6, \mathfrak{e}_7, \mathfrak{e}_8\,.
\end{equation}
Of these single factors, the only ones that contain $\gsm$ 
as a subgroup are
$E_8,E_7$, and $E_6$. For a product of two gauge factors, the
possible algebras are
\begin{equation}
\label{eq:review:rigid-product}
    \mathfrak{su}(2)\oplus \mathfrak{su}(2),
\hspace*{0.1in} \mathfrak{su}(3)\oplus \mathfrak{su}(2),
\hspace*{0.1in} \mathfrak{su}(3)\oplus \mathfrak{su}(3),
\hspace*{0.1in} \mathfrak{g}_2\oplus \mathfrak{su}(2),
\hspace*{0.1in} \mathfrak{so}(7)\oplus \mathfrak{su}(2)\,.
\end{equation}
In particular, this includes the non-abelian part of $\gsm$ but it is hard to incorporate the remaining
$\U(1)$ in a rigid way \cite{MartiniTaylorSemitoric, WangU1s}. In this thesis, we focus on the case of a
single gauge factor that contains $\gsm$. 
Formalizing the heuristic picture of \eqref{eq:review:local-expansion},
the
presence of a given rigid gauge factor can be easily determined by the
following analysis
\cite{MorrisonTaylor4DClusters}: we define the following divisors on
$\Sigma$ (not the base $B$)
\begin{align}
\label{eq:review:FkGl}
    F_k &= -4K_\Sigma+(4-k)N_\Sigma\,, \nonumber \\
    G_l &= -6K_\Sigma+(6-l)N_\Sigma\,.
\end{align}
We then determine the minimum values of $(k,l)$ such that $F_k,G_l$
are effective.  Any Weierstrass model is then forced to have vanishing
orders of at least $(k,l)$ on $\Sigma$.  When $\Sigma$ is near or
intersecting other divisors with sufficiently negative normal bundles,
this can cause a further enhancement of the gauge group factor over
$\Sigma$; for example, this effect arises in the 6D case where an
isolated curve of self-intersection $-3$ supports a rigid $SU(3)$
gauge factor, but a pair of intersecting curves with
self-intersections $(-3, -2)$ support a rigid $G_2 \times SU(2)$ group
with a minimum amount of jointly charged matter (which is insufficient
to Higgs the group down to a smaller subgroup)
\cite{MorrisonTaylorClusters}.

Rigid gauge groups are much more generic than the tuned ones in the
landscape for various reasons.  First, tuned gauge factors require
fine-tuning of moduli over any given base, while we get rigid gauge
groups automatically when the base contains divisors with reasonably
negative normal bundles.  Second, as mentioned above, most bases
contain many rigid gauge factors, so such factors are clearly
ubiquitous in the landscape.  Third, since many divisors already
support rigid gauge groups, on a generic base few (or even no)
divisors are available for tuning additional gauge factors; this
effect becomes increasingly strong as $h^{1, 1}(B)$ increases and the
number of complex structure moduli $h^{3, 1}(\hat{Y})$ (for a threefold base)
decreases. Therefore, from a statistical point of view (such as in \ \cite{AshokDouglas,DenefDouglas}), in the absence of other
considerations not yet understood, we may expect that it is much more
likely for $\gsm$ to arise from rigid gauge groups than from simply
fine tuning over a set of divisors that do not support rigid gauge
groups, over a base such as a weak Fano threefold.

It is natural then to consider classes of models in which the SM gauge group $\gsm$ arises from a rigid gauge factor $E_6,E_7$,
or $E_8$.  While the precise abundance of these three gauge groups in
the landscape is not fully understood, it is clear that each of them
arises as a rigid gauge factor over a vast set of bases, both for 6D
and 4D F-theory models.  This abundance is most clearly understood for
6D F-theory models, where the toric bases for such models have been
completely classified \cite{MorrisonTaylorToric} and there is also
some understanding of the full set of allowed non-toric bases,
particularly at large $h^{2, 1}(\hat{Y})$
\cite{MartiniTaylorSemitoric,TaylorWangNon-toric}.  In particular,
among the 61539 toric base twofolds (including toric bases with $-9,
-10$ and $-11$ curves, which support rigid $E_8$ gauge factors and
contain (4, 6) points that must be blown up for a smooth base), 26958,
36698, 37056 bases have rigid $E_6,E_7,E_8$ factors respectively, so
more than half of all bases support each of $E_7$ and $E_8$ groups,
and 55332 ($\sim$ 90\%) contain a divisor that supports either an
$E_7$ or $E_8$ factor. 
At least for large Hodge numbers, the structure of non-toric bases is
similar, and toric bases form a good representative sample
\cite{TaylorWangNon-toric}, although it is plausible that at small Hodge numbers
non-toric bases with fewer large exceptional groups dominate.
On the other
hand, the total number of toric bases alone for 4D F-theory models is
$\mathcal O(10^{3000})$
\cite{TaylorWangMC,HalversonLongSungAlg,TaylorWangLandscape}, or $\mathcal O(10^{90})$
if counting only base polytopes instead of their triangulations \cite{Taylor:2025gnp}, which is much
too large for explicit analysis. It is expected that $E_8$ is (much)
more generic than the other exceptional group factors for elliptic
CY fourfolds with toric bases, but there is no
good measure of the relative abundance between $E_7$ and $E_6$.  One
estimate of these abundances from a partial statistical analysis of
toric bases comes from a Monte Carlo analysis on blowups of $\mathbb
P^3$, without rigid $E_8$ factors or codimension-two $(4,6)$
singularities \cite{TaylorWangMC}. It is estimated that 18\% of the
bases in this study contain rigid $E_7$ factors and 26\% of them
contain rigid $E_6$ factors, but the errors in these estimates may be
large.  In general we expect that the two gauge groups have similar
relative abundance, while the overall fractions may get %slightly
smaller when $E_8$'s are included. This is the case for 6D F-theory
models: there are 24483 toric bases without rigid $E_8$'s, of which
18276 (75\%) contain rigid $E_7$ factors and 13843 (57\%) contain
rigid $E_6$ factors. 

The above estimates are focused on toric bases only, so it is clearly desirable to
have some better estimates of how common rigid exceptional groups are
in the broader landscape of elliptic \cy fourfolds with
both toric and non-toric bases.  For \cy fourfolds with threefold bases there are
also questions of how to statistically weight the sets of possible
fluxes and different triangulations of the base, each of which can
give exponentially large factors \cite{Wang:2020gmi,Taylor:2025gnp} (see also
\cite{Demirtas:2020dbm} on related issues). Nevertheless, it is clear in any case that 
the rigid $E_6, E_7,$ and $E_8$ factors arise
on a vast class of F-theory bases $B$.

\subsection{Matter from codimension-2 singularities}
\label{subsec:review:matter}

We now turn to the matter content
in F-theory models.  Matter fields in the low-energy theory can
arise from both localized and global features in the gauge divisor
$\Sigma$.  When a gauge divisor intersects another component of the
discriminant locus over a codimension-2 locus $C$, in general the fiber singularity
is enhanced over $C$, resulting in matter multiplets in the low-energy theory.
In the resolved geometry these enhancements result in additional
$\P^1$ components in the fibers over $C$.
In the M-theory picture, the matter multiplets are associated with
M2-branes wrapping these fibral curves.  When a non-abelian gauge
divisor intersects another non-abelian gauge divisor the resulting
matter is charged under both gauge groups, while intersections with
the residual discriminant locus over components not carrying a gauge
group (like the $I_1$ locus where $f, g \neq 0,\Delta = 0$) give
matter that is only charged under the single gauge factor.  There is also
``bulk'' matter in the adjoint representation supported over the full
divisor $\Sigma$.  

In many situations
the matter representations $R$ over a matter curve $C$
can be 
determined in a relatively simple way
 directly from the singular geometry, known as the Katz-Vafa method \cite{KatzVafa}. First,  one
determines the vanishing orders on $C$ and associates them with a
(naive) Kodaira type, hence a larger non-abelian group $\tilde G$. 
 The adjoint representation of $\tilde G$ can then be decomposed
into
representations of the original gauge group $G$. Apart from the
adjoint of $G$ supported on the bulk of $\Sigma$, this also
includes some new representations and some singlets. These are
the matter representations supported on $C$. We denote $C_R$ as the matter
locus supporting representation $R$. Note that unlike the Kodaira classification
in Table \ref{tab:Kodaira}, the Katz-Vafa method may not work in some
edge cases and one must study the fiber structure directly.

Here is a concrete example for an $\SU(3)$ gauge group, continuing the example in 
\S\ref{sec:review:basics:tate}. From \eqref{eq:review:SU3discriminant}, the discriminant
can be factorized into $\Delta=\Delta_0s^3$, where $\{s=0\}$ is the gauge divisor $\Sigma$. Intersecting $\Sigma$ with the locus $\{\Delta_0=0\}$, we see that there are
two subloci where the vanishing orders of $f,g,\Delta$ increase, hence the singularities
are enhanced.
\begin{align}
\label{eq:review:SU3matterLoci}
    C_1&=\{s=a_1=0\}\,,\quad&\left({\rm ord}(f),{\rm ord}(g),{\rm ord}(\Delta)\right)=(2,2,4)\,,\\
    C_2&=\{s=a_1a_{2,1}a_{3,1}^2 - a_{3,1}^3 - a_1^2 a_{3,1} a_{4,2} + a_1^3 a_{6,3}=0\}\,,\quad&\left({\rm ord}(f),{\rm ord}(g),{\rm ord}(\Delta)\right)=(0,0,4)\,.
\end{align}
From Table \ref{tab:Kodaira}, we see that the singularity on $C_1$ is still $\SU(3)$,
so there is no additional matter on $C_1$.\footnote{Strictly speaking, it is also 
because the coefficient of $s^2$ in $g$ is a 
perfect square, so the monodromy condition for $\SU(3)$ singularity is satisfied.} 
On the other hand, the singularity on $C_2$
is enhanced from $\SU(3)$ to $\SU(4)$. Decomposing the $\SU(4)$ adjoint
\begin{equation}
\label{eq:review:SU4toSU3}
    \mathbf{15}\rightarrow \mathbf{8}\oplus\mathbf{3}\oplus\mathbf{\bar 3}\oplus\mathbf{1}\,.
\end{equation}
Therefore, $C_2$, now denoted as $C_\mathbf{3}$, supports fundamental and 
antifundamental matter of $\SU(3)$. From \eqref{eq:review:SU3discriminant}, its 
homology class is $C_\mathbf{3}=-\Sigma\cdot(9K_B+3\Sigma)$.

In the next chapters, we will need the following results for $E_7,E_6$ gauge groups.
For rigid $E_7$ models, the matter locus $C_\mathbf{56}=-\Sigma\cdot (4K_B+3\Sigma)$,
if nontrivial, supports half multiplets of fundamental matter $\mathbf{56}$. For rigid
$E_6$ models, the matter locus $C_\mathbf{27}=-\Sigma\cdot(3K_B+2\Sigma)$ supports
fundamental matter $\mathbf{27}$ and antifundamental matter $\mathbf{\bar{27}}$.

While determining the representations is straightforward, determining the matter 
multiplicities is much harder in general. In 6D F-theory, the matter loci are a set
of points, and the matter multiplicities are simply the numbers of these points, given
by intersection numbers on the base. On the other hand, in 4D F-theory, the multiplicities
depend on both the geometry and flux data, and the exact dependence is still not
fully understood. Fortunately, the calculation of chiral indices
(i.e., the difference between the numbers of chiral and anti-chiral
multiplets) has been well established (and is reviewed in
\S\ref{subsec:review:G4fluxes}).  The computation of the number of vector-like
chiral/anti-chiral pairs is much more subtle
\cite{Bies:2014sra,Bies:2021nje,Bies:2021xfh,Bies:2022wvj,Bies:2023sfm}; a particularly
simple case of this computation is done in \S\ref{subsec:e7:vectorlike}.
In addition, as studied in the next chapters, the multiplicities also change when the geometric gauge
group $G$ itself is broken by flux to a smaller group $G' \subset G$.
Matter can appear in various representations of $G'$ that are contained
within the representations of $G$ that may arise geometrically in the
unbroken theory. Remarkably, chiral matter
can arise for $G'$ even when there are no allowed chiral
representations of $G$ (such as for $G = E_7$).

In this thesis, we generally avoid situations where the degrees of a codimension-2 singularity reach $(4,6)$
or higher, where the above picture breaks down, signaling the
presence of strongly coupled sectors \cite{HeckmanMorrisonVafa,Apruzzi:2018oge}.
This situation happens when there is, for example, an $E_8$ gauge group with nontrivial
codimension-2 singularities. These singularities are still not fully understood 
especially in 4D F-theory, and remain as open and important problems for understanding
matter spectra in F-theory.

\subsection{Yukawa couplings from codimension-3 singularities}
\label{subsec:review:yukawa}

When three matter loci intersect (some of the loci may be the same, i.e., 
self-intersecting), they can give rise to codimension-3 singularities where the degrees of
singularity are enhanced again. Instead of having more $\mathbb P^1$ components in the
fiber, the fibral curves on one matter locus split and join into the fibral curves on
other matter loci at codimension-3 singularities. Accordingly, M2-branes wrapping these
fibral curves can homologically deform into each other at these singularities. 
From the low-energy perspective, these singularities give rise to Yukawa couplings but
not additional matter (unless the degrees of singularity reach $(4,6)$, to be discussed
in \S\ref{sec:Pheno}). The homological relations between the fibers 
\begin{equation}
\label{eq:review:YukawaFiber}
    \mathbb P^1_{\mathbf R_1}=\mathbb P^1_{\mathbf R_2}+\mathbb P^1_{\mathbf R_3}\,,
\end{equation}
correspond to the
constraints that Yukawa couplings must be gauge invariant
\begin{equation}
\label{eq:review:YukawaRep}
    \mathbf{\bar R}_1\times \mathbf R_2\times \mathbf R_3\rightarrow \mathbf 1\,.
\end{equation}

Unlike codimension-1 and -2 singularities, there are no simple methods to read out the
types of Yukawa couplings, and one must study the fiber structure directly. To 
illustrate this analysis, we continue the $\SU(3)$ example in 
\S\ref{sec:review:basics:tate}. We see that there is a codimension-3 singularity along
the locus
\begin{equation}
\label{eq:review:SU3YukawaLocus}
    \{s=a_1=a_{3,1}=0\}\,,\quad\left({\rm ord}(f),{\rm ord}(g),{\rm ord}(\Delta)\right)=(2,3,6)\,.
\end{equation}
The exceptional divisors $D_1,D_2$ restricted on the locus $\{a_1=a_{3,1}=0\}$ become
\begin{align}
    D_1&:\quad \{e_1=e_2=0\}\,,\\
    D_2&:\quad \{e_2=e_1(x^3+a_{2,1}sx^2+a_{4,2}s^2x+a_{6,3}s^3)=0\}\,.
\end{align}
We see that $D_2$ splits into multiple components at the singularity. The first 
component $\{e_1=e_2=0\}$ is just the same as $D_1$, while the remaining becomes cubic
and splits into three irreducible components, all of which support (anti)fundamental
matter. Therefore, the codimension-3 singularity supports the baryonic coupling $\mathbf 3^3$ \cite{Lawrie:2012gg}.

In general, the matter involved in the Yukawa couplings can come from either 
codimension-2 loci $C$ or the adjoints on the bulk of gauge divisor $\Sigma$. Therefore,
there are three possible types of Yukawa couplings: between three fields on the bulk of 
$\Sigma$ (denoted by $\Sigma\Sigma\Sigma$), between one field on the bulk of $\Sigma$
and two fields on codimension-2 loci (denoted by $\Sigma CC$), and between three fields
on codimension-2 loci (denoted by $CCC$). Their properties and the distinctions between
these couplings will be further discussed in \S\ref{sec:Pheno}.

\subsection{Toric geometry}
\label{subsec:review:toric}

So far, we have been working at the level of general Weierstrass models without assuming
any specific bases. While this is a powerful approach, in many cases it is useful to 
construct explicit internal manifolds $Y$. In particular, one of the approaches to
study the string landscape is to construct a large set of possible $Y$ and investigate
their properties. Toric geometry is a framework for constructing these manifolds and
computing their topological data such as intersection numbers and Hodge numbers with
well-established combinatorial algorithms, many of which are available in \texttt{Sage}
\cite{sagemath}. Here we review the basics of toric geometry and its use in string theory, 
following \cite{HuangTaylorLargeHodge,Jefferson:2022ssj}. For simplicity, we only focus
on smooth toric varieties.

Toric geometry generalizes the notion of projective spaces. In toric geometry, a 
$n$-dimensional toric variety $X$ is defined by its fan $\Sigma$ (not to be confused 
with gauge divisors), which is a set of cones in the space $N\otimes \mathbb R$, where 
$N=\mathbb Z^n$ is a lattice. The cones are spanned by a set of one-dimensional cones
known as rays given by vectors $\vec v_i\in N$. The vertices of $\vec v_i$ form a 
lattice polytope $\nabla$ in $N$. Each ray represents a homogeneous coordinate $z_i$
corresponding to a divisor $D_i=\{z_i=0\}$, and higher-dimensional cones spanned by 
rays correspond to intersections of the divisors with multiple vanishing homogeneous 
coordinates. Now, $X$ is defined as
\begin{equation}
\label{eq:review:toricDef}
    X=\frac{\mathbb C^n\backslash SRI}{G}\,,
\end{equation}
where the Stanley-Reisner ideal $SRI$ contains subsets of $\vec v_i$ that do not share
a higher-dimensional cone, i.e., their homogeneous coordinates cannot vanish simultaneously. $G$ is a group of scaling relations given by
\begin{equation}
\label{eq:review:toricScaling}
    \sum_i a_i\vec v_i=0\quad\Leftrightarrow\quad z_i\rightarrow\lambda^{a_i}z_i\,.
\end{equation}

As a simple example, the projective space $\mathbb P^2$ is given by the rays
\begin{equation}
\label{eq:review:toricP2}
    \vec v_i=(1,0),(0,1),(-1,-1)\,.
\end{equation}
Note that this is not the unique choice of rays, and different choices are related by
${\rm SL}(n,\mathbb Z)$ transformations. Let the homogeneous coordinates be $x,y,z$.
We see that the three rays together do not span a cone in 3D, so $SRI$ contains the
origin $(x,y,z)=(0,0,0)$. Moreover, we observe that $\vec v_1+\vec v_2+\vec v_3=0$,
which corresponds to the scaling $(x,y,z)\rightarrow(\lambda x,\lambda y,\lambda z)$.
This is indeed the description of $\mathbb P^2$.

Alternatively, we can interpret $X$ as a compactification by gluing different patches
of algebraic tori $(\mathbb C^*)^r$ together, known as stratification. The origin of 
the fan corresponds to the bulk $(\mathbb C^*)^n$, and each $(n-r)$-dimensional cone
corresponds to a patch of $(\mathbb C^*)^r$. For $\mathbb P^2$, the origin corresponds
to the patch of nonvanishing $x,y,z$, the ray $\vec v_1$ corresponds to the patch $x=0$
but nonvanishing $y,z$, and so on.

Various topological data of $X$ can be computed with known formulas. For example, the
total Chern class is
\begin{equation}
\label{eq:review:totalChern}
    c(X)=\prod_i(1+D_i)\,.
\end{equation}
In particular, the anticanonical class of $X$ is the sum of $D_i$. For $\mathbb P^2$,
all three $D_i$ are the same as the hyperplane class $H$, and we indeed get $-K_X=3H$.

Intersection numbers are also easy to compute for 2D toric varieties. Adjacent rays
$\vec v_i,\vec v_{i+1}$ in the fan corresponds the intersection number $D_i\cdot D_{i+1}=1$, 
since they span a 2D cone. The self-intersection numbers are $D_i^2=-m$, where $m$
satisfies $m\vec v_i=\vec v_{i-1}+\vec v_{i+1}$. All the other intersection numbers 
vanish. In higher dimensions, however, 
intersection numbers are not only determined by the lattice polytope $\nabla$. More precisely, $\nabla$ does not fully
determine the fan $\Sigma$ in higher dimensions since triangulations of polytopes, 
which determine the set of higher-dimensional cones, are
far from unique in higher dimensions.

As a more nontrivial example, we consider the Hirzebruch surface $\mathbb F_n$, which
is a $\mathbb P^1$-bundle over another $\mathbb P^1$ with twist $n$. The rays are
\begin{equation}
\label{eq:review:Hirzebruch}
    \vec v_i=(1,0),(0,1),(-1,-n),(0,-1)\,.
\end{equation}
We see that $\vec v_1,\vec v_3$ correspond to the base $\mathbb P^1$, and 
$\vec v_2,\vec v_4$ correspond to the fiber $\mathbb P^1$. The scaling relations are
$\vec v_1+\vec v_3+n\vec v_2=0$ and $\vec v_2+\vec v_4=0$. The divisor classes and
homological equivalence can also be seen from the coefficients in the scaling 
relations. Here, $\vec v_1,\vec v_3$ have the same coefficients and $D_1,D_3$ are in the
same class $F$. We further have $D_4=S$ and $D_2=S+nF$. Using the above formulas, the
intersection numbers are $F^2=0,S\cdot F=1,S^2=-n$.

In F-theory, toric geometry can be used to describe not only the bases, but also CY 
manifolds by taking the anticanonical hypersurfaces of toric varieties. CY hypersurfaces
are constructed using reflexive polytopes \cite{batyrev1993dual}; see a brief 
construction of hypersurfaces of general classes in Appendix \ref{appendix:e7}.
A polytope $\nabla$ 
containing the origin as its only interior point is reflexive if its dual polytope $\Delta$ defined as
\begin{equation}
\label{eq:review:dualPolytope}
    \Delta=\{\vec m \in M\;|\;\vec m\cdot\vec v\geq -1,\forall \vec v\in \nabla\}\,,
\end{equation}
is also a lattice polytope, where $M={\rm Hom}(N,\mathbb Z)$ is the dual lattice of $N$.
Each lattice point in $\Delta$ corresponds to a monomial
\begin{equation}
\label{eq:review:dualPolytopeMonomial}
    p(\vec m_i)=\prod_j z_j^{\vec m_i\cdot \vec v_j+1}\,.
\end{equation}
We recall that $z_j$ is the homogeneous coordinate associated with the ray $\vec v_j$.
Then, the CY hypersurface equation is
\begin{equation}
\label{eq:review:dualPolytopeCY}
    \sum_i c_i p(\vec m_i)=0\,,
\end{equation}
where $c_i$ are coefficients. While the resulting CY manifold and its intersection
numbers depend on the triangulation of $\nabla$, its Hodge numbers depend on $\nabla$
only and can be computed using Batyrev's formulas \cite{batyrev1993dual}.

There are 16 distinct reflexive polytopes in 2D for elliptic curves, one of which is the space
$\mathbb P^{2,3,1}$ used in the Weierstrass model. Continuing our example, $\mathbb P^2$
can also be used to construct elliptic curves. The dual polytope of \eqref{eq:review:toricP2} 
is a lattice polytope with vertices $(-1,-1),(-1,2),(2,-1)$,
and the CY hypersurface equation is the generic cubic polynomial in $x,y,z$. In higher 
dimensions, all 4319 3D reflexive polytopes for constructing K3 surfaces \cite{Kreuzer:1998vb} and 
473,880,776 4D reflexive polytopes for constructing CY threefolds 
\cite{Kreuzer:2000xy,Kreuzer:2000qv} have been classified
in the Kreuzer-Skarke (KS) database. 5D reflexive polytopes for constructing CY fourfolds
are still far from being fully classified, although the maximal polytopes for constructing 
reflexive polytopes have been classified in the KS database \cite{Scholler:2018apc}.

Among the reflexive polytopes for constructing CY hypersurfaces, a useful subclass of polytopes that 
closely resemble the Tate models are the standard $\mathbb P^{2,3,1}$-fibered
polytopes. First, the rays of $\mathbb P^{2,3,1}$ are $(-1,0),(0,-1),(2,3)$ satisfying
$2\vec v_1+3\vec v_2+\vec v_3=0$. Then, a $\mathbb P^{2,3,1}$-fibration over an $n$-dimensional base $B$
is given by the following rays. There are three rays associated with the fiber
\begin{equation}
\label{eq:review:P231polytopeFiber}
    \vec v_x=(\vec 0,-1,0)\,,\quad \vec v_y=(\vec 0,0,-1)\,,\quad \vec v_z=(\vec 0,2,3)\,,
\end{equation}
where $\vec 0$ is $n$-dimensional. In addition, each ray $\vec v_i^B$ of the base $B$
corresponds to the ray
\begin{equation}
\label{eq:review:P231polytopeBase}
    \vec v_i=(\vec v_i^B,2,3)\,.
\end{equation}
One can show that the monomials from the dual polytope are exactly the ones in \eqref{eq:review:Tate}.

The above polytopes represent smooth Tate models without any gauge groups. To construct
gauge groups, we need to set certain monomials to vanish by removing points in the dual
polytope $\Delta$, which is the same as adding rays into the polytope $\nabla$. These
additional rays, known as tops, correspond to exceptional divisors and have the form
\begin{equation}
\label{eq:review:top}
    (l\vec v_i^B,\vec p_j)\,,
\end{equation}
where $l$ is a parameter called level, $\vec v_i^B$ is the ray associated with the gauge
divisor $\Sigma=D_i$ on the base, and
\begin{equation}
    \vec p_j=(2,3),(1,2),(1,1),(0,1),(0,0),(-1,0),(0,-1)\,,
\end{equation}
are the seven lattice points in the polytope of $\mathbb P^{2,3,1}$. The polytope 
with the tops directly gives the resolved CY manifold $\hat Y$, with different 
triangulations of the polytope corresponding to different routes of resolution of $Y$.
The required tops for different gauge groups have been classified in 
\cite{Candelas:1996su,HuangTaylorLargeHodge}. For example, an $\SU(3)$ gauge group is
constructed by adding the tops $(\vec v_i^B,1,2),(\vec v_i^B,1,1)$, corresponding to the
two exceptional divisors. Note that this construction requires that the gauge divisor
is associated with only one ray on the base. In fact, when the gauge divisor is
associated with multiples of the base rays, the resulting CY manifold after resolution
may no longer be a toric hypersurface in general.

\subsection{$G_4$ fluxes}
\label{subsec:review:G4fluxes}

Apart from the geometry of the elliptic fibration, 
further data is needed to fully
define a 4D F-theory model and determine its gauge group and
matter content. 
The structure of this extra data
is most easily understood in the
dual M-theory picture, where  the 3-form potential
$C_3$ and its field strength $G_4=dC_3$
provide extra parameters associated with a compactification. 
The degrees of freedom of $C_3$ contain continuous degrees of freedom
when $h^{2, 1} (\hat{Y})$ is nontrivial; completely incorporating the
effects of these degrees of freedom is necessary
to determine the exact matter
spectrum, which is a complex task with the current technologies, as
reviewed in \cite{WeigandTASI}. Fortunately in this thesis, $G_4$ flux is sufficient to determine the gauge group
and the matter content, and the tools for analyzing these aspects of the
theory are well developed.

In general, $G_4$ is a discrete flux that takes values in the fourth
cohomology $H^4(\hat Y,\mathbb R)$.
The quantization condition on $G_4$ is slightly subtle and is given by
\cite{Witten:1996md}
\begin{equation}
\label{eq:review:fluxquant}
    G_4+\frac{1}{2}c_2(\hat Y) \in H^4(\hat Y,\mathbb Z)\,,
\end{equation}
where $c_2(\hat Y)$ is the second Chern class of $\hat Y$. In
general, $c_2(\hat Y)$ can be odd (i.e., non-even), in which case the discrete
quantization of $G_4$ contains a half-integer shift.
In particular, this implies that in some cases we are
forced to turn on some flux that may break the gauge groups, see Appendix \ref{appendix:e7} for an example.

Next, to preserve the minimal amount of supersymmetry (SUSY) in 4D, $G_4$
must live in the middle cohomology i.e. $G_4\in H^{2,2}(\hat Y,\mathbb
R)\cap H^4(\hat Y,\mathbb Z/2)$. 
Supersymmetry also imposes the condition of primitivity \cite{Becker:1996gj,Gukov:1999ya}:
\begin{equation}
\label{eq:review:primitivity}
    J\wedge G_4=0\,,
\end{equation}
where $J$ is the K\"ahler form of $\hat Y$. This is
automatically satisfied when the geometric gauge group is not broken, but
not obviously satisfied when the gauge group is broken by (vertical) flux. The
interpretation of this condition is that it stabilizes some (but not all)
K\"ahler moduli; stabilizing these moduli within the K\"ahler cone
imposes additional flux constraints. This will be explained
further in Section \ref{subsec:fluxbreaking:primitive}.

We also have the D3-tadpole condition \cite{Sethi:1996es} that must be
satisfied for a consistent vacuum solution:
\begin{equation}
\label{eq:review:tadpole}
    \frac{\chi(\hat Y)}{24}-\frac{1}{2}\int_{\hat Y} G_4\wedge G_4=N_{D3}\in \mathbb Z_{\geq 0}\,,
\end{equation}
where $\chi(\hat Y)$ is the Euler characteristic of $\hat Y$,
and $N_{D3}$ is the number of D3-branes, or M2-branes in the
dual M-theory. To preserve SUSY and stability, we
forbid the presence of anti-D3-branes i.e. $N_{D3}\geq 0$. The
integrality of $N_{D3}$ is guaranteed by
\eqref{eq:review:fluxquant}. This condition has an immediate
consequence on the sizes of fluxes. Recall the topological
formulae for CY fourfolds (see e.g. \cite{Klemm_1998}):
\begin{align}
\label{eq:review:CYhodge}
    \chi &= 6(8+h^{1,1}-h^{2,1}+h^{3,1})\,, \nonumber \\
    h^{2,2} &= 44+4h^{1,1}-2h^{2,1}+4h^{3,1}\,,
\end{align}
where $h^{i,j}$ are the Hodge numbers. It is then clear that
$h^{2,2}>2\chi/3\gg \chi/24$. Therefore, if we randomly turn on
flux in the whole middle cohomology such that the tadpole constraint
is satisfied, a generic flux
configuration vanishes or has small magnitude in most of the
$h^{2,2}$ independent directions.

There are more flux constraints on the \emph{vertical} part
of $G_4$, such that $G_4$ dualizes to a consistent F-theory
background that preserves Poincar\'{e} invariance, which we return to below.
%This vertical flux is also our key to analyze
To analyze
flux breaking and chiral matter it is helpful
to consider
the orthogonal decomposition of the middle cohomology
\cite{Braun:2014xka}:
\begin{equation}
\label{eq:review:decomp}
    H^{4}(\hat Y,\mathbb C)=H^{4}_\mathrm{hor}(\hat Y,\mathbb C)\oplus H^{2,2}_\mathrm{vert}(\hat Y,\mathbb C)\oplus H^{2,2}_\mathrm{rem}(\hat Y,\mathbb C)\,.
\end{equation}
The horizontal subspace comes from the complex structure
variation of the holomorphic 4-form $\Omega$ associated with
the CY fourfold. 
Flux in these directions has the effect of inducing a superpotential and
stabilizing complex structure moduli \cite{Gukov:1999ya}. The vertical subspace is spanned by products of harmonic
$(1,1)$-forms (which are Poincar\'e dual to divisors, denoted
by $[D_I]$)
\begin{equation}
    H^{2,2}_\mathrm{vert}(\hat Y,\mathbb C)=\mathrm{span}\left(
    H^{1,1}(\hat Y,\mathbb C)\wedge H^{1,1}(\hat Y,\mathbb C)\right)\,.
\label{eq:review:vertsubspace}
\end{equation}
Finally, there may be components that do not belong to the
horizontal or vertical subspaces; these are referred to as
remainder flux.  While there are various types of remainder
flux, we will need the following type in
the analysis below. Consider a
curve $C_{\mathrm{rem}}\in H_{1,1}(\Sigma,\mathbb Z)$  in
$\Sigma$, such that its pushforward $\iota_*C_\mathrm{rem}\in H_{1,1}(B,\mathbb Z)$ is trivial, where
$\iota:\Sigma\rightarrow B$ is the inclusion map.
While such a
curve  cannot be realized on toric divisors on toric bases, it has
been suggested that such curves do exist on ``typical'' bases
\cite{Braun:2014xka}, so toric geometry may be insufficiently
generic for this class of constructions. In the following chapters, we will 
construct examples of non-toric bases that support $C_\mathrm{rem}$.
In any case, we
now restrict each $D_i$ onto $C_\mathrm{rem}$. Its Poincar\'e dual $[D_i|_{C_\mathrm{rem}}]$ is a $(2,2)$-form, but since $C_\mathrm{rem}$ cannot be obtained by intersections of base divisors, we must have
\begin{equation}
    \left[D_i|_{C_\mathrm{rem}}\right]\in H^{2,2}_\mathrm{rem}(\hat Y,\mathbb C)\,.
\end{equation}

Here we explain more about vertical flux. Combining (\ref{eq:review:vertsubspace}) with  (\ref{eq:review:fluxquant}) gives the
integral vertical subspace $H^{2,2}_\mathrm{vert}(\hat Y,\mathbb
R)\cap H^4(\hat Y,\mathbb Z)$. 
We focus primarily here on the vertical subspace spanned by integer
multiples of forms $[D_{I}] \wedge[D_J]$
\begin{equation}
    H^{2,2}_\mathrm{vert}(\hat Y,\mathbb Z):=\mathrm{span}_{\mathbb Z}\left(
    H^{1,1}(\hat Y,\mathbb Z)\wedge H^{1,1}(\hat Y,\mathbb
    Z)\right)\,.
\label{eq:review:vertZ}
\end{equation}
While this subspace does not necessarily include all lattice points in
the full vertical cohomology $H^{2, 2}_\mathrm{vert} (\hat{Y},\C)\cap
H^4 (\hat{Y},\Z)$ of the same dimension, 
this subspace provides access to much of the interesting
physics, including the production of chiral matter and the flux
breaking mechanism we study in this paper.
The full intersection pairing on $H^4 (\hat{Y},\Z)$ is unimodular, and
in many cases there are elements of this lattice that have  components in the full vertical subspace
(\ref{eq:review:vertsubspace}) that do not lie in (\ref{eq:review:vertZ}).
Some
of these quantization issues have recently been discussed in, e.g.,
\cite{CveticEtAlQuadrillion, Jefferson:2021bid}, but various questions remain
outstanding regarding the full characterization of this quantization,
which is complicated further in connection with the possibility of non-even values of
$c_2 (\hat{Y})$.  We avoid these complications in this thesis and
focus primarily on the space (\ref{eq:review:vertZ}) and, when
possible, on even $c_2 (\hat{Y})$. This will suffice for the examples
that we explore explicitly in this thesis.

Now we set up some notations for   vertical fluxes. We expand
\begin{equation}
    G_4^\mathrm{vert}=\phi_{IJ} [D_I]\wedge [D_J]\,,
\label{eq:review:g4vert}
\end{equation}
and work with integer (or possibly half-integer if $c_2$ is
odd) flux parameters $\phi_{IJ}$. Note that the
expansion depends on the choice of basis of base divisors,
which we will specify depending on context. We denote the
integrated flux as \cite{Grimm:2011fx}
%\begin{equation}
%    \Theta_{\Lambda\Gamma}=\int_{\hat Y} G_4\wedge [D_\Lambda]\wedge
%          [D_\Gamma]\,,
%\label{eq:integrated-flux}
%\end{equation}
%where $D_\Lambda,D_\Gamma$ are linear combinations of $D_I$.
%Clearly $\Theta_{\Lambda\Gamma}$ is also spanned by
 %$\Theta_{IJ}$.
\begin{equation}
    \Theta_{\Lambda\Gamma}=\int_{\hat Y} G_4\wedge [\Lambda]\wedge
          [\Gamma]\,,
\label{eq:review:intflux}
\end{equation}
where $\Lambda,\Gamma$ are arbitrary linear combinations of $D_I$; subscripts $0, i, \alpha, \ldots$ refer to the basis divisors $D_0, D_i, D_\alpha, \ldots$.
Using the intersection
numbers on $\hat Y$, studying these
objects is turned into simple linear algebra problems, as we will see in \S\ref{subsec:review:intersections}.

Now we are ready to write down the remaining flux constraints.
To preserve Poincar\'e symmetry after dualizing, we require
\cite{Dasgupta:1999ss}
\begin{equation}
\label{eq:review:Poincare}
    \Theta_{0\alpha}=\Theta_{\alpha\beta}=0\,.
\end{equation}
(Recall that Greek  indices $\alpha, \ldots$ correspond to divisors that are
pullbacks from the base, while Roman indices $i$ correspond to Cartan
divisors, and the index 0 refers to the global zero section of the elliptic
fibration.)
Next, if the whole geometric gauge symmetry is preserved, a necessary condition is that
\begin{equation}
\label{eq:review:gaugepres}
    \Theta_{i\alpha}=0\,,
\end{equation}
for all $i,\alpha$, 
otherwise the gauge group is broken. This condition is not
sufficient when there is nontrivial remainder flux.
Note that the condition
(\ref{eq:review:Poincare}) for Poincar\'{e} symmetry
is unchanged even when (\ref{eq:review:gaugepres}) is violated.

Much of the above discussion on vertical flux extends naturally to the
type of remainder flux $G_4^\mathrm{rem}$ we need. Similarly, we expand
\begin{equation}
    G_4^\mathrm{rem}=\phi_{ir} \left[D_i|_{C_\mathrm{rem}}\right]\,,
\end{equation}
and work with integer $\phi_{ir}$. In this thesis, we always turn on 
remainder flux with a single $C_\mathrm{rem}$ only, so we do not specify
the choice of $C_\mathrm{rem}$ in the flux parameters; instead we just 
label them by ``$r$''.

Apart from breaking the gauge group, the \emph{vertical} flux
can also induce chiral matter. The famous
index formula states that for a weight $\beta$ in
representation $R$, its chiral index $\chi_\beta$, i.e.,
the difference in multiplicities between weights $\beta$ and $-\beta$,
is \cite{Braun_2012,Marsano_2011,KRAUSE20121,Grimm:2011fx}
\begin{equation} \label{eq:review:ordinarychi}
    \chi_\beta=\int_{S(\beta)}G_4^\mathrm{vert}\,,
\end{equation}
where $S(\beta)$ is called the matter surface of $\beta$. When
$R$ is localized on a matter curve $C_R$, $S(\beta)$ is the
fibration of the blowup $\mathbb P^1$ corresponding to $\beta$
over $C_R$. The procedure of finding the matter surfaces will be
discussed in \S\ref{subsec:review:intersections}.
When the gauge group $G$ is not broken, the vanishing of all
$\Theta_{i\alpha}$ guarantees that all $\beta$ in $R$ give the
same $\chi_R$. This is no longer true if $G$ is broken by fluxes; see
\S\ref{subsec:fluxbreaking:matter} for more detailed discussions.

\subsection{Intersection theory on fourfolds}
\label{subsec:review:intersections}

In \cite{Jefferson:2021bid}, a unified approach was developed for
organizing the relevant components of the intersection numbers on
$\hat{Y}$ into a resolution-independent structure
that conceptually simplifies the analysis of symmetry
constraints, flux breaking, and chiral matter.
The basic idea is that the intersection numbers 
\begin{equation}
M_{IJKL} = \int_{\hat{Y}} [D_I] \wedge [D_J] \wedge [D_K] \wedge [D_L] 
\end{equation}
can be organized into a matrix
\begin{equation}
M_{(IJ)(KL)} = M_{IJKL} = S_{IJ} \cdot S_{KL} \,,
\end{equation}
where the formal surface $S_{IJ} = D_I\cap D_J$ is equivalent to an element of vertical homology
$H_{2, 2} (\hat{Y},\Z)$, and ``dots'' denote the intersection
product.
In terms of this matrix, the equations
(\ref{eq:review:intflux}-\ref{eq:review:gaugepres}), as well as the expressions
for chiral matter multiplicities in terms of $G_4$ can be expressed
simply in terms of linear algebra.

A key aspect of this perspective is that the basis  of formal surfaces
$S_{IJ}$ is redundant \cite{LinWeigandG4, Bies_2017}.  There are various equivalences between these
surfaces in homology; for example the set of such formal surfaces
$S_{\alpha \beta}$ associated with pullbacks of intersections of
divisors on the base naively has $h_{1, 1} (B) (h_{1, 1} (B) + 1)/2$
elements, whereas by Poincar\'{e} duality the number of homologically
independent curves on the base is $h_{2, 2} (B) = h_{1, 1} (B)$, so
there are at least  $h_{1, 1} (B) (h_{1, 1} (B) -1)/2$ redundant
formal surfaces $S_{\alpha \beta}$.  Such homological equivalences
between the $S_{IJ}$ correspond to null vectors of the matrix $M$.
Removing all such homological equivalences $\sim$ gives a reduced
matrix $M_{\rm red}$, which encodes the intersection product on middle
vertical homology/cohomology.  One of the key observations of
\cite{Jefferson:2021bid} is that this intersection matrix seems to
always be resolution invariant even though the quadruple intersection
numbers $M_{IJKL}$ are resolution dependent.

For an elliptic CY fourfold with a single non-abelian gauge factor, in
many cases\footnote{For most gauge groups this is the form of $M_{\rm
    red}$ when the gauge group is associated with an isolated
  singularity over a divisor in the base and there are no further
  enhanced singularities on curves intersecting that divisor.  The situation
  becomes more complicated when there are, e.g., further gauge factors
  on intersecting divisors, although codimension two enhanced
  singularities on curves can also arise in the absence of further
  gauge factors.  In a few other situations where the geometry has
  codimension three (4, 6) loci, including cases with the gauge group
  $E_7$, as well as other groups such as $SU(7)$ and $SO(12)$, there
  are extra homologically independent surfaces $S_{i j}$ that support
  flux but not conventional chiral matter fields; we avoid turning
  on these fluxes in this thesis.} where there is no chiral matter the matrix $\mred$
takes the simple form
\begin{equation}
        M_\text{red} =  \begin{pmatrix}
               D_{\alpha'} \cdot K \cdot D_\alpha  & D_{\alpha'}  \cdot D_{\alpha} \cdot D_\beta  & 0 \\
              D_{\alpha' } \cdot D_{\beta'} \cdot D_{\alpha }  &0 &0 \\
            0 & 0& -\kappa^{ij}\Sigma \cdot D_\alpha \cdot D_{\alpha'} 
        \end{pmatrix} \,
\label{eq:review:Mred-simple}
\end{equation}
 in a basis of independent vertical homology classes
$S_{0 \alpha},
S_{\alpha \beta}, S_{i \alpha}$.
Here $\kappa^{ij}$ is the inverse Killing metric of the
gauge algebra, which is also the Cartan matrix $C^{ij}$ for ADE
groups.
Note that the products here are taken in the base;
for convenience, in general we will not mention explicitly the space where
the products are taken, as the space ($\hat{Y}$ or $B$)
is already clear from context.
This is clearly resolution independent; indeed, the quadruple intersection
numbers involved in this matrix have long been  known in these cases
for general bases and gauge factors (see, e.g., \cite{Grimm:2011sk}).
The above also straightforwardly generalizes
to remainder flux by replacing the triple intersection on the base with
the intersection of two $C_\mathrm{rem}$'s on $\Sigma$.

Perhaps more remarkably, this resolution invariance of $M_{\rm red}$
up to a choice of basis
appears to hold even when there are homologically
nontrivial surfaces $S_{ij}$, which usually
(with the exceptions of the cases mentioned in the preceding footnote)
correspond to ``matter surfaces'' that
can support chiral matter. In this case, the general form of $M_{\rm
  red}$ in a basis of independent classes
$S_{0 \alpha},
S_{\alpha \beta}, S_{i \alpha}, S_{ij}$ is
\begin{equation}
        M_\text{red} =  \begin{pmatrix}
               D_{\alpha'} \cdot K \cdot D_\alpha  & D_{\alpha'}  \cdot D_{\alpha} \cdot D_\beta  & 0 & 0 \\
              D_{\alpha' } \cdot D_{\beta'} \cdot D_{\alpha }  &0 &0 & * \\
            0 & 0& -\kappa^{ij}\Sigma \cdot D_\alpha \cdot D_{\alpha'}& * \\
        0 & * &  * & *
        \end{pmatrix} \,.
\label{eq:mr-matter}
\end{equation}
While naively the elements marked with $*$ are resolution-dependent,
it was observed in \cite{Jefferson:2021bid} that up to an integer change
of basis, the matrices $M_{\rm red}$ given by (\ref{eq:mr-matter}) for
 distinct resolutions $\hat{Y}, \hat{Y}'$ of a singular
geometry $Y$ are equivalent in many classes of examples, and it was
conjectured that this resolution-independence always holds.
Furthermore,
given the form (\ref{eq:mr-matter}) there is a rational change
of basis under which $M_{\rm red}$ can be put in a canonical form
\begin{equation}
U^{\rm t}        M_\text{red} U =  \begin{pmatrix}
               D_{\alpha'} \cdot K \cdot D_\alpha  & D_{\alpha'}  \cdot D_{\alpha} \cdot D_\beta  & 0 & 0 \\
              D_{\alpha' } \cdot D_{\beta'} \cdot D_{\alpha }  &0 &0 & 0 \\
            0 & 0& -\kappa^{ij}\Sigma \cdot D_\alpha \cdot D_{\alpha'}& 0 \\
        0 & 0 &  0 &  
 \frac{M_{\rm phys}}{(\det \kappa)^2} 
        \end{pmatrix} \,.
\label{eq:review:Mred-matter}
\end{equation}
Here, $M_{\rm phys}$ is a matrix that in general
encodes the relations between fluxes and chiral matter; for any
particular choice of gauge group $G$, $M_{\rm phys}$ can be expressed
in terms of characteristic data of the gauge divisor and canonical
class of the base.  Note that because the transformation matrix
$U$ is generally rational, the appropriate lattice on which this
canonical form acts for physical flux configurations
is generally a finite index sublattice of $\Z^{N}$, where $N = h^{2,
  2}_{\rm vert} (\hat{Y})$.

One application of this framework is that we can very easily analyze
the constraint equations (\ref{eq:review:Poincare}), (\ref{eq:review:gaugepres}) and
chiral matter from $G$-flux in a simple linear algebraic framework
using $\mred$.  In this matrix notation we can write
(\ref{eq:review:intflux}) in the form
\begin{equation}
\Theta_{IJ} = M_{(IJ)(KL)} \phi_{KL} \,,
\label{eq:review:theta-matrix}
\end{equation}
where $\phi_{KL}$ is a vector of integers parameterizing the
$G$-flux as in (\ref{eq:review:g4vert}).  Restricting to an independent basis of (co-)homology
cycles, for example  (\ref{eq:review:Poincare}) and (\ref{eq:review:gaugepres}) become
simple linear constraints on the flux $\phi$. 
In particular, because of the block-diagonal form of the matrix
$\mred$, (\ref{eq:review:Poincare}) simply imposes the condition $\phi_{0
  \alpha} = \phi_{\alpha \beta} = 0$, and (\ref{eq:review:gaugepres}) imposes the
condition that $\phi_{i \alpha} = 0$ for all $i, \alpha$ 
when $M_{\rm red}$ is given by (\ref{eq:review:Mred-simple}) and/or there are
no nonzero flux parameters $\phi_{ij}$.
 We will use this same
framework here to analyze fluxes throughout the next chapters.  
Note that while the conjectured resolution-independence of
(\ref{eq:mr-matter}) has not been generally proven, we do not rely in
any significant way in this thesis on the validity of this conjecture; the
analysis for the group $E_7$ relies only on the form
(\ref{eq:review:Mred-simple}), which is manifestly resolution-independent, and
for the $E_6$ analysis we use a specific resolution and associated
forms (\ref{eq:mr-matter}) and (\ref{eq:review:Mred-matter}).
\chapter{Gauge symmetry breaking with fluxes and natural Standard Model structure from exceptional GUTs in F-theory}
\label{chap:fluxbreaking}

In this chapter, we give a general description of gauge symmetry breaking using
vertical and remainder fluxes in 4D F-theory models.
The fluxes can break a geometric gauge group to a smaller group and induce chiral
matter, even when the larger group admits no chiral matter
representations.  We focus specifically on applications to
realizations of the Standard Model gauge group and chiral
matter spectrum through breaking of rigid
exceptional gauge groups $E_7, E_6$, which are ubiquitous in
the 4D F-theory landscape. Supplemented by an intermediate
$\SU(5)$ group, these large classes of models give
natural constructions of Standard Model-like theories with small
numbers of generations of matter in F-theory. The results in this chapter
are based on \cite{Li:2021eyn,Li:2022aek}.

\section{Introduction}
\label{sec:fluxbreaking:intro}

String theory provides a consistent framework for a unified theory
that combines gravity with the other fundamental forces described by
quantum field theory.  To describe the real world, however,
ten-dimensional string theory must be compactified on a real
six-dimensional manifold, and various further objects like branes,
flux, and orientifolds must be incorporated.  Such constructions give
an enormous number (perhaps on the order of something like
$10^{272000}$ \cite{TaylorWangVacua}) of string theory vacua, known as
the string landscape.  Despite this large number, so far it has not
been clear which low-energy theories can be UV-completed and realized
in the string landscape.  The investigation of this question, known as
the Swampland program \cite{VafaSwamp,OoguriVafaSwamp}, has been a
rapidly evolving research area.

Here we focus on another related main challenge in string
phenomenology.  Despite decades of work, it is not yet clear whether
the well-established Standard Model of particle physics (SM) can be
realized in the string landscape, including all details of observed
phenomenology; for a recent review of work in this direction, see
\cite{Cvetic:2022fnv}.  Beyond the simple existence of such a solution, it is
perhaps even more important to understand the extent to which the
Standard Model can arise as a \emph{natural} solution in string
theory.  In other words, we would like to understand the extent to
which solutions like the Standard Model are widespread in the string
landscape or require extensive fine-tuning.  Constructing the detailed
Standard Model requires many elements such as the gauge group, the
matter content including both chiral matter and the Higgs, the Yukawa
couplings, a supersymmetry (SUSY)-breaking mechanism, values of the 19
free parameters, and possibly some room to address beyond-SM problems
as well as cosmological aspects such as the density of dark energy.
Unfortunately, the current available string theory techniques are far
from enough to compute all these features precisely. Although there is
some recent development on finding the exact matter spectrum
\cite{Bies:2014sra,Bies:2021nje,Bies:2021xfh,Bies:2022wvj,Bies:2023sfm}
in F-theory, in this chapter we only
focus on the gauge group and chiral matter content, where the
techniques have been well developed.  The general philosophy is that
if we can identify a natural class of models that realize the Standard
Model gauge group and chiral matter fields, these structures may
naturally correlate with certain other features of SM or beyond SM
physics.

These aspects have been long-standing and primary goals in string
phenomenology, and there has been a great amount of work on them in
the last two decades, starting from heterotic string compactifications,
which naturally carry $E_8$ gauge groups that can be broken down to
the standard model gauge group.  Recently, F-theory
\cite{VafaF-theory,MorrisonVafaI,MorrisonVafaII} has become the most
promising framework for studying string compactifications and
phenomenology, as it provides a global description of a large
connected class of supersymmetric string vacua. (See
\cite{WeigandTASI} for a review.) In particular, F-theory gives 4D
$\mathcal N=1$ supergravity models when compactified on 
elliptically fibered Calabi-Yau (CY) fourfolds, corresponding to
non-perturbative compactifications of type IIB string theory on general
(non-Ricci flat) complex K\"ahler threefold base manifolds $B$.  The
number of such threefold geometries $B$ alone seems to be on the order
of $10^{3000}$
\cite{TaylorWangMC,HalversonLongSungAlg,TaylorWangLandscape}, without
even considering the exponential multiplicity of fluxes possible for
each geometry.  F-theory is also known to be dual to many other types
of string compactifications such as heterotic models.  Briefly,
F-theory is a strongly coupled version of type IIB string theory with
non-perturbative configurations of 7-branes balancing the curvature of
the compactification space. The non-perturbative brane physics is
encoded geometrically into the elliptically fibered manifold, which
can be analyzed using powerful tools from algebraic geometry. The
gauge groups and matter content supported on these branes can then be
easily determined when combined with flux data. Applying these
techniques, here we construct a novel class of F-theory models that
naturally give the SM gauge group and chiral matter content.
Note that in this chapter we focus exclusively on 4D models with ${\cal
  N} = 1$ supersymmetry (4 supercharges).  While  low-scale supersymmetry has not been
observed in nature, supersymmetry provides additional symmetry
structure that enables systematic analytic study of a broad class of
vacua; since some structure such as typical rigid gauge groups are
similar between 6D theories with 8 supercharges and 4D theories with 4
supercharges, we have some optimism that some of the structure of
typical geometric gauge groups and chiral matter content may persist
from 4D  ${\cal N} = 1$ theories to theories with broken supersymmetry.

There have been many attempts in the literature to construct
supersymmetric
models of compactified string theory
with the SM gauge group $\gsm=\SU(3)\times \SU(2)\times \U(1)/\mathbb
Z_6$.
(As noted in \cite{TaylorTurnerGeneric,TaylorTurner321}, for the gauge
group $\SU(3) \times\SU(2) \times\U(1)$
without the quotient by the $\Z_6$ center, the SM chiral
matter content is highly non-generic and involves a great deal of fine
tuning; we proceed under the assumption that the gauge group of
the Standard Model is really $\gsm$.)
The results of these efforts suggest that the (supersymmetric)
landscape may contain a wide variety of SM-like models. The constructions of such
models in F-theory can be loosely classified in the following ways: As in field
theory approaches, one can directly build models with $\gsm$, or start
with grand unified theories (GUTs) and break the larger gauge group
down to $\gsm$ in various ways. There are also two essentially
distinct types of geometric gauge groups in F-theory. On the one hand,
one can tune a desired gauge group by fine-tuning many complex
structure moduli. In contrast, most F-theory compactification bases
contain divisors with very negative normal bundle. The strong
curvature of these geometries forces singularities in the elliptic
curve over these divisors, giving rigid (a.k.a \emph{geometrically
  non-Higgsable} \cite{MorrisonTaylor4DClusters}) gauge symmetries,
which are present throughout the whole branch of moduli space and
ubiquitous in the F-theory landscape
\cite{TaylorWangMC,HalversonLongSungAlg,TaylorWangLandscape}. Below we
comment on each type of approach:

\begin{itemize}
    \item \textbf{Directly tuned $\gsm$:}
      These models do not require any symmetry
      breaking mechanisms except the usual Higgs. Recently significant
      progress on these has been gained. In
      \cite{CveticEtAlQuadrillion}, $10^{15}$ explicit solutions of
      directly tuned $\gsm$ with three generations of SM chiral
      matter (a ``quadrillion Standard Models''), have been
      constructed, based on the ``$F_{11}$'' fiber of
      \cite{KleversEtAlToric};. It has also been shown that the SM
      matter representations generically appear when $\gsm$ is
      directly tuned, in the sense that these matter representations
      are included among those that require the least amount of moduli
      fine-tuning given the gauge group \cite{TaylorTurnerGeneric},
      and a universal Weierstrass model for such tunings has been
      constructed \cite{Raghuram:2019efb}, which includes those of
      \cite{CveticEtAlQuadrillion} in one particular subclass.  All
      these constructions include the presence of the
 $\mathbb Z_6$
      quotient in $\gsm$.
    
    \item \textbf{Directly tuned GUT:}
    These models have been studied for over a decade, starting
    from
    \cite{Donagi:2008ca,BeasleyHeckmanVafaI,BeasleyHeckmanVafaII,DonagiWijnholtGUTs}. Most
    of the work on these models has focused on the GUT
    group of $\SU(5)$ and its $\U(1)$ extensions
    \cite{Blumenhagen:2009yv,Marsano:2009wr,Grimm:2009yu,KRAUSE20121,Braun:2013nqa},
    while there has also been some study of
    $\SO(10)$ GUTs \cite{Chen:2010ts}. (See \cite{HeckmanReview} for a
    review) Most of these constructions break the GUT group using the
    so-called hypercharge flux further discussed in
    \cite{Mayrhofer:2013ara,Braun:2014pva}, which is a kind of ``remainder''
    flux \cite{Braun:2014xka} breaking the gauge group into the commutant of broken
    directions, including the $\U(1)$'s of these directions \cite{Buican:2006sn}.
    
    \item \textbf{Rigid $\gsm$:}
    Despite the success of the above models, they cannot be
    the most generic or natural SMs in the landscape, as
    extensive fine-tuning is generally required to get the directly
    tuned $\gsm$ or a non-rigid (tuned)
GUT group such as  $\SU(5)$ (see
    e.g.\ \cite{BraunWatariGenerations}).
    Moreover, the presence of rigid gauge groups forbids
    tuning additional gauge factors like $\gsm$ on most bases.
    Finding a rigid $\gsm$ seems to be a more natural way.
    Nevertheless, while the non-abelian $\SU(3)\times \SU(2)$
    parts of $\gsm$ can easily
be realized as a rigid structure \cite{GrassiHalversonShanesonTaylor},
    constructing the $\U(1)$ is much more subtle, and bases that support
    non-Higgsable $\U(1)$ factors are rather rare
    \cite{MartiniTaylorSemitoric, WangU1s}.
    
    \item \textbf{Rigid GUT:} The rigid gauge groups that contain
      $\gsm$ as a subgroup are $E_8,E_7,E_6$
      \cite{MorrisonTaylor4DClusters}, and these rigid groups are
      ubiquitous in the F-theory landscape.  Of these, it seems that
      in 4D (as well as in 6D),
      $E_8$ appears most frequently in the landscape, while $E_7$ and
      $E_6$ are also quite abundant
      \cite{TaylorWangMC,HalversonLongSungAlg,TaylorWangLandscape}. Starting
      with one of these rigid exceptional groups and breaking down to
      $\gsm$ is in principle the most natural way to construct SM-like
      models, from the point of view of prevalence in the F-theory
      landscape, and this is the approach taken in this chapter. On the other hand,
      SM-like models using these groups bring other
      challenges. Undesired exotic matter can be easily induced by
      such large gauge groups. While $E_6$ has been one of the
      traditional GUT groups 
(see, e.g., \cite{Gursey:1975ki,Achiman:1978vg,Barbieri:1980vc}, and
\cite{Chen:2010tg,Callaghan:2012rv,Callaghan:2013kaa}
for realizations in F-theory and further references)
, $E_7$ and $E_8$ do not themselves
      support chiral matter and have not received as much attention as
      GUT groups.  These groups, especially $E_8$,
are often associated with high
      degrees of singularity in the elliptic fibration (i.e.,
      codimension two  (4, 6) loci), that involve
      strongly coupled sectors that are poorly
      understood \cite{HeckmanMorrisonVafa,Apruzzi:2018oge}; the
      constructions we consider here avoid these issues.
    
\end{itemize}

To address some of the above challenges, we propose a general class of
SM-like models using a rigid (or even tuned) $E_7$ GUT group in
F-theory, with an intermediate $\SU(5)$ group. These models enjoy the advantages of being natural and
requiring little fine-tuning, and address some of the above
challenges.  Specifically, fluxes can be used to break the
geometric $E_7$ group in an F-theory construction in a way that is not
transparent in the low-energy field theory, but
%we can go beyond the field theory limit and
%use $E_7$ as GUT group, 
gives the correct SM gauge group and some chiral matter.
Although in many cases the breaking leads to exotic chiral
matter, there are large families of models in which the
correct SM chiral matter representations are obtained through
an intermediate $\SU(5)$.  The number of generations
can easily be small and we have demonstrated that three
generations can naturally arise in many of these models.  In this chapter, we present
the general formalism and various technical subtleties, describe the
$E_7$ models in detail, and generalize the construction to
other groups such as $E_6$. In particular, we give a fully
explicit example of our SM-like models, incorporating both
vertical and remainder fluxes.  These constructions open large new
regions of the landscape for string phenomenology. Note that for
various reasons explained below, we do not include $E_8$ GUTs,
although it is the most frequent exceptional gauge factor in the landscape.

The central tool we use to construct these models is gauge
symmetry breaking by flux living in \emph{vertical} and \emph{remainder} cohomologies (we
use the name ``flux breaking'' from now on; vertical and remainder cohomologies are reviewed in
\S\ref{subsec:review:G4fluxes}). This is an economic
way to deal with some of the above challenges. 
By imposing
simple linear constraints, we can break the larger GUT group
down to $\gsm$ without extra $\U(1)$'s. At the same time, the \emph{vertical} flux
induces chiral matter regardless of whether the original group
supports chiral matter.
%As we will show below, there are many
%gauge-breaking flux parameters contributing to a single chiral
%index.
The resulting chiral index has a linear Diophantine
structure related to the geometry of the F-theory base that generically allows any small number of
generations; sometimes three is the most
preferred number of generations. Remarkably, no highly tuned geometry
or nontrivial quantization condition on
the manifold is needed to achieve
 this structure. Certainly the idea of vertical
flux breaking is not new, but below we develop it to some
depth so that only a relatively simple calculation is needed to find the
chiral index. The calculation is based on the techniques in
\cite{Jefferson:2021bid}, which provide a conjecturally resolution-independent
description of the mathematical structure needed to compute chiral
indices for a fixed gauge group structure on a general base.

This chapter is organized as follows. we describe the formalism of
flux breaking in \S\ref{sec:fluxbreaking:formalism}. There we write down the
flux constraints for gauge breaking and the formula for chiral
indices. We also describe various technical points such as determining
matter surfaces, primitivity and K\"ahler moduli stabilization. To
demonstrate how the formalism works, we work out several simple
$\SU(N)$ examples focusing on anomaly cancellation.

In \S\ref{sec:fluxbreaking:SM} we present the construction of natural SM-like models from $E_7$ flux breaking. 
We first discuss different embeddings of $\gsm$ into
$E_7$, which induce SM chiral matter or various exotic matter.
Then we write down the class of SM-like models in general,
without assuming a specific base.

The same method can be
straightforwardly generalized to other large gauge groups such as
$E_6$.  We discuss these applications in Section
\ref{sec:fluxbreaking:othergroups}. There we also discuss some obstructions to
applying the same formalism to $E_8$. As a useful example, in \S\ref{sec:fluxbreaking:example} we 
work out an explicit construction on a particular
base that can give three generations of SM chiral
matter as the minimal and preferred chiral spectrum.
This construction is the simplest example
that we are aware of
where all the ingredients in our class of SM-like models can be
realized. Note that as mentioned above,
these SM-like models are far from complete to really describe our
Universe. In \S\ref{sec:fluxbreaking:conclusion} we finally conclude and
discuss further questions in these directions. We address several
technical points in Appendix~\ref{chap:appendix3}.

The necessary background from F-theory relevant to these constructions is reviewed in Chapter~\ref{chap:review}.

\section{Formalism of flux breaking}
\label{sec:fluxbreaking:formalism}

We now present a general formalism for
describing flux breaking in F-theory.  While the basic ideas
underlying this process have been understood previously in the
literature \cite{WeigandTASI}, 
explicit examples of this phenomenon in F-theory
have to date not been studied in detail.  We describe here flux
breaking in a general way that makes possible a simple analysis of a
wide range of flux breaking scenarios over general bases.  We discuss
various technical points that are worth extra attention, and give some
simple examples. We will then apply the results of this section in
\S\ref{sec:fluxbreaking:SM} to build our SM-like models.

\subsection{Flux breaking}
\label{subsec:fluxbreaking:breaking}

The mechanism of gauge breaking using fluxes is
certainly not a new idea. In general, both vertical and
remainder fluxes are involved in flux breaking, giving
qualitatively different breaking patterns. The vertical flux,
at the same time, can also induce chiral matter.

Let us first study vertical flux.
Consider a non-abelian group $G$ with
its Cartan directions labelled by $i=1,2,...,\mathrm{rank}(G)$,
corresponding to the exceptional divisors $D_i$. It is well
known \cite{WeigandTASI} that if we turn on nonzero flux
\begin{equation}
    G_4^\mathrm{vert}=\sum_i \phi_{i\alpha} [D_i]\wedge[D_\alpha]\,,
\end{equation}
for a single $\alpha$ (in an arbitrary basis), $G$ is broken
into the commutant of $T=\phi_{i\alpha} T_i$ within $G$, where
the Cartan generators $T_i$ are associated with the simple
roots $\alpha_i$ i.e. in the co-root basis. The
commutant can be factorized into
$G'=H\times \U(1)^{\mathrm{rank}(G)-\mathrm{rank}(H)}$, where
$H$ does not contain any $\U(1)$ factors. The
remaining $\U(1)$'s, however, are also generically broken since the flux induces
masses to the corresponding gauge bosons through the St\"uckelberg mechanism
\cite{Grimm:2010ks,Grimm:2011tb} (see Appendix \ref{sec:Stuckelberg}).  Below we rephrase this procedure in
a more efficient language.

Recall that preserving the whole geometric gauge symmetry
requires $\Theta_{i\alpha}=0$ for
all $i,\alpha$. Now we violate some of these conditions by
turning on
some nonzero parameters $\phi_{i\alpha}$. Consider a generator $e_\beta$
corresponding to the root $\beta=-b_i\alpha_i$. We then
compute\footnote{Indices appearing twice are summed over; other
  summations are indicated explicitly.}
\begin{equation}
    [T,e_\beta]=-\phi_{i\alpha} C^{ij} b_j e_\beta\propto-\sum_j b_j\left<\alpha_j,\alpha_j\right>\kappa^{ij}\phi_{i\alpha} e_\beta\,,
\end{equation}
where $\left<.,.\right>$ is the inner product of root
vectors. The commutator vanishes, hence the generator is
preserved, only when
\begin{equation} \label{rootCondition}
    \sum_i b_i\left<\alpha_i,\alpha_i\right>\Theta_{i\alpha}=0\,,
\end{equation}
for all
$\alpha$. By Appendix
\ref{sec:Stuckelberg}, the corresponding linear combination of
Cartan generators
\begin{equation}
    \sum_i b_i\left<\alpha_i,\alpha_i\right> T_i\,,
\end{equation}
is also preserved. These
generators form the non-abelian group $H$ after breaking.
Below we will focus on ADE groups, so
$\left<\alpha_i,\alpha_i\right>$ are the same for all $i$ and
Eq.\ (\ref{rootCondition}) simply becomes
$b_i\Theta_{i\alpha}=0$ for all $\alpha$.
% \drop{If we turn on nonzero
%$\phi_{i\alpha}$ for multiple
%distinct $\alpha$'s, for simplicity we require the ratio
%$\phi_{i\alpha}/\phi_{i\alpha'}$ for nonzero
%$\phi_{i\alpha},\phi_{i\alpha'}$ to be the same for all $i$ (for each
%pair $\alpha, \alpha'$),
%such that they all give the same gauge group $H$. As we will see,
%this condition also follows from primitivity.}

The simplest example of vertical flux breaking is that we turn on $\Theta_{i'\alpha}\neq 0$
for some set of Dynkin indices $i'\in I'$ and some $\alpha$, in a
generic way such that Eq.\ (\ref{rootCondition}) is satisfied only when $b_{i'}=0$ for
all $i' \in I'$. Then
$H$ is given by removing the corresponding
nodes in the Dynkin diagram of $G$. The simple roots of $H$ are
directly descended from $G$ and are given by $\alpha_{i\notin I'}$.  We will focus on this kind of breaking below.
\begin{comment}
If $\Theta_{i'\alpha}\neq 0$ for some
$i',\alpha$, the corresponding Cartan gauge bosons will get
St\"uckelberg masses (see Appendix \ref{sec:Stuckelberg}) and the gauge group $G$ breaks to $H$. The
non-abelian part of $H$ is given by removing the corresponding
nodes in the Dynkin diagram of $G$. The simple roots of $H$ are
directly descended from $G$ and are given by $\alpha_{i\notin i'}$. From now on we denote the broken directions as $i'$.
The correspondence between the two descriptions is as
follows. Consider the commutator
\begin{equation}
    [T,e_j]=\sum_i \phi_{i\alpha} C_{ij} e_j
    \propto\sum_i \kappa^{-1}_{ij}\phi_{i\alpha}e_j\,,
\end{equation}
which vanishes if $\Theta_{j\alpha}=0$ for all $\alpha$,
otherwise equals to $\Theta_{j\alpha'}e_j$ for some $\alpha'$.
Here $e_j$ is the generator corresponding to the simple root
$\alpha_j$. Therefore, the commutant of
$T$ in $G$ has roots spanned by $\alpha_j$ with
$\Theta_{j\alpha}=0$ for all $\alpha$.
\end{comment}

The statements for remainder flux are similar. If we turn on
\begin{equation}
    G_4^\mathrm{rem}=[c_i D_i|_{C_\mathrm{rem}}]\,,
\end{equation}
for some $C_\mathrm{rem}$ satisfying the property mentioned
in \S\ref{subsec:review:G4fluxes}, $G$ is broken into the commutant of
$T=c_i T_i$ within $G$. The difference is that the remainder
flux does not turn on any $\Theta_{i\alpha}$, so there is no
St\"uckelberg mechanism and all the $\U(1)$ factors in 
the commutant are preserved. In other words, breaking using
remainder flux never decreases the rank of the gauge group,
while breaking using vertical flux always decreases the rank.
In general, when both types of fluxes are turned on, only the
intersection of the two commutants are preserved. As a
result, a wide variety of breaking patterns can be
constructed using combinations of these fluxes. Note that
when $G$ is a rigid gauge group, $\Sigma$ is a rigid divisor
and supports remainder flux breaking only when embedded into
a non-toric base.
This follows because for a toric base $B$, toric divisors span the
cone of effective divisors, so any rigid effective divisor $\Sigma$ is
toric, and toric curves in a toric $\Sigma$ span $h^{1,1} (\Sigma)$.

So far we have focused on the non-abelian part of the broken
gauge group, while there can also be $\U(1)$ factors remaining. There
are two ways to get $\U(1)$'s in our formalism. The first way
is, obviously, breaking $G$ with remainder flux in which all
the $\U(1)$'s in the commutant are preserved. It is also
possible to get $\U(1)$'s with vertical flux. By imposing
$p_i\Theta_{i\alpha}=0$ for all $\alpha$ and some $p_i$, the
linear combination of Cartan generators $T_p=p_i T_i$ is
preserved. By choosing $p_i$ such that $p_i \alpha_i$ (modulo the preserved roots) is not
along a root of $G$, there is no additional root to be preserved, so $T_p$
corresponds to an extra $\U(1)$ factor instead of a part of
$H$. Note that the $\U(1)$'s induced by vertical flux are
always ``exotic'': a $\U(1)$ that coincides with a root of $G$
must be obtained through remainder instead of vertical flux,
otherwise the $\U(1)$ enhances to a part of $H$.
\begin{comment}
On the
other hand, if there is a linear relation between
$\Theta_{i'\alpha}$'s with different $i'$ i.e.
$c_{i'}\Theta_{i'\alpha}=0$ for all $\alpha$, the corresponding
Cartan direction $c_{i'}T_{i'}$ becomes massless again and is a
part of $H$. In practice, we choose the non-abelian part of $H$
by choosing the Dynkin nodes to be removed, and the abelian
part of $H$ by imposing these linear relations.
\end{comment}

There is an additional 
subtlety from vertical flux breaking. Let the
$\alpha$'s giving homologically independent $S_{i\alpha}$ be
$\alpha_1,\alpha_2, \ldots,\alpha_r$.
\begin{comment}
By basic linear algebra,
no matter how we turn on $\phi_{i\alpha}$, there must be at
least $\left(\mathrm{rank}(G)-r\right)$ linear relations in the form of
Eq.\ (\ref{rootCondition}). Therefore, we must have
\end{comment}
From the above breaking rules, we see that the
difference $\mathrm{rank}(G)-\mathrm{rank}(G')$ is
given by the rank of the  ($r$  $\times$
$\mathrm{rank}(G)$)
matrix $\Theta_{(\alpha_a)(i)}$ (where $a$ and $i$
are the indices for rows and columns respectively).
As we will show in \S\ref{subsec:fluxbreaking:primitive},
to satisfy primitivity the rank of the matrix is constrained to be
 at most $r-1$.
Therefore, we get a lower bound on $r$ for given $G$ and $G'$:
\begin{equation} \label{rlowerbound}
    r\geq\mathrm{rank}(G)-\mathrm{rank}(G')+1\,.
\end{equation}
In particular, we must have a sufficiently large number $r$
of $\alpha$'s giving independent cycles $S_{i \alpha}$
(associated with independent curves in $\Sigma$ that are also
independent in $B$) in order
to get a desired $G'$. This condition imposes constraints on
the possible geometries that support a given vertical flux breaking.
\begin{comment}
Apart from the linear relations
we impose, we may be able to construct some \emph{accidental}
ones that exist but are not imposed intentionally. In other
words, the St\"uckelberg mass matrix (see Appendix
\ref{sec:Stuckelberg}) has rank at most $h^{1,1}(\Sigma)$,
which may be smaller than our desired number of massive
directions. This leads to extra $\U(1)$'s and makes $H$
deviate from our desired one. To avoid so, we must require
\begin{equation}
    \mathrm{rank}(G)-\mathrm{rank}(H)\geq h^{1,1}(\Sigma)\,.
\end{equation}
\end{comment}

\subsection{Chiral matter and matter surfaces}
\label{subsec:fluxbreaking:matter}

Apart from breaking the gauge group, the \emph{vertical} flux
can also induce chiral matter. The famous
index formula states that for a weight $\beta$ in
representation $R$, its chiral index $\chi_\beta$ is \cite{Braun_2012,Marsano_2011,KRAUSE20121,Grimm:2011fx}
\begin{equation} \label{ordinarychi}
    \chi_\beta=\int_{S(\beta)}G_4^\mathrm{vert}\,,
\end{equation}
where $S(\beta)$ is called the matter surface of $\beta$. When
$R$ is localized on a matter curve $C_R$, $S(\beta)$ is the
fibration of the blowup $\mathbb P^1$ corresponding to $\beta$
over $C_R$. When $G$ is not broken, the vanishing of all
$\Theta_{i\alpha}$ guarantees that all $\beta$ in $R$ give the
same $\chi_R$. When $G$ is broken to $G'$, $R$ decomposes into
different irreducible representations $R'$ in $G'$ and the above is no
longer true. Instead, we need that all $\beta'$ in $R'$ give
the same $\chi_{R'}$.

This can be seen as follows. Since weights differ by roots,
given a weight $\beta$ in $R$ of $G$, it is useful to
expand
$\beta=-b_i\alpha_i$. Hence we can decompose its matter surface
$S(\beta)$ as \cite{WeigandTASI}
\begin{equation} \label{Sdecompose}
    S(\beta)=S_0(R)+b_i\left.D_i\right|_{C_R}\,,
\end{equation}
where $S_0$ only depends on $R$ but not $\beta$. We will prove
this decomposition below. When $G$ is not broken, $\chi_R$ is
calculated using $S_0$. The Poincare dual $[S_0(R)]$ is the corresponding
%matter surface
flux that gives chiral matter without breaking
$G$ when $G$ supports chiral matter.  As we will see more explicitly below, $S_0 (R)$ and its Poincar\'{e} dual correspond to the last row/column of (\ref{eq:review:Mred-matter}). We will now focus on the
second term of (\ref{Sdecompose}) and determine $S_0$ later. Matter curves in general
can be written as
\begin{equation}
    C_R=\Sigma\cdot(p_R K_B+q_R \Sigma)\,,
\end{equation}
where $p_R,q_R$ are some (integer) coefficients. Then,
\begin{equation} \label{partialchiR'}
    \int_{S(\beta)}G_4^\mathrm{vert}=\int_{S_0(R)}G_4^\mathrm{vert}+b_i\int_{\hat Y} G_4^\mathrm{vert}\wedge[D_i]\wedge\pi^*[p_R K_B+q_R \Sigma]\,.
\end{equation}
The second term is a linear combination of $\Theta_{i\alpha}$
and we can replace the $i$ summation with $i' \in I'$ since the other terms vanish. Since the
weights of $R'$ differ by combinations of $\alpha_{i\notin I'}$
only, each set of $b_{i'}$ gives a representation $R'$, and Eq.\ (\ref{partialchiR'}) is the same for all weights of $R'$. In
general, different $b_{i'}$ and different $R$ can give  rise to the same irreducible representation
$R'$. We must sum over these contributions to get the complete $\chi_{R'}$.
Applying Eq.\ (\ref{ordinarychi}), we get
\begin{equation} \label{fullchiR'}
    \chi_{R'}=\sum_R \sum_{b_{i'}} \left(\int_{S_0(R)}G_4^\mathrm{vert}+b_{i'}\left(p_R \Theta_{i'K_B}+q_R \Theta_{i'\Sigma}\right)\right)\,.
\end{equation}
This is our main tool to calculate chiral indices in models
with flux breaking. An important feature that can be seen here
is that $\chi_{R'}$ for complex $R'$ can be nontrivial even if
$R$ is non-complex. In other words, there can be chiral matter
after flux breaking even if $G$ does not support chiral matter.
This formula passes several consistency checks, such as $\chi_{\bar R'}=-\chi_{R'}$, since taking the conjugate representation flips
all contributions in Eq.\ (\ref{fullchiR'}) to opposite signs.
Moreover, in all examples we will see, anomaly cancellation is
preserved after the breaking as long as the flux constraints
are satisfied.

So far, we have been focusing on matter localized on curves. On
the other hand, adjoint matter lives on the bulk of $\Sigma$
and matter curves or surfaces for this representation are not well-defined.
Nevertheless, it has been shown that adjoint matter can also
become chiral after flux breaking, and the chiral indices are
given by setting $S_0(\mathrm{Adj})=0$ and replacing $C_R$ by
$K_\Sigma$ \cite{Bies:2017fam}. By the adjunction formula,
$K_\Sigma=\Sigma\cdot(K_B+\Sigma)$ and we should set
$p_\mathrm{Adj}=q_\mathrm{Adj}=1$.

It may sound strange that $K_{\Sigma}$ directly appears in
$\chi$, while in 6D F-theory models it is well-known that the
number of adjoint hypermultiplets is the genus
$g=\left(K_{\Sigma}+2\right)/2$ \cite{Witten:1996qb}. Should there also be such
a shift in the 4D formula? In fact, we should compare the formula
with the Dirac index of adjoint matter in 6D instead. In 6D
$\mathcal{N}=1$ SUSY, each vector multiplet contains two
$(0,1/2)$ spinors, while each hypermultiplet
(two half-hypermultiplets) contains two $(1/2,0)$ spinors.
Since there is one vector multiplet and there are $g$ hypermultiplets,
the Dirac index in 6D is indeed $2g-2=K_{\Sigma}$.

Now we return to the determination of $S_0$. Since the
nontrivial $\Theta$ are $\Theta_{ij}$ and $\Theta_{i\alpha}$,
we only need the $S_{ij}$ and $S_{i\alpha}$ components in
$S_0$. The $S_{ij}$ components, if they exist, give chiral matter
even when $G$ is not broken.  A useful indirect procedure to determine such
components has been established,  through the
matching of Chern-Simons (CS) terms in M/F-theory duality \cite{Grimm:2011sk,Grimm:2011fx,Cvetic:2012xn}. To
be precise, in the 3D M-theory dual, $\Theta_{ij}$ are the
classical CS couplings appearing in the 3D effective action. These
match with the one-loop corrected CS couplings in the 4D
F-theory when compactified (additionally) on a circle. The
charged fermions running through the loop relate the couplings
to chiral indices. As a result, we can establish relations in
the form of $\chi_R=x_R^{ij}\Theta_{ij}$, where $x_R^{ij}$ are
some coefficients.
We refer to \cite{Jefferson:2021bid} for more details. Note that
this determines the $S_{ij}$ components in $S_0$, which are
sufficient when $G$ is not broken. To include the $S_{i\alpha}$
components for the broken case, we make the following ansatz:
\begin{equation}
    S_0(R)=x_R^{ij}D_i\cdot D_j+D_i\cdot D_R^i\,,
\end{equation}
where $D_R^i$ is some linear combination of $D_\alpha$. Now we
determine $D^i_R$. First, we choose a base divisor $D'$ such
that it intersects $C_R$ only once i.e. $C_R\cdot D'=1$. Then
by definition, the fibral curve $C_\beta$ corresponding to
weight $\beta$ is
\begin{equation}
    C_\beta=S(\beta)\cdot D'=S_0(R)\cdot D'+b_i\mathbb P^1_i\,,
\end{equation}
where $\mathbb P^1_i$ is the fibral curve in $D_i$. Now we must have
\begin{equation}
    D_i\cdot C_\beta=\beta_i\,,
\end{equation}
where $\beta_i=-C_{ij} b_j$ are the components of $\beta$ in a basis of
fundamental weights.
By $D_i\cdot \mathbb P^1_j=-C_{ij}$, we see that the
second term in Eq.\ (\ref{Sdecompose}) gives all the weights,
and the condition reduces to simply
\begin{equation}
    S_0(R)\cdot D_i\cdot D'=0\,.
\end{equation}
All intersection numbers in the above involve triple
intersections of $\Sigma,D'$, and some other classes on the
base. Since we have the freedom to choose $D'$ as long as it is
properly normalized, the $\mathrm{rank}(G)$ constraints
determine $\Sigma\cdot D^i_R$ in terms of other known
classes, namely $\Sigma^2,\Sigma\cdot K_B$. This is equivalent to
determining $D_i\cdot D^i_R$ since only $\Sigma\cdot D^i_R$
appears in its intersection numbers. Therefore, $S_0(R)$ has
been fixed. Notice that these constraints also mean that $S_0(R)$
must live in the directions of $M_{\mathrm{phys}}$, confirming that this surface and the associated Poincar\'{e} dual flux correspond to the final row/column of (\ref{eq:review:Mred-matter}) as asserted above; in fact this conclusion can also be arrived at directly from the observation that the Poincar\'{e} dual $[S_0 (R)]$ is the only flux direction that preserves Poincar\'{e} and gauge symmetry. The
block-diagonal form of $M_{\mathrm{red}}$ then implies that we
can always separate the chiral indices into contributions
preserving $G$, and those induced by flux breaking. These
correspond to the two terms in Eq.\ (\ref{Sdecompose}), hence
give a resolution-independent description of matter surfaces.
This also recovers the statement that chiral matter in $G$ is
induced by flux along the Poincar\'e dual $[S_0(R)]$ \cite{Marsano_2011}.
We will explicitly demonstrate these  relations in \S\ref{subsec:fluxbreaking:e6}.

It is useful to have a simple result from the above procedure.
For non-complex $R$, it is clear that the procedure gives
trivial $S_0(R)$. In particular, for $G$ not supporting chiral
matter, $S_0(R)$ is always absent.

\subsection{Primitivity}
\label{subsec:fluxbreaking:primitive}

The gauge-breaking flux must also satisfy various flux
constraints discussed in \S\ref{subsec:review:G4fluxes}. Interestingly, the \emph{vertical} flux
we turn on does not automatically
satisfy primitivity ($J\wedge G_4=0$). Extra attention must be paid and we will see
that primitivity leads to additional flux constraints.

It is useful to first review the basics of the K\"ahler form $J$.
The volume of a complex $d$-dimensional submanifold $\mathcal M^d$ in $\hat Y$ is given by
\begin{equation}
    \mathrm{vol}(\mathcal M^d)=\int_{\mathcal M^d} J^d\,.
\end{equation}
The K\"ahler cone is then the cone of $J$ giving positive volumes. We can expand $J$ of $\hat Y$ in the K\"ahler cone as
\begin{equation}
    J=t^0[D_0]+t^\alpha[D_\alpha]+t^i[D_i]\,,
\end{equation}
where the K\"ahler moduli $t$ are
restricted to the positive K\"ahler cone.
%positive and we have chosen the
%K\"ahler basis on the base for $D_\alpha$.
So far we have focused
on the resolved manifold $\hat Y$, which is on the M-theory
Coulomb branch where $G$ is broken into
$\U(1)^{\mathrm{rank}(G)}$. To take the F-theory limit and
restore the whole $G$, we need to shrink the fibers to zero
volume, while keeping the (pullbacks of the) base divisors at finite
volumes. Note that $t^0$ and $t^i$ measure the elliptic and
exceptional fiber volumes respectively. Therefore, we need to
send $t^0$ and $t^i$ to zero and scale up $t^\alpha$. To be
precise, the limit can be done by the following rescaling \cite{Grimm:2010ks,Bonetti:2011mw}:
\begin{equation}
    t^0\rightarrow \epsilon t^0\,,\quad t^\alpha\rightarrow \epsilon^{-1/2} t^\alpha\,,\quad t^i\rightarrow \epsilon^{3/2} t^i\,,
\end{equation}
where the limit is now $\epsilon\rightarrow 0$. Therefore, we
only need to consider $J\rightarrow\pi^* J_B=t^\alpha[D_\alpha]$ when studying primitivity.

First we recall how primitivity is satisfied when $G$ is not
broken. By Eq.\ (\ref{eq:review:Poincare}) and (\ref{eq:review:gaugepres}), we
have $\Theta_{I\alpha}=0$ for all $\alpha$. This already
guarantees $J\wedge G_4=0$ and primitivity is automatically
satisfied. It is also clear that nonzero $\Theta_{i\alpha}$
breaks the above argument and primitivity is not always
satisfied. In particular, generically we have
\begin{equation}
    \int_{\hat Y} [D_i]\wedge J\wedge G_4=t^\alpha \Theta_{i\alpha}\neq 0\,.
\end{equation}
The above vanishes only for specific values of $t^\alpha$. The
interpretation is that by turning on gauge-breaking flux, some
K\"ahler moduli are stabilized (but not all, as an overall
rescaling of $t^\alpha$ also satisfies the constraint).

On the other hand, not all choices of nonzero
$\Theta_{i\alpha}$ can stabilize the K\"ahler moduli within the
K\"ahler cone. As a first step, one necessary condition for consistent
stabilization is that the flux should give a positive tadpole
$\int_{\hat Y}G_4\wedge G_4>0$, which is already not always
true for gauge-breaking vertical flux. From the form of
$M_{\mathrm{red}}$, the sign of the tadpole is determined by the
triple intersection form $\Sigma\cdot D_\alpha\cdot D_\beta$
on the base. If this is positive semidefinite, the flux always
gives a nonpositive tadpole, hence is not ever consistent. Although
the intersection forms for most geometries of $\Sigma$ have both
positive and negative directions, $\Sigma$ cannot be as simple
as $\mathbb P^2$. When the
tadpole can be positive, during vertical flux breaking we must turn on
some gauge-breaking flux along negative
directions in the intersection form of $\Sigma$.

We must stress again that having  a positive tadpole is not a
sufficient condition for primitivity. Here we show  how
primitivity leads to additional flux and geometric constraints. We consider
\begin{equation}
\label{eq:fluxbreaking:intPrimitivity}
    t^\alpha \Theta_{i\alpha}=0\Rightarrow t^\alpha \Sigma\cdot D_\alpha \cdot D_\beta \phi_{i\beta}=0\,,
\end{equation}
for all $i$.  
%\drop{This is satisfied by the condition that
%$\phi_{i\alpha}/\phi_{i\alpha'}$ is the same for all
%$i$, but} 
Notice that the solutions of
$t^\alpha$ live in the left null space of the matrix
$\Theta_{(\alpha_a)(i)}$. Therefore to have
nontrivial solutions of $t^\alpha$, the rank of the
matrix must be less than $r$, i.e., at most $r-1$,
where we recall that $r$ is the number of  $\alpha$'s giving
homologically independent cycles $S_{i \alpha}$.
This leads to Eq.\ (\ref{rlowerbound}).  The
positivity of $t^\alpha$ is more subtle.
%For an arbitrary $i$,
%we now decompose the above into the Mori basis $l_\alpha$ on
%the base (which is dual to the K\"ahler basis $D_\alpha$).
In the simplest cases where the Mori cone (dual to the K\"ahler cone) is generated by $h^{1,1} (B)$ basis curves $l_\alpha$, we can decompose the above into this basis.
That is, for any $i$ we have $\phi_{i\beta}\Sigma\cdot D_\beta=m_{i\alpha} l_\alpha$
for some coefficients $m_{i\alpha}$. Then the primitivity constraint
is simply
\begin{equation}
    t^\alpha m_{i\alpha}=0\,.
\end{equation}
Therefore to have positive $t^\alpha$, we must have at least a
pair of $m_{i\alpha}$ with opposite signs. This imposes some sign
constraints on $\phi_{i\alpha}$ as demonstrated below in some specific
cases.

\subsection{Simple $\SU(N)$ models}
\label{subsec:fluxbreaking:SUNex}

So far we have presented the general formalism of flux
breaking. To see how it works, it is useful to  illustrate with some simple
examples involving vertical flux breaking of $G=\SU(N)$ to
$G'=\SU(N-1)$ with no extra $\U(1)$.
We focus on checking the formalism using anomaly cancellation, which is automatically
achieved in all examples below as a consequence of (\ref{fullchiR'}). Interestingly, this is a
result from nontrivial cancellation between the matter
representations in $G$. To focus on the effect of flux
breaking, we do not include any chiral matter in the unbroken
models. In other words, we focus on the second term in Eq.\ (\ref{fullchiR'}), which should satisfy anomaly cancellation
on its own as discussed.
In each case, we turn on $\Theta_{i' \alpha}$ for $i' = N -1$, to
break the Dynkin diagram $A_{N -1}$ to $A_{N -2}$.

\begin{itemize}
    \item $\SU(3)\rightarrow \SU(2)$
\end{itemize}

Since $\SU(2)$ does not support any chiral matter, all chiral
indices should vanish. Indeed, all $\SU(2)$ representations
come from pairs of opposite $b_{i'}$. For example, the $\SU(3)$
adjoint $\mathbf 8$ gives two copies of the $\SU(2)$
fundamental representation
$\mathbf 2$ with $b_2=\pm 1$. The $\SU(3)$ fundamental $\mathbf 3$ gives a $\mathbf 2$ with $b_2=1/3$ which is nonzero, but we
also have the $\SU(3)$ antifundamental $\bar{\mathbf 3}$ giving
another $\mathbf 2$ with $b_2=-1/3$. Eq.\ (\ref{fullchiR'})
then implies that $\chi_{\mathbf 2}=0$. In general, such
a cancellation holds for any non-complex $R'$. Note that a
single $R$ may have nonzero contribution to $\chi_{R'}$,
although it must get cancelled. This shows that only the total
$\chi_{R'}$ is a physical quantity.

\begin{itemize}
    \item $\SU(4)\rightarrow \SU(3)$
\end{itemize}

$\SU(3)$ has complex representations such as $\mathbf 3$, but a
generic $\SU(3)$ F-theory model only contains $\mathbf 8,\mathbf 3,\bar{\mathbf 3}$, and $\chi_{\mathbf 3}=0$ is
required by anomaly cancellation. Interestingly, this is more
nontrivial from the $\SU(4)$ perspective. Consider a generic
$\SU(4)$ model, which contains the representations
$\mathbf{15},\mathbf{6},\mathbf{4}$ and the conjugates. For
the latter two, the matter curves are
$C_{\mathbf{6}}=-\Sigma\cdot K_{B}$ and
$C_{\mathbf{4}}=-\Sigma\cdot\left(8K_{B}+4\Sigma\right)$. The
three representations all break to $\mathbf{3}$ with
$b_3=1,-1/2,1/4$ respectively. Be careful here to recall that
$C_{\mathbf{6}}$ actually contains two copies of $\mathbf{6}$.
Now we have
\begin{equation}
    \sum_R b_3^R C_R=\Sigma\cdot\left(1\cdot\left(K_{B}+\Sigma\right)+2\cdot\left(-\frac{1}{2}\right)\cdot\left(-K_{B}\right)+\frac{1}{4}\cdot\left(-8K_{B}-4\Sigma\right)\right)=0\,.
\end{equation}
Eq.\ (\ref{fullchiR'}) then implies that $\chi_{\mathbf 3}=0$.
We see that anomaly cancellation has become a cancellation
%between
that involves both weights and classes of matter curves.

\begin{itemize}
    \item $\SU(5)\rightarrow \SU(4)$
\end{itemize}
The calculation is similar. A generic $\SU(5)$ model contains
$\mathbf{24},\mathbf{10},\mathbf{5}$ and the conjugates. The
matter curves are $C_{\mathbf{10}}=-\Sigma\cdot K_{B}$ and
$C_{\mathbf{5}}=-\Sigma\cdot\left(8K_{B}+5\Sigma\right)$. The
three representations all break to $\mathbf{4}$ with
$b_4=1,-3/5,1/5$ respectively. Then,
\begin{equation}
    \sum_{R}b_4^RC_{R}=\Sigma\cdot\left(1\cdot\left(K_{B}+\Sigma\right)+\left(-\frac{3}{5}\right)\cdot\left(-K_{B}\right)+\frac{1}{5}\cdot\left(-8K_{B}-5\Sigma\right)\right)=0\,.
\end{equation}
Therefore, $\chi_{\mathbf{4}}=0$ as required by anomaly cancellation.

\begin{itemize}
    \item $\SU(6)\rightarrow \SU(5)$
\end{itemize}
This is a more interesting example since $G'$ now supports
chiral matter. We show that flux breaking can induce chiral
matter satisfying anomaly cancellation, even if there is no
chiral matter in the unbroken phase. A generic $\SU(6)$
model contains $\mathbf{35},\mathbf{15},\mathbf{6}$. The
matter curves are $C_{\mathbf{15}}=-\Sigma\cdot K_{B}$ and
$C_{\mathbf{6}}=-\Sigma\cdot\left(8K_{B}+6\Sigma\right)$. All
three representations break to $\mathbf{5}$ with
$b_5=1,-2/3,1/6$ respectively, while $\mathbf{15}$ also breaks
to $\mathbf{10}$ with $b_5=1/3$. Using Eq.\ (\ref{fullchiR'}),
we get
\begin{equation}
    \chi_{\mathbf{5}}=-\chi_{\mathbf{10}}=\frac{1}{3}\Theta_{iK_{B}}\,.
\end{equation}
Therefore, the chiral indices become nontrivial, and the
anomaly cancellation condition $\chi_{\mathbf{5}}=-\chi_{\mathbf{10}}$
is satisfied. Note that despite the presence of  the factor of
$1/3$, the flux constraints must guarantee integer chiral
indices.

\section{Standard Model structure from $E_7$ flux breaking}
\label{sec:fluxbreaking:SM}

We are now ready to discuss the breaking $E_7\rightarrow\gsm$,
which leads to the SM gauge group and exact chiral spectrum from a
gauge group ubiquitous in the landscape. This is the main
result of \cite{Li:2021eyn}, but here we provide more details. In
particular, we discuss more about different embeddings of
$\gsm$ into $E_7$ and solve the flux constraints more
generally. As shown in Section
\ref{sec:fluxbreaking:othergroups}, all these results can be
easily generalized to $E_6$ flux breaking.

\subsection{Embeddings of the gauge group $\gsm$}
\label{subsec:fluxbreaking:embedding}

As a first step, here is a general picture of $E_7$ flux
breaking. A generic $E_7$ model contains the adjoint
$\mathbf{133}$ and the fundamental $\mathbf{56}$, with matter
curve $C_{\mathbf{56}}=-\Sigma\cdot(4K_B+3\Sigma)$. In
particular, there are models with only $\mathbf{133}$ but no
$\mathbf{56}$, so the two representations should satisfy
anomaly cancellation separately under flux breaking, unlike
the $\SU(N)$ models. The number of independent sets of chiral
matter induced by vertical flux breaking depends on the embedding of
$\gsm$ into $E_7$. Usually, the number allowed by anomaly
cancellation is (much) less than the number of independent flux
parameters we
turn on. Therefore, many flux parameters contribute to a
single chiral index, naturally leading to a small number of
generations as demonstrated below. More surprisingly, the
chiral matter induced may not realize all the independent sets
allowed by anomaly cancellation, unlike the situation for generic
chiral matter representations in universal tuned $\gsm$ models without
flux breaking \cite{Jefferson:2021bid}.

Below we see two examples of $\gsm$ embeddings into $E_7$. The first one
gives SM chiral matter as the only allowed matter spectrum.
The second one gives exotic chiral matter (defined below)
which is only part of the spectrum allowed by anomaly
cancellation.

\subsubsection{Standard Model chiral matter}
\label{ssubsec:fluxbreaking:SMchiral}

\begin{figure}[t]
\centering
\includegraphics[width=0.5\columnwidth]{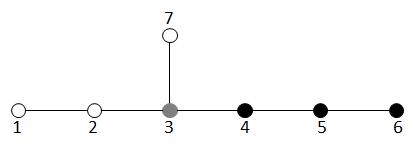}
\caption{The Dynkin diagram of $E_7$. The Dynkin node labelled
$i$ corresponds to the exceptional divisor $D_i$. The solid
nodes are the ones we break to get the Standard Model gauge
group and chiral matter. Node 3 (in gray) is broken by
remainder flux while the others are broken by vertical flux.}
\label{dynkine7}
\end{figure}

Here we consider embeddings of $\gsm$ that lead to SM chiral
matter. First, the embedding of the non-abelian part i.e.
$\SU(3)\times\SU(2)$ is unique up to $E_7$ automorphisms,
when we restrict to root embeddings,
see
Appendix \ref{sec:embeddingcount}.  Without loss of
generality, we put the non-abelian part in nodes $1,2,7$ in
the Dynkin diagram, see Figure \ref{dynkine7}. Then, there
are 4 choices of $\U(1)$ (with generator $T_Y=Y_i T_i$) in $\gsm$ that
play the role of hypercharge and give only SM chiral matter (see Appendix \ref{sec:embeddingcount}):
\begin{align} \label{hyperchargeChoice}
    Y_i=&-(1/3,2/3,1,0,0,0,1/2)\,,\quad-(1/3,2/3,1,1,0,0,1/2)\nonumber \\
    &-(1/3,2/3,1,1,1,0,1/2)\,,\quad-(1/3,2/3,1,1,1,1,1/2)\,.
\end{align}
In fact, however, these are also equivalent under
automorphisms, and
for all choices $Y_{i=3,4,5,6}$ coincides with a
root of $E_7$ that expands the group to $SU(5)$, so the hypercharge must be obtained through
remainder flux. Moreover, vertical flux is also necessary for chiral
matter. For simplicity, we will focus on the
first choice of $Y_i$, while other choices give similar
results. Therefore, the proposal is that we first break $E_7$
down to an intermediate $\SU(5)$ with vertical flux, then
obtain $\gsm$ using hypercharge flux, in parallel with
earlier work on tuned SU(5) GUT
models \cite{Donagi:2008ca,BeasleyHeckmanVafaI,BeasleyHeckmanVafaII,DonagiWijnholtGUTs,Blumenhagen:2009yv,Marsano:2009wr,Grimm:2009yu,KRAUSE20121,Braun:2013nqa}. As mentioned in
\S\ref{subsec:fluxbreaking:breaking}, the construction can be done on
typical but non-toric bases supporting rigid $E_7$ factors.

Following this approach, we first break $E_7$ down to $\SU(5)$
by turning on nonzero $\Theta_{i'\alpha}$ for $i'=4,5,6$ and
some $\alpha$, see Figure \ref{dynkine7}. Then we further
break $\SU(5)$ down to $\gsm$ by turning on the hypercharge flux:
\begin{equation} \label{hyperchargeflux}
    G_4^\mathrm{rem}=\left[D_Y|_{C_\mathrm{rem}}\right]\,,
\end{equation}
for some $C_\mathrm{rem}$, where $D_Y=2D_1+4D_2+6D_3+3D_7$ is
the exceptional divisor corresponding to the hypercharge
generator from the first choice of $Y_i$. This remainder flux
further breaks node 3 in Figure \ref{dynkine7} and gives $\gsm$.

\begin{comment}
For
simplicity, we only focus on the case where the non-abelian
part of $\gsm$ is produced by breaking the Dynkin nodes
$i'=3,4,5,6$, see Figure \ref{dynkine7}. Hence we turn on nonzero $\Theta_{i'\alpha}$ for $i'=3,4,5,6$.
\end{comment}

Since only the vertical flux induces chiral matter, we can analyze
the matter content by breaking $E_7 \rightarrow SU(5)$, where
the $\mathbf{56}$ breaks into a combination of $\mathbf{5},
\mathbf{10}$, uncharged singlets and conjugate representations, and $\mathbf{133}$
includes these as well as the adjoint $\mathbf{24}$.  Since the
adjoint is non-chiral, the only chiral representations we expect for
$\gsm$ after the final breaking by remainder flux
are the Standard Model representations 
\begin{equation}
    \left(\mathbf{3},\mathbf{2}\right)_{1/6}\,,\quad
    \left(\mathbf{3},\mathbf{1}\right)_{2/3}\,,\quad
    \left(\mathbf{3},\mathbf{1}\right)_{-1/3}\,,\quad
    \left(\mathbf{1},\mathbf{2}\right)_{1/2}\,,\quad
    \left(\mathbf{1},\mathbf{1}\right)_{1}\,.
\end{equation}

As mentioned above, we expect anomaly cancellation separately from the
matter arising from the $\mathbf{56}, \mathbf{133}$ of $E_7$.
Using Eq.\ (\ref{fullchiR'}), we
indeed get SM chiral matter from vertical flux and the $\mathbf{56}$ with
\begin{equation} \label{chi56}
    \chi_{\left(\mathbf{3},\mathbf{2}\right)_{1/6}}^{\mathbf{56}}=\frac{1}{2}\left(3\Theta^{\mathbf{56}}_4+2\Theta^{\mathbf{56}}_5+\Theta^{\mathbf{56}}_6\right)\,,
\end{equation}
where $\Theta^{\mathbf{56}}_i=-4\Theta_{iK_{B}}-3\Theta_{i\Sigma}$.
Similarly,
$\mathbf{133}$ also gives SM chiral matter with
\begin{equation} \label{chi133}
    \chi_{\left(\mathbf{3},\mathbf{2}\right)_{1/6}}^{\mathbf{133}}=-\left(3\Theta^{\mathbf{133}}_4+2\Theta^{\mathbf{133}}_5+\Theta^{\mathbf{133}}_6\right)\,,
\end{equation}
where $\Theta^{\mathbf{133}}_i=\Theta_{iK_{B}}+\Theta_{i\Sigma}$. We see that only certain linear combinations of
$\Theta_{i'\alpha}$ appear in the chiral indices.

\subsubsection{Exotic matter}
\label{ssubsec:fluxbreaking:exotic}

To get 
SM chiral matter but not other representations $R'$ of $\gsm$, in the above
procedure, it is important to choose the right embedding. For
directly tuned $\gsm$, it has been argued in \cite{TaylorTurnerGeneric} that the
model generically contains the SM matter fields and the representations
$\left(\mathbf{3},\mathbf{1}\right)_{-4/3},\left(\mathbf{1},\mathbf{2}\right)_{3/2},\left(\mathbf{1},\mathbf{1}\right)_{2}$,
while constructing representations $R'$ other than these (defined as
 exotic matter representations) requires extensive amounts of fine-tuning. Here
we will see that this is no longer the situation in the case of vertical
flux breaking. As described in \S\ref{subsec:fluxbreaking:breaking}, we
can get exotic $\U(1)$'s from simple vertical flux
constraints, leading
to many possible exotic representations $R'$. Below we give such an example. Note that
when $R'$ other than SM matter representations
are involved, the flux breaking may not realize all
independent sets of chiral matter allowed by anomaly cancellation.

The directly tuned $\gsm$ models containing generic matter, which
includes the Standard Model representations, can be naturally unHiggsed into
$\SU(4)\times \SU(3)\times \SU(2)$ models \cite{Raghuram:2019efb}. It is interesting
that the converse cannot be achieved from the perspective of
$E_7$ flux breaking. As an example, we consider a flux breaking
pattern from vertical flux that can be associated with the breaking
route $E_{7}\rightarrow \SU\left(4\right)\times
\SU\left(3\right)\times \SU\left(2\right)\rightarrow
\SU\left(3\right)^{2}\times
U\left(1\right)/\mathbb{Z}_{3}\rightarrow\gsm$.
Note that in this example we do not use remainder flux; all the
breaking comes from vertical fluxes.
This time we put the non-abelian part in nodes $1,4,5$, so
we turn on nonzero $\Theta_{i'\alpha}$ for
$i'=2,3,6,7$. Then we find that the $\U(1)$ charge is
%\begin{equation}
%    q=\frac{3}{2}b_{2}+\frac{1}{3}b_{3}-\frac{4}{3}b_{6}-2b_{7}\,,
%\end{equation}
given by the generator
\begin{equation}
    T=\frac{1}{2}T_{1}+T_{2}-\frac{1}{3}T_{4}-\frac{2}{3}T_{5}-T_{6}-T_{7}\,.
\end{equation}
Therefore, we further impose $\Theta_{2\alpha}=\Theta_{6\alpha}+\Theta_{7\alpha}$ for all
$\alpha$. This condition does not coincide with any root of
$E_7$, so it really induces an exotic $\U(1)$.

As above, we analyze the breaking of $\mathbf{56}$ and $\mathbf{133}$
separately. The first observation is that $\mathbf{56}$ does not break
into the generic matter representations that appear in directly tuned $\gsm$ models. Instead, it
breaks into the representations
\begin{equation} \label{E7to432fund}
    \left(\mathbf{3},\mathbf{2}\right)_{1/6},\left(\mathbf{3},\mathbf{1}\right)_{5/3},\left(\mathbf{3},\mathbf{1}\right)_{-1/3},\left(\mathbf{1},\mathbf{2}\right)_{1/2},\left(\mathbf{1},\mathbf{1}\right)_{1},\left(\mathbf{3},\mathbf{1}\right)_{-4/3},\left(\mathbf{1},\mathbf{2}\right)_{3/2},\left(\mathbf{1},\mathbf{1}\right)_{2}\,.
\end{equation}
That is, the right-handed up quark is replaced by the exotic
$\left(\mathbf{3},\mathbf{1}\right)_{5/3}$, and there are various
exotic representations that appear. There are three independent sets of chiral matter from anomaly cancellation. The chiral indices from flux breaking, however, only realize two of them:
\begin{equation}
    \chi^{\mathbf{56}}=\Theta^{\mathbf{56}}_{3}\left(1,0,-3,2,-3,1,-1,2\right)+\frac{1}{2}\left(\Theta^{\mathbf{56}}_{6}+\Theta^{\mathbf{56}}_{7}\right)\left(3,-1,-6,3,-7,1,-2,5\right)\,.
\end{equation}
Here the components of the vectors of fields correspond to the matter representations in Eq.\ (\ref{E7to432fund}) in the same order. The first set of
of anomaly-canceling
chiral matter fields only contains the generic matter representations
that appear in directly tuned
$\gsm$ models, while the second set involves the exotic
$\left(\mathbf{3},\mathbf{1}\right)_{5/3}$.

The analysis for $\mathbf{133}$ is  similar. Under the prescribed
breaking route it breaks into
\begin{gather}
    \left(\mathbf{3},\mathbf{2}\right)_{1/6},\left(\mathbf{3},\mathbf{2}\right)_{7/6},\left(\mathbf{3},\mathbf{2}\right)_{-11/6},\left(\mathbf{3},\mathbf{1}\right)_{2/3},\left(\mathbf{3},\mathbf{1}\right)_{-1/3},\left(\mathbf{3},\mathbf{1}\right)_{-4/3},\nonumber\\
    \left(\mathbf{3},\mathbf{1}\right)_{5/3},\left(\mathbf{1},\mathbf{2}\right)_{1/2}, \left(\mathbf{1},\mathbf{2}\right)_{3/2},\left(\mathbf{1},\mathbf{1}\right)_{1},\left(\mathbf{1},\mathbf{1}\right)_{2},\left(\mathbf{1},\mathbf{1}\right)_{3}\,.
\end{gather}
This gives many more exotic matter representations than $\mathbf{56}$,
which can have nontrivial chiral indices for generic fluxes.
Again, only two of
the allowed independent sets of chiral matter are realized.
The chiral indices are
\begin{align}
    \chi^{\mathbf{133}}=&\Theta^{\mathbf{133}}_{3}\left(-1,0,0,0,0,1,1,1,0,1,-1,0\right)\nonumber\\
    &+\left(\Theta^{\mathbf{133}}_{3}+\Theta^{\mathbf{133}}_{6}+\Theta^{\mathbf{133}}_{7}\right)\left(0,2,1,-3,3,-4,2,-6,1,4,-2,1\right)\,,
\end{align}
following the above order. The first set contains only those $R'$ from
both $\mathbf{56}$ and $\mathbf{133}$.
Note that while the $\mathbf{56}$ alone does not generate all states
in the Standard Model spectrum, the missing states are supplied by the
$\mathbf{133}$.
On the other hand, there is no choice of fluxes that gives only SM
matter and no exotics: the second set of representations from
$\mathbf{133}$ is the only place that some exotic matter fields like
$\left(\mathbf{1},\mathbf{1}\right)_{3}$ appear, so the fluxes
generating this family must cancel in the absence of exotic matter.
But this is the only combination that includes the field
$\left(\mathbf{3},\mathbf{1}\right)_{2/3}$, so we cannot get the full
Standard Model chiral matter spectrum from this construction without
at least some exotics.

In conclusion,
while the flux constraints are equally simple in many constructions,
choosing  the incorrect embedding of $\gsm$ into $E_7$ can lead to a
variety of exotic chiral matter fields. These matter representations
are generically present since there are many more than 4
choices of $\U(1)$ in $\gsm$ that can be embedded into $E_7$.
For many of these choices, unlike the case just analyzed here, there
may actually be no resulting fields in some of the Standard Model
representations.  In others, like this one, all of the SM
representations may appear along with some exotics; while in this
specific case we can show that no flux combination is possible that
gives just SM matter without exotics, it is possible that for other
U(1) choices, a judicious tuning of fluxes may cancel the chiral
multiplicities of all exotic matter fields, still allowing for an SM
construction with the expected matter fields and no exotics, but we
leave a full consideration of this question for further research.
%In principle, one can count
%all such embeddings and study the matter spectrum given by
%each embedding. Although the analysis can be tedious, the
%genericity between different embeddings remains an
%interesting open question.

\subsection{Solving the flux constraints}
\label{subsec:fluxbreaking:solve}

Now we turn back to the construction of  the
Standard Model gauge group and
chiral matter from \S\ref{ssubsec:fluxbreaking:SMchiral}. Although we have obtained the chiral index in
terms of $\Theta_{i'\alpha}$, we still need to solve the flux
constraints and express everything in terms of flux
parameters $\phi$.

There is a subtlety before breaking $E_7$. Most $E_7$
models have codimension-3 singularities with degree
$(4,6,12)$. Such singularities can no longer be simply
interpreted as Yukawa couplings. The fiber becomes non-flat
at these points and supports an extra vertical flux.
It also seems to correspond to extra strongly coupled
(chiral) degrees of freedom, possibly M5-branes wrapping
non-flat fibers \cite{Candelas:2000nc,Lawrie:2012gg,Jefferson:2021bid}. Although $E_7$ itself does not support any
chiral matter, after flux breaking the extra flux may induce
more chiral matter which is not covered by our formalism.
This will be studied in \cite{46}. For realistic SM-like models, we
simply set such extra flux to vanish, so all the flux we
consider is for flux breaking.

It is now straightforward to solve the flux constraints by
considering independent $S_{i\alpha}$. Recall from \eqref{eq:review:theta-matrix} that
$\Theta_{i\alpha}=-\Sigma\cdot D_\alpha\cdot D_\beta \kappa^{ij} \phi_{j\beta}$.
For independent $S_{i\alpha}$,
the triple intersection form $M^B_{\alpha\beta}=\Sigma\cdot D_\alpha\cdot D_\beta$ on $B$ is invertible. The solution to
$\Theta_{1\alpha}=\Theta_{2\alpha}=\Theta_{3\alpha}=\Theta_{7\alpha}=0$ is simply
%\begin{gather}
%    \phi_{1\alpha}=2n_\alpha\,,\quad\phi_{2\alpha}=4n_\alpha\,,\quad\phi_{3\alpha}=6n_\alpha\,,\nonumber\\
%    \phi_{4\alpha}=5n_\alpha\,,\quad\phi_{5\alpha}=\phi_{5\alpha}\,,\quad\phi_{6\alpha}=\phi_{6\alpha}\,,\quad\phi_{7\alpha}=3n_\alpha\,,
%\end{gather}
\begin{gather}
    \phi_{1\alpha}=2n_\alpha\,,\quad\phi_{2\alpha}=4n_\alpha\,,\quad\phi_{3\alpha}=6n_\alpha\,,\quad
    \phi_{4\alpha}=5n_\alpha\,,\quad\phi_{7\alpha}=3n_\alpha\,,
    \label{eq:phi-n}
\end{gather}
with $\phi_{5 \alpha}, \phi_{6 \alpha}$ arbitrary,
but we pick sufficiently generic
$\phi_{5 \alpha},\phi_{6 \alpha}$ such that the
resulting gauge group
does not get further enhanced.
These fluxes give
\begin{equation}
    \Theta_{4\alpha}=M^B_{\alpha\beta}(\phi_{5\beta}-4n_\beta)\,,\quad\Theta_{5\alpha}=M^B_{\alpha\beta}(5n_\beta-2\phi_{5\beta}+\phi_{6\beta})\,,\quad\Theta_{6\alpha}=M^B_{\alpha\beta}(\phi_{5\beta}-2\phi_{6\beta})\,.
\end{equation}
%\drop{The
%condition that the ratio
%$\phi_{i\alpha}/\phi_{i\alpha'}$ is independent of $i$ is
%satisfied as long as the  components $n_\alpha,\phi_{5\alpha},\phi_{6\alpha}$
%satisfy it separately.} 
The flux quantization condition is satisfied by integer $\phi_{i\alpha}$, hence integer $n_\alpha$ when $c_2(\hat Y)$ is even. The D3-tadpole condition is satisfied when $\phi_{i\alpha}$ are sufficiently small. Now Eq.\ (\ref{chi56}) and (\ref{chi133}) give
\begin{equation} \label{generalGSMchi}
    \chi_{(\mathbf 3,\mathbf 2)_{1/6}}=\Sigma\cdot(6K_B+5\Sigma)\cdot D_\alpha n_\alpha\,.
\end{equation}
This is one of the main results in this chapter.
The independence of the chiral multiplicity from the parameters $\phi_{5 \alpha}, \phi_{6 \alpha}$ can be understood from the fact that these fluxes do not hit the roots of the preserved part of the gauge group.
Note that
$-(6K_B+5\Sigma)$ is the class of the coefficient of $s^5 z^6$ in the $E_7$ Tate model \cite{BershadskyEtAlSingularities,KatzEtAlTate}. Intersecting it with
$C_{\mathbf{56}}$ gives the codimension-3 singularities.

We see that in a generic basis for the base divisors $D_\alpha$, there are $r$ (see \S\ref{subsec:fluxbreaking:breaking}) quantized flux parameters 
contributing to a single
chiral index, and the chiral index has a linear
Diophantine structure. This is unlike the case in directly
tuned $\gsm$ models, where the chiral index is controlled by
a single flux parameter with a large constant factor, and
either specific geometries must be chosen, or a better understanding of the quantization conditions discussed in \S\ref{subsec:review:G4fluxes} must be achieved, to make
the chiral index as small as 3.
In our case, generically the
intersection numbers $\Sigma\cdot(6K_B+5\Sigma)\cdot D_\alpha$ have no common factors, and the chiral index can be
any integer. A generic flux configuration has both positive
and negative $n_\alpha$ with small magnitudes
(due to the large number of flux directions that can contribute to the
tadpole as discussed in \S\ref{subsec:review:G4fluxes}), making the terms in
Eq.\ (\ref{generalGSMchi}) cancel and naturally leading to a small
chiral index. Heuristically, if we sample $\chi_{(\mathbf 3,\mathbf 2)_{1/6}}$ throughout the landscape, we expect a
distribution peaking at $\chi=0$ and decaying as $\chi$
becomes large \cite{Andriolo:2019gcb}.
Therefore, $\chi=3$ is a natural solution although it may not
be the most preferred. In conclusion, Eq.\ (\ref{generalGSMchi}) is favored by phenomenology.

There may be also some rare cases where the triple
intersection numbers have a common factor. Most probably the
common factor forbids the possibility of $\chi=3$, but if the
common factor is 3, interestingly $\chi=3$ becomes both the
minimal and natural nontrivial chiral spectrum.

The appearance of a nontrivial minimal multiplicity, and other aspects
of multiplicity quantization, can be understood in terms of
intersection theory on the base $B$, combined with the structure of
the $E_7$ lattice.  The intersection product $C = \Sigma \cdot (6 K_B +5 \Sigma)$
is a curve in integer homology of the base $B$.  For generic choices of characteristic
data, we expect that this curve will be primitive, in which case
Poincar\'{e} duality asserts that there is a divisor $D' = D_\alpha
n'_\alpha$ with $C \cdot D' = 1, n'_\alpha \in\Z$.  This is the generic
case described above where there are no common factors and the chiral
index can be any integer.  Thus, in some sense the flux associated
with the chiral index can be characterized by a single parameter
$\lambda$, with $n_\alpha = \lambda n'_\alpha$.  On the other hand, a
full treatment of the proper basis for fluxes would involve
identifying flux directions with minimal tadpole contribution, which
we do not investigate further here.
When $C = m C'$ is not primitive but is an integer multiple of a
primitive curve $C'$, this corresponds to the situation where there
are common factors and there is a non-unit minimal multiplicity for
$\chi$; the case where $\chi = 3$ is minimal corresponds to the
situation where $m = 3$.

It is interesting that more
generally we can consider fluxes in $H^4 (\hat{Y},\Z)$ that
may lead to fractional values of $n_\alpha$.
Since  $H^4 (\hat{Y},\Z)$ has a unimodular intersection form, we
expect that there may be fluxes in $H^4 (\hat{Y},\Z)$ with fractional vertical components $n_\alpha$,
when the fluxes $\phi_{i \alpha}$ lie
in the dual of the root lattice (i.e., the weight lattice) of $E_7$.
From the form of the
$E_7$ lattice, the only non-integer fluxes allowed have half-integer
entries for $\phi_{4 \alpha}, \phi_{6 \alpha}, \phi_{7 \alpha}$.
This results in half-integer $n_\alpha$ in Eq.\ (\ref{eq:phi-n}) and
naively appears to lead to half-integer
multiplicities in Eq.\ (\ref{generalGSMchi}). The multiplicities should be, however, guaranteed to be integers from the structure of $H^4 (\hat{Y},\Z)$.
%Thus, either such
%situations are not possible, or the necessary horizontal flux that
%must be included to give an integer flux vector stabilize moduli to
%change the geometry, or something else fixes up the physics of these
The explicit form of $H^4 (\hat{Y},\Z)$ is not yet fully elucidated,
as discussed in  \S\ref{subsec:review:G4fluxes}, which makes these issues a bit
subtle,
and we leave a more detailed investigation of such situations
to further work.

\subsection{A simplified example of small number of generations}
\label{ssubsec:fluxbreaking:F1example}

%We are now ready to write down
%explicit F-theory constructions for our class of SM-like models. This can be

%The construction for our
This class of SM-like models,
combining vertical and remainder fluxes, 
%is more complicated but 
can be %done
realized on a large class of
bases with rigid (or tuned) $E_7$ but cannot be
%done
constructed completely in
simple toric geometries. Here for simplicity, to illustrate the
multiplicity of generations, we focus on {\it vertical} flux 
breaking to SU(5)
and give
an oversimplified example, using the Hirzebruch surface $\mathbb{F}_1$ as
the gauge divisor $\Sigma$.\footnote{This
oversimplification also leads to exotic $\U(1)$
gauge factors along with the $\SU(5)$. Here as a mere demonstration, we
focus on the $\SU(5)$ representations and ignore
the $\U(1)$ charges when calculating chiral
indices.}
When
further breaking to $\gsm$ through remainder flux is possible 
on more complicated surfaces,
this further breaking does not affect multiplicities.
%the same
%number of generations results.
%\footnote{Using this gauge
%divisor may lead to exotic $\U(1)$ extensions of $\gsm$
%\cite{Li:2022aek}, which we ignore here. The issue can be %easily avoided
%using more complicated $\Sigma$.} \kobe{Changed it to $\mathbb F_1$ on general bases}

$\mathbb{F}_1$ is a $\mathbb{P}^1$-bundle over another
$\mathbb{P}^1$. We denote $S$ as the section $\mathbb{P}^1$
and $F$ as the fiber $\mathbb{P}^1$. Then the intersection
numbers are $F^{2}=0,F\cdot S=1,S^{2}=-1$. Its anticanonical
class is $-K_\Sigma=2S+3F$. Now embed $\mathbb{F}_1$ into $B$ with normal bundle
$N_\Sigma=-aS-bF$. Let $F_S,F_F$ be divisors with
$\Sigma\cdot F_S,\Sigma\cdot F_F$ being pushforwards of
$S$ and $F$ into $B$ respectively. Without {\it remainder}
flux, we can assume $\Sigma\cdot F_S,\Sigma\cdot F_F$ are independent.
By choosing $N_{\Sigma}=-8S-7F$,
\begin{align}
    F_k=-4K_\Sigma+(4-k)N_\Sigma\,,\nonumber\\
    G_l=-6K_\Sigma+(6-l)N_\Sigma\,,
\end{align}
are both effective only when $k\geq 3,l\geq 5$, so we have a rigid $E_7$
supported on $\Sigma$ \cite{MorrisonTaylor4DClusters}. The nonzero
intersection numbers are then
$\Sigma\cdot F_S\cdot F_F=1,\Sigma^2\cdot F_F=-8,\Sigma\cdot
F_S^{2}=-1,\Sigma^2\cdot F_S=1,\Sigma^3=48$.

We claim that all the above constraints on {\it vertical}
flux can be solved inside the
K\"ahler cone by turning on nonzero but sufficiently small integer
$\phi_{iF_S}$ and $\phi_{iF_F}$ with opposite signs.
We require the
ratio $\phi_{iF_S}/\phi_{iF_F}$ to be the same for all $i$. To break
the gauge group, we turn on integer
$\phi_{5F_S},\phi_{6F_S}$ freely and
\begin{equation}
    \left(\phi_{1F_S},\phi_{2F_S},\phi_{3F_S},\phi_{4F_S},\phi_{7F_S}\right)=(2,4,6,5,3)n_S\,,    
\end{equation}
and similarly for $\phi_{iF_F}$, where
$n_S,n_F$ are integers with opposite signs. Eq. (\ref{fullchiR'})
then gives a simple formula for the number of
generations of $\SU(5)$ GUT matter:
\begin{equation}
    \chi_{\mathbf{10}}=-\chi_{\mathbf{5}}=-7n_S-4n_F\,.
\end{equation}
This is a linear Diophantine equation and the number of
generations can be any sufficiently small integer.
%The fourfold $\hat Y$ has
%$h^{1,1}=11,h^{2,1}=0,h^{3,1}=8497$, hence $h^{2,2}=34076$
%and $\chi(\hat Y)=51096$.
As explained above, it is natural
to consider small $\phi_{iF_S}$ and $\phi_{iF_F}$. The
minimal flux configuration has $n_S=-1$ and
$n_F=1$, hence $\chi=3$ appears to be preferred.
This is an example of an
F-theory model with exactly three
generations of chiral matter, with minimal
fine-tuning.
%Such a construction can be done on many
%(non-toric) bases with the same divisor that supports
%hypercharge flux \cite{Braun:2014pva,Braun:2014xka}, and %the same normal bundle.

The above is the most general {\it vertical} flux we can turn on
given the flux constraints and conditions on
$\Theta_{i\alpha}$. All other $\phi_{i\alpha}$ are
equivalent to a combination of $\phi_{iF_S}$ and
$\phi_{iF_F}$ by homology relations.

The above construction can be easily generalized to
incorporate hypercharge flux by using more complicated $\Sigma$'s on
non-toric bases \cite{Braun:2014pva,Braun:2014xka}
(with different multiplicities in each case).
%It makes no qualitative difference at low energies except the gauge
%group.
We provide such explicit constructions in \S\ref{sec:fluxbreaking:example}.

\section{Breaking other gauge groups}
\label{sec:fluxbreaking:othergroups}

So far we have focused on the breaking $E_7\rightarrow\gsm$,
while among the rigid gauge groups, $E_6$ and $E_8$ can also
be broken into $\gsm$. $E_6$ has been one of the traditional
GUT groups, so its breaking is less novel than $E_7$'s. 
Flux breaking of (non-rigid) $E_6$ F-theory models has been described
in the dual heterotic framework in \cite{Chen:2010tg}.
On
the other hand, $E_6$ also appears in a significant portion of the
landscape, and we therefore generalize our construction to $E_6$
for completeness. $E_8$ is clearly the most abundant
exceptional gauge
group in the landscape, but unfortunately our formalism does
not work for $E_8$ for several reasons. Below we discuss
these two gauge groups separately.

\subsection{$E_6$}
\label{subsec:fluxbreaking:e6}

It is clear that the above construction for $E_7$ also works
for $E_6$, since we can first break $E_7$ down to $E_6$ in
our breaking by vertical flux. The generalization to $E_6$, however, is more
nontrivial since $E_6$ itself supports chiral matter without
flux breaking. There are more flux parameters $\phi_{ij}$ to
turn on, and both terms in Eq.\ (\ref{fullchiR'}) contribute
to the chiral indices. Although as discussed before the two
terms in Eq.\ (\ref{fullchiR'}) are independent contributions
controlled by different flux parameters, the flux
configuration itself becomes more nontrivial due to flux
quantization. This will be explained below.

\begin{figure}[t]
\centering
\includegraphics[width=0.4\columnwidth]{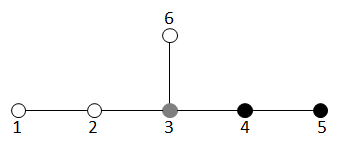}
\caption{The Dynkin diagram of $E_6$. The Dynkin node labelled
$i$ corresponds to the exceptional divisor $D_i$. The solid
nodes are the ones we break to get the Standard Model gauge
group and chiral matter. Node 3  (gray) is broken by
remainder flux while the others are broken by vertical flux.}
\label{dynkine6}
\end{figure}

Although we expect the
middle intersection form on vertical fluxes (\ref{eq:review:Mred-matter}) as
well as the
physics to be resolution-independent,
in practice it is useful to work with a certain resolution,
and extract resolution-independent information from the
results. Here we choose the resolution studied in
\cite{Esole:2017kyr,Bhardwaj_2019,Jefferson:2021bid},  which
for completeness we review in Appendix \ref{sec:resolution}. The exceptional
divisors and the broken directions are described as in Figure \ref{dynkine6}.
The result in \cite{Jefferson:2021bid} shows that when the gauge group is
unbroken, there is a chiral index for the $E_6$ fundamental
$\mathbf{27}$, given by
\begin{equation}
    \chi_{\mathbf{27}}=\Theta_{24}\,.
\end{equation}
Following the procedure in \S\ref{subsec:fluxbreaking:matter}, we find that the
matter surface $S_0(\mathbf{27})$ is
\begin{equation}
    S_0(\mathbf{27})=D_2\cdot
    D_4+\frac{1}{3}\pi^*(3K_B+2\Sigma)\cdot(-D_1+D_2+2D_4+D_5)\,,
\label{eq:e6-s0}
\end{equation}
where the class $(3K_B+2\Sigma)$ appears in the matter curve
$C_{\mathbf{27}}=-\Sigma\cdot(3K_B+2\Sigma)$.

Recall that both the flux configuration and chiral indices
can be separated into parts that preserve or break the gauge
group. We now decompose $G_4=G_4^{p}+G_4^{b}$, where $G_4^p$
lives in the directions in $M_{\mathrm{phys}}$ and preserves
the gauge group, while $G_4^b$ is the gauge-breaking flux.
Correspondingly we define the flux parameters $\phi^p,\phi^b$
and chiral indices $\chi^p,\chi^b$. By solving the flux
constraints, $G_4^p$ is given by
\begin{equation}
    G_4^p=\phi_{24}^p\left[S_0(\mathbf{27})\right]\,,
\end{equation}
inducing
\begin{equation}
\label{e6chib4breaking}
    \chi^p_{(\mathbf 3,\mathbf 2)_{1/6}}=\chi_{\mathbf{27}}=\frac{1}{3}\Sigma\cdot(3K_B+2\Sigma)\cdot(6K_B+5\Sigma)\phi_{24}^p\,.
\end{equation}
As expected, this chiral index is controlled by a single flux
parameter $\phi^p_{24}$.  We discuss the detailed quantization
condition on this parameter below, but note that $K_B \cdot \Sigma
\cdot \Sigma$ is always even, by the Hirzebruch-Riemann-Roch theorem
for surfaces, so the chiral multiplicity is always an integer when
$\phi_{24}^p$ is a multiple of 3/2.

Now we turn to $G_4^b$. Here we follow the breaking route
used in \S\ref{subsec:fluxbreaking:solve}. In principle, we can
apply the same procedure as in \S\ref{sec:fluxbreaking:SM} to
obtain $G_4^b$ and $\chi^b$, but there is a faster way using
the result for $E_7$. We can first break $E_7$ down to
$E_6$ by removing node 6 in Figure \ref{dynkine7}. Note that
$\phi_{6\alpha}$ in the $E_7$ model are completely
independent of other flux parameters and do not contribute to
the chiral indices. Therefore, $G_4^b$ for
$E_6$ is the same as that for $E_7$ by ignoring node 6 in
$E_7$. That is
\begin{gather}
    \phi^b_{1\alpha}=2n_\alpha\,,\quad\phi^b_{2\alpha}=4n_\alpha\,,\quad\phi^b_{3\alpha}=6n_\alpha\,,\nonumber\\
    \phi^b_{4\alpha}=5n_\alpha\,,
    %\quad\phi_{5\alpha}=\phi_{5\alpha}\,,
    \quad\phi^b_{6\alpha}=3n_\alpha\,,
    \label{eq:e6fluxbreaking}
\end{gather}
where $\phi^b_{5 \alpha}$ is arbitrary.
Therefore, the chiral index from flux breaking is
\begin{equation}
\label{e6chibreaking}
    \chi^b_{(\mathbf 3,\mathbf 2)_{1/6}}=\Sigma\cdot(6K_B+5\Sigma)\cdot D_\alpha n_\alpha\,,
\end{equation}
which is exactly the same as Eq.\ (\ref{generalGSMchi}). The
total chiral index is then $\chi=\chi^p+\chi^b$. Despite the
extra $\chi^p$, with the inclusion of $\chi^b$ it is
qualitatively the same as that in $E_7$ models.

So far $\phi^p$ and $\phi^b$ are totally separated, but it
becomes more interesting when flux quantization is
considered. First, in $E_6$ models it is unavoidable to have
a non-even $c_2(\hat Y)$. Using the techniques in \cite{Jefferson:2021bid}, we find
that for our choice of resolution (in terms of independent surfaces)
\begin{align}
    \left[c_2(\hat Y)\right]=&\left[c_2(B)\right]+11\pi^{*}K_{B}^{2}+\left(-12D_{0}+17D_{1}+27D_{2}+30D_{3}+24D_{4}+11D_{5}+14D_{6}\right)\cdot\pi^{*}K_{B} \nonumber \\
    &+\left(6D_{1}+8D_{2}+6D_{3}+6D_{4}+2D_{5}+2D_{6}\right)\cdot\pi^{*}\Sigma+D_2\cdot D_4\,.
\end{align}
Note that $\left[c_2(B)\right]+\pi^{*}K_{B}^{2}$ is always
even \cite{Collinucci:2010gz}. 
%If the gauge group is unbroken i.e. $G_4^b=0$, we
%see that flux quantization generically requires 
%$\phi_{24}^p=3(2k+1)/2$
%for some integer $k$. The situation, however, is different if
%both $G_4^p$ and $G_4^b$ are present. The crucial point is
%that flux quantization only applies to the total flux $G_4$,
%allowing more flux configurations if we look at one of the
%sectors only. In particular, now $\phi_{24}^p$ can be any
%half-integer, as long as appropriate \emph{fractional}
%$\phi_{i\alpha}$ are turned on such that the total flux is
%correctly quantized. 
Note also from the form of Eq.\ (\ref{eq:e6-s0}) that the part of $c_2$
that is odd precisely contributes to a half-integer contribution to $G_4^p$
 and does not necessitate breaking of the $E_6$.
If the gauge group is unbroken i.e. $G_4^b=0$, we
see that flux quantization generically requires 
$\phi_{24}^p=3(2k+1)/2$
for some integer $k$. 
(As noted above, this always gives integer chiral multiplicity since  $K_B \cdot \Sigma \cdot \Sigma$ is
always even.)
The situation, however, is different if
both $G_4^p$ and $G_4^b$ are present. The crucial point is
that flux quantization only applies to the total flux $G_4$,
allowing more flux configurations if we look at one of the
sectors only. In particular, now $\phi_{24}^p$ can be any
half-integer, as long as appropriate \emph{fractional}
$\phi_{i\alpha}$ are turned on such that the total flux is
correctly quantized. Therefore, the presence of
gauge-breaking flux enlarges the possibilities of matter
surface flux, although they contribute to chiral indices
independently.

\subsection{$E_8$}
\label{subsec:fluxbreaking:e8}

It is tempting to apply our formalism to $E_8$ models.
Nevertheless, these models have very different physics from
$E_7,E_6$ models, and the direct construction of Standard Model-like
vacua from flux breaking of rigid $E_8$ factors fails for various
reasons.

The first reason is that an $E_8$ model generically contains
codimension-2 $(4,6)$ singularities. While this type of
singularity in 4D has not been completely understood, it is
believed to be parallel to the story in 6D F-theory models.
There, a simple physical interpretation of these singularities is
obtained by blowing up the locus into the tensor branch till
the singularities are within the minimality bound i.e. degree
$<(4,6,12)$. The origin of  the tensor branch corresponds to
shrinking the resulting exceptional divisors to zero volume,
giving a strongly coupled limit of the model. D3-branes
wrapping these exceptional divisors also become tensionless
strings in the low-energy theory. All these signal the
presence of strongly coupled superconformal sectors
\cite{HeckmanMorrisonVafa,DelZotto:2014hpa,Apruzzi:2018oge}. These
extra degrees of freedom, called conformal matter, are not
covered by our formalism for analyzing flux breaking.

Still, there are $E_8$ models without these kinds of
singularities and naively our formalism should work in such cases.
 The second reason, however, that these geometries are problematic is
 that the condition for the absence of
conformal matter is that the codimension-2 singularity has
trivial homology class i.e. $\Sigma\cdot(6K_B+5\Sigma)=0$.
Surprisingly, using Eq.\ (\ref{generalGSMchi}) this immediately
implies that no chiral matter can be induced, even if we
break $E_8\rightarrow\gsm$. It remains interesting to find a
reason behind this apart from direct computations.

All this seem to suggest that the class of SM-like models we
have constructed may still not be the largest class in the
landscape. In particular, the F-theory geometry with the most
flux vacua contains many factors of $E_8$, but
no factors of $E_7,E_6$ and does not
support our formalism \cite{TaylorWangVacua}. In principle, the most generic SM
matter should come from the strongly coupled matter in $E_8$.
Some initial investigation into studying the 4D spectrum from strongly
coupled $E_8$ matter
in the context of E-string
theory is described in \cite{TianWangEString}.

\section{An explicit example}
\label{sec:fluxbreaking:example}

The above construction of SM gauge group and chiral matter can be done on a large class of bases
containing rigid or tuned $E_7,E_6$ factors. For the rigid case, we need
a non-toric base such that there can be a rigid divisor
supporting hypercharge flux. In this section, we
provide an explicit example of such a construction,
with three generations of SM chiral matter as the
minimal and preferred chiral matter content. Since
the $E_7$ models require a more involved
construction with $r\geq 4$, below we construct
rigid $E_6$ with $r=3$. As shown below, the gauge divisor is a del
Pezzo surface $dP_4$, so the model has a limit where gravity is
fully decoupled \cite{BeasleyHeckmanVafaII}.

We choose a base $B$ through the following procedure: first consider
$A=\mathbb P^1\times \mathbb F_1$ where $\mathbb F_1$ is the Hirzebruch
surface. Then $B$ is a certain hypersurface in an
ambient space $X$ that is a $\mathbb P^1$-bundle over $A$
with a certain normal bundle. This example is
a generalization of an example of a geometry supporting remainder flux in \cite{Braun:2014pva}; more
generally we can similarly analyze any hypersurface $B$ in a toric
fourfold $X$ as long as $B$ is ample.
In the explicit example we
consider here,
both vertical and remainder fluxes are
incorporated, and all flux constraints can be explicitly
solved.

Let us first construct the ambient space $X$. To
construct a model with $r\geq 3$ (see
\S\ref{subsec:fluxbreaking:breaking}), we need to start with a threefold
$A$ with $h^{1,1}\geq 3$.
 As an example with $h^{1,1}(A)=3$, we choose $A$ to be
$\mathbb P^1\times \mathbb F_1$. Within $\mathbb F_1$,
we denote $s$ as the
$\mathbb P^1$ section and $f$ as the $\mathbb P^1$ fiber. Then
the intersection numbers are $f^{2}=0,f\cdot s=1,s^2=-1$.
Now on $A$, we denote $\sigma$ as the
$\mathbb{F}_{1}$ section and $S,F$ as the
$\mathbb P^1$ product with $s$ and $f$ respectively. Then
the anticanonical class of $A$ is
$-K_{A}=2\sigma+2S+3F$.
\begin{comment}
We denote $\sigma$ as the $dP_2$ section and
$S_1,S_2,H$ as the $\mathbb P^1$ product with the
two exceptional divisors and the hyperplane of
$dP_2$ respectively. Note that the effective
divisors are spanned by $\sigma,S_1,S_2$, and
$F=H-S_1-S_2$. The anticanonical class of $A$ is $-K_A=2\sigma+2S_1+2S_2+3F$.
\end{comment}
The nonzero intersection numbers are:
\begin{equation} \label{p1crossdp2}
    \sigma\cdot S\cdot F=1\,,\;\sigma\cdot S^2=-1\,.
\end{equation}
Finally we can
describe $X$ as a $\mathbb P^1$-bundle over $A$. We denote
$\sigma_A$ as the section and $F_\sigma,F_S,F_F$ as the
fibers along $\sigma,S,F$ respectively. Let the normal bundle
be $N_A=-a\sigma-bS-cF$ where $a,b,c\in\mathbb Z_{\geq0}$.
Then its anticanonical class is
$-K_X=2\sigma_A+(a+2)F_\sigma+(b+2)F_S+(c+3)F_F$. The
intersection numbers can be calculated using Eq.\ (\ref{p1crossdp2}) and the relations
$\sigma_A\cdot(\sigma_A+aF_\sigma+bF_S+cF_F)=0$. Note that with the below choice of $N_A$, $X$ is a
smooth, projective toric variety with a unique triangulation.

We then choose the base as a hypersurface in $X$ with irreducible class
$B=\sigma_A+(a+1)F_\sigma+(b+1)F_S+(c+2)F_F$. By abuse of notation, we use
$B$ to denote both the base and its divisor class in $X$. By
adjunction we have $-K_B=B\cdot(\sigma_A+F_\sigma+F_S+F_F)$.
As shown below, $B$ is strictly inside the K\"ahler cone of $X$, so $B$ is ample in $X$.  By Lefschetz's hyperplane theorem, we then have the isomorphism
$H^{1,1}\left(B,\mathbb{Z}\right)\cong H^{1,1}\left(X,\mathbb{Z}\right)$. In other words, divisors
on $B$ are spanned by the intersections $B\cdot\sigma_A,B\cdot F_\sigma,B\cdot F_{S},B\cdot F_{F}$.
The intersection numbers relevant to our purpose are
\begin{gather}
    B\cdot\sigma_A\cdot F_\sigma^2=0\,,\;B\cdot\sigma_A\cdot F_\sigma\cdot F_S=1\,,\;B\cdot\sigma_A\cdot F_\sigma\cdot F_F=1\nonumber \\
    B\cdot\sigma_A\cdot F_S\cdot F_F=1\,,\;B\cdot\sigma_A\cdot F_S^2=-1\,,\;B\cdot\sigma_A\cdot F_F^2=0\,.
    \label{Sigmaintersections}
\end{gather}

\begin{comment}
By choosing $N_{\Sigma}=-8S-7F$, there is a rigid $E_7$
supported on $\Sigma$ (see \S\ref{sec:review:basics:rigid}) and an even $c_2(\hat Y)$. The nonzero intersection numbers are then
$\Sigma\cdot F_{S}\cdot F_{F}=1,\Sigma^{2}\cdot F_{F}=-8,\Sigma\cdot F_{S}^{2}=-1,\Sigma^{2}\cdot F_{S}=1,\Sigma^{3}=48$.
\end{comment}

Now consider the gauge divisor
$\Sigma=B\cdot\sigma_A=\sigma_A\cdot(F_\sigma+F_S+2F_F)$.
To determine the rigid gauge group on $\Sigma$, we
calculate
\begin{gather}
    -K_{\Sigma}=B\cdot\sigma_A\cdot \left(F_\sigma+F_S+F_F\right)=\sigma_A\cdot\left(2 F_\sigma\cdot F_S+3F_\sigma\cdot F_F+2F_S\cdot F_F\right)\,, \\
    N_{\Sigma}=B\cdot\sigma_A^2
    =-\sigma_A\cdot\left((a+b)F_\sigma\cdot F_S+(2a+c)F_\sigma\cdot F_{F}+(b+c)F_S\cdot F_F\right)\,.
\end{gather}
By the conditions in
\S\ref{sec:review:basics:rigid}, we choose $(a,b,c)=(3,3,3)$ such that $\Sigma$
is a rigid divisor
supporting a rigid $E_6$. Note that with this choice of
$N_A$, $C_\mathbf{27}$ is trivial and all
the matter is in the  $E_6$ adjoint $\mathbf{78}$ before flux breaking.
Note also that with this choice $N_\Sigma = 3K_\Sigma$ is divisible by 3, so the curve $C =\Sigma \cdot
(6K_B + 5 \Sigma)$ appearing in Eq.\ (\ref{e6chibreaking}) is not primitive
but takes the form $C = 3C'$ as discussed in \S\ref{subsec:fluxbreaking:solve}, and
we expect chiral multiplicities that are multiples of 3.
Note further that we have not ruled out fully the possibility of increased
enhancement over curves in the base, which might in principle give
rise to additional surfaces.  Even if this occurs,
however, it should not be relevant for our construction as we can
simply keep any associated additional fluxes that may arise to
vanish.  For future work and more general constructions, however, it
would be useful to develop more completely the methodology for
analyzing the structure of hypersurface bases of this general kind.

We can solve the flux constraints after determining the
geometry. First we focus on vertical flux. Let us analyze the constraint from
primitivity. The
independent $S_{i\alpha}$ are $S_{i(B\cdot F_\sigma)},S_{i(B\cdot F_S)},S_{i(B\cdot F_F)}$, while $S_{i\Sigma}$ is a linear
combination of the former three. Using the
intersection numbers, we see that the triple intersection
form $M^B_{\alpha\beta}$ has one positive and two negative
directions, so primitivity can be satisfied.
As explained in \S\ref{subsec:fluxbreaking:primitive}, we focus on
the K\"ahler form of the base $J_B$, which can be expanded using a
basis of base divisors (recall that in this case as noted above,
$H^{1,1}\left(B,\mathbb{Z}\right)\cong
H^{1,1}\left(X,\mathbb{Z}\right)$):
\begin{equation}
    \left[J_{B}\right]=B\cdot\left(t_{1}F_F+t_{2}\left(F_S+F_{F}\right)+t_{3}F_\sigma+t_{4}\left(\sigma_A+3F_\sigma+3F_S+3F_F\right)\right)\,,
\end{equation}
where $t_{1},t_{2},t_{3},t_{4}$ are linear
combinations of K\"ahler moduli,
%on $B$
 and may be negative
inside the K\"ahler cone of $B$
in general. While determining the
exact K\"ahler
cone of a hypersurface in a toric variety can be subtle, the K\"ahler cone of $B$ must contain that of $X$
 \cite{Demirtas:2018akl}.
For simplicity, we look for a solution of the primitivity constraints in
the K\"ahler cone of $X$ only.
First, the
K\"ahler cone of $X$ can be obtained from the Mori cone, which is
spanned by $F_\sigma\cdot F_S\cdot F_F,\sigma_A\cdot F_\sigma\cdot F_S,\sigma_A\cdot F_\sigma\cdot F_F,\sigma_A\cdot F_S\cdot F_F$. By computing the
dual cone, we see that the interior of the K\"ahler
cone of $X$ corresponds to $t_{1},t_{2},t_{3},t_{4}>0$. Now primitivity implies that for all $i$
\begin{equation}
    t_1(\phi_{i\sigma}+\phi_{iS})+t_2(2\phi_{i\sigma}+\phi_{iF})+t_3(\phi_{iS}+\phi_{iF})=0\,.
\end{equation}
To determine the chiral matter spectrum, we first
focus on $n_\alpha$. There must be a pair of
coefficients of $t_a$ with opposite signs, which
places constraints on the possible $n_\alpha$.

\begin{comment}
There is another constraint on $n_\alpha$ from flux
quantization. Interestingly, our example has
a non-even $c_2(\hat Y)$. Using the techniques in \cite{Jefferson:2021bid} and the choice of resolution in Appendix \ref{sec:resolution}, we find
that for models with a single $E_7$: (in terms of homologically independent surfaces)
\begin{align}
    \left[c_2(\hat Y)\right]=&\left[c_2(B)\right]+11\pi^{*}K_{B}^{2}\nonumber \\
    &+\left(-12D_{0}+14D_{1}+30D_{2}+48D_{3}+41D_{4}+28D_{5}+17D_{6}+27D_{7}\right)\cdot\pi^{*}K_{B} \nonumber \\
    &+\left(2D_{1}+6D_{2}+12D_{3}+12D_{4}+8D_{5}+6D_{6}+8D_{7}\right)\cdot\pi^{*}\Sigma\,.
\end{align}
Note that $\left[c_2(B)\right]+\pi^{*}K_{B}^{2}$ is always
even \cite{Collinucci:2010gz}. Since $\Sigma.K_B=K_\Sigma-N_\Sigma$, Eq.\ (\ref{eq:review:fluxquant})
\end{comment}

Now we consider the chiral index. First, $\phi^p$ and Eq.\ (\ref{e6chib4breaking}) vanish since $C_\mathbf{27}$ is
trivial. Eq.\ (\ref{e6chibreaking}) then gives
\begin{equation}
    \chi_{\left(\mathbf{3},\mathbf{2}\right)_{1/6}}=-3(2n_\sigma+n_S+2n_F)\,,
\end{equation}
where $n_I=n_{(B\cdot F_I)}$. 
As discussed before, it
is natural to consider small $\phi_{i\alpha}$. To
fulfill flux quantization as in \S\ref{subsec:fluxbreaking:e6}, we turn on  integer $n_\alpha$.
One of the minimal flux configurations satisfying the flux
constraints has $(n_\sigma,n_S,n_F)=(-1,1,0)$, hence
gives $\chi=3$ as the minimal and preferred chiral spectrum.

We now turn to remainder flux. It can be shown that $\Sigma$ is a del
Pezzo surface $dP_4$ and supports remainder flux. First notice that
$\Sigma$ is a hypersurface in $A$
with class $\sigma+S+2F$. In other words, $\Sigma$ is the vanishing
locus
\begin{equation} \label{fluxbreaking:Sigmalocus}
    xP+yP'=0\,,
\end{equation}
in $A$, where $P,P'$ are sections of $\mathcal O_A(S+2F)$, and $x,y$
are the homogeneous coordinates of the $\mathbb P^1$ in $A$. For generic
points in the $\mathbb F_1$, Eq. (\ref{fluxbreaking:Sigmalocus}) has a unique
solution, representing a single point in $\mathbb P^1$. On the other hand,
there are $(s+2f)^2=3$ points in $\mathbb F_1$ such that $P=P'=0$,
and Eq. (\ref{fluxbreaking:Sigmalocus}) represents the whole $\mathbb P^1$.
Therefore, the geometry of $\Sigma$ is $\mathbb F_1$ blown up in 3
generic points i.e. a $dP_4$, where the projection $A \rightarrow
\F_1$ gives the blow-down map
$\Sigma \rightarrow \F_1$.

To construct the remainder flux, notice that the three exceptional
curves on $\Sigma$ from blowing up $\mathbb F_1$ (denoted by
$e_1,e_2,e_3$) are all $\mathbb P^1$
fibers in $A$, which have class $S\cdot F$. Under the
inclusion map $\iota:\Sigma\rightarrow B$, we then have $\iota_* e_i=\sigma_A\cdot F_S\cdot F_F$
for all $i=1,2,3$.  Therefore, we can choose e.g. $C_{\mathrm{rem}}=e_1-e_2$ and turn on the remainder flux
\begin{equation} \label{eq:e6remainderflux}
    G_4^\mathrm{rem}=\left[\left(D_Y+\phi_{4r}D_4+\phi_{5r}D_5\right)|_{C_\mathrm{rem}}\right]\,,
\end{equation}
where the first term is the hypercharge flux in Eq.\ (\ref{hyperchargeflux}) and the other two terms with free flux
parameters $\phi_{4r},\phi_{5r}$ do not affect the gauge group. Notice
that when $\phi_{4r}=4$, it is known that this flux removes the exotic
vector-like $(\mathbf 3,\mathbf 2)_{-5/6}$ \cite{BeasleyHeckmanVafaII},
avoiding a number of phenomenological inconsistencies such as part of proton decay. Importantly,
this removal can be achieved without the complication of using fractional line bundles,
which must be used in traditional $\SU(5)$ models, due to more flux
parameters. This is parallel to how the Diophantine structure in vertical flux enables much more possibilities for chiral indices.
These further phenomenological features of our models will be studied
in 
a future publication. In the analysis below, we keep this choice of $\phi_{4r}$.

For consistency, we still need to study the tadpole condition, see Eq.\ (\ref{eq:review:tadpole}). First we specify the remaining flux parameters
$\phi_{5\alpha},\phi_{5r}$. As an example with small tadpole, we choose
$(\phi_{5F_\sigma},\phi_{5F_S},\phi_{5F_F},\phi_{5r})=(-3,2,1,2)$. The
total flux is then given by these four parameters, the vertical flux in
Eq. (\ref{eq:e6fluxbreaking}), and the remainder flux in Eq.
(\ref{eq:e6remainderflux}) with $\phi_{4r}=4$. The
vertical flux consistently stabilizes the K\"ahler moduli at
$2t_1=2t_2=t_3$. The total tadpole is
\begin{equation}
    \frac{1}{2}\int_{\hat Y}G_4^\mathrm{vert}\wedge G_4^\mathrm{vert}+\frac{1}{2}\int_{\hat Y}G_4^\mathrm{rem}\wedge G_4^\mathrm{rem}=13+6=19\,.
\end{equation}
Here we have naturally extended Eq. (\ref{eq:review:Mred-simple}) to remainder
flux, since the intersections are all localized on $\Sigma$. 
Using the
technique in \cite{Esole:2017kyr}, we find that $\chi(\hat Y)=2088$.
Since $\chi(\hat Y)/24=87>19$, the tadpole condition is satisfied.

We would like to emphasize that although we have chosen an
explicit global example here, the analysis is purely local.
The same breaking pattern and matter spectrum are expected
whenever there is a $\Sigma$ in $B$ with the same geometry and normal bundle. This is analogous to 6D
F-theory models, in which any curve of self-intersection $-6$
supports a rigid $E_6$ \cite{MorrisonTaylorClusters}. We expect that many of the F-theory threefold
bases contain the above local structure. Moreover, there are
lots of local structures throughout the landscape that
support the same flux breaking. Therefore, our construction
provides a large class of models with SM gauge group and
chiral matter, with 
much less fine-tuning that is needed for other known constructions.

\section{Conclusion and further questions}
\label{sec:fluxbreaking:conclusion}

\subsection{Summary of results}
\label{subsec:fluxbreaking:summary}

In this chapter, we have described a large class of  Standard
Model-like
models with the right gauge group and chiral matter spectrum,
using the framework of F-theory compactifications. These
models originate from rigid $E_7$ (covered briefly in \cite{Li:2021eyn}) or $E_6$ gauge symmetries,
which are ubiquitous in the string landscape and do not
require any fine-tuning of moduli. In particular, the UV
physics of string theory allows us to use $E_7$, in
addition to the traditional $E_6$, as a GUT group. The same
construction can be  carried out on many (but non-toric) F-theory threefold
bases that contain rigid $E_7$ or $E_6$ local structures. Due to
such genericity, we expect that this is a natural way for 
the Standard Model
to arise in the landscape. Although we do not have an exact
quantification, we believe these models should be more
generic than tuned SM-like models in the landscape.

Remarkably, these models also enjoy the advantage of typically having
small chiral indices. While the chiral indices in tuned
SM-like models are usually too large unless very specific geometries are considered, or the subtle flux quantization issues discussed in \S\ref{subsec:review:G4fluxes} are managed,
the chiral indices in our models
have a linear Diophantine structure that naturally leads to
small integers for typical geometries. As a result, three generations of SM chiral
matter can be easily realized in our models. In particular, a
subset of them have $\chi=3$ as the minimal or preferred
matter content. This is favored by phenomenology. We hope
that this large class of SM-like constructions can shed some
light on where our Universe sits in the string landscape, and
whether it is a \emph{natural} solution in the landscape.

The main tool we have used to achieve the above results is gauge
symmetry breaking with both vertical and remainder fluxes. This
is an efficient way to build models, as it breaks the
gauge group and induces chiral matter at the same time. While
this idea is not new, we have developed it here in depth to give a
systematic procedure to describe the flux breaking from any
$G$ to any $G'$ on almost any base, and calculate the chiral
spectrum induced by the vertical flux. All these calculations can be done
using simple formulas and give results that are manifestly
resolution-independent, with the base geometry and group
theory data as the only input. A remarkable fact from this
procedure is that even if $G$ does not support any chiral
matter, generically a chiral spectrum is still induced if $G'$
supports chiral matter. This is why we can use $G=E_7$ in our
SM-like models. The only exception we find is
$G=E_8$. The procedure developed here is a byproduct of our study of SM-like
models, and should be useful for other types of F-theory
model building in the future.

\subsection{Further questions}
\label{subsec:fluxbreaking:questions}

As mentioned at the beginning of this chapter, although the models we have
constructed here
have the right gauge group and chiral matter spectrum, they
are far from complete in realizing the full details of the
 Standard Model in string theory. 
More work is needed to understand the full matter spectrum including
vector-like fields, Yukawa couplings, questions related to proton
decay, etc.
Many other more general
questions can also be asked, regarding
both theoretical and phenomenological aspects of these models. Examples include:

\begin{itemize}
    \item One interesting feature of our formalism of flux
    breaking is that it intrinsically relies on the
    non-perturbative physics of F-theory. The gauge-breaking
    flux we turn on does not have any immediately obvious
    description in the low-energy theory. In particular, the
    approach of inducing chiral matter with the flux cannot be
    realized in the framework of field theory in any known way. Although the
    broken gauge group and chiral matter spectrum are
    certainly low-energy observables, they do not give full
    information on the flux configuration, and the original $E_7$ or
    $E_6$ gauge group does not seem to be apparent in any clear way in
    the low-energy theory. To gain a more
    complete picture, it would be interesting to understand the
    structure of these models better from the low-energy perspective
    and/or in the dual heterotic  framework.
    
    \item In a string compactification compatible with observations,
      the moduli must be stabilized. In F-theory, the stabilization of
      complex structure moduli is done by turning on horizontal flux,
      inducing a superpotential for the moduli. This flux is
      orthogonal to vertical flux and does not affect the matter
      spectrum. In models with tuned gauge groups, however, some
      complex structure moduli must be fixed and this complicates the
      problem of computing the period vectors, hence superpotential,
      when combined with these tunings. On the other hand, our models rely on
      rigid gauge groups and there is no constraint on complex
      structure moduli. Therefore, the stabilization can be done
      independently without affecting the gauge sectors.  This
      promises, in principle, to make the calculation of moduli
      stabilization easier,
      and opens up an interesting possibility of finding SM-like
      models with moduli stabilized, along the lines of
      \cite{Demirtas:2021nlu,Demirtas:2021ote} and related
      work.\footnote{We thank Manki Kim for discussion on this.}
    
    \item Our construction of SM-like models is
      base-independent.
It is thus possible to apply our
    construction to a large number of explicit F-theory
    threefold bases and perform statistical analysis. 
 There are several distinct such statistical problems of interest.  On
 the one hand, for a given local geometry that supports this
 construction, it will be useful to know  what portion of
 flux configurations can break the rigid gauge group
    down to $\gsm$, and/or give three generations of SM
    chiral matter. 
At the same time it would be desirable to have a better understanding
of the global space of threefold bases that support 4D elliptic
Calabi-Yau spaces, and how ubiquitous the presence of rigid $E_6$ or
$E_7$ gauge factors is in this space.
In particular,
while the current list of F-theory threefold
    bases is far from complete, the large ensembles of toric bases
    considered in
    \cite{TaylorWangMC,HalversonLongSungAlg,TaylorWangLandscape}
suggest that $E_6$ and $E_7$ factors occur frequently.  The naive
expectation would be that this is similarly true for non-toric bases,
although it would be important to initiate some systematic survey of
non-toric bases (perhaps, e.g., general hypersurfaces in toric
fourfolds), to confirm or contradict that hypothesis.
Such a survey would also give insight into whether the cycles needed
for remainder fluxes are indeed typical, as suggested in \cite{Braun:2014xka}.
For a given fourfold geometry, with multiple rigid gauge factors,
we
    can apply our construction to any rigid $E_7$ or $E_6$
    factor (while other gauge factors can serve as hidden
    sectors such as dark matter \cite{MorrisonTaylor4DClusters,Halverson:2016nfq}). We
    can then count the
    configurations of gauge-breaking flux explicitly, while
estimating the number of horizontal flux configurations
    using statistical methods \cite{DenefLesHouches}.
This can give a sense of the statistical likelihood of realizing the
Standard Model using the construction presented here for a given
geometry.  Combining these global and local analyses of large
classes of models in a systematic way could give a more precise
framework for characterizing the extent to which the construction
presented here is ``natural'' in the string landscape.
    
    \item We have focused here on the chiral part of the matter
    spectrum only, while the full matter spectrum also
    includes vector-like matter like the Higgs. Analyzing
    the
    vector-like spectrum requires explicit cohomology data
    from topologically nontrivial $C_3$ potential
    backgrounds.  These  are usually much harder to
    compute than $G_4$ flux, although recently
    analytical tools have been developed for some special
    cases of these
    \cite{Bies:2014sra,Bies:2021nje,Bies:2021xfh,Bies:2022wvj,Bies:2023sfm}; such analysis goes beyond the scope of this chapter.
    On the other hand, we have a qualitative picture of the
    vector-like spectrum. Since we have started with a gauge
    group $G$ much larger than $\gsm$, generically there
    would be a large amount of vector-like matter, coming from the
    adjoint of $G$. It has been shown in
    \cite{BeasleyHeckmanVafaII} that it is
    impossible to remove all the vector-like exotics
    when the GUT group is $\SO(10)$ or higher\footnote{The argument
      given in \cite{BeasleyHeckmanVafaII} appears in a context where
 the gauge
      divisor is del Pezzo but the same argument holds whenever
the
 gauge divisor has an effective anti-canonical class, and has
      vanishing $h^{2,0}$.}, but
    it may be possible to remove the overly dangerous ones completely from the spectrum, such as $(\mathbf 3,\mathbf 2)_{-5/6}$ as demonstrated in \S\ref{sec:fluxbreaking:example}. We also expect the remaining
    vector-like matter to get large
    masses and lift from the low-energy theory. From this
    point of view, it has not been clear how the Higgs sector
    can be obtained with the right mass within  F-theory or any other
    approach for supersymmetric compactification of string
    theory. It is
    important to address this question if we want to fully
    realize the Standard Model in string theory.
    
    \item It is natural to consider $\U(1)$ extensions to our
    SM-like models, as extra $\U(1)$ factors can be easily
    constructed using the formalism of flux breaking. First,
    recall that some Cartan gauge bosons become massive due
    to vertical flux. In fact, they are still associated with
    global $\U(1)$ symmetries, although we expect that these
    symmetries are further (slightly) broken by other
    effects such as instantons \cite{Banks:2010zn}. Moreover, while $\U(1)$ gauge factors
    usually originate from a
    nontrivial Mordell-Weil group of rational sections in the global
    elliptic geometry
    \cite{AspinwallMorrisonNonsimply,MorrisonVafaII}, the
    $\U(1)$ factors from fluxes only depend on the local
    geometry on $\Sigma$, hence do not constrain the global
    geometry much. The resulting charges can easily be large,
    as shown in \S\ref{ssubsec:fluxbreaking:exotic}. Including these
    $\U(1)$'s in the models presented here can lead to extra selection
    rules and help resolve the puzzles in GUTs
    such as proton decay \cite{Marsano:2009wr,Grimm:2010ez},
    and is important in further studies of these SM-like
    models. In addition, it is interesting to explore the
    possibilities of large $\U(1)$ charges in 4D F-theory
    models from (vertical) flux breaking. (See e.g.
    \cite{Raghuram:2018hjn} for such an analysis in 6D F-theory
    models)
    
    \item Comparing with other tuned SM-like or GUT models,
    the origin of Yukawa couplings in our models is less
    clear. In the tuned models, only matter localized on
    curves $C$ is chiral and the Yukawa couplings are
    between
    three fields on $C$ ($CCC$), which are well understood by
    studying codimension-3 singularities (see, e.g.,
    \cite{WeigandTASI} for a review and further references). In contrast, chiral
    matter in our models may live on both the bulk of
    $\Sigma$ and on matter curves. Hence there are three
    possible types of Yukawa couplings: couplings between three
    fields on the bulk of $\Sigma$ ($\Sigma\Sigma\Sigma$),
    couplings between two fields on the matter curve $C$ and one
    field on $\Sigma$ ($CC\Sigma$), and the above $CCC$
    couplings \cite{BeasleyHeckmanVafaI}. It is
    natural to realize the Higgs on the bulk of $\Sigma$ since
    $\Sigma$ supports much vector-like matter, while a generic matter curve only supports chiral matter \cite{Bies:2020gvf}. The SM Yukawa couplings, which are between two
    chiral fields and the Higgs, thus should
correspond to $\Sigma\Sigma\Sigma$ and $CC\Sigma$ couplings.
    Nevertheless, rigid gauge groups %are 
can be
realized on
    $\Sigma$ with effective
    $-K_\Sigma$
(and therefore also
    $h^{2,0}(\Sigma)=0$), 
as in the explicit example of \S\ref{sec:fluxbreaking:example},
where $\Sigma\Sigma\Sigma$
    couplings are absent by the logic of
    \cite{BeasleyHeckmanVafaI}. Therefore to have
    the correct Yukawa couplings in this situation, extra tuning on
    fluxes must be done such that the chiral matter
    is localized on $C$ only.  While the tuning can
    be easily done in general, it is not possible in
    the example in \S\ref{sec:fluxbreaking:example} since
    $C_\mathbf{27}$ is trivial. Excluding
    this issue, we see no obstruction to having the Standard Model Yukawa
    couplings, but a rigorous construction is still lacking.
    There is a second issue specifically for $E_7$ models:
    as mentioned before,  codimension-3 singularities can arise in
    $E_7$ models with degrees $(4,6,12)$, which cannot be simply
    interpreted as $CCC$ couplings. This fact can also be seen from
    group theory, since $\mathbf{56}^3$ does not contain any
    singlets.
For a complete understanding of rigid $E_7$ flux breaking, the role of
fluxes through extra cycles associated with these singularities should
be better understood.
\end{itemize}

We hope to address some of these issues in future studies.

\chapter{Towards natural and realistic $E_7$ GUTs in F-theory}
\label{chap:e7}

In this chapter, we consider phenomenological aspects of a natural class of
Standard Model-like supersymmetric F-theory vacua realized through
flux breaking of rigid $E_7$ gauge factors.  
%It is fairly likely to 
%realize the 
Three
generations of Standard Model matter
are realized in many of these vacua. 
We further
find that many other Standard Model-like features
are naturally %can easily be
 compatible
with these constructions. For example, dimension-4 and 5 terms associated
with proton decay are ubiquitously suppressed. Many of these features are
due to the %peculiar group theoretical and F-theory geometrical
           %structure 
group theoretical structure
of $E_7$ and associated F-theory geometry. In particular, a set of approximate global symmetries descends
from the $E_7$ group,
leading to exponential suppression of undesired
couplings. The results in this chapter are based on \cite{Li:2024uwf}.

\section{Introduction}
\label{sec:Intro}

String theory provides a consistent framework for a unified theory
that combines gravity with the other fundamental forces described by
quantum field theory.  To describe the real world, however,
ten-dimensional string theory must be compactified on a real
six-dimensional manifold, and various further objects like branes,
fluxes, and orientifolds must be incorporated.  Such constructions give
an enormous number (perhaps on the order of something like
$10^{272000}$ \cite{TaylorWangVacua}) of string theory vacua, known as
the string landscape. As part of the program to realize our Universe
in string theory, it has been a long-standing and primary goal
to find the  structure of the Standard Model (SM) of particle
physics within the string landscape.
%Despite decades of work, 
While many (supersymmetric) string vacua have been identified that share
 many of the principal features of the Standard Model,
there is as yet no single vacuum known in the string landscape that reproduces
all the observed phenomenological details of our world;
% can be 
%realized within
%the string landscape
 for recent reviews of work in this direction, see
\cite{Cvetic:2022fnv,Marchesano:2022qbx}.  

Beyond the simple question of the
  existence of a vacuum matching observed physics, it is
perhaps even more important to understand the extent to which the
physical features of the
Standard Model arise  \emph{naturally} in string
theory.  In other words, we would like to understand the extent to
which solutions like the Standard Model are widespread in the string
landscape or require extensive fine-tuning.  
This is a principal focus of this work and the associated research
program: we take a top-down perspective on the global set of string
vacua and attempt to identify realizations of the Standard Model that
are compatible with the most typical structures arising in string
theory.  %It is natural to 
We
use
F-theory \cite{VafaF-theory,MorrisonVafaI,MorrisonVafaII} to study
these
 questions, as this approach gives a global and nonperturbative
picture of the largest currently understood set of string vacua.
For reviews of F-theory and applications to Standard Model
constructions, see \cite{WeigandTASI,HeckmanReview}.
This chapter describes some more detailed phenomenological aspects of
SM constructions
presented in Chapter~\ref{chap:fluxbreaking}
that are realized through flux breaking of
rigid $E_6$ and $E_7$ gauge factors,
which are relatively common features in F-theory geometries.

Constructing the detailed
Standard Model requires many elements such as the gauge group, the
matter content including both chiral matter and the Higgs, the Yukawa
couplings, a supersymmetry (SUSY)-breaking mechanism, values of the 19
free parameters, and possibly some room to address beyond-SM problems
as well as cosmological aspects such as the density of dark energy. 
Unfortunately, the current available string theory techniques are far
from enough to compute all these features precisely. Among the above SM 
features, string theory techniques for constructing the
gauge group and the chiral spectrum are well-established. While there is
some recent progress on the Higgs sector \cite{Bies:2021nje,Bies:2021xfh,Bies:2022wvj,Bies:2023jqg, Bies:2023sfm} and the Yukawa
couplings \cite{Cvetic:2019sgs} in a large class of F-theory models,
so far no fully precise statement on the realization of these features
in a way that matches observed physics
has been made in this context.
On the other hand, incorporating these established features
with e.g. SUSY breaking is far beyond our current techniques.  
Although at 
this moment no complete  realization of the Standard Model
has been constructed in any version of string theory, if we can identify a 
natural class of models that realize a decent portion of
the coarsest features of the SM, 
these structures may
naturally correlate with certain other features of the SM or beyond SM
physics. We will explore this philosophy in this chapter.

One obvious way in which the
models studied here (and elsewhere in much of the string theory literature)
differ from observed physics is that we focus on solutions with
supersymmetry.  Supersymmetry has not yet been observed at low (TeV or
below) energies in nature, but as a theoretical tool it increases our
level of analytic control.  By studying solutions with supersymmetry,
we can gain some perspective on global aspects of the string
landscape.  Of course, eventually we need to understand
non-supersymmetric solutions to match observed physics.  One
possibility is that the physics we see is in a broken-symmetry phase
of a theory with supersymmetry at energies beyond the TeV scale.
Even if supersymmetry is broken at the Planck or string scale,
many insights gained by exploring the space of supersymmetric vacua may
be relevant to the less controlled
non-supersymmetric vacua.

\subsection{Natural vs. tuned  features}
\label{subsec:natural}

Before describing our results, it is worth clarifying the concept of
\emph{naturalness} used in this chapter.
To obtain vacua with all the SM features considered in this chapter, 
quite a few specific choices must be made in the
construction of vacua.  A
list of such choices in the context of the models studied in this
chapter
 is summarized in \S\ref{subsec:overview}, and the mathematical
 conditions imposed for such choices are given at the beginning of
\S\ref{sec:ExplicitConstruction}. 
%Many of the choices are natural in
%different extents, 
The extent to which these different kinds of choices are natural varies,
within a hierarchy of naturalness/tuning.
Roughly speaking, each of the choices made in constructing a specific
class of string vacua can be characterized as
belonging to one of the following categories:

\begin{enumerate}
\item Physical constraints: These constraints come from string theory itself
and must be satisfied in all string compactifications. These constraints
ensure physically sensible vacua that have, e.g., Poincar\'{e} (or AdS)
invariance. Examples include tadpole cancellation and primitivity of fluxes.

\item Ubiquitous/common conditions: Let us consider a reasonably large
but presumably finite set of 
string vacua or compactification geometries, such as ${\cal N} = 1$ 4D
F-theory vacua or the associated set of topologically distinct
 complex
threefold bases that support elliptic Calabi-Yau fourfolds.
A condition is common if it holds for
an ${\cal O} (1)$ fraction
of the set of vacua or geometries, considered with a simple counting
measure. In particular, the condition is ubiquitous if it holds for a
substantial majority. As examples, the existence of rigid
$E_8$ gauge factors in (known) F-theory base geometries is ubiquitous, and
that of $E_7$ gauge factors appears to be common. (See, e.g.,
\cite{HalversonLongSungAlg,TaylorWangLandscape,Taylor:2025gnp} and discussions below)

\item Fairly likely conditions: Sometimes, there are a family of
  similar conditions, such as possible values of a discrete
  parameter. Each possible value may
only hold for a relatively small fraction of vacua
  within the above set, so that none of the conditions are
  ubiquitous. We refer to a condition as being fairly likely if the
 fraction of vacua or geometries with this property
 is considerably higher than for most of the
  other possibilities. 
As an illustration, we would say that rolling a sum of 6 on a pair of
six-sided dice is ``fairly likely,'' although rolling a sum of 7 is slightly
more likely.
%In other words, 
As another example, 
%preferred conditions are represented by
  points near the peaks in a distribution of some discrete parameters
  correspond to fairly likely conditions.  See
  \cite{Andriolo:2019gcb} for more discussions along these lines. As
%  an
a further example, \cite{Li:2021eyn,Li:2022aek} argued that three
  generations of chiral matter is fairly likely in this sense in
  our $E_7$ model (although, for example, zero generations may be  more
  likely). (See also, e.g., \cite{BraunWatariGenerations}.)

\item Natural choices: These are  choices for discrete parameters having
many possible values that are not (obviously)
preferred in any way, but 
imposing a particular chosen value does not require 
exponential amounts of tuning. Such choices may
hold at the level of, e.g., 0.1\% of the given set of vacua.
%These choices are
%necessary conditions
Such choices may be needed for obtaining some \emph{qualitative}
features of observed phenomenology in some constructions.  For
example, obtaining the SM gauge group and matter representations from
flux breaking of $E_7$ involves some choices of fluxes given by mild
linear constraints, which seem to be natural in this sense, although
they do not seem to be preferred in any particular way over other
choices that would give a variety of other possible groups and
representations.

\item Fine-tuning: These are choices involving setting one or more
continuous variables to take specific values, 
or making an exponentially rare choice among discrete possibilities.
Vacua based on such choices
 are increasingly non-generic 
in the landscape as the number of such tunings increases. In some 
constructions of string vacua, such
choices are needed to obtain  certain \emph{qualitative} features of
observed phenomenology.
For example, a tuned $\SU(5)$ or SM
gauge group in F-theory involves extensively fine-tuning many moduli to
specific values
\cite{BraunWatariGenerations} (unless these moduli are somehow
automatically tuned by a specific class of flux choices). 
Notably, it seems that
no such fine-tuning is involved in our $E_7$
models.

\item Technical choices: To facilitate analytic control of the vacua
  and make some particular calculations manageable, in some cases
  technical choices are made by restricting attention to some specific
  relatively simple choices of vacua. These choices are not necessary
  either for physical or phenomenological consistency, but are made to
  illustrate specific examples as simply as possible. The features of
  the models chosen in this way should be representative of some
  larger class of vacua or geometries.  
In some situations, technical choices can be made just to simplify
calculations that are in principle possible and expected to give
qualitatively similar results for all other choices.  In other cases,
technical choices are made where it is not clear how to do the
computation explicitly in general, and/or whether a completely general
choice will give qualitatively similar results. If not, some choices or
tuning of one of the above types may be
implicitly involved.  For the specific technical choices made here,
we have some confidence that
  similar results should also hold for a broader  class of vacua
  without those technical choice. Nevertheless, some qualitative
  simplifications occur based on these choices, thus
more
  explicit further studies are required to understand the extent to which these technical choices
  are relevant for phenomenologically interesting features. Examples of 
technical
  choices include picking some certain topological types for the
  compactification, which we do in this chapter
(specifically by choosing  models where the gauge divisor is a del
  Pezzo surface and the matter curve is a $\P^1$)
%for explicit  computations in some examples.
to facilitate and simplify the analysis; these technical choices made
here fall in the latter category above
that may implicitly involve some more or less natural choice or
tuning,
 as they may affect qualitative
aspects of the low-energy physics.
\end{enumerate}

While the term ``natural'' is used widely in many different ways in
the literature, we attempt to use the above classification to be slightly
more precise about the types of choices involved in the construction of
our models and the realization of phenomenological features.  This is a  coarse characterization, however, as choices and
tunings can occur across a broad spectrum, and
we do not attempt to make any precise
division between the gradations of ``common,'' ``fairly likely,'' and
``natural'' conditions.  In particular,
we do not have a perfect understanding of the class of string
geometries or F-theory compactifications, so any attempt at classification of
this type is necessarily quite imperfect given the current state of knowledge.
Moreover, the
measure problem on the landscape is not at all understood, so we
really do not have any good sense of the proper probability measure to
use on the landscape.  Nonetheless, in the absence of any known or
conjectured dynamical mechanism that would modify these considerations,
features that seem to require exponentially large amounts of fine-tuning under a
simple counting measure seem likely to occur less frequently in a
large string multiverse than  features that are ubiquitous,
fairly likely, or even natural in the preceding terminology.  In principle, even without solving the
measure problem, this may give us some insight into the extent to
which the Standard Model may be realized naturally in string theory,
and what BSM physics may be most naturally associated with those SM structures.

\subsection{Review of previous work}
\label{subsec:e7:review}

In recent years, F-theory
 has become a particularly
promising framework for studying many aspects of
 string compactifications and
phenomenology, as it provides a global description of a large
connected class of supersymmetric string vacua. (See
\cite{WeigandTASI} for a review.) In particular, F-theory gives 4D
$\mathcal N=1$ supergravity models when compactified on 
elliptically fibered Calabi-Yau (CY) fourfolds $Y$, corresponding to
non-perturbative compactifications of type IIB string theory on general
(non-Ricci flat) complex K\"ahler threefold base manifolds $B$.  The
number of such threefold geometries $B$  seems to already be on the order
of $10^{3000}$ for toric bases $B$
\cite{TaylorWangMC,HalversonLongSungAlg,TaylorWangLandscape}, without
even considering the exponential multiplicity of fluxes possible for
each geometry, although the number of flop equivalence classes of bases is
somewhat smaller
%, on the order of $10^{50}$
\cite{Taylor:2025gnp}.  F-theory is also known to be dual to many other types
of string compactifications such as heterotic models.  Briefly,
F-theory is a strongly coupled version of type IIB string theory with
non-perturbative configurations of 7-branes balancing the curvature of
the compactification space. The non-perturbative brane physics is
encoded geometrically into the elliptically fibered manifold, which
can be analyzed using powerful tools from algebraic geometry. The
gauge groups and 
chiral
matter content supported on these branes can then be
easily determined when combined with flux data.

Applying the above
techniques, many SM-like constructions of 4D F-theory models with the gauge group $\gsm=\SU(3)\times \SU(2)\times \U(1)/\mathbb
Z_6$ have been
achieved in the literature. The early literature, starting from 
\cite{Donagi:2008ca,BeasleyHeckmanVafaI,BeasleyHeckmanVafaII,DonagiWijnholtGUTs}, focused on the breaking of GUT groups of $\SU(5)$ and its $\U(1)$
extensions
\cite{Blumenhagen:2009yv,Marsano:2009wr,Grimm:2009yu,KRAUSE20121,Braun:2013nqa}, 
while there has also been some study of $\SO(10)$ 
\cite{Chen:2010ts} and $E_6$ 
\cite{Chen:2010tg,Callaghan:2012rv,Callaghan:2013kaa} GUTs. (See \cite{WeigandTASI,HeckmanReview} for more extensive reviews.) These constructions break the GUT group using the
so-called hypercharge flux further discussed in
\cite{Mayrhofer:2013ara,Braun:2014pva}, which is a kind of ``remainder''
flux \cite{Buican:2006sn,Braun:2014xka} to be reviewed below. Some later constructions 
tried to construct $\gsm$ directly without any symmetry breaking, with the recent culmination of finding $10^{15}$ explicit solutions of
directly tuned $\gsm$ with three generations of SM chiral
matter (a ``quadrillion Standard Models'' \cite{CveticEtAlQuadrillion}), 
based on the ``$F_{11}$'' fiber in \cite{KleversEtAlToric}. These 
constructions are further generalized in \cite{Raghuram:2019efb,Jefferson:2022yya}. Although these models nicely
capture some of the most important phenomenological features, they face
one common issue: 
In terms of the notions discussed in \S\ref{subsec:natural}
the gauge groups in these constructions are
highly ``fine-tuned'', namely they are
obtained by setting specific values for many complex structure
moduli. 
Furthermore, on
most F-theory bases such tuning of $\gsm$ is forbidden due to the presence
of rigid gauge groups (to be discussed shortly). Even if the tuning is
available, it may not be compatible with moduli stabilization by fluxes and/or nonperturbative effects.

A more natural class of SM-like constructions in F-theory comes from
rigid gauge groups such as $E_7,E_6$ \cite{MorrisonTaylorClusters,MorrisonTaylor4DClusters}. These are gauge groups enforced by
strong curvature (to be more precise, very negative normal bundle) on the
base, and are present throughout the whole branch of moduli space
over that base, hence 
avoid the issue of tuning moduli. Moreover, statistical studies on (toric)
F-theory bases have suggested that these rigid gauge groups are
%ubiquitous
fairly common
in the landscape.
While the specific base naively associated with the most flux vacua
\cite{TaylorWangVacua}
does not contain $E_7$ or $E_6$ factors,
 these gauge factors arise in a substantial fraction of F-theory base
 geometries enumerated by a simple counting measure (which may or may
 not distinguish bases related by a flop).
The fraction of toric bases for 6D F-theory models that contain
rigid $E_7$ and $E_6$ factors is more than 50\% \cite{MorrisonTaylorToric}.
The statistics of $E_7$ and $E_6$ factors in threefold bases for 4D
F-theory models is less well understood; one study found
$E_7$ factors in $\sim 20$\% of a limited simple of bases
\cite{TaylorWangMC}, and a more detailed analysis of the prevalence of
such factors is studied in \cite{Taylor:2025gnp}.
Nonetheless, breaking these gauge groups to $\gsm$ should
give us a very large set of SM-like constructions. In Chapter~\ref{chap:fluxbreaking}, 
we have proposed a general class of
SM-like models using rigid $E_7,E_6$ GUT groups in
F-theory, with an intermediate $\SU(5)$ group. These models 
enjoy the advantages of being natural
and involving little or no fine-tuning.
  Specifically, a combination of ``vertical''
and ``remainder'' fluxes can be used to break the rigid gauge groups
in a way that is not
transparent in the low-energy field theory, but
gives the correct SM gauge group and some chiral matter.
Although in many cases the breaking leads to exotic chiral
matter, there are large families of models in which the
correct SM chiral matter representations are obtained through
an intermediate $\SU(5)$.  The number of generations
can easily be small and we have demonstrated that three
generations are fairly likely in many of these models. In particular,
a fully global explicit construction of such an $E_6$ model has been given.

\subsection{Overview of results}
\label{subsec:overview}

As discussed before, 
%we may hope to naturally find more SM features in this
%large set of SM-like constructions. 
one might hope that
there are string realizations of the Standard Model in which most or
all of the features observed in nature arise in a relatively natural
way.
In this chapter, we show that apart from
the SM gauge group and chiral spectrum, several
additional
 important SM features can be easily obtained in the $E_7$ (but not 
$E_6$) models, with some additional but mild tuning on the geometry and 
flux background. Specifically, we obtain the following features, 
%by making 
%the following choices,
each of which depends upon making choices with
 various extents of naturalness:
\begin{itemize}
    \item As mentioned above, rigid $E_7$ factors are 
%ubiquitous
quite common in the F-theory landscape, and may be natural or likely
depending on the proper vacuum measure. 
For generic (non-toric) bases,
    the $E_7$ gauge group can be broken down to $\gsm$ by some natural choices
    of vertical and remainder fluxes.
\item 
For the models with flux breaking of $E_7 \rightarrow\gsm$, a set of
approximate global $\U(1)$ symmetries descend from the $E_7$, leading to
exponential suppression of certain couplings.
    \item With the $E_7 \rightarrow\gsm$ gauge-breaking fluxes, it 
appears to be fairly likely to have 3 generations of SM chiral
matter (although 0, 1, or 2 generations may be more likely), and
fairly likely that the
    exotic $(\mathbf 3,\mathbf 2)_{-5/6}$ representation is removed
from the spectrum.
    \item Due to the use of $E_7$, there are always candidate Higgs 
    sectors with a string
    theory origin different from that of chiral matter. Such a structure
    automatically leads to distinct dynamics between the Higgs and chiral
    matter, and gives rise to unsuppressed SM Yukawa couplings.
    \item Under this setup, dimension-4 and 5 proton decay is
      ubiquitously suppressed to phenomenologically safe levels.
\end{itemize}

The distinction between the Higgs and chiral matter, the appearance of the approximate global $\U(1)$ symmetries, and the
ubiquitous suppression of dimension-4 and 5 proton decay are the
strongest features of these constructions, in which desirable
properties associated with observed physics arise essentially
automatically.  Most of
the remaining features we explore generally require
small amounts of discrete tuning. They may involve common, fairly
likely or natural choices and do not arise automatically,
but do not seem to require extensive fine-tuning.

\begin{itemize}
    \item There is some automatic splitting between the doublet and triplet masses, although the amount of splitting and the exact masses are unknown.
    \item 
%Despite the presence of additional fields, 
%we can realize the correct set of Yukawa couplings with some further
%natural choices of fluxes. 
Although there are extra charged vector-like exotics in the spectrum, the Yukawa couplings
between most of these fields (all besides the triplet Higgs) and the
SM matter are exponentially suppressed through the
above-mentioned approximate symmetries. We call these fields \emph{inert}
vector-like exotics.

\item It is plausible that there
    is some hierarchy in the SM
Yukawa couplings, but the exact values are unknown.
    \item With the setup so far, the model contains three right-handed
    neutrinos with masses lower than the string/GUT scale. It is plausible but not fully
    clear that the seesaw mechanism occurs.
    %\item Gauge coupling unification is achieved by assuming the presence of several inert
    %vector-like exotics at TeV scales, which are well within the current
    %observational bounds but may be observable in the near
    %future.
\end{itemize}

 To facilitate the discussions and calculations in this chapter, we technically choose the gauge divisor to be a del Pezzo surface, and the
    matter curve to be a $\mathbb P^1$.  Although we expect similar
    results for many other choices, these choices do lead to some
    qualitative simplifications in the analysis, and further work is
    needed to determine whether  low-energy models
with similar structure arise for a broader
    class of gauge divisors and matter curves, and/or to determine how
    natural or fine-tuned these geometric choices may be.

As an 
example, we  work out an explicit global construction of the $E_7$ models that 
%explicitly
 realizes all of the above SM features. We emphasize that many
of these phenomenological advantages are specific for the $E_7$ models,
and may be (much) harder to realize in other types of SM-like constructions
in F-theory. Some of the above features are inherited from the group 
structure of $E_7$ itself, regardless of the string theory physics. To the
authors' knowledge, however, these group theoretical features
have not been noticed in the field theory literature, probably because $E_7$
itself does not support any chiral matter, if there is no additional input
like fluxes from the UV.

While
 the $E_7$ models considered in this chapter
 have quite a few phenomenological advantages
 over some other stringy realizations of the Standard Model, 
we
note
that these models potentially still suffer from the following
issues, in light of which extra care must be taken when interpreting  the
results presented here. First, these models contain many vector-like exotics that cannot be
removed by fluxes (except the most dangerous $(\mathbf 3,\mathbf
2)_{-5/6}$, which is fairly likely
to be absent).  In particular, these exotics
include other copies of the Higgs field. From the effective field theory
perspective, we generically expect these exotics to get heavy masses
near the GUT/string scale such that they do not affect the low-energy
phenomenology. On the other hand, this expectation in general may not
be true in string theory, and it is important to develop further
techniques to ensure the right masses.  Although it may be possible,
we do not see any reason in these models why one of the Higgs doublet
pairs
should get much
lower masses than the other copies. In other words, there is no
totally clear solution to the $\mu$-problem in our setup. Next, these
$E_7$ models have codimension-3 $(4,6)$ singularities on the base,
which correspond to an extra family of flux and may be associated to
an extra sector of strongly coupled superconformal and chiral matter
\cite{Candelas:2000nc,Lawrie:2012gg,Achmed-Zade:2018idx,Jefferson:2021bid,46}. We
can easily control the flux such that this sector is non-chiral, but
since we understand very little about this sector, further studies are
needed to ensure that this sector does not affect phenomenology.

\subsection{Outline of chapter}

This chapter is organized as follows. We first extend the flux breaking
formalism developed in Chapter~\ref{chap:fluxbreaking} to compute the
vector-like spectrum in \S\ref{subsec:e7:vectorlike}.
Despite the difficulty of computing the vector-like spectrum in general,
the vector-like spectrum can be completely determined in some special 
cases. We discuss how our $E_7$ models easily fit into these cases, so that
we can fully compute the matter spectrum in our models.

To initiate our discussions on semi-realistic $E_7$ GUTs in F-theory, 
in \S\ref{sec:E7Review} we
first summarize the $E_7$ models proposed in Chapter~\ref{chap:fluxbreaking}. We describe the geometry and 
fluxes needed to get the SM gauge group and three generations of SM chiral
matter from a rigid $E_7$.

In \S\ref{sec:Pheno}, we discuss various phenomenological aspects of the $E_7$ 
models proposed in Chapter~\ref{chap:fluxbreaking}, namely the vector-like matter, 
Yukawa couplings, proton decay, the Higgs sector, the neutrino sector, and
gauge coupling unification. One of our main tools is the St\"uckelberg mechanism \cite{DonagiWijnholtModelBuilding,Grimm:2010ks,Grimm:2011tb} used
in our flux breaking of $E_7$ to $\gsm$, which leaves several approximate
$\U(1)$ global symmetries. We discuss these symmetries in detail and 
study how they constrain the couplings and mass 
terms\footnote{Throughout this chapter, ``mass terms'' refer
to the $\mu$-term and other similar terms in the superpotential, which involve two different fields.} in the low-energy 
theory. These constraints, plus some additional tuning on the fluxes, lead
to many of the above phenomenological advantages, especially the 
ubiquitous suppression of proton decay. We also discuss the vector-like 
matter that appears in the spectrum. We demonstrate how to easily remove the
most dangerous $(\mathbf 3,\mathbf 2)_{-5/6}$ vector-like exotic, and 
discuss why most other vector-like exotics are \emph{inert}. Although we 
cannot make
any precise statements, we discuss various possible origins of the 
vector-like (including the Higgs) masses. Based on such discussions, we
make some brief comments about the Higgs sector, the neutrino sector, and
gauge coupling unification.
%and suggest some estimates on
%the mass scales based on a rough calculation on gauge coupling unification.
%In this process, we will see that it is unavoidable to have several inert
%vector-like exotics at TeV scales, although we do not explain their 
%origins. These exotics include both the Higgs and the other SM 
%representations, but they do not violate current experimental bounds.

After describing the recipe of getting semi-realistic $E_7$ GUTs F-theory,
in \S\ref{sec:ExplicitConstruction} we write down an explicit global construction of a $E_7$ model
that achieves all the above phenomenological features. This example
demonstrates the fact that these features can indeed be obtained through 
some mild tuning on the geometry and the fluxes, but %not any
without the necessity of fine-tuning any
 moduli.
Therefore, it is reasonable to regard these features as being natural in
the string landscape. To emphasize various advantages and disadvantages of
the $E_7$ models, in \S\ref{sec:Comparison} we briefly compare our models with other
SM-like F-theory constructions in the literature. We finally conclude in \S\ref{sec:Conclusion}. 
In Appendix \ref{appendix:e7}, we discuss several technical tools that are useful
in the construction in \S\ref{sec:ExplicitConstruction}.

\section{Vector-like matter from flux breaking}
\label{subsec:e7:vectorlike}

The flux breaking formalism developed in Chapter~\ref{chap:fluxbreaking}
induces vector-like matter apart from chiral matter. In this section,
we describe how vector-like matter arises in these models, and how the 
spectrum can be computed in some special cases.

To fully understand the phenomenology of the $E_7$ models, it is important to study
the vector-like spectrum 
%apart from
in addition to the chiral spectrum. One of the main reasons 
for this is that in
a realistic model we need to realize the Higgs sector, while avoiding
dangerous vector-like exotics.  While the techniques for computing the
chiral spectrum are already at hand, computing the
vector-like spectrum in general requires not only the $G_{4}$ flux,
but the full information of $C_{3}$ in terms of line bundles on $\Sigma$
and $C$.\footnote{In general, these are described by sheaves when
  there are more severe singularities on $\Sigma$ and/or $C$. In this
  chapter, we only consider completely smooth geometry on the bases, so
  the description by line bundles is sufficient.} In many cases, these
things are hard to compute and some of the relevant technology has
only been developed fairly recently
\cite{Bies:2021nje,Bies:2021xfh,Bies:2022wvj,Bies:2023jqg}. Fortunately,
our models admit several important simplifications such that the
$G_{4}$ flux itself already determines the vector-like spectrum.

Let us first focus on vector-like matter that lives on the bulk of $\Sigma$.
We follow the formalism in \cite{Bies:2014sra,Bies:2017fam}. At least in most cases, the full $C_3$ can be captured by an
algebraic complex 2-cycle $\mathcal{A}$ in the Chow group $\mathrm{CH}^{2}(\hat{Y})$
(algebraic cycles modulo  rational equivalence
instead of homological equivalence),
with homology class is $\left[G_{4}\right]$ \cite{Braun_2012}. We consider the restriction of $\mathcal{A}$
onto a Cartan divisor $D_{i}$, given by the intersection product
$\mathcal{A}\cdot D_{i}\in\mathrm{CH}^{2}\left(D_{i}\right)$. Its
projection onto $\Sigma$, given by $\pi_{*}\left(\mathcal{A}\cdot D_{i}\right)\in\mathrm{CH}^{1}\left(\Sigma\right)$,
is a curve on $\Sigma$ associated with the line bundle
\begin{equation}
L_{i}=\mathcal{O}_{\Sigma}\left(\pi_{*}\left(\mathcal{A}\cdot D_{i}\right)\right)\,.
\end{equation}
Now for each weight $\beta=-b_{i}\alpha_{i}$ of the adjoint,
we define the line bundle
\begin{equation}
L_{\beta}=\otimes_{i}L_{i}^{b_{i}}\,.
\end{equation}
Then the chiral and anti-chiral multiplicities for $\beta$ are counted
by the following sheaf cohomologies \cite{Donagi:2008ca,BeasleyHeckmanVafaI}
\begin{align}
\mathrm{chiral}&:\quad H^{0}\left(\Sigma,L_{\beta}\otimes K_{\Sigma}\right)\oplus H^{1}\left(\Sigma,L_{\beta}\right)\oplus H^{2}\left(\Sigma,L_{\beta}\otimes K_{\Sigma}\right)\,, \\
\mathrm{anti-chiral}&:\quad H^{0}\left(\Sigma,L_{\beta}\right)\oplus H^{1}\left(\Sigma,L_{\beta}\otimes K_{\Sigma}\right)\oplus H^{2}\left(\Sigma,L_{\beta}\right)\,.
\end{align}

Notice that the sheaf cohomologies for chiral and anti-chiral
matter are related by Serre duality. To calculate their dimensions,
we apply the following two simplifications \cite{BeasleyHeckmanVafaI}. First, for
fluxes with \emph{nontrivial} $L_\beta$ and satisfying primitivity
(i.e. preserving SUSY), we have $H^{0}\left(\Sigma,L_{\beta}\right)=H^{2}\left(\Sigma,L_{\beta}\otimes K_{\Sigma}\right)=0$.
Next, we %restrict
assume that
 $\Sigma$ is a rational surface with effective
$-K_{\Sigma}$,\footnote{As shown below, the condition that
$-K_\Sigma$ is effective is a reasonable simplifying assumption in the
context of rigid gauge groups. All toric surfaces are rational and
have effective $-K_\Sigma$.
We assume these conditions on $\Sigma$ in the
rest of this chapter; in much of this chapter we restrict attention to the
special case where $\Sigma$ is a (generally non-toric)
del Pezzo surface.
Further work would be needed to understand the detailed structure of
the resulting models when these technical conditions are relaxed.} %then
in this case, we have $H^{2}\left(\Sigma,L_{\beta}\right)=H^{0}\left(\Sigma,L_{\beta}\otimes K_{\Sigma}\right)=0$.
Therefore, the exact multiplicities $n_{\beta}$ and $n_{-\beta}$ are fully
determined by $h^{1}\left(\Sigma,L_{\beta}\right)$ and $h^{1}\left(\Sigma,L_{\beta}\otimes K_{\Sigma}\right)$
respectively. Since only $H^1$ is nontrivial, the multiplicities are also captured by the topological Euler
characteristics $\chi\left(\Sigma,L_{\beta}\right)$ and $\chi\left(\Sigma,L_{\beta}\otimes K_{\Sigma}\right)$.
These are fully determined by $c_{1}\left(L_{\beta}\right)$, given
by the Hirzebruch-Riemann-Roch theorem: \cite{Donagi:2008ca,BeasleyHeckmanVafaI,Blumenhagen:2008zz}
\begin{align}
\label{eq:HirzebruchRiemannRoch}
n_{\beta} & = -\chi\left(\Sigma,L_{\beta}\right) \nonumber \\
& =\frac{1}{2}\left[c_{1}\left(L_{\beta}\right)\right]\cdot K_{\Sigma}-\frac{1}{2}\left[c_{1}\left(L_{\beta}\right)\right]^{2}-1\nonumber \\
 & =\frac{1}{2}\chi_{\beta}-\frac{1}{2}\left[c_{1}\left(L_{\beta}\right)\right]^{2}-1\,,
\end{align}
where $\chi_{\beta}$ is the chiral index for $\beta$ given in \S\ref{subsec:fluxbreaking:matter}, and
the Poincar\'e dual is taken with respect to $\Sigma$.
The expression for $n_{-\beta}$ is the same except that the sign of the
first term is flipped; indeed we get back $n_\beta-n_{-\beta}=\chi_\beta$.
In our models where only gauge-breaking fluxes are turned on, we can
read off $c_{1}\left(L_{\beta}\right)$ from \eqref{eq:review:Mred-simple}:
\begin{equation}
\label{eq:c1ofL}
\left[c_{1}\left(L_{\beta}\right)\right]=-b_{i}\kappa^{ij}\left(\phi_{j\alpha}\Sigma\cdot D_\alpha+\phi_{jr}C_{\mathrm{rem}}\right)\,.
\end{equation}
Combining Eqs. (\ref{eq:HirzebruchRiemannRoch}) and (\ref{eq:c1ofL}), 
this gives us a formula to compute the exact matter multiplicities from
the bulk of $\Sigma$ in terms of the vertical and remainder flux
parameters.

Now we turn to vector-like matter localized on matter curves. Similarly for a weight $\beta\in R$ supported on $C_R$,
we consider the pullback of $\mathcal A$ onto a matter surface $S(\beta)$
given by $\mathcal A\cdot S(\beta)\in \mathrm{CH}^{2}\left(S(\beta)\right)$. 
Its projection onto $C_R$, given by $\pi_*(A\cdot S(\beta))\in \mathrm{CH}^{1}\left(C_R\right)$ defines a line bundle for $\beta$:
\begin{equation}
L_\beta=\mathcal{O}_{C_R}\left(\pi_{*}\left(\mathcal{A}\cdot S(\beta)\right)\right)\,.
\end{equation}
Then the chiral and anti-chiral multiplicities for $\beta$ are counted
by the following sheaf cohomologies
\begin{align}
\label{eq:mattercurvecohomologies}
\mathrm{chiral}&:\quad H^{0}\left(C_R,L_{\beta}\otimes \sqrt{K_{C_R}}\right)\,, \\
\mathrm{anti-chiral}&:\quad H^{1}\left(C_R,L_{\beta}\otimes \sqrt{K_{C_R}}\right)\,,
\end{align}
where $\sqrt{K_{C_R}}$ is the spin bundle on $C_R$. These sheaf 
cohomologies are more subtle than those for the bulk of $\Sigma$. While
they are well understood when the matter curve has genus 0 or 1, for
irreducible curves with higher genus, these cohomologies have complicated dependence on moduli,
and their dimensions can jump at special points in the moduli space. For
reducible curves, there can also be vector-like pairs between 
different irreducible components of the curves, if the total chiral index
is split into different components accordingly. Instead of running into
all these subtleties, below we just focus on a special case where the
matter curve is simply a $\mathbb P^1$. In this case, there can never be
any vector-like pairs from the matter curve, since for $\mathbb P^1$ only
one of the $H^0,H^1$ is nontrivial, depending on the sign of the line 
bundle. 
This means in particular that the vector-like spectrum is independent of the choice
of spin bundle, significantly simplifying the analysis.
As shown below, this geometry is 
not hard to achieve in our $E_7$ models, and we leave the generalizations
to more complicated matter curves in future work.

So far, we have discussed the vector-like multiplicities for each
weight separately. On the other hand, we recall that weights from 
different $b_{i'}$ and $R$ can contribute to the same chiral $R'$ in Eq.\ 
(\ref{fullchiR'}) during flux breaking. Similarly, these different weights
can form vector-like pairs after flux breaking, even if each weight is 
purely chiral. This effect
can occur on both matter curves and the bulk of $\Sigma$. Such vector-like
matter has qualitatively different behavior from that obtained from sheaf
cohomologies, and has interesting phenomenological implications. More 
details will be discussed in later sections.

\section{Review of $E_7$ GUTs in F-theory}
\label{sec:E7Review}

With the above extension of the flux breaking formalism, now we are ready to discuss the
$E_7$ models in more detail. For completeness, first we briefly summarize the $E_7$ models
in Chapter~\ref{chap:fluxbreaking}, namely
%getting
describing how the SM gauge
group and chiral spectrum can be realized in a
 natural way through flux breaking of a rigid $E_7$ factor.

\subsection{Flux breaking of rigid $E_7$ factors}

Recall that gauge groups in F-theory arise from sufficiently high degrees
of fiber singularities on a gauge divisor $\Sigma$. For an $E_7$ gauge 
group, the (singular) elliptic \cy fourfold $Y$
is described by a certain form of Weierstrass model 
\cite{Kodaira,Neron,BershadskyEtAlSingularities}. Treating the elliptic curve as the
\cy hypersurface in $\mathbb{P}^{2,3,1}$ with
homogeneous coordinates $[x:y:z]$, $Y$ is given by the
locus of
\begin{equation}
y^2 = x^3 + s^3 f_3xz^4 + s^5 g_5 z^6 \,,
\label{eq:e7}
\end{equation}
where $s, f_3, g_5$ are sections of line bundles ${\cal O} (\Sigma),{\cal O} (-4K_B-3
\Sigma),{\cal O} (-6K_B-5 \Sigma)$ on the base $B$,
and the gauge divisor
$\Sigma$
 supporting the $E_7$ factor  is
given by $s=0$. There is adjoint matter $\mathbf{133}$ arising from
excitations localized around the bulk of
$\Sigma$. There is also fundamental matter $\mathbf{56}$ localized on
the curve $s= f_3=0$, or $C_{\mathbf{56}}=-\Sigma \cdot (4K_B+3\Sigma)$
in terms of the intersection product, when the curve is nontrivial in
homology. When $\Sigma$ has a sufficiently negative normal bundle 
$N_\Sigma$, singularities of the elliptic fibration are enforced on $\Sigma$, and the 
Weierstrass model for $Y$ is automatically restricted to the form 
 (\ref{eq:e7}). A rigid $E_7$ is then realized on $\Sigma$. To be precise, we can consider the following divisors
on $\Sigma$ (not on $B$) \cite{MorrisonTaylor4DClusters}:
\begin{align}
    F_k &= -4K_\Sigma+(4-k)N_\Sigma\,, \nonumber \\
    G_l &= -6K_\Sigma+(6-l)N_\Sigma\,,
\label{eq:nonHiggsable}
\end{align}
where $k,l$ are integers. Then there is a rigid $E_7$ on $\Sigma$ if 
$F_k,G_l$ are effective for $k\geq 3,l\geq 5$ only. A simple way to 
satisfy this condition is to consider effective $-K_\Sigma$ and
$-N_\Sigma$, such that $-3K_\Sigma+N_\Sigma$ is not effective but 
$-4K_\Sigma+N_\Sigma$ is effective. As discussed
% before
previously, the natural choice of effective $-K_\Sigma$ 
matches nicely with the simplification we made in \S\ref{subsec:e7:vectorlike} for computing
vector-like spectrum, and we assume that this condition holds.

After setting up the geometry, we now turn on the flux background.
We break $E_7$ to $\gsm$ in two steps. Since remainder flux preserves
$\U(1)$'s along the roots but vertical flux does not, we first break 
$E_7$ to $\SU(5)$ with vertical flux, then break $\SU(5)$ to $\gsm$ with
remainder flux. The latter flux is very similar to the hypercharge flux
in traditional $\SU(5)$ GUTs in F-theory. To perform the first step
of breaking, we turn on nonzero $\Theta_{i'\alpha}$ for some $\alpha$ and
$i'=4,5,6$ subject to the flux constraints listed in \S\ref{subsec:review:G4fluxes}, 
see Figure \ref{dynkine7}. In terms of flux parameters $\phi_{i\alpha}$, we turn on
\begin{equation}
    \phi_{1\alpha}=2n_\alpha\,,\quad\phi_{2\alpha}=4n_\alpha\,,\quad\phi_{3\alpha}=6n_\alpha\,,\quad
    \phi_{4\alpha}=5n_\alpha\,,\quad\phi_{7\alpha}=3n_\alpha\,.
    \label{eq:e7:phi-n}
\end{equation}
The values of $\phi_{5 \alpha}, \phi_{6 \alpha}$, if sufficiently 
generic, do not affect the gauge group, but they will be fixed by other
flux and phenomenological constraints. Here we define a new set of flux parameters
$n_\alpha$, which can be integers or half-integers depending on the parity
of $c_2(\hat Y)$. 
At this point the gauge group has been broken to SU(5).
To perform the second step of breaking, we similarly
turn on the remainder flux
\begin{equation}
    \phi_{1r}=2n_r\,,\quad\phi_{2r}=4n_r\,,\quad\phi_{3r}=6n_r\,,\quad
    \phi_{7r}=3n_r\,,
    \label{eq:phi-r}
\end{equation}
and $\phi_{4r},\phi_{5r},\phi_{6r}$ plays the same role as $\phi_{5 \alpha}, \phi_{6 \alpha}$. Here $n_r$ is always integer. Under the
construction of rigid $E_7$, we require a non-toric base $B$ to ensure
the existence of $C_\mathrm{rem}$, hence this remainder flux.
After the remainder flux breaking, the remaining unbroken gauge group
is
\begin{equation}
\gsm=\SU(3)\times \SU(2)\times \U(1)/\mathbb
Z_6\,.
\label{eq:SMgaugegroup}
\end{equation}

\subsection{Chiral spectrum in flux-broken $E_7$ models}

It is straightforward to calculate the chiral spectrum given the above
fluxes. Since only the vertical flux induces chiral matter, we can analyze
the matter content by breaking $E_7 \rightarrow SU(5)$, where
the $\mathbf{56}$ breaks into a combination of $\mathbf{5},
\mathbf{10}$, uncharged singlets and conjugate representations, and $\mathbf{133}$
includes these as well as the adjoint $\mathbf{24}$.  Since the
adjoint
of SU(5)
 is non-chiral, the only chiral representations we expect for
$\gsm$ after the whole breaking
are the Standard Model representations,
which descend from the  $\mathbf{5},
\mathbf{10}$ of SU(5),
\begin{equation}
\label{eq:e7:SMreps}
    Q=\left(\mathbf{3},\mathbf{2}\right)_{1/6}\,,\quad
    \bar U=\left(\bar{\mathbf{3}},\mathbf{1}\right)_{-2/3}\,,\quad
    \bar D=\left(\bar{\mathbf{3}},\mathbf{1}\right)_{1/3}\,,\quad
    L=\left(\mathbf{1},\mathbf{2}\right)_{-1/2}\,,\quad
    \bar E=\left(\mathbf{1},\mathbf{1}\right)_{1}\,.
\end{equation}
Using 
Eq.\ (\ref{fullchiR'}), we
indeed get the anomaly-free combination of SM chiral matter from vertical
flux. It will be useful to separate the contributions from $\mathbf{56}$
and $\mathbf{133}$ to the total chiral index, i.e. $\chi_{\left(\mathbf{3},\mathbf{2}\right)_{1/6}}=\chi_{\left(\mathbf{3},\mathbf{2}\right)_{1/6}}^{\mathbf{56}}+\chi_{\left(\mathbf{3},\mathbf{2}\right)_{1/6}}^{\mathbf{133}}$, where each contribution is anomaly-free by itself. The fundamental $\mathbf{56}$ gives
\begin{equation}
    \chi_{\left(\mathbf{3},\mathbf{2}\right)_{1/6}}^{\mathbf{56}}=\Sigma\cdot (4K_B+3\Sigma)\cdot D_\alpha n_\alpha\,,
\end{equation}
and the adjoint $\mathbf{133}$ gives
\begin{equation}
    \chi_{\left(\mathbf{3},\mathbf{2}\right)_{1/6}}^{\mathbf{133}}=2\Sigma\cdot (K_B+\Sigma)\cdot D_\alpha n_\alpha\,.
\end{equation}
Note that the total chiral indices only depend on $n_\alpha$ but not
$\phi_{5 \alpha}, \phi_{6 \alpha}$. This is no longer true when
we look at the chiral indices for each weight in the phenomenological
analysis below. An important feature of these chiral indices is that
they have a linear Diophantine structure in the quantized flux parameters,
with the coefficients not being very large.
If we randomly pick some small values of $n_\alpha$ (bounded by the 
tadpole constraint (\ref{eq:review:tadpole})), 
%most probably
generically different terms in the
chiral indices will cancel each other, resulting in small chiral indices.
Therefore, small chiral indices are preferred in these models,
and it is not hard to achieve three generations of SM chiral matter.

In most cases, the above Weierstrass model also has codimension-3 singularities at the 
locus $s=f_3=g_5=0$. Traditionally, codimension-3 singularities are
interpreted as Yukawa couplings in the low-energy theory. In $E_7$ models,
however, these are so-called non-minimal singularities (with degrees 
$(4,6)$ or higher) where the fiber becomes non-flat, i.e. its dimension
jumps. Such singularities can no longer be interpreted as Yukawa 
couplings;\footnote{It was pointed out in \cite{Achmed-Zade:2018idx} that
these singularities may give rise to quartic or higher order couplings. 
Nevertheless, the singularities in \cite{Achmed-Zade:2018idx} have an 
unusual local geometry where a curve intersects another curve three times.
We do not see any such intersections or any evidence of such higher order 
couplings in our models.} this is also manifest by noticing that $\mathbf{56}^3$ does not
contain any singlets, hence cannot form any gauge-invariant 
couplings.
%\footnote{From the perspective after the breaking of $E_7$,
%Eq.\ (\ref{eq:branchingrules56}) tells us that all fields descended from 
%$\mathbf{56}$ carry half-integer additional $\U(1)$ charges. Hence we 
%cannot form invariant couplings from any three of those fields.}
Instead, there is an extra family of vertical flux associating to the 
non-flat fiber with nontrivial $\phi_{ij}$ components 
\cite{Jefferson:2021bid}. Analogous to codimension-2 $(4,6)$ 
singularities in 6D F-theory models \cite{HeckmanMorrisonVafa,Apruzzi:2018oge}, there has been
evidence that this flux switches on an extra sector of strongly coupled
superconformal \emph{chiral} matter, given by M2-branes wrapping curves
on the non-flat fiber \cite{46}. For our phenomenological purpose, we can always
set this flux to zero i.e. $\phi_{ij}=0$ for all $i,j$, such that the
extra sector becomes non-chiral and probably does not affect the Standard
Model sector. We should warn readers, however, that without further
studies on these extra sectors, we cannot precisely rule out the 
possibility that these sectors ruin the desired phenomenology.

\section{Phenomenology of $E_7$ GUTs}
\label{sec:Pheno}

So far, we have studied the gauge group and the chiral spectrum in the above class of $E_7$ models. In this
section, we start to analyze the phenomenological aspects of these models
in more detail. 
 The presence of approximate global symmetries descending from the
 underlying $E_7$ group suppresses certain couplings, with significant
 implications for phenomenology of these models;
in particular, we show that proton decay is automatically suppressed.
%Specifically, we
We focus further on the Higgs sector and the
interactions in these models.
We also discuss vector-like exotics in these models and %argue how
%they can still
the circumstances under which they can
 be phenomenologically safe. We consider the extent
to which the various features of the $E_7$ flux-broken models possibly, or even
naturally, match with observed phenomenology. The analysis in this
section, together with an explicit
construction of an example model in \S\ref{sec:ExplicitConstruction}, are the main
results of this chapter.
These results provide evidence
%They provide strong evidence towards
that natural and realistic $E_7$ GUTs can be realized in F-theory. 
On the other hand, due to %lack of
limits on existing
technologies %to compute
for computing detailed aspects of % continuous data in
 F-theory models (such as the specific values of couplings), most of the
analysis in this section is purely qualitative.

\subsection{Approximate global symmetries}
\label{subsec:approxu1}

Our starting point is 
based on considering
 approximate $\U(1)$ global symmetries that arise in the $E_7$  flux-broken
models.  
These approximate 
symmetries directly originate from vertical flux breaking, and
 control the structure of interactions and mass terms in the
low-energy theory,  as well as leading to many of the phenomenologically attractive
features of these models such as suppression of proton decay. 
 Recall that during vertical
flux breaking from $E_7$ to $\gsm$, the Cartan gauge bosons along the generators $T_4,T_5,T_6$ get masses from the St\"uckelberg
mechanism \cite{DonagiWijnholtModelBuilding,Grimm:2010ks,Grimm:2011tb}. 
These masses explicitly break the corresponding Cartan $U(1)$'s of the 
nonabelian gauge  symmetry,
but the matter interactions descending from the unbroken
gauge group still respect the global parts of the broken $U(1)$ gauge
symmetries (at least for the symmetries without mixed anomalies with the remaining gauge 
group).  These
$U(1)$ global symmetries are broken only by D3/M5-instanton
effects, which are exponentially suppressed in K\"ahler moduli \cite{Witten:1996bn, Blumenhagen:2006xt,Ibanez:2006da,Blumenhagen:2007zk}. In the low-energy theory, these effects turn on exponentially
suppressed mass terms and interactions that violate the global symmetries.
Since these effects are small, the symmetries still remain as approximate
global symmetries in the theory. This scenario is consistent with
the No Global Symmetries Conjecture \cite{Banks:2010zn,Harlow:2018tng}. Note that without more details of
the model, we cannot quantitatively specify %
the extent to which
 a certain quantity is
%exponentially
 suppressed by these symmetries. Throughout this chapter, we
only adopt the qualitative picture of exponential suppression, and leave
efforts towards explicit calculations of these quantities
for future work.

To study the implications of these $U(1)$ symmetries, it is important to understand
the branching rules from $E_7$ to $\gsm$  in the presence of these additional $U(1)$
charges. Below we use the basis $\left(Y,b_{4},b_{5},b_{6}\right)$
for the $U(1)$ charges, where $Y$ is the SM hypercharge.\footnote{One 
direction of the additional $\U(1)$ symmetries actually has mixed 
anomalies with the SM gauge group (see also, e.g.\ \cite{Ibanez:2006da},
for a similar situation). In the example below, the anomaly-free
directions are spanned by $-4b_4/5+b_5$ and $2b_4/5+b_6$. It turns out 
that excluding the anomalous symmetry does not affect the  selection
rules below at
all, so for completeness we still use all the $\U(1)$ charges 
%as the 
%basis
in labeling the charges.} The
branching rules are
\begin{align}
\label{eq:branchingrules56}
\mathbf{56} & \rightarrow\left(\mathbf{1},\mathbf{1}\right)_{0,5/2,2,3/2}+\left(\mathbf{1},\mathbf{1}\right)_{0,5/2,2,1/2}+\left(\mathbf{1},\mathbf{1}\right)_{0,5/2,1,1/2}+\left(\mathbf{1},\mathbf{1}\right)_{1,3/2,1,1/2}\nonumber \\
 & +\left(\mathbf{3},\mathbf{2}\right)_{1/6,3/2,1,1/2}+\left(\bar{\mathbf{3}},\mathbf{1}\right)_{-2/3,3/2,1,1/2}+\left(\bar{\mathbf{3}},\mathbf{1}\right)_{1/3,1/2,1,1/2}+\left(\bar{\mathbf{3}},\mathbf{1}\right)_{1/3,1/2,0,1/2}\nonumber \\
 &
+\left(\bar{\mathbf{3}},\mathbf{1}\right)_{1/3,1/2,0,-1/2}+\left(\mathbf{1},\mathbf{2}\right)_{-1/2,1/2,1,1/2}+\left(\mathbf{1},\mathbf{2}\right)_{-1/2,1/2,0,1/2}+\left(\mathbf{1},\mathbf{2}\right)_{-1/2,1/2,0,-1/2}
\nonumber \\
&+\mathrm{conjugates}\,,
\end{align}
\begin{align}
\label{eq:branchingrules133}
\mathbf{133} & \rightarrow\left(\mathbf{8},\mathbf{1}\right)_{0,0,0,0}+\left(\mathbf{1},\mathbf{3}\right)_{0,0,0,0}+4\times\left(\mathbf{1},\mathbf{1}\right)_{0,0,0,0}\nonumber \\
 &
+\left[\left(\mathbf{1},\mathbf{1}\right)_{0,0,0,1}+\left(\mathbf{1},\mathbf{1}\right)_{0,0,1,0}+\left(\mathbf{1},\mathbf{1}\right)_{0,0,1,1}+\left(\mathbf{1},\mathbf{1}\right)_{1,-1,0,0}+\left(\mathbf{1},\mathbf{1}\right)_{1,-1,-1,0}
\right.\nonumber \\
 &\hspace*{0.1in} +\left(\mathbf{1},\mathbf{1}\right)_{1,-1,-1,-1}+\left(\mathbf{3},\mathbf{2}\right)_{-5/6,0,0,0}+\left(\mathbf{3},\mathbf{2}\right)_{1/6,-1,0,0}+\left(\mathbf{3},\mathbf{2}\right)_{1/6,-1,-1,0}+\left(\mathbf{3},\mathbf{2}\right)_{1/6,-1,-1,-1}\nonumber \\
&\hspace*{0.1in}  +\left(\bar{\mathbf{3}},\mathbf{1}\right)_{-2/3,-1,0,0}+\left(\bar{\mathbf{3}},\mathbf{1}\right)_{-2/3,-1,-1,0}+\left(\bar{\mathbf{3}},\mathbf{1}\right)_{-2/3,-1,-1,-1}+\left(\bar{\mathbf{3}},\mathbf{1}\right)_{1/3,-2,-1,0}\nonumber \\
&\hspace*{0.1in}  +\left(\bar{\mathbf{3}},\mathbf{1}\right)_{1/3,-2,-1,-1}+\left(\bar{\mathbf{3}},\mathbf{1}\right)_{1/3,-2,-2,-1}+\left(\bar{\mathbf{3}},\mathbf{1}\right)_{1/3,3,2,1}+\left(\mathbf{1},\mathbf{2}\right)_{-1/2,-2,-1,0}\nonumber \\
&\hspace*{0.1in} \left. +\left(\mathbf{1},\mathbf{2}\right)_{-1/2,-2,-1,-1}+\left(\mathbf{1},\mathbf{2}\right)_{-1/2,-2,-2,-1}+\left(\mathbf{1},\mathbf{2}\right)_{-1/2,3,2,1}+\mathrm{conjugates}\right]\,.
\end{align}
It is then straightforward to apply the rule
that only terms  with all net
$U(1)$ charges vanishing are %not
unsuppressed in the superpotential
of the low-energy theory.
Notice that there are three copies of 
$\bar{D} =\left(\bar{\mathbf 3}, \mathbf 1\right)_{1/3}$
and $L = \left(\mathbf 1,\mathbf 2\right)_{-1/2}$ (or the $\SU(5)$ 
fundamental before remainder flux breaking) in each of
the decompositions $\mathbf{56},\mathbf{133}$, which are distinguished
by having different
%with different additional
$U(1)$ charges.
Without further information or inputs, 
 three families of SM chiral matter arising in a given model
 in general may be distributed
within the three copies. As shown below, such a distribution may lead to 
phenomenological inconsistencies, and some extra tuning must be done
% in the
%model
 to avoid those issues.

\subsection{Vector-like exotics}
\label{subsec:vectorexotics}
 
As in all GUT models, the $E_7$ models face the issue of having (many)
vector-like exotics that have not been observed in experiments. Although
we have chosen a $\mathbb P^1$ matter curve to ensure
that there are
 no vector-like 
pairs on the matter curve, generically there are many vector-like
matter
fields
on the bulk of $\Sigma$. Therefore, all representations in Eq. 
(\ref{eq:branchingrules133}), except the SM adjoint representations, 
generically have nontrivial vector-like matter multiplicities. 
Interestingly, these bulk vector-like fields involve the usual MSSM
 Higgs 
fields $H_u, H_d$, which indeed play the role of
a SM Higgs sector in the discussion below.
This feature does not seem to happen in models with smaller GUT groups.
On the other hand, there are also inert Higgs fields $H_u', H_d'$, which
have the same representation $(\mathbf 1,\mathbf 2)_{\pm1/2}$ under
the SM
gauge group, but do not have the right additional $\U(1)$ charges to form
unsuppressed Yukawa couplings with SM matter (see 
\S\ref{subsec:Yukawalist}). There are also similar sets of fields for the
triplet Higgs $T_u,T_d,T_u',T_d'$, as well as vector-like fields in other
exotic representations, namely $(\mathbf 3,\mathbf 2)_{-5/6},(\mathbf 3,\mathbf 2)_{1/6},(\bar{\mathbf 3},\mathbf 1)_{-2/3},(\mathbf 1,\mathbf 1)_{1}$.
In particular, the exotic $(\mathbf 3,\mathbf 2)_{-5/6}$ ruins
phenomenology by causing proton decay and spoiling gauge coupling unification (see also \S\ref{subsec:unification}), and must be removed
from the spectrum. We will also discuss the phenomenological safety of
other vector-like exotics as we proceed in later sections.

As seen
in \S\ref{subsec:e7:vectorlike}, the multiplicities of these vector-like fields are controlled
by the fluxes. Unfortunately, it has been shown in 
\cite{BeasleyHeckmanVafaII} that for GUT groups higher or equal to $\SO(10)$,
it is impossible to remove all the vector-like exotics by tuning the 
fluxes. Nevertheless, it was pointed out in \cite{Li:2022aek} that it is easy to remove the most dangerous
$(\mathbf 3,\mathbf 2)_{-5/6}$.

Now we show that in the $E_7$
models, this representation is %preferred
reasonably likely
to be removed from the spectrum, at least for certain kinds of gauge
divisor $\Sigma$.
First, we notice that the representation and its conjugate have
$(b_3,b_4,b_5,b_6)=(\pm1,0,0,0)$. Recall that the vertical flux we turn
on breaks directions $4,5,6$. We then see that $\chi_{(\mathbf
  3,\mathbf 2)_{-5/6}}=0$,
since $\Theta_{i \alpha} \neq 0$ only for $i = 4, 5, 6$ in
\eqref{rootCondition},
 and the vertical flux does not contribute in Eq.
(\ref{eq:c1ofL}). In other words, the multiplicity is purely controlled
by the remainder flux in \eqref{eq:phi-r}, given by\footnote{When $\phi_{4r}=5n_r$, $L_\beta$ becomes trivial and Eq.\ (\ref{eq:HirzebruchRiemannRoch}) no longer applies.}
\begin{equation}
    n_{(\mathbf 3,\mathbf 2)_{-5/6}}=-\frac{1}{2}(5n_r-\phi_{4r})^2 C_{\mathrm{rem}}^2-1\,.
\end{equation}
Therefore, $n_{(\mathbf 3,\mathbf 2)_{-5/6}}=0$ if $5n_r-\phi_{4r}=\pm1$
and $C_\mathrm{rem}^2=-2$. Interestingly, some choices of the remainder flux with the 
smallest tadpole %satisfies
satisfy these conditions. Consider the tadpole
\begin{equation}
    \frac{1}{2}[G_4^\mathrm{rem}]^2=-\frac{1}{2}C_\mathrm{rem}^2\kappa^{ij}\phi_{ir}\phi_{jr}\,.
\end{equation}
As demonstrated in \S\ref{sec:ExplicitConstruction}, in many cases $\Sigma$ is a del Pezzo surface
and the available $C_\mathrm{rem}$ with the least negative
self-intersection has $C_\mathrm{rem}^2=-2$. 
Going forward we assume these technical conditions, which also imply the
condition discussed earlier that $-K_\Sigma$ is effective.
Further work would be needed to generalize the analysis to the
situation when  these technical conditions are relaxed.
In this situation, $\kappa^{ij}\phi_{ir}\phi_{jr}$ is minimized by e.g.
\begin{equation}
\label{eq:remainderfluxchoice}
    \phi_{ir}=(2,4,6,4,2,1,3)\,,
\end{equation}
which indeed leads to $n_{(\mathbf 3,\mathbf 2)_{-5/6}}=0$. From now on,
we always assume this choice of remainder flux, with tadpole
\begin{equation}
    \frac{1}{2}[G_4^\mathrm{rem}]^2=4\,.
\end{equation}

How about other vector-like exotics? Unlike the above, vertical flux also
contributes to the multiplicities
of other vector-like exotics. Comparing to remainder flux, vertical flux
satisfies more constraints like primitivity. We also need vertical flux
to get the right chiral spectrum and, to be discussed below, the right 
interactions. After fulfilling these more important requirements, 
we find no more room to remove the remaining vector-like exotics; i.e.,
generically there is a nontrivial or even large contribution to $n_\beta$ from vertical
flux. Since remainder flux is orthogonal to vertical flux, there is
no remainder flux that can cancel the contribution from
vertical flux. Therefore, we expect that all the other vector-like
exotics are present in our models. Fortunately, below we will show that
these vector-like exotics, including the triplet Higgs $(\bar{\mathbf 3},\mathbf 1)_{1/3}$, which potentially mediates dangerous proton decay, 
can still be phenomenologically acceptable if some additional assumptions
and tunings are made.

\subsection{List of Yukawa couplings}
\label{subsec:Yukawalist}

With the selection rules \S\ref{subsec:approxu1}, we can now list the Yukawa couplings that
are not suppressed by the approximate global symmetries. Recall that in
a general 4D F-theory model, there are three types
of Yukawa couplings on a gauge divisor $\Sigma$ \cite{BeasleyHeckmanVafaI}. First, there are Yukawa
couplings between three fields on the bulk of $\Sigma$ (denoted by 
$\Sigma\Sigma\Sigma$), but it has been shown in 
\cite{BeasleyHeckmanVafaI} that these couplings are all absent when 
$-K_\Sigma$ is effective, which is assumed in our $E_7$ models
as discussed in \S\ref{subsec:e7:vectorlike}. The 
second type of Yukawa couplings are between one field on the bulk of $\Sigma$ and two fields on matter curves (denoted by $\Sigma CC$). These
couplings are generically present, and we will simply assume that
all 
 couplings of this type that satisfy the symmetry constraints are present. Finally, there
are Yukawa couplings between three fields on matter curves (denoted by
$CCC$). These couplings are characterized by codimension-3 singularities
of the fibration. Nevertheless as discussed in last section, the 
codimension-3 singularities in the $E_7$ models cannot be interpreted as
Yukawa couplings. In conclusion, there are only $\Sigma CC$-type
couplings in our theory.\footnote{This coupling structure, from the low-energy perspective, can also be understood as a kind of R-symmetry, where
the fields on the bulk of $\Sigma$ have R-charge $1$, and those on matter
curves have R-charge $1/2$. We thank Jesse Thaler for this comment.} Note
that this UV structure of Yukawa couplings
%are vastly
is quite different from that of other % any
 previous SM-like constructions in 
F-theory, and we expect new phenomenological features in the IR to arise from this structure.

We now 
investigate how
%try to reproduce
 the SM Yukawa couplings can arise from
% with
 $\Sigma CC$-type couplings. In principle, we can
localize the vector-like Higgs on matter 
curves more general than $\mathbb P^1$. If we make such a choice, however,
the general (including both diagonal and off-diagonal) Yukawa couplings 
will require the non-existent $CCC$- and/or $\Sigma\Sigma C$-type 
couplings apart from $\Sigma CC$-type couplings. Therefore, to reproduce the SM Yukawa couplings
with mixing between all three generations, it is necessary to localize the Higgs on the bulk of $\Sigma$, and
all SM chiral matter on the matter 
curve $C_\mathbf{56}$.
 This choice of localization also matches with the fact 
that, from the discussion of \S\ref{subsec:e7:vectorlike}, 
generically
there are  many 
vector-like fields on the bulk of $\Sigma$, but no such pairs on $\mathbb P^1$ matter
curves.
We also notice that 
this choice of Higgs is only 
available when the
GUT group is as large as $E_7$, such that the adjoint includes the Higgs
after breaking. On the other hand, the chiral matter induced
by vertical flux breaking descends from both $\mathbf{133}$ and 
$\mathbf{56}$, and can be localized on both $\Sigma$ and $C_\mathbf{56}$. Therefore to reproduce
the SM Yukawa couplings, we need to impose the flux constraint
\begin{equation}
\label{eq:nochifrom133}
     \chi_{\left(\mathbf{3},\mathbf{2}\right)_{1/6}}^{\mathbf{133}}=2\Sigma\cdot (K_B+\Sigma)\cdot D_\alpha n_\alpha=0\,.
\end{equation}
Below we
will see that this constraint can be easily satisfied. It is worth
emphasizing that this choice of localization automatically implies very
different low-energy physics between the Higgs and chiral matter, due to
their distinct geometric origins.

Now, assuming that
the SM chiral spectrum is supported on $C_{\mathbf{56}}$, we can easily
list all couplings that do not violate the approximate global
symmetries. For simplicity, here we 
first ignore the couplings involving
uncharged singlets under the SM gauge group; %while
 these singlets may play the
role of right-handed neutrinos and will be studied in \S\ref{subsec:neutrino}.
We then have the SM Yukawa couplings:\footnote{Similar to mass terms, the couplings described here are 
terms in the superpotential $W$; Yukawa couplings between one boson
and two fermions come as usual from the contributions to the
potential $V$ of the form $ (\partial^2 W/\partial \phi_i \partial
\phi_j) \psi_i \psi_j$ and its conjugate.}
\begin{align}
H_uQ\bar U&:\quad\left(\mathbf 1,\mathbf 2\right)_{1/2,-3,-2,-1}\times\left(\mathbf 3,\mathbf 2\right)_{1/6,3/2,1,1/2}\times\left(\bar{\mathbf 3},\mathbf 1\right)_{-2/3,3/2,1,1/2}\,,\nonumber \\
H_dQ\bar D&:\quad
\begin{cases}
\left(\mathbf 1,\mathbf 2\right)_{-1/2,-2,-2,-1}\times\left(\mathbf 3,\mathbf 2\right)_{1/6,3/2,1,1/2}\times\left(\bar{\mathbf 3},\mathbf 1\right)_{1/3,1/2,1,1/2}\,, \\
\left(\mathbf 1,\mathbf 2\right)_{-1/2,-2,-1,-1}\times\left(\mathbf 3,\mathbf 2\right)_{1/6,3/2,1,1/2}\times\left(\bar{\mathbf 3},\mathbf 1\right)_{1/3,1/2,0,1/2}\,, \\
\left(\mathbf 1,\mathbf 2\right)_{-1/2,-2,-1,0}\times\left(\mathbf 3,\mathbf 2\right)_{1/6,3/2,1,1/2}\times\left(\bar{\mathbf 3},\mathbf 1\right)_{1/3,1/2,0,-1/2}\,,
\end{cases} \nonumber \\
H_d L\bar E&:\quad
\begin{cases}
\left(\mathbf 1,\mathbf 2\right)_{-1/2,-2,-2,-1}\times\left(\mathbf 1,\mathbf 2\right)_{-1/2,1/2,1,1/2}\times\left(\mathbf 1,\mathbf 1\right)_{1,3/2,1,1/2}\,, \\
\left(\mathbf 1,\mathbf 2\right)_{-1/2,-2,-1,-1}\times\left(\mathbf 1,\mathbf 2\right)_{-1/2,1/2,0,1/2}\times\left(\mathbf 1,\mathbf 1\right)_{1,3/2,1,1/2}\,, \\
\left(\mathbf 1,\mathbf 2\right)_{-1/2,-2,-1,0}\times\left(\mathbf 1,\mathbf 2\right)_{-1/2,1/2,0,-1/2}\times\left(\mathbf 1,\mathbf 1\right)_{1,3/2,1,1/2}\,,
\end{cases}
\label{eq:SMYukawa}
\end{align}
where the first representation in each product is the up and down
Higgs $H_u,H_d$. We also have %the following
a number of other exotic couplings. 
There are %also
 couplings involving the triplet Higgs $\left(\mathbf 3,\mathbf 1\right)_{-1/3}$:
\begin{align}
T_uQQ&:\quad\left(\mathbf 3,\mathbf 1\right)_{-1/3,-3,-2,-1}\times\left(\mathbf 3,\mathbf 2\right)_{1/6,3/2,1,1/2}\times\left(\mathbf 3,\mathbf 2\right)_{1/6,3/2,1,1/2}\,,\nonumber \\
T_u\bar U\bar E&:\quad\left(\mathbf 3,\mathbf 1\right)_{-1/3,-3,-2,-1}\times\left(\bar{\mathbf 3},\mathbf 1\right)_{-2/3,3/2,1,1/2}\times\left(\mathbf 1,\mathbf 1\right)_{1,3/2,1,1/2}\,,\nonumber \\
T_dQL&:\quad
\begin{cases}
\left(\bar{\mathbf 3},\mathbf 1\right)_{1/3,-2,-2,-1}\times\left(\mathbf 3,\mathbf 2\right)_{1/6,3/2,1,1/2}\times\left(\mathbf 1,\mathbf 2\right)_{-1/2,1/2,1,1/2}\,, \\
\left(\bar{\mathbf 3},\mathbf 1\right)_{1/3,-2,-1,-1}\times\left(\mathbf 3,\mathbf 2\right)_{1/6,3/2,1,1/2}\times\left(\mathbf 1,\mathbf 2\right)_{-1/2,1/2,0,1/2}\,, \\
\left(\bar{\mathbf 3},\mathbf 1\right)_{1/3,-2,-1,0}\times\left(\mathbf 3,\mathbf 2\right)_{1/6,3/2,1,1/2}\times\left(\mathbf 1,\mathbf 2\right)_{-1/2,1/2,0,-1/2}\,,
\end{cases} \nonumber \\
T_d\bar U\bar D&:\quad
\begin{cases}
\left(\bar{\mathbf 3},\mathbf 1\right)_{1/3,-2,-2,-1}\times\left(\bar{\mathbf 3},\mathbf 1\right)_{-2/3,3/2,1,1/2}\times\left(\bar{\mathbf 3},\mathbf 1\right)_{1/3,1/2,1,1/2}\,,\\
\left(\bar{\mathbf 3},\mathbf 1\right)_{1/3,-2,-1,-1}\times\left(\bar{\mathbf 3},\mathbf 1\right)_{-2/3,3/2,1,1/2}\times\left(\bar{\mathbf 3},\mathbf 1\right)_{1/3,1/2,0,1/2}\,,\\
\left(\bar{\mathbf 3},\mathbf 1\right)_{1/3,-2,-1,0}\times\left(\bar{\mathbf 3},\mathbf 1\right)_{-2/3,3/2,1,1/2}\times\left(\bar{\mathbf 3},\mathbf 1\right)_{1/3,1/2,0,-1/2}\,.
\end{cases}
\label{eq:tripletHiggsYukawa}
\end{align}
These
couplings are always present together with the SM Yukawa couplings, but
the ones with triplet Higgs mediate dimension-5 proton decay and need
extra attention. Note that there are unique sets of additional $\U(1)$ 
charges for $H_u,T_u$.
This uniquely identifies these fields in the decomposition
 (\ref{eq:branchingrules133}) as
\begin{equation}
H_u=\left(\mathbf{1},\mathbf{2}\right)_{1/2, -3, -2, -1}\,,\quad
T_u=\left(\bar{\mathbf{3}},\mathbf{1}\right)_{-1/3, -3, -2, -1}\,.
\label{eq:Higgsrepresentations}
\end{equation}
On the other hand, there are three possible %sets
fields with distinct approximate U(1) charges for each of $H_d,T_d$,
each of which couples to $\bar{D}, L$ in one of the three possible
copies of the SU(5) fundamentals. The
choices of these charges will be discussed below.

There are also couplings involving
other types of vector-like exotics, namely $(\mathbf 3,\mathbf 2)_{1/6},(\mathbf 3,\mathbf 1)_{2/3},(\mathbf 1,\mathbf 1)_1$:
\begin{gather}
\left(\mathbf 3,\mathbf 2\right)_{1/6,-1,-1,-1}\times\left(\bar{\mathbf 3},\mathbf 1\right)_{1/3,1/2,1,1/2}\times\left(\mathbf 1,\mathbf 2\right)_{-1/2,1/2,0,1/2}\,,\nonumber \\
\left(\mathbf 3,\mathbf 2\right)_{1/6,-1,-1,-1}\times\left(\bar{\mathbf 3},\mathbf 1\right)_{1/3,1/2,0,1/2}\times\left(\mathbf 1,\mathbf 2\right)_{-1/2,1/2,1,1/2}\,,\nonumber \\
\left(\mathbf 3,\mathbf 2\right)_{1/6,-1,-1,0}\times\left(\bar{\mathbf 3},\mathbf 1\right)_{1/3,1/2,1,1/2}\times\left(\mathbf 1,\mathbf 2\right)_{-1/2,1/2,0,-1/2}\,,\nonumber \\
\left(\mathbf 3,\mathbf 2\right)_{1/6,-1,-1,0}\times\left(\bar{\mathbf 3},\mathbf 1\right)_{1/3,1/2,0,-1/2}\times\left(\mathbf 1,\mathbf 2\right)_{-1/2,1/2,1,1/2}\,,\nonumber \\
\left(\mathbf 3,\mathbf 2\right)_{1/6,-1,0,0}\times\left(\bar{\mathbf 3},\mathbf 1\right)_{1/3,1/2,0,1/2}\times\left(\mathbf 1,\mathbf 2\right)_{-1/2,1/2,0,-1/2}\,,\nonumber \\
\left(\mathbf 3,\mathbf 2\right)_{1/6,-1,0,0}\times\left(\bar{\mathbf 3},\mathbf 1\right)_{1/3,1/2,0,-1/2}\times\left(\mathbf 1,\mathbf 2\right)_{-1/2,1/2,0,1/2}\,,\nonumber \\
\left(\bar{\mathbf 3},\mathbf 1\right)_{-2/3,-1,-1,-1}\times\left(\bar{\mathbf 3},\mathbf 1\right)_{1/3,1/2,1,1/2}\times\left(\bar{\mathbf 3},\mathbf 1\right)_{1/3,1/2,0,1/2}\,,\nonumber \\
\left(\bar{\mathbf 3},\mathbf 1\right)_{-2/3,-1,-1,0}\times\left(\bar{\mathbf 3},\mathbf 1\right)_{1/3,1/2,1,1/2}\times\left(\bar{\mathbf 3},\mathbf 1\right)_{1/3,1/2,0,-1/2}\,,\nonumber \\
\left(\bar{\mathbf 3},\mathbf 1\right)_{-2/3,-1,0,0}\times\left(\bar{\mathbf 3},\mathbf 1\right)_{1/3,1/2,0,1/2}\times\left(\bar{\mathbf 3},\mathbf 1\right)_{1/3,1/2,0,-1/2}\,,\nonumber \\
\left(\mathbf 1,\mathbf 1\right)_{1,-1,-1,-1}\times\left(\mathbf 1,\mathbf 2\right)_{-1/2,1/2,1,1/2}\times\left(\mathbf 1,\mathbf 2\right)_{-1/2,1/2,0,1/2}\,,\nonumber \\
\left(\mathbf 1,\mathbf 1\right)_{1,-1,-1,0}\times\left(\mathbf 1,\mathbf 2\right)_{-1/2,1/2,1,1/2}\times\left(\mathbf 1,\mathbf 2\right)_{-1/2,1/2,0,-1/2}\,,\nonumber \\
\left(\mathbf 1,\mathbf 1\right)_{1,-1,0,0}\times\left(\mathbf 1,\mathbf 2\right)_{-1/2,1/2,0,1/2}\times\left(\mathbf 1,\mathbf 2\right)_{-1/2,1/2,0,-1/2}\,.
\label{eq:exoticYukawa}
\end{gather}
These couplings induce, e.g., additional proton decay and may not be 
compatible with phenomenology. On the other hand, all these couplings
%exchange
mix
distinct copies of $(\bar{\mathbf 3},\mathbf 1)_{1/3}$ and/or 
$(\mathbf 1,\mathbf 2)_{-1/2}$ with different U(1) charges, while 
%each
the couplings in Eqs. 
(\ref{eq:SMYukawa}) and (\ref{eq:tripletHiggsYukawa}) 
relate $H_d, T_d$ in a given copy with matter fields within the same corresponding
copy.  As
shown in \S\ref{subsubsec:Yukawastructure}, extra tuning of the model is available such that the
couplings in Eq.\ (\ref{eq:exoticYukawa}) are absent, and we assume
such absence throughout this chapter. All the other couplings without
uncharged singlets and not listed above, if present, are exponentially
suppressed by the approximate global symmetries.

\subsection{Proton decay}
\label{subsec:protondecay}

Before building semi-realistic Higgs sector and Yukawa couplings, there
is an important issue to be resolved: as in every GUT model, 
%there must
%be a mechanism to suppress proton decay. 
there is a possibility of couplings that give a proton decay rate that
exceeds experimental limits.
In particular, since the triplet Higgs cannot
be removed from the spectrum, the couplings in last subsection naively seem to
suggest that the $E_7$ models will suffer from
excess %too large
 proton decay mediated by
 dimension-5 operators.  We
now show that, fortunately, both dimension-4 and 5 proton decay 
are ubiquitously suppressed in the $E_7$ models.
This feature in some sense ``comes for free'' with the construction of
these models.

First, dimension-4 proton decay in the MSSM is driven by the R-parity
violating terms in the superpotential:
\begin{equation}
\label{eq:dim4protondecay}
    W\supset \alpha_1QL\bar D+\alpha_2LL\bar E+\alpha_3\bar D\bar D\bar U\,,
\end{equation}
where we have used the notation in \eqref{eq:e7:SMreps}.
These are
couplings between three chiral fields, which all descend from 
$\mathbf{56}$ under the assumption of Eq.\ (\ref{eq:nochifrom133}). Hence
from the geometric perspective, dimension-4 proton decay requires 
$CCC$-type couplings, which are absent in the $E_7$ models. 
The absence of these couplings is also %complemented
natural from
% by
 the form of the fields in the
symmetry-broken theory: since the approximate global charges $b_4,b_6$ of
all fields (including all copies of $\bar D,L$) appearing in Eq.\ 
(\ref{eq:dim4protondecay}) are half-integers, none of the interactions
of this type
have vanishing net $b_4,b_6$ charges, so all such interactions violate
one of the approximate
global symmetries.
%On the other
%hand, it is
%not clear whether flux breaking would induce new but exponentially small
%$CCC$-type couplings; the interplay between codimension-3 $(4,6)$ 
%singularities and flux breaking will be studied in the future. 
%This
Thus, dimension-4 proton decay is automatically absent in the $E_7$ 
models (with the caveat that we do not completely understand the $(4,6)$
singularities; it is not fully clear whether the interplay between these
singularities and flux breaking would modify this conclusion). In fact, it was already pointed out in \cite{Tatar:2006dc} that
dimension-4 proton decay can be eliminated in this way only when the
GUT group is $E_7$ or $E_8$; %such  elimination was also realized in
this suppression arises in
heterotic and M-theory
models as well as in F-theory.

Now we turn to dimension-5 proton decay. Such decay in conventional
supersymmetric GUTs comes from the following terms
\begin{equation}
\label{eq:conventionaldim5}
W\supset \lambda_1 T_uQQ+\lambda_2 T_dQL+MT_{u}T_{d}\,,
\end{equation}
where $T_u,T_d$ are the triplet Higgs, and $M$ is a large mass close to 
the GUT scale $M_\mathrm{GUT}$. Integrating
out $T_{u},T_{d}$, we then get the dimension-5 operator $QQQL/M$, which is
only suppressed by $1/M$ and 
leads to
an unacceptable rate of proton decay. Nevertheless, the $E_7$ models are different
from %the
 conventional GUTs in the following sense: although the first two terms
in Eq.\ (\ref{eq:conventionaldim5}) are not suppressed, we see from Eq.\ 
(\ref{eq:tripletHiggsYukawa}) that $T_u,T_d$ never have opposite additional $\U(1)$
charges, hence the mass term in Eq.\ (\ref{eq:conventionaldim5}) is 
exponentially suppressed. Instead, $T_u,T_d$ have their
own vector-like partners, denoted by $T_d',T_u'$, with opposite
additional $\U(1)$ charges. These ``primed'' fields
are \emph{inert}, i.e.\ their Yukawa interactions with SM chiral matter are
exponentially suppressed, but they give $T_u,T_d$ large masses by the
conventional mass terms. Now the superpotential schematically has the form of
\begin{equation}
W\supset \lambda_1 T_uQQ+\lambda_2 T_dQL+MT_{u}T_{d}'+MT_{u}'T_{d}+mT_{u}T_{d}+mT_{u}'T_{d}'\,,
\end{equation}
where $m$ is exponentially suppressed compared to $M$.
The operator we get by integrating out the triplet Higgs and the vector-like partners is roughly
$\left(m/M\right)QQQL/M$, which is indeed further exponentially 
suppressed by the factor $m/M$. Therefore, as long as $M$ is
sufficiently
large (probably close to $M_{GUT}$) and $m/M$ is sufficiently small, the
$E_7$ models are safe from overly dangerous proton decay.

Regarding dimension-6 proton decay mediated by gauge bosons along the 
broken directions of the gauge group, we do not see an obvious mechanism
of suppression.
 The gauge bosons typically have masses around the KK/GUT scale, %and 
which
may be 
sufficiently high to evade the current experimental bounds \cite{Workman:2022ynf}. Moreover,
fluxes on $\Sigma$ can make the ground state wavefunction of the gauge
bosons more localized, and suppress its wavefunction overlap
(hence the coupling) to the chiral matter on $C_\mathbf{56}$
\cite{Ibanez:2012zg}. Note that such suppression may not be exponential 
\cite{Hebecker:2014uaa}, but may already be sufficient for our purposes due
to the high GUT scale.
Despite all these heuristic arguments, more techniques and explicit
calculations are still required to %figure
determine the exact rate of proton decay,
which is essential for realistic model building.

\subsection{Higgs and Yukawa sectors}
\label{subsec:HiggsAndYukawa}

Now we turn to the Higgs and Yukawa sectors in the $E_7$ models. Even
with the Yukawa couplings in Eq.\ (\ref{eq:SMYukawa}) that have the right 
representations, it is not
guaranteed that those couplings resemble the structure of Higgs and Yukawa sectors
in the Standard Model, due to various differences of the $E_7$ models
from the conventional Standard Model. Below we explain each of these
differences, and write down the necessary conditions for realizing the
SM Higgs and Yukawa sectors.

\subsubsection{Doublet-triplet splitting and the Higgs masses}
\label{subsubsec:Higgs}

Apart from proton decay, one of the important questions in general GUT
models is the doublet-triplet splitting problem, or why the masses of 
doublet and triplet Higgs are separated %or
by many orders of magnitude.
In F-theory GUTs,  this splitting in principle can be explained by
the presence of hypercharge flux \cite{BeasleyHeckmanVafaII}. For
the tuned $\SU(5)$ GUTs, however, the explicit realization of such 
splitting can be difficult; see \S\ref{sec:Comparison} for further
discussion.  In contrast, the doublet and triplet
Higgs in the $E_7$ model live on the bulk of $\Sigma$ and always receive
mass splitting from hypercharge flux, which is also localized on a
(remainder) surface on $\Sigma$. Therefore, the doublet and triplet
masses are automatically split once we break $E_7$ to $\gsm$, although
the amount of splitting is still unknown and new techniques
must be developed for finding out the Higgs mass spectrum.

\begin{comment}
Although
the amount of splitting is still unknown, in principle the hypercharge 
flux can induce exponential
separation between the two masses as follows.
It was demonstrated in \cite{BeasleyHeckmanVafaII} that when $\Sigma$ is
a del Pezzo surface as in our models, depending on
the choice of flux, the wavefunctions involved in the mass terms can repel
one another, hence give exponentially suppressed mass terms. The ranges 
of such choice are different for doublet and triplet Higgs. Therefore, 
there may be a choice of hypercharge flux such that the mass term for
doublet Higgs is exponentially suppressed by wavefunction repulsion, 
while that for triplet Higgs is not. Without explicit
geometric and flux curvature data of
the model, however, it is hard to check whether this scenario can be 
realized.
\end{comment}

Still, what controls the mass terms before the splitting by
hypercharge flux? Similar to the triplet Higgs in
\S\ref{subsec:protondecay}, the conventional $\mu$-term
i.e. $\mu H_u H_d$ is exponentially suppressed. This suppression, however,
does not mean that the $\mu$-problem is solved, since the Higgs can still
get large masses from terms $H_u H_d', H_d H_u'$, when 
$H_u,H_d$ have their own vector-like partners $H_d',H_u'$. On the other 
hand, it means that there is some vector-like matter with light masses
when such vector-like partners do not exist. 
%Such 
Indeed, such a scenario generically 
%happens
arises for $H_d$, and the essence of this effect lies in
vertical flux breaking: although we have imposed that the total chiral
index of fields arising
from $\mathbf{133}$ vanishes, there can still be nontrivial chiral
surpluses for each of the three copies of doublet and triplet Higgs in
this representation
(see \S\ref{subsec:approxu1}). Suppose we have the following spectrum
for the three copies of Higgs fields:
\begin{align}
    (\mathbf 1,\mathbf 2)_{-1/2,-2,-2,-1}&:\,n_1\,,\quad(\mathbf 1,\mathbf 2)_{1/2,2,2,1}:\,n_1'\,,\nonumber\\
    (\mathbf 1,\mathbf 2)_{-1/2,-2,-1,-1}&:\,n_2\,,\quad(\mathbf 1,\mathbf 2)_{1/2,2,1,1}:\,n_2'\,,\nonumber\\
    (\mathbf 1,\mathbf 2)_{-1/2,-2,-1,0}&:\,n_3\,,\quad(\mathbf 1,\mathbf 2)_{1/2,2,1,0}:\,n_3'\,,\label{eq:Higgs-n-spectrum}
\end{align}
where $n_i,n_i'$ denote the multiplicities. Eq. (\ref{eq:nochifrom133})
implies that $n_1+n_2+n_3=n_1'+n_2'+n_3'$, hence the spectrum is non-chiral
under the SM gauge group. On the other hand, generically we have 
$n_i\neq n_i'$ and the spectrum would be
% is
 chiral if the additional 
$\U(1)$'s were gauge symmetries, i.e. the St\"uckelberg mechanism was absent.
If $n_i\neq n_i'$ for some $i$, there must be a
field direction in the $i$-th copy that cannot acquire any mass terms 
within the same copy. It can only get mass terms from fields in other 
copies. Since they do not have opposite additional $\U(1)$ charges,
the resulting mass terms are exponentially suppressed, leading
to a light doublet Higgs $H_d$.

It is tempting to use the above mechanism to solve the $\mu$-problem.
Unfortunately, the $\mu$-problem cannot be solved in this way %because
                                %of
for
two reasons. First, only $H_d$, but not $H_u$, has three copies in the
branching rule. In other words, %the same
this mechanism for producing light
$H_d$ cannot produce a light $H_u$. 
Second, the above mechanism relies on vertical flux breaking,
which only breaks $E_7$ to $\SU(5)$ instead of $\gsm$. This means that
whenever a light $H_d$ is produced in this way, there must also be a
light $T_d$. Although
there may still be doublet-triplet splitting from hypercharge flux, the
mass of the $T_d$ is still exponentially suppressed. Such $T_d$ directly interacts with SM chiral matter and
ruins the argument in \S\ref{subsec:protondecay}, i.e. there is still too much dimension-5
proton decay even with the exponential suppression in \S\ref{subsec:protondecay}. In this sense, we should even avoid
any light $H_d$ or $T_d$ produced in this way. As discussed in \S\ref{subsubsec:Yukawastructure}, we will
arrange the fluxes such that only one copy of $H_d$ interacts with SM
chiral matter. Without loss of generality, let us pick the copy with
$(b_4,b_5,b_6)=(2,2,1)$. Then avoiding light $H_d$ and $T_d$ coming from the above mechanism is 
achieved by the flux constraint
\begin{equation}
\label{eq:133samecopy}
    \chi^\mathbf{133}_{(\bar{\mathbf 3},\mathbf 1)_{1/3,-2,-2,-1}}=0\,.
\end{equation}
This is another linear constraint on the flux parameters, similar to the
ones for breaking the gauge group or inducing three generations of chiral
matter. This new constraint, however, is the first constraint that
involves the previously unused flux parameters $\phi_{5\alpha},\phi_{6\alpha}$. Therefore, given the gauge group and total chiral spectrum, 
there is always still some room in the $E_7$ models for
satisfying this new constraint.

The above still does not explain the origin of light masses in the SM
Higgs sector. Sadly, in the current construction of our $E_7$ models,
there is still no obvious solution to the $\mu$-problem. This is 
understandable, however, since the Higgs masses in F-theory are very 
complicated quantities to calculate. Traditionally, the Higgs masses come from the vevs of some fields
localized on divisors other than $\Sigma$ but intersecting with $\Sigma$.
These fields behave as singlets and couple to the vector-like matter
on $\Sigma$. Nevertheless, the vevs or potential of these fields depends on
many factors, including %by
but not limited to the detailed couplings between
these fields, the D-term potential, the nonperturbative superpotential,
and most importantly, soft SUSY breaking \cite{Palti:2016kew}. Therefore without understanding
more basic issues like moduli stabilization and SUSY breaking in F-theory, no precise
statements on these vector-like masses can be made. On the other hand,
given such a complicated origin of the Higgs masses, it is reasonable to
expect that some hierarchy is generated and brings some of the Higgs
to light scales. At the same time, we should
not allow more than one pair
of Higgs to be at the electroweak scale, although generically there are
many vector-like fields with the same representation. This is because
when more than one Higgs field couples to SM chiral matter in the same
way, the flavor basis generically does not align with the Higgs mass
basis. Such misalignment produces tree-level flavor-changing neutral
currents (FCNCs), which are not observed in experiments.\footnote{We
thank Jesse Thaler for pointing out this issue.} In conclusion, to
reproduce the SM Higgs sector, it is far from clear how to realize
exactly one pair of light Higgs doublets among all the Higgs fields. This is a major
%disadvantage 
shortcoming
 of the $E_7$ models, and we hope to give a better explanation
for this Higgs hierarchy in the future.

As a remark, there are still many \emph{inert} Higgs fields in the other
two copies. In particular, there can be multiple light inert Higgs fields,
coming from pairing chiral surpluses between the copies or other ways. 
Fortunately since they are inert, there is no tight constraint on these
fields. We note that the 
current experimental lower bound on $H_u',H_d'$ masses is around 100 GeV \cite{Workman:2022ynf}.

\subsubsection{Structure of Yukawa couplings}
\label{subsubsec:Yukawastructure}

One of the most dangerous features in the $E_7$ model is that there are three copies
of $(\bar{\mathbf 3},\mathbf 1)_{1/3}$ and $(\mathbf 1,\mathbf 2)_{-1/2}$
in the branching rules in \S\ref{subsec:approxu1}, with different
additional $\U(1)$ charges. The
generic case where the three generations of chiral matter are distributed
in all the copies is not phenomenologically acceptable for various 
reasons. First, generically there are chiral %surpluses
differences for each of the
copies,
as in (\ref{eq:Higgs-n-spectrum}).
While
these add up to the three generations in the total chiral spectrum, as
demonstrated in \S\ref{subsubsec:Higgs} they can also form light vector-like exotics between
different copies. Next, having different copies in the chiral spectrum
turns on an unsuppressed set of exotic couplings in Eq.\ (\ref{eq:exoticYukawa}),
including additional proton decay. More seriously, multiple light Higgs fields are
required to generate unsuppressed SM Yukawa couplings for all the copies, see Eq.\ (\ref{eq:SMYukawa}).\footnote{While there is indeed some hierarchy between
Yukawa couplings in the observed Standard Model, we do not expect
the hierarchy to be as large as the exponential suppression from the 
approximate global symmetries. Therefore, if we only use one Higgs for 
multiple copies with exponentially suppressed couplings, very probably it will not give
the right flavor structure.} As discussed in \S\ref{subsubsec:Higgs},
such a Higgs sector
again leads to unacceptable FCNCs. Therefore to avoid all the above 
issues, we must arrange all three generations of chiral matter to be
within the same copy. For the choice of light Higgs with
$(b_4,b_5,b_6)=(2,2,1)$ in \S\ref{subsubsec:Higgs}, for example,
we should
choose the corresponding copy of chiral matter with $(b_4,b_5,b_6)=(1/2,1,1/2)$ and impose the following
flux constraints:
\begin{equation}
\label{eq:56samecopy}
    \chi^\mathbf{56}_{(\bar{\mathbf 3},\mathbf 1)_{1/3,1/2,0,1/2}}=\chi^\mathbf{56}_{(\bar{\mathbf 3},\mathbf 1)_{1/3,1/2,0,-1/2}}=0\,.
\end{equation}
Again, these constraints are mild tuning on the remaining flux parameters
$\phi_{5\alpha},\phi_{6\alpha}$, which is
generically achievable. Nevertheless, it will be interesting to see
whether there is a more fundamental reason that leads to such choice of
fluxes.

After this choice of chiral matter, the remaining couplings in the 
low-energy theory are just the SM Yukawa couplings and their counterparts
with the triplet Higgs. It would be even more informative if we can get 
the values of the Yukawa couplings. Although calculating those
values is beyond our current F-theory technologies, we can gather some of
their qualitative features. Unlike the conventional F-theory models with
$CCC$-type couplings, the use of $\Sigma CC$-type couplings means that the
Yukawa couplings are supported on the whole $C_\mathbf{56}$ instead of
points on it. If the Higgs wavefunction is nearly uniform on 
$C_\mathbf{56}$, the Higgs will interact with all three 
generations of chiral matter in the same way, thus the Yukawa couplings
will be undesiredly close to an identity matrix.\footnote{We thank Jonathan Heckman for pointing this out.}
Nevertheless, especially with the presence of bulk fluxes, we expect 
the
Higgs wavefunction to be non-uniform and peak in some smaller
region. A simple %generic
 scenario would be that the region intersects with  
$C_\mathbf{56}$ in a connected small but finite range.
% in.
This scenario is then similar to
the case of a single Yukawa point studied in e.g. \cite{Heckman:2008qa,Cecotti:2009zf}, where the small nonperturbative
correction is now due to the finite size of the interaction region.
In this way, the 
Yukawa hierarchy is generated as in the $\SU(5)$ F-theory
GUTs. On the other hand, to really compute the Yukawa couplings, we first need to
understand the Higgs wavefunction profile and its possible
correlations with the exponentially low Higgs mass. Once we understand
these issues, we may be able to use the ultra-local approach developed in
\cite{Font:2012wq,Font:2013ida,Marchesano:2015dfa,Carta:2015eoh} to
computing Yukawa couplings within the intersecting region, but understanding those 
issues remains very challenging. In %less generic
more complicated scenarios where there are
multiple disconnected interaction regions, they are similar to the case of
multiple Yukawa points. The arguments in \cite{Cvetic:2019sgs} then 
suggest that there is also some Yukawa hierarchy, although the methods in
\cite{Cvetic:2019sgs} do not straightforwardly generalize to our models due to
the use of flux breaking. In summary, it is possible that there is
 some hierarchy between the Yukawa couplings.
This hierarchy may match with the observed Yukawa hierarchy, but 
explicitly computing the Yukawa matrix in our models will be an important
future step for realistic model building.

\subsection{Neutrino sector}
\label{subsec:neutrino}

Here we turn to the neutrino sector and make some brief comments. From Eq.
(\ref{eq:branchingrules56}), we see that $\mathbf{56}$ gives three copies
of singlets that can be right-handed neutrinos. Since in the
above we have restricted the leptons into one copy, only one copy of the
singlets $(\mathbf 1,\mathbf 1)_{0,5/2,1,1/2}$ have unsuppressed Yukawa
couplings with SM chiral matter:
\begin{gather}
    (\mathbf 1,\mathbf 2)_{1/2,-3,-2,-1}\times (\mathbf 1,\mathbf 2)_{-1/2,1/2,1,1/2}\times (\mathbf 1,\mathbf 1)_{0,5/2,1,1/2}\,,\nonumber \\
    (\mathbf 3,\mathbf 1)_{-1/3,-3,-2,-1}\times (\bar{\mathbf 3},\mathbf 1)_{1/3,1/2,1,1/2}\times (\mathbf 1,\mathbf 1)_{0,5/2,1,1/2}\,.
\end{gather}
The other two copies of singlets have nontrivial multiplicities but belong
to inert matter.

We can obtain the multiplicity of right-handed neutrinos from its ``chiral
index''. It sounds strange to calculate a ``chiral index'' for non-chiral
matter; the correct interpretation is that the right-handed neutrinos
carry additional $\U(1)$ charges, hence ``would be'' chiral matter if we ignore the St\"uckelberg
mechanism.
Since there are no vector-like exotics on the matter curve, the chiral
index is the same as the exact multiplicity, which remains unchanged
under the symmetry breaking. Now from the flux constraints we imposed in
previous sections, we see that the chiral index is fixed to be
\begin{equation}
    \chi_{(\mathbf 1,\mathbf 1)_{0,5/2,1,1/2}}=3\,.
\end{equation}
Therefore, we have three right-handed neutrinos, which favorably combine
with the left-handed ones to give three Dirac neutrinos and a square PMNS
matrix. The above Yukawa couplings then give the usual Dirac mass terms
after electroweak symmetry breaking. This scenario is more or less the
same as conventional GUTs with $\SO(10)$ gauge group or above.

There are also Majorana mass terms from the $\Sigma CC$-type couplings
involving two right-handed neutrinos and a bulk singlet. Since
all right-handed neutrinos have the same additional $\U(1)$ charges,
the Majorana masses are always exponentially suppressed (compared
to string/GUT scale) by the additional $\U(1)$ symmetries. In fact, similar suppression was already used in some 
early type II \cite{Blumenhagen:2006xt,Ibanez:2006da} and F-theory 
\cite{BeasleyHeckmanVafaII} SM-like model. It was estimated in those 
references that the exponential suppression factor might be around 
$10^{-6}$ to $10^{-4}$, which is much more mild than the electroweak 
hierarchy. This is not incompatible
 %Therefore, this 
%suppression %seems to match 
with the observational constraints on the 
seesaw mechanism. We emphasize that, however, these numerical estimates
are very crude, and without more explicit
computations of the masses and couplings, we cannot make fully precise
statements on how the left-handed neutrinos get very small masses.

\subsection{Gauge coupling unification}
\label{subsec:unification}

Here we briefly comment on the possibility of gauge coupling unification
in our models. Despite the use of $E_7$ in the construction of models,
whether %the gauge coupling unification is preserved 
gauge coupling unification is present
in any useful sense
is far from obvious.
From the point of view of the world-volume theory on the
IIB 7-branes supporting the $E_7$ gauge
theory (as in, e.g., \cite{BeasleyHeckmanVafaI}), it should be possible to find a classical %(though
%nonperturbative)
 description of flux breaking through turning on flux (T-dual to
 turning on an adjoint scalar as in, e.g., \cite{Taylor:1996ik}).
From this perspective, at sufficiently high energies
 the world-volume
$E_7$ gauge symmetry would be effectively restored, and the expected extra gauge bosons %should
would
become  relatively light, so there is some sense in which gauge coupling unification
might be expected.  
Note, however, that the quantization of flux means that the background
flux will give a mass scale $m_\mathrm{KK}=1/l_{\rm KK}$, where $l_{{\rm KK}}$ is
the compactification scale, so that this unification only occurs much above
the KK scale.  Furthermore,
in the nonperturbative F-theory regime, where there is no
weakly coupled description, it is not clear that the 7-brane
world-volume theory can be meaningfully separated from string theory
in the bulk space.
Thus, we do not necessarily expect unification even at the
compactification scale.
To understand some of the issues, we first clarify the meaning of gauge coupling
unification in our string theory context.

There are two separate aspects. First at the GUT scale
$M_\mathrm{GUT}$,\footnote{In the string theory context, $M_\mathrm{GUT}$ may be around the KK scale or string scale depending on model details.} the gauge
couplings in our models are clearly unified if flux breaking is absent.
The coupling is given by the volume of the gauge divisor:
\begin{equation}
\label{eq:divisorvol}
\frac{1}{g^{2}}\simeq\mathrm{vol}\left(\Sigma\right)\,.
\end{equation}
It is estimated from observations that $1/\alpha\simeq 24$ at 
$M_\mathrm{GUT}$ \cite{Martin:1997ns}; we simply assume that the divisor volume is stabilized 
to this particular value by certain mechanisms. On the other hand,
the remainder flux breaks $\SU(5)$ to $\gsm$ and 
induces some splitting of gauge couplings at $M_\mathrm{KK}$. 
It is 
then important to understand such splitting and determine its
size. Such
splitting has been understood in type IIB models \cite{Blumenhagen:2008aw,Mayrhofer:2013ara}:
the splitting between the $\SU(3)$ and $\SU(2)$ gauge couplings is
\begin{equation}
\label{eq:splitting}
\frac{1}{\alpha_{2}\left(M_\mathrm{GUT}\right)}-\frac{1}{\alpha_{3}\left(M_\mathrm{GUT}\right)}\simeq-\frac{1}{10g_s}\left[c_{1}\left(L_{3}\right)\right]^{2}=\frac{1}{5g_s}(n_{Q'}+1)\,,
\end{equation}
where $g_s$ is the string coupling, and $n_{Q'}$ is the number of vector-like pairs in the exotic 
representation $(\mathbf 3,\mathbf 2)_{-5/6}$; recall that we have set $n_{Q'}=0$ in previous sections. There is also a unification-like relation
\begin{equation}
\label{eq:unificationlike}
    \frac{1}{\alpha_{Y}\left(M_\mathrm{GUT}\right)}=\frac{5}{3}\frac{1}{\alpha_{1}\left(M_\mathrm{GUT}\right)}=\frac{1}{\alpha_{2}\left(M_\mathrm{GUT}\right)}+\frac{2}{3}\frac{1}{\alpha_{3}\left(M_\mathrm{GUT}\right)}\,.
\end{equation}
All the above, however, cannot be directly applied to F-theory models,
especially when the models, like our $E_7$ models, are intrinsically 
strongly coupled and have no type IIB limit.
This is because the axio-dilaton varies over the internal space and the
meaning of the $1/g_s$ correction is no longer clear. The worldvolume 
theory, which was used to derive the type IIB result, also needs to be
reconsidered in F-theory setups. In addition, there may be large
stringy threshold corrections to the gauge kinetic functions
due to the strong coupling nature of these models. All these subtleties imply that the splitting at 
$M_\mathrm{GUT}$ may not be small even if we set $n_{Q'}=0$.

Next, at scales lower than $M_\mathrm{GUT}$, the RG flow
of the SM gauge couplings are affected by the vector-like
exotics. The RG flow depends on both the representations
and the masses of the vector-like exotics. Since the remainder flux
already breaks the GUT group at $M_\mathrm{GUT}$, it is possible that
some vector-like exotics are light and do not form GUT multiplets. These
exotics seriously alter the RG flow and may ruin gauge coupling 
unification. The existence of such exotics, however, depends crucially on
uncontrolled aspects of the models such as SUSY breaking. It is also
possible that the presence of these exotics compensates the above splitting
at $M_\mathrm{GUT}$ and makes the couplings apparently unified from the
bottom-up perspective. Therefore, so far we cannot make any definite
statement on how the vector-like spectrum may affect the RG flow.

In conclusion, the gauge couplings in our models are affected by a
number of uncontrolled aspects, thus gauge coupling unification is not
guaranteed in our models. From this perspective, the unification of 
the observed gauge couplings in ordinary MSSM looks like an accident if
our models really describe our Universe. Nevertheless, this conclusion
mainly comes from our inability to compute non-topological details of our
models. A more careful string theory analysis in the future may reveal 
that the observed unification is in fact not an accident at all.

\section{Explicit global constructions of $E_7$ GUTs}
\label{sec:ExplicitConstruction}

In all the above sections, we have written down many necessary constraints
on the geometry and fluxes for constructing semi-realistic $E_7$ GUTs in
F-theory. It remains important to see whether all these constraints can
be satisfied simultaneously within a 4D F-theory model. In this section,
we provide an explicit global construction of such a model, using the
tools of toric hypersurfaces. The construction here is a 
generalization of that in \cite{Braun:2014pva}. It is also the first
explicit example of a rigid $E_7$ GUT
% instead of the 
 (rigid $E_6$ GUTs were presented in 
\cite{Li:2022aek}). Although we only present a single example here,
the same construction can be generalized to large class of F-theory
compactifications. Before writing down such an explicit model, it is
useful to first review the geometric and flux constraints we want to
achieve:
\begin{itemize}
\item $\Sigma$ as a del Pezzo surface supporting both rigid $E_{7}$ 
(with effective $-K_\Sigma$)
and
hypercharge flux, and $C_{\mathbf{56}}=-\Sigma\cdot\left(4K_{B}+3\Sigma\right)$
as a $\mathbb{P}^{1}$, to enable explicit computations and interesting
phenomenology. The first requirement demands that $\Sigma$ is a rigid
divisor on a non-toric base.
\item The general flux constraints in \S\ref{subsec:review:G4fluxes}: flux quantization, primitivity for
vertical flux, and tadpole cancellation. In particular, we should
look for flux configurations with minimal tadpole.
\item A vertical flux breaking $E_{7}\rightarrow\SU(5)$ and a remainder
(hypercharge) flux breaking $\SU(5)\rightarrow G_{\mathrm{SM}}$.
In particular we need $r\geq4$, see \S\ref{subsec:fluxbreaking:breaking}.
\item $\chi_{\left(\mathbf{3},\mathbf{2}\right)_{1/6}}^{\mathbf{133}}=0$ and $\chi_{\left(\mathbf{3},\mathbf{2}\right)_{1/6}}^{\mathbf{56}}=3$ for the total chiral
spectrum.
\item All three families of chiral $\left(\bar{\mathbf 3},\mathbf 1\right)_{1/3}$ and $\left(\mathbf 1,\mathbf 2\right)_{-1/2}$
coming from the same copy, i.e. Eq.\ (\ref{eq:56samecopy}),
to avoid exotic vector-like spectrum and couplings.
\item The copy of bulk $\left(\bar{\mathbf 3},\mathbf 1\right)_{1/3}$ and $\left(\mathbf 1,\mathbf 2\right)_{-1/2}$
that interacts with the chiral matter
being itself non-chiral, i.e. Eq.\ (\ref{eq:133samecopy}), to avoid light vector-like exotics.
\item $\left[c_{1}\left(L_{3}\right)\right]^{2}=-2$ for hypercharge flux,
to remove the exotic $\left(\mathbf 3,\mathbf 2\right)_{-5/6}$.
\end{itemize}
As always, we set the flux associated to the non-flat fiber to zero.

Now we write
down an explicit F-theory model that satisfies all the above constraints.
As in \cite{Li:2022aek}, we choose the base $B$ through the following
procedure. We start with an auxilliary toric threefold $A$ with $h^{1,1}(A)=4$.
Then the ambient fourfold $X$ is a $\mathbb P^1$-bundle over $A$ with a
certain normal bundle, and $B$ is a certain hypersurface in $X$. The
geometry of $B$ can be analyzed using the techniques in Appendix \ref{appendix:Hypersurfaces}.
With appropriate choices in the above procedure, we can construct $B$
containing a rigid $\Sigma$ with $r=4$ and nontrivial remainder flux.

\begin{table}[t]
\centering
\begin{tabular}{|c|c|}
\hline 
Toric ray & Divisor \tabularnewline
\hline
$(1,0,0,0)$ & $F_{E_1}+F_F$ \tabularnewline
\hline
$(0,1,0,0)$ & $F_{E_2}+F_F$ \tabularnewline
\hline
$(-1,-1,-1,0)$ & $F_F$ \tabularnewline
\hline
$(-1,0,-1,-3)$ & $F_{E_1}$ \tabularnewline
\hline
$(0,-1,-1,0)$ & $F_{E_2}$ \tabularnewline
\hline
$(0,0,-1,-4)$ & $F_\sigma$ \tabularnewline
\hline
$(0,0,1,0)$ & $F_\sigma+F_{E_1}+F_{E_2}+F_F$ \tabularnewline
\hline
$(0,0,0,-1)$ & $\sigma_A$ \tabularnewline
\hline
$(0,0,0,1)$ & $\sigma_A+4F_\sigma+3F_{E_1}$ \tabularnewline
\hline 
\end{tabular}
\caption{The toric rays and the corresponding divisors in the toric
construction of the ambient fourfold $X$.}
\label{tableoftoricrays}
\end{table}

Let us first construct the ambient space $X$. 
We choose $A$ to be a 
$\mathbb{P}^{1}$-bundle over the del Pezzo surface $dP_{2}$, which has
a toric description.
Let us first introduce the notations. Within $dP_{2}$ i.e. blowup of
$\mathbb{P}^{2}$ at two generic points, let $e_{1},e_{2}$ be the
exceptional curves from the blowup, and $h=f+e_{1}+e_{2}$ be the
hyperplane. The intersection numbers are $f^{2}=e_{1}^{2}=e_{2}^{2}=-1,f\cdot e_{1}=f\cdot e_{2}=1,e_{1}\cdot e_{2}=0$.
Now on $A$, we denote $\sigma$ as the $dP_{2}$ section and $E_{1},E_{2},F$
as the $\mathbb{P}^{1}$-fibers along $e_{1},e_{2},f$ respectively.
We choose the normal bundle of the $\mathbb{P}^{1}$-bundle to be $N_{\sigma}=-h$,
and the anticanonical class is $-K_{A}=2\sigma+3E_{1}+3E_{2}+4F$.
The intersection numbers on $A$ follow straightforwardly from those
on $dP_{2}$ and the relation $\sigma\cdot\left(\sigma+F+E_{1}+E_{2}\right)=0$.
Finally we let the fourfold $X$ be a $\mathbb{P}^{1}$-bundle over $A$
with normal bundle $N_{A}=-4\sigma-3E_{1}$. We denote $\sigma_{A}$ as the section
and $F_{I}$ be the fiber along $I\in\left\{ \sigma,E_{1},E_{2},F\right\} $.
The anticanonical class is $-K_{X}=2\sigma_{A}+6F_{\sigma}+6F_{E_{1}}+3F_{E_{2}}+4F_{F}$.
Again, the intersection numbers follow from those on $A$ and the
relation $\sigma_{A}\cdot\left(\sigma_{A}+4F_{\sigma}+3F_{E_{1}}\right)=0$. Note that with these choices of normal bundles, there is
a unique triangulation such that $X$ is a 
smooth and projective toric variety. The toric rays of $X$ are listed in Table \ref{tableoftoricrays}.

We now choose the threefold base $B$ as a hypersurface in $X$ with
irreducible class $B=\sigma_{A}+5F_{\sigma}+5F_{E_{1}}+2F_{E_{2}}+3F_{F}$.
By abuse of notation, we use $B$ to denote both the base and its divisor
class in $X$.
By adjunction $-K_{B}=B\cdot\left(\sigma_{A}+F_{\sigma}+F_{E_{1}}+F_{E_{2}}+F_{F}\right)$.
Using the techniques in Appendix \ref{appendix:Hypersurfaces}, one can
check that $h^{1,1}(B)=h^{1,1}(X)=5$. 
%In other words,
In particular, in this situation
 the divisors of $B$ are spanned by intersections in $X$. The
intersection numbers
of these divisors relevant to our purpose are
\begin{equation}
B\cdot\sigma_{A}\cdot F_{I}\cdot F_{J}=\left(\begin{array}{cccc}
-2 & 1 & 1 & 0\\
1 & -1 & 0 & 1\\
1 & 0 & -1 & 1\\
0 & 1 & 1 & -1
\end{array}\right)\,.
\end{equation}

Now we consider the divisor $\Sigma=B\cdot\sigma_{A}$. It is also
the hypersurface in $A$ with class $\sigma+2E_{1}+2E_{2}+3F$. This
class is irreducible and does not have any base locus (considered as a
hypersurface in $A$),
 so it is a
well-defined %rigid
irreducible gauge divisor. We compute
\begin{align}
-K_{\Sigma}
& =
B\cdot\sigma_{A}\cdot\left(F_{\sigma}+F_{E_{1}}+F_{E_{2}}+F_{F}\right)\nonumber \\
 & =\sigma_{A}\cdot\left(2F_{\sigma}\cdot F_{E_{1}}+2F_{\sigma}\cdot F_{E_{2}}+3F_{\sigma}\cdot F_{F}+3F_{E_{1}}\cdot F_{F}\right)\,,\\
N_{\Sigma} & =B\cdot\sigma_{A}^{2}\nonumber \\
 & =-\sigma_{A}\cdot\left(7F_{\sigma}\cdot F_{E_{1}}+4F_{\sigma}\cdot F_{E_{2}}+8F_{\sigma}\cdot F_{F}+3F_{E_{1}}\cdot F_{F}\right)\,.
\end{align}
Therefore by \eqref{eq:review:FkGl}, we see that $\Sigma$ is indeed a rigid divisor supporting
$E_{7}$. The matter curve is
\begin{align}
C_{\mathbf{56}}  & =\Sigma \cdot (-4K_B - 3 \Sigma)\\
& =B\cdot\sigma_{A}\cdot\left(\sigma_{A}+4F_{\sigma}+4F_{E_{1}}+4F_{E_{2}}+4F_{F}\right)\nonumber \\
 & =B\cdot\sigma_{A}\cdot\left(F_{E_{1}}+4F_{E_{2}}+4F_{F}\right)\,.
\end{align}
Notice that the divisor $E_{1}+4E_{2}+4F$ in $A$ is also
irreducible and does not have any base locus. Therefore, $C_{\mathbf{56}}$
is also irreducible with genus
\begin{equation}
g=1+\frac{1}{2}C_{\mathbf{56}}\cdot\left(C_{\mathbf{56}}+K_{\Sigma}\right)=0\,,
\end{equation}
which means that the matter curve is simply a $\mathbb{P}^{1}$.
%On $\Sigma$, this matter curve self-intersects at $C_\mathbf{56}^2=7$ points,
%so it can support a rank-3 Yukawa matrix without nonperturbative effects.

We can now study the constraints on vertical flux. First, we study
primitivity by expanding the K\"ahler
form of $B$ using a basis of base divisors:
\begin{align}
\left[J_{B}\right]= & B\cdot(t_{1}\left(F_{E_{1}}+F_{F}\right)+t_{2}\left(F_{E_{2}}+F_{F}\right)+t_{3}\left(F_{E_{1}}+F_{E_{2}}+F_{F}\right)\nonumber \\
 & +t_{4}\left(F_{\sigma}+F_{E_{1}}+F_{E_{2}}+F_{F}\right)+t_{5}\left(\sigma_{A}+4F_{\sigma}+4F_{E_{1}}+4F_{E_{2}}+4F_{F}\right)\,,
\end{align}
where $t_{1},t_{2},t_{3},t_{4},t_5$ are linear
combinations of K\"ahler moduli,
 and may be negative
inside the K\"ahler cone of $B$
in general. While determining the
exact K\"ahler
cone of a hypersurface in a toric variety can be subtle, the K\"ahler cone of $B$ must contain that of $X$
 \cite{Demirtas:2018akl}.
For simplicity, we look for a solution of the primitivity constraints in
the K\"ahler cone of $X$ only.
By a direct toric computation, one can check that the K\"ahler cone of
$X$ is
given by $t_{1},t_{2},t_{3},t_{4},t_{5}>0$. The independent $S_{i\alpha}$
are $S_{i\sigma},S_{iE_{1}},S_{iE_{2}},S_{iF}$, where we have simplified
the notation and denoted $S_{i\left(B\cdot F_{I}\right)}$ as $S_{iI}$.
The primitivity condition is then
\begin{equation}
\label{eq:explicitprimitivity}
t_{1}\left(\phi_{iE_{2}}+\phi_{i\sigma}\right)+\left(t_{2}+3t_{5}\right)\left(\phi_{iE_{1}}+\phi_{i\sigma}\right)+\left(t_{3}+t_{5}\right)\left(\phi_{iF}+2\phi_{i\sigma}\right)+t_{4}\left(\phi_{iF}+\phi_{iE_{1}}+\phi_{iE_{2}}\right)=0\,,
\end{equation}
for all $i$. A necessary but not sufficient condition for satisfying
primitivity is
that there must be
some coefficients in Eq.\ (\ref{eq:explicitprimitivity}) with
opposite signs for each $i$.
Below we will find explicit solutions to primitivity and check that the
solutions are within the K\"ahler cone.

Next, we require the total chiral indices to be
\begin{gather}
\chi^{\mathbf{133}}_{(\mathbf 3,\mathbf 2)_{1/6}}=-2\left(n_{F}+n_{E_{1}}+n_{E_{2}}\right)=0\,,\\
\chi^{\mathbf{56}}_{(\mathbf 3,\mathbf 2)_{1/6}}=-n_{F}-3n_{E_{1}}-5n_{\sigma}=3\,.
\end{gather}
Here the $n_I$ parameterize the fluxes through Eq. (\ref{eq:e7:phi-n}).
To understand what values of $n_{I}$ we should turn on to get the right
total chiral spectrum,
we should first look at flux quantization, since for $E_{7}$ models
$c_{2}(\hat{Y})$ is not necessarily even. We can calculate $c_2(\hat Y)$
using the techniques in \cite{Jefferson:2021bid}, which involve picking
a particular resolution of the $E_7$ models, but the parity of 
$c_2(\hat Y)$ is expected to be resolution-independent. We outline the
procedure in Appendix \ref{appendix:Quantization}, while here we only apply the result, which
tells us that we can turn on half-integers $n_{E_{1}},n_{F}$ and integers
$n_{E_{2}},n_{\sigma}$ (similarly for $\phi_{6\alpha}$) to guarantee flux
quantization. Note that these may not be the only choices of $n_I$, 
since the structure of $H^4(\hat Y,\mathbb Z)$ is subtle and may 
include elements with fractional coefficients. Also note that these choices of $n_I$
do not necessarily mean that the vertical flux does not belong to $H^{4}(\hat{Y},\mathbb{Z})$,
since as we will see its tadpole is still integer. Now with these choices
of $n_I$, we see that
an almost minimal flux configuration $\left(n_{\sigma},n_{E_{1}},n_{E_{2}},n_{F}\right)=\left(0,-3/2,0,3/2\right)$
already gives the above two chiral indices and is consistent with primitivity.

There are more flux conditions that constraint the values of $\phi_{5\alpha},\phi_{6\alpha}$.
First, by flux quantization we should turn on half-integers $\phi_{6E_{1}},\phi_{6F}$
and integers for the remaining parameters. Primitivity still constraints
their values nontrivially. Moreover, to put the chiral matter into
the copy $\left(b_{4},b_{5},b_{6}\right)=\left(1/2,1,1/2\right)$,
we need to impose Eq.\ (\ref{eq:56samecopy}), or in terms of flux parameters
\begin{equation}
-\phi_{5F}-3\phi_{5E_{1}}-5\phi_{5\sigma}=12\,,\quad-\phi_{6F}-3\phi_{6E_{1}}-5\phi_{6\sigma}=6\,.
\end{equation}
We also need to avoid light vector-like exotics in the bulk
copy $\left(b_{4},b_{5},b_{6}\right)=\left(2,2,1\right)$. Eq.\ (\ref{eq:133samecopy}) then leads
to
\begin{equation}
\phi_{5F}+\phi_{5E_{1}}+\phi_{5E_{2}}=0\,.
\end{equation}
These are mild linear constraints on the flux parameters $\phi_{5\alpha},\phi_{6\alpha}$.
Although there are multiple solutions to these linear constraints, we should seek for solutions with minimal
tadpole. By a brute force search, we find that one of the optimal
solutions is

\begin{equation}
\left(\phi_{5\sigma},\phi_{5E_{1}},\phi_{5E_{2}},\phi_{5F},\phi_{6\sigma},\phi_{6E_{1}},\phi_{6E_{2}},\phi_{6F}\right)=\left(0,-5,2,3,1,-\frac{7}{2},1,-\frac{1}{2}\right)\,,
\end{equation}
which consistently stabilizes the K\"ahler moduli at $t_{1}=t_{2}+3t_{5}=t_{3}+t_{5}=3t_{4}$.
Together with $n_{I}$, this vertical flux gives a tadpole
\begin{equation}
\frac{1}{2}\left[G_{4}^{\mathrm{vert}}\right]\cdot\left[G_{4}^{\mathrm{vert}}\right]=32\,.
\end{equation}
As a comparison, if we do not impose Eqs. 
(\ref{eq:133samecopy}) and (\ref{eq:56samecopy}) i.e. primitivity is the only constraint on $\phi_{5\alpha},\phi_{6\alpha}$,
the minimal tadpole is
\begin{equation}
\frac{1}{2}\left[G_{4}^{\mathrm{vert}}\right]\cdot\left[G_{4}^{\mathrm{vert}}\right]=20\,,
\end{equation}
given by e.g.
\begin{equation}
\left(\phi_{5\sigma},\phi_{5E_{1}},\phi_{5E_{2}},\phi_{5F},\phi_{6\sigma},\phi_{6E_{1}},\phi_{6E_{2}},\phi_{6F}\right)=\left(0,-4,-1,5,1,-\frac{5}{2},-1,\frac{3}{2}\right)\,.
\end{equation}
Therefore, we see that the vertical flux we need to turn on is slightly
non-generic.

Let us now turn to remainder flux. It can be shown that $\Sigma$ is a del
Pezzo surface $dP_6$ and supports remainder flux. First recall that
$\Sigma$ is a hypersurface in $A$
with class $\sigma+2E_1+2E_2+3F$. In other words, $\Sigma$ is the vanishing
locus
\begin{equation} \label{Sigmalocus}
    xP+yP'=0\,,
\end{equation}
in $A$, where $P,P'$ are sections of $\mathcal O_A(2E_1+2E_2+3F),\mathcal O_A(E_1+E_2+2F)$ respectively, and $x,y$
are the homogeneous coordinates of the $\mathbb P^1$ in $A$. For generic
points in the $dP_2$, Eq.\ (\ref{Sigmalocus}) has a unique
solution, representing a single point in $\mathbb P^1$. On the other hand,
there are $\left(2e_{1}+2e_{2}+3f\right)\cdot\left(e_{1}+e_{2}+2f\right)=4$ points in $dP_2$ such that $P=P'=0$,
and Eq.\ (\ref{Sigmalocus}) represents the whole $\mathbb P^1$.
Therefore, the geometry of $\Sigma$ is $dP_2$ blown up in 4
generic points i.e. a $dP_6$.

To construct the remainder flux, notice that the four exceptional
curves on $\Sigma$ from blowing up $dP_2$ (denoted by
$e_3$ to $e_6$) are all $\mathbb P^1$
fibers in $A$, hence all have the same class in $B$.  Therefore, we can
choose e.g. $C_{\mathrm{rem}}=e_3-e_4$ (or any difference $e_i-e_j$ for 
distinct $i,j=3,4,5,6$) with $C_{\mathrm{rem}}^2=-2$, and turn on the 
remainder flux specified in \S\ref{subsec:vectorexotics}. Therefore, we need a total tadpole of
36 to satisfy all the
flux constraints. Using the techniques in \cite{Jefferson:2021bid}, we 
find that $\chi(\hat{Y})=1176$
and $\chi(\hat{Y})/24=49>36$, so tadpole cancellation
is satisfied. Unfortunately, there seems to be not much room to achieve
full moduli stabilization, but the situation should improve if we
consider more complicated geometries.

Having the full flux configuration, it is now straightforward to also
compute the vector-like spectrum. For simplicity again we ignore the
uncharged singlets. Since $C_{\mathbf{56}}$ is a $\mathbb{P}^{1}$, all vector-like
pairs comes from the bulk of $\Sigma$. Using the
formula for $n_{\beta}$ in \S\ref{subsec:e7:vectorlike}, we get the vector-like spectrum as in
Table \ref{tableofvector}. It is clear that there are too many doublet
and triplet Higgs that are not inert, and it is important to understand
how the mass hierarchy is produced, such that we only see one doublet
Higgs (pair) at the electroweak scale. There are also a number of light and inert vector-like exotics from the table.

\begin{table}[t]
\centering
\begin{tabular}{|c|c|c|c|}
\hline 
$\left(\mathbf 1,\mathbf 2\right)_{-1/2,3,2,1}$ & 18 & $\left(\mathbf 1,\mathbf 2\right)_{1/2,-3,-2,-1}$ & $\mathbf{18}$\tabularnewline
\hline 
$\left(\mathbf 1,\mathbf 2\right)_{-1/2,-2,-2,-1}$ & $\mathbf{4}$ & $\left(\mathbf 1,\mathbf 2\right)_{1/2,2,2,1}$ & 4\tabularnewline
\hline 
$\left(\mathbf 1,\mathbf 2\right)_{-1/2,-2,-1,-1}$ & 19 & $\left(\mathbf 1,\mathbf 2\right)_{1/2,2,1,1}$ & 16\tabularnewline
\hline 
$\left(\mathbf 1,\mathbf 2\right)_{-1/2,-2,-1,0}$ & 24 & $\left(\mathbf 1,\mathbf 2\right)_{1/2,2,1,0}$ & 27\tabularnewline
\hline 
$\left(\bar{\mathbf 3},\mathbf 1\right)_{1/3,3,2,1}$ & 21 & $\left(\mathbf 3,\mathbf 1\right)_{-1/3,-3,-2,-1}$ & $\mathbf{21}$\tabularnewline
\hline 
$\left(\bar{\mathbf 3},\mathbf 1\right)_{1/3,-2,-2,-1}$ & $\mathbf{3}$ & $\left(\mathbf 3,\mathbf 1\right)_{-1/3,2,2,1}$ & 3\tabularnewline
\hline 
$\left(\bar{\mathbf 3},\mathbf 1\right)_{1/3,-2,-1,-1}$ & 16 & $\left(\mathbf 3,\mathbf 1\right)_{-1/3,2,1,1}$ & 13\tabularnewline
\hline 
$\left(\bar{\mathbf 3},\mathbf 1\right)_{1/3,-2,-1,0}$ & 21 & $\left(\mathbf 3,\mathbf 1\right)_{-1/3,2,1,0}$ & 24\tabularnewline
\hline 
$\left(\bar{\mathbf 3},\mathbf 1\right)_{-2/3,-1,-1,-1}$ & 6 & $\left(\mathbf 3,\mathbf 1\right)_{2/3,1,1,1}$ & 3\tabularnewline
\hline 
$\left(\bar{\mathbf 3},\mathbf 1\right)_{-2/3,-1,-1,0}$ & 1 & $\left(\mathbf 3,\mathbf 1\right)_{2/3,1,1,0}$ & 4\tabularnewline
\hline 
$\left(\bar{\mathbf 3},\mathbf 1\right)_{-2/3,-1,0,0}$ & 19 & $\left(\mathbf 3,\mathbf 1\right)_{2/3,1,0,0}$ & 19\tabularnewline
\hline 
$\left(\mathbf 3,\mathbf 2\right)_{1/6,-1,-1,-1}$ & 5 & $\left(\bar{\mathbf 3},\mathbf 2\right)_{-1/6,1,1,1}$ & 2\tabularnewline
\hline 
$\left(\mathbf 3,\mathbf 2\right)_{1/6,-1,-1,0}$ & 0 & $\left(\bar{\mathbf 3},\mathbf 2\right)_{-1/6,1,1,0}$ & 3\tabularnewline
\hline 
$\left(\mathbf 3,\mathbf 2\right)_{1/6,-1,0,0}$ & 16 & $\left(\bar{\mathbf 3},\mathbf 2\right)_{-1/6,1,0,0}$ & 16\tabularnewline
\hline 
$\left(\mathbf 3,\mathbf 2\right)_{-5/6,0,0,0}$ & 0 & $\left(\bar{\mathbf 3},\mathbf 2\right)_{5/6,0,0,0}$ & 0\tabularnewline
\hline 
$\left(\mathbf 1,\mathbf 1\right)_{1,-1,-1,-1}$ & 6 & $\left(\mathbf 1,\mathbf 1\right)_{-1,1,1,1}$ & 3\tabularnewline
\hline 
$\left(\mathbf 1,\mathbf 1\right)_{1,-1,-1,0}$ & 1 & $\left(\mathbf 1,\mathbf 1\right)_{-1,1,1,0}$ & 4\tabularnewline
\hline 
$\left(\mathbf 1,\mathbf 1\right)_{1,-1,0,0}$ & 15 & $\left(\mathbf 1,\mathbf 1\right)_{-1,1,0,0}$ & 15\tabularnewline
\hline 
\end{tabular}
\caption{The representations and multiplicities of vector-like matter
originated from the adjoint $\mathbf{133}$ on the bulk of gauge divisor.
Only the bold multiplicities correspond to fields interacting with the
SM chiral matter without exponential suppression. All the other fields
are inert vector-like exotics. Note that there are 
nontrivial linear relations between these numbers implied
by the formulas in \S\ref{subsec:e7:vectorlike}.}
\label{tableofvector}
\end{table}

In conclusion, we have obtained an explicit F-theory model with the SM gauge
group from rigid $E_{7}$, three families of SM chiral
matter with qualitatively standard Yukawa couplings and suppressed proton decay,
and excess numbers of heavy Higgs with some doublet-triplet splitting.
The flux configuration requires some but not too much fine-tuning. We
emphasize again that most analysis in this section depends on the local
geometry only. We expect that many of the F-theory threefold bases 
contain local geometries that are the same or similar
to the above, so this explicit
 construction can be easily generalized to large class of 4D F-theory
compactifications.

\section{Comparing with other F-theory constructions in the literature}
\label{sec:Comparison}

As we have pointed out a number of times in the previous sections, the
$E_7$ models have many features that are distinct from previous SM-like
constructions in F-theory. This distinction makes the $E_7$ models a
new interesting class of models to be studied in depth in the future.
In this section, we %elaborate the distinction in details. We 
explain in  more detail
some of the specific differences between
the models presented here and the F-theory models with tuned $\gsm$ or
$\SU(5)$ reviewed briefly in \S\ref{sec:Intro}, as well as the rigid $E_6$
GUTs.

\subsection{Tuned models}
\label{subsec:comparetuned}

There have been many SM-like F-theory constructions using tuned gauge
groups such as $\gsm$ or $\SU(5)$ (again, for reviews see
\cite{WeigandTASI,HeckmanReview,Cvetic:2022fnv,Marchesano:2022qbx}). The
most obvious difference between those
constructions and the $E_7$ models presented here has been discussed
in \S\ref{sec:Intro}: namely, fine-tuning of many complex structure
moduli is required to obtain $\gsm$ or $\SU(5)$ geometrically, while
the presence of rigid $E_7$ only depends on the normal bundle of
$\Sigma$ instead of any moduli.  It seems that rigid $E_7$ factors are relatively abundant
in the landscape. Although the measure on the landscape has never been
clear, a naive counting measure on the (singular) geometries suggests
a large exponential dominance of geometries supporting rigid $E_7$ factors over those supporting tuned gauge
factors.
Note, however, that it is possible that certain flux choices may in
some situations force
complex structure moduli to a tuned locus with an enhanced gauge
group; further investigation of this possibility is needed to clarify
the level of tuning really involved in geometrically tuned
constructions beyond the level suggested by the analysis of e.g., \cite{BraunWatariGenerations}.
Due to the  moduli-independence of the relevant gauge group, it may also be easier to
incorporate  a full analysis of moduli stabilization
 in the rigid models than the tuned
ones.

Another significant difference regards the Yukawa couplings. In many
SM-like F-theory constructions, some selection rules 
are required to get rid of exotic couplings. The Yukawa couplings
in tuned $\gsm$ or $\SU(5)$ models come from $CCC$-type couplings, and all the
matter fields are localized on matter curves. Therefore, the selection
rules are usually obtained by engineering a set of multiple matter curves
where different types of matter localize separately, or some additional
$U(1)$'s by tuning the global geometry. In the $E_7$ models,  selection
rules that remove exotic couplings
 automatically follow from the use of $\Sigma CC$-type couplings,
and easily separate the chiral matter from vector-like matter including the Higgs. There
are also approximate $\U(1)$'s from the St\"uckelberg mechanism that
arrive
without
additional tuning. Therefore,  the
selection rules needed to match expectations from observed physics
 are more easily
realized in the $E_7$ models than in other constructions. An example is the proton decay suppression
described in \S\ref{subsec:protondecay}.

The means of realizing the Higgs in the two classes of models is also
qualitatively different. In tuned $\gsm$ or $\SU(5)$, the Higgs comes
from some
vector-like matter on matter curves. Such a construction requires explicit
specification of the sheaf cohomology groups in Eq.\ 
(\ref{eq:mattercurvecohomologies}), which are in general very hard to
compute since they are moduli-dependent quantities. More exotic tools 
like root bundles \cite{Bies:2021nje,Bies:2021xfh,Bies:2022wvj,Bies:2023jqg, Bies:2023sfm} may also be needed in the construction.
 In many cases, there
is no Higgs in the low-energy theory unless some further tuning is done.
In $\SU(5)$, we also need the Higgs matter surfaces to have remainder
components, such that the hypercharge flux is present on the Higgs curves
and doublet-triplet splitting can be achieved. Generic matter surfaces,
however, are purely vertical 
unless further tuning on moduli is done, and
global examples of such scenarios are
 rare in the literature (see e.g. \cite{Braun:2014pva}). In contrast, in the $E_7$ models we can instead
realize the Higgs as bulk vector-like matter, which generically has
nonzero multiplicities that are easily calculated from the fluxes. 
In this situation,
there are 
already Higgs fields with some doublet-triplet splitting without any further tuning, but the issue becomes
having too many instead of too few Higgs fields. It is less clear how to
make one pair of the Higgs exponentially lighter in the $E_7$ models, 
while  in the tuned models there can be exactly one pair of Higgs, and thus
the way to obtain the Higgs hierarchy may be clearer.

Because of the use of flux breaking and $E_7$, there are many further
differences between these two classes of models in terms of computational abilities. First, in tuned $\gsm$ 
or $\SU(5)$ the total chiral spectrum is controlled by one flux parameter
only. In many cases the chiral indices contain large prefactors, and 
three generations of chiral matter cannot easily be obtained using integer 
fluxes, unless more nontrivial (and less completely
understood)
quantization conditions are used, as in
\cite{CveticEtAlQuadrillion,Jefferson:2022yya}. In the $E_7$ 
models, however, many flux parameters from vertical flux breaking 
contribute to the chiral indices, giving a linear Diophantine structure.
As a result, it is natural to get three generations of 
chiral matter just by generic integer fluxes. 

Specifically for $\SU(5)$, the removal of exotic $(\mathbf 3,\mathbf
2)_{-5/6}$ appears to be harder.  This is because the form of hypercharge flux
is completely fixed to be $\phi_{ir}\propto (2,4,6,3)$ and there is no
free flux parameter like $\phi_{4r}$ as in the $E_7$ models. Therefore,
there must be a factor of 5 in $c_1(L_3)$, and we need to use fractional
line bundles to satisfy the condition $[c_1(L_3)]^2=-2$ for removing the
exotic $(\mathbf 3,\mathbf 2)_{-5/6}$. In contrast, as in \S\ref{subsec:vectorexotics},
this condition in $E_7$ is already satisfied by a fairly likely choice of
integer remainder flux. On the other hand, there can be some controlled
scenario in $\SU(5)$ where all the vector-like exotics are removed, while
in $E_7$ we cannot remove most of the vector-like exotics; we can at best
arrange them into inert fields.

In conclusion, we see that the $E_7$ models are not only more natural in
the landscape, but also possess a number of phenomenological advantages
over the tuned models. These models demonstrate how using naturalness
as the guiding philosophy can help us discover more semi-realistic 
features in the landscape. On the other hand, these models still have 
their own shortcomings especially regarding the heavy mass spectrum, due
to the lack of computational technologies.

\subsection{Rigid $E_6$ GUTs}
\label{subsec:compareE6}

In \cite{Li:2021eyn,Li:2022aek}, it was proposed that the rigid 
construction of SM-like models works equally well for both $E_7$ and 
$E_6$, since these two gauge groups are similarly abundant in the 
landscape. While the two GUT groups share some features such as the 
naturalness of the gauge group and three generations of chiral matter, 
$E_6$ behaves differently when coming to Yukawa couplings. While $E_7$
models do not have any $CCC$-type couplings, in $E_6$ models the gauge 
group only gets enhanced to $E_8$ at codimension-3 singularities, which
are well-defined Yukawa points giving $CCC$-type couplings. Moreover,
the branching rules from $E_6$ to $\gsm$ including the additional $\U(1)$ charges
$(b_4,b_5)$ are
\begin{align}
\mathbf{27} & \rightarrow\left(\mathbf 1,\mathbf 1\right)_{0,5/3,4/3}+\left(\mathbf 1,\mathbf 1\right)_{0,5/3,1/3}+\left(\mathbf 1,\mathbf 1\right)_{1,2/3,1/3}\nonumber \\
 & +\left(\mathbf 3,\mathbf 2\right)_{1/6,2/3,1/3}+\left(\bar{\mathbf 3},\mathbf 1\right)_{-2/3,2/3,1/3}+\left(\bar{\mathbf 3},\mathbf 1\right)_{1/3,-1/3,1/3}+\left(\bar{\mathbf 3},\mathbf 1\right)_{1/3,-1/3,-2/3}\nonumber \\
 & +\left(\bar{\mathbf 3},\mathbf 1\right)_{1/3,4/3,2/3}+\left(\mathbf 1,\mathbf 2\right)_{-1/2,-1/3,1/3}+\left(\mathbf 1,\mathbf 2\right)_{-1/2,-1/3,-2/3}+\left(\mathbf 1,\mathbf 2\right)_{-1/2,4/3,2/3}\,,\\
\mathbf{78} & \rightarrow\left(\mathbf 8,\mathbf 1\right)_{0,0,0}+\left(\mathbf 1,\mathbf 3\right)_{0,0,0}+3\times\left(\mathbf 1,\mathbf 1\right)_{0,0,0}\nonumber \\
 & +[\left(\mathbf 1,\mathbf 1\right)_{0,0,1}+\left(\mathbf 1,\mathbf 1\right)_{1,-1,0}+\left(\mathbf 1,\mathbf 1\right)_{1,-1,-1}+\left(\mathbf 3,\mathbf 2\right)_{-5/6,0,0}+\left(\mathbf 3,\mathbf 2\right)_{1/6,-1,0}+\left(\mathbf 3,\mathbf 2\right)_{1/6,-1,-1}\nonumber \\
 & +\left(\bar{\mathbf 3},\mathbf 1\right)_{-2/3,-1,0}+\left(\bar{\mathbf 3},\mathbf 1\right)_{-2/3,-1,-1}+\left(\bar{\mathbf 3},\mathbf 1\right)_{1/3,-2,-1}+\left(\mathbf 1,\mathbf 2\right)_{-1/2,-2,-1}+\mathrm{conjugates]}\,.
\end{align}
Unlike $E_7$, one can check that there is no suitable field on the bulk
of $\Sigma$ that can play the role of Higgs, so the Higgs must be localized
on the matter curve. The Yukawa couplings in $E_6$ models are more similar to the tuned models and many results in \S\ref{sec:Pheno} do not apply to $E_6$ models, while
$E_{6}$ models also suffer from many vector-like exotics. In this
sense, $E_{7}$ models are fundamentally different from any other F-theory
GUT models.

\section{Conclusion}
\label{sec:Conclusion}

In this chapter, we have studied various phenomenological aspects of $E_7$
GUTs in 4D F-theory compactifications. These models were proposed in
\cite{Li:2021eyn,Li:2022aek} as a large class of natural SM-like
constructions, since rigid $E_7$ gauge factors
are moduli independent and common in the F-theory landscape. Vertical and remainder fluxes
are used to break $E_7$ to the SM gauge group, and appear fairly likely to induce three
generations of SM chiral matter. Here we have shown that the use of $E_7$
and flux breaking also naturally implies several more phenomenologically favorable
features, including suppression of proton decay, doublet-triplet 
splitting, and Higgs candidates with the right structure of
SM Yukawa couplings due to approximate global symmetries descending
from the $E_7$ Cartan generators. 
%The model also predicts the existence of several 
%inert vector-like exotics around TeV scale, which is a valid prediction 
%on experiments. 
For the first time, we have
written down an explicit global construction of such $E_7$ models that
achieve all the above features. The construction of these features is
qualitatively distinct from other SM-like constructions in the previous 
F-theory literature. In particular, only mild tuning on the discrete data
of the geometry and the flux background is involved in the construction.
The results in this chapter give us strong hints towards
SM constructions in F-theory that are both realistic and natural. In
other words, these results appear to be compatible with the hypothesis
% the philosophy
 that our Universe 
can be described as a
 \emph{natural} solution in the string landscape.

These results, on the other hand, are still far from complete in realizing
the full details of the Standard Model in string theory. Since the $E_7$
models are inherently strongly coupled and there is extensive use of 
fluxes, the machinery for computing continuous parameters, or at least
the moduli dependence of continuous parameters, is very limited. While we
can more or less fully specify the discrete data in the $E_7$ models, our
arguments are at best qualitative when it comes to questions on Yukawa
couplings, mass scales, etc. Such limitations lead to a number of
shortcomings of the $E_7$ models, especially the unavoidable presence of
(mostly inert) vector-like exotics with masses not determined. The Higgs
hierarchy problem, i.e. the $\mu$-problem,
also remains unsolved in our models.

There are many challenges to
% is a long way of studies before
 answering these important questions.
First, we need to develop new tools beyond the ultra-local approach \cite{Font:2012wq,Font:2013ida,Marchesano:2015dfa,Carta:2015eoh} to
compute the moduli dependence of various quantities. After that, we still
need to understand more fundamental questions like the realization of
moduli stabilization and SUSY breaking in F-theory, which by themselves
are essential components of realistic model building. Solutions to these questions also
involve tackling some open problems such as computing the K\"ahler
potential in F-theory. All these tasks are particularly challenging when
there is no weakly coupled type IIB limit for the $E_7$ models.
Nonetheless, some insight into aspects like gauge coupling unification may be
possible by considering the 8D world-volume theory on the $E_7$
7-branes, where flux breaking should have a classical (if
nonperturbative) description.

There are also several extensions of the $E_7$ models presented in this
chapter that should be investigated further. Throughout this chapter, we have made several strong assumptions on
the geometry, such as restricting the matter curve to be $\mathbb P^1$,
to enable more interesting computations on the discrete data. It will
be interesting, although technically challenging, to relax these 
assumptions and explore more behavior of the $E_7$ models. It is also
important to study the statistics of these $E_7$ models in the F-theory
landscape. By scanning through a large set of F-theory bases and flux
configurations, we can quantitatively analyze the genericity of different
features of the $E_7$ models, which further sheds light on where our
Universe sits in the landscape.

We hope to address some of these questions in future studies.

\chapter{Large $\U(1)$ charges from flux breaking in 4D F-theory models}
\label{chap:u1}

In this chapter, we study the massless charged spectrum of $\U(1)$ gauge
fields in F-theory that arise from flux breaking of a nonabelian
group.  The $\U(1)$ charges that arise in this way can be very large.
In particular, using vertical flux breaking, we construct an
explicit 4D F-theory model with a $\U(1)$ decoupled from other gauge
sectors, in which the massless/light fields have charges as large as
657. This result greatly exceeds prior results in the literature. We
argue heuristically that this result may provide an upper bound on
charges for light fields under decoupled $\U(1)$ factors in the F-theory landscape. We
also show that the charges can be even larger when the $\U(1)$ is
coupled to other gauge groups. The results in this chapter are based 
on \cite{Li:2023rqf}.

\section{Introduction}
\label{sec:u1:intro}

It is well-known that string theory, when
compactified on manifolds in various dimensions,
gives a vast range of vacuum solutions known as the
string landscape. The low-energy physics of
these vacuum solutions can be
described by quantum field theories coupled to
gravity, with a wide range of
different gauge groups and matter
content. Nevertheless, there are strong constraints
from string theory, or quantum gravity in general,
on the low-energy theories that have a consistent
UV completion with gravity. 
Such constraints have been a key component in the analysis of string
theory since the early days of the subject, when Green and Schwarz
identified the strong conditions imposed by anomaly cancellation on
quantum theories of gravity in ten dimensions \cite{Green:1984sg},
leading to the identification of the heterotic string theory
\cite{Gross:1984dd}; later work has shown that indeed all consistent
theories of quantum gravity in ten dimensions with supersymmetry are
those that come from string theory (at least at the level of massless
spectra) \cite{Adams:2010zy,Kim:2019vuc}.

In lower space-time dimensions, particularly in 4D, the relationship
is much less clear
between the set of theories that can be realized in string theory and
those that appear consistent from the point of view of low-energy EFT
coupled to gravity.  In recent years, observations of general features
of string vacua and black hole behavior have led to a number of
speculations regarding quantum gravity constraints on low-energy
physics that are often referred to as  \emph{Swampland
  Conjectures} \cite{VafaSwamp,OoguriVafaSwamp} (see \cite{vanBeest:2021lhn} for a recent
review). These constraints have the potential
not only to shed light on
the general structure of string theory and quantum
gravity, but also may have phenomenological
implications leading to insights into physics
beyond the Standard Model.

One concrete set of questions regarding consistent quantum gravity
theories and the string landscape addresses bounds on the complexity
of the gauge and matter fields that are possible.
For example, while the rank of the gauge group in 6D or 4D
supersymmetric
string
vacua can be very high (see, e.g., \cite{Candelas:1997eh,Aspinwall:1997ye,MorrisonTaylorToric,Wang:2020gmi}), there is believed to be a
finite bound.  Similarly,
explicit string constructions of vacua in four and
higher dimensions give massless or light matter
representations of bounded complexity for nonabelian gauge groups
(see, e.g., \cite{Dienes:1996yh,KleversEtAlExotic,Cvetic:2018xaq}).
In this chapter we focus on the question of what kinds of charges are
possible for massless or light fields charged under a U(1) gauge group
in  a 4D string vacuum constructed from F-theory  \cite{VafaF-theory,MorrisonVafaI,MorrisonVafaII}.

One of the most
widely accepted swampland-style conjectures is the
\emph{Completeness Hypothesis} \cite{Polchinski:1998rr,Banks:2010zn}, which states that
in a gauge theory coupled to gravity, all gauge
charges (consistent with charge quantization) must
be realized by some physical states. 
This conjecture has been proven in the context of quantum gravity in
AdS space with a holographic dual description \cite{Harlow:2018tng}.
Here the
physical states can be massless, or massive
including black holes. 
On the other hand, the situation is less well understood if we
consider
only massless or light\footnote{By massless or light fields in the F-theory context, we mean states coming from
branes wrapping cycles with vanishing volume. In 4D, these
include both chiral fields, which are truly massless, and
vector-like fields, which are kinematically massless but get
some light masses (relative to black hole masses) in the low-energy
theory from interactions in the
superpotential. We discuss both cases in our examples.}
fields. We may expect upper bounds on the gauge
charges that can be realized by massless fields in
the landscape, but it is not clear how large the
upper bounds are or whether the bounds even exist.
This is particularly unclear for $\U(1)$ charges
since, as explained below, it is very hard to
geometrically engineer $\U(1)$ gauge groups with
even moderately large charges (i.e., $q > 3$)
for massless states in string theory.

It is natural to look for such upper bounds using
the framework of F-theory, since this approach provides a
global description of the largest 
connected class of supersymmetric string vacua that is currently understood (see
\cite{WeigandTASI} for a review). 
F-theory gives 4D
$\mathcal N=1$ supergravity models when
compactified on elliptically fibered Calabi-Yau
(CY) fourfolds $Y$, corresponding to
non-perturbative compactifications of type IIB
string theory on general
(non-Ricci flat) complex K\"ahler threefold base
manifolds $B$. F-theory is also known to contain many vacua that are dual to many other types
of string compactifications, such as heterotic
models. The power of F-theory comes from
geometrizing the non-perturbative 7-brane
backgrounds in type IIB string theory into
elliptically fibered manifolds, which can be
analyzed using well-established tools in algebraic
geometry. Therefore, F-theory allows us to explore
the strongly coupled regime of the string landscape.
Charge completeness in the
context of F-theory is shown in \cite{Morrison:2021wuv} to follow from
some standard assumptions regarding the physical interpretation of the
F-theory geometry, for 6D theories and corresponding gauge sectors of
4D theories.

In the F-theory framework, nonabelian
gauge groups arise from singularities on divisors (algebraic
codimension-one loci) on $B$. In six or more space-time dimensions,
the form of the nonabelian part of the gauge group and corresponding
massless matter content can be easily determined using the local
geometry \cite{Kodaira,Neron,BershadskyEtAlSingularities,KatzVafa},
which is easy to study.  In contrast, $\U(1)$ gauge factors in 6D and 8D
F-theory models, as well as in many 4D models, arise from the global geometry. To
be precise, these abelian factors in the gauge group arise
 from a Mordell-Weil group of rational sections
with nonzero rank in the elliptic fibration
\cite{MorrisonVafaII,AspinwallMorrisonNonsimply,Aspinwall:2000kf,Grimm:2010ez}. It
is much harder to engineer these models, and surprisingly few explicit
F-theory constructions have been found with any but the simplest
charged matter structure. The best-understood class of models with
a single $\U(1)$, known as the Morrison-Park model
\cite{MorrisonParkU1}, 
gives a universal form of Weierstrass model with U(1) gauge group and
 massless (absolute values of)
charges\footnote{Throughout the chapter, we normalize the nonzero
  charges such that they are all integers with the greatest common
  divisor being 1.} $q=1,2$.  Explicit models with $q=3,4,5$
have been constructed in \cite{KleversEtAlToric,Raghuram34,Knapp:2021vkm}
respectively, while models with $q=6$ are inferred from the type IIB
limit in \cite{Cianci:2018vwv}, and a procedure for constructing these
charges explicitly from universal flops is given in
\cite{Collinucci:2019fnh}. It has also been argued that $q$ can be
as large as 21 in 6D F-theory models using implicit Higgsing arguments
\cite{Raghuram:2018hjn}, 
and an algorithm for computing general U(1) charges from the form of a given
Weierstrass model has been developed in
\cite{Raghuram:2021wvx},
but explicit models with $q>6$ are still
  lacking. 
On the other hand, it was argued in
  \cite{Taylor:2018khc} that there is an infinite swampland of massless $\U(1)$
  charge spectra  in 6D supergravity theories.
In \cite{Raghuram:2020vxm}, a systematic criterion was proposed for
ruling out most of this infinite swampland, as F-theory constructions
of these models generally lead to an ``automatic enhancement'' of the
gauge group, and some low-energy arguments for this automatic enhancement
were put forth in \cite{Cvetic:2021vsw}.

Note that we primarily focus in this chapter on charges of massless or light
fields under isolated $\U(1)$ factors only; more complicated charge
structures can arise when there are also nonabelian gauge factors
and there are fields that  have both $\U(1)$ and nonabelian charges,
as discussed in \S\ref{sec:u1:coupling}.

The preceding discussion has focused primarily on 6D F-theory
models.
While 4D F-theory models  can be constructed with similar charges
using the same kinds of Weierstrass models described above
(Morrison-Park, etc.\ for charges up to $q = 6$), there
are also some qualitatively
different possibilities in 4D due to the inclusion of flux backgrounds,
which can affect the gauge groups and matter content.
In particular, with the power of fluxes it becomes
possible to build $\U(1)$ gauge groups from the
local geometry, which enables us to construct a much
larger class of $\U(1)$ models with larger $q$.
Indeed, it was noticed in \cite{Li:2022aek} that large $\U(1)$
charges can easily arise through breaking of
nonabelian gauge groups using  so-called
\emph{vertical} flux (referred to as  ``vertical flux
breaking''  henceforth), which will be described
below. In this chapter, we take this approach.  We describe the general
framework of F-theory models with $\U(1)$ factors from flux breaking,
construct some examples with large charges ($q\gg 6$), and try
to identify a plausible upper bound for $q$ in the
4D F-theory landscape.

The strategy is as follows: We first identify
nonabelian models that support vertical flux
breaking down to a single decoupled $\U(1)$. We can
choose an arbitrarily exotic linear combination of
the Cartan $\U(1)$'s to be preserved, as long as
appropriate vertical flux satisfying all relevant
constraints is turned on. This exotic
$\U(1)$ is the source of large $q$.
%\footnote{It has
%been observed that in 6D, the same construction of
%$\U(1)$ without fluxes leads to automatic
%enhancement to nonabelian gauge groups \cite{Raghuram:2020vxm,Cvetic:2021vsw}. Here the
%nonabelian gauge group is explicitly broken by
%supersymmetric fluxes, and we do not expect such enhancement to
%happen.} 
As the combination becomes more exotic,
more flux is needed to satisfy flux quantization \cite{Witten:1996md},
and the flux configuration finally hits the tadpole
bound \cite{Sethi:1996es}. These are the only constraints that lead to
an upper bound of $q$ for a given geometry. 
We describe the general framework for this flux breaking and analyze
some specific models that give particularly large values of $q$.
To
maximize $q$, we should maximize the tadpole bound,
which is fixed by the Euler characteristic
$\chi(\hat Y)$ of the resolved elliptic Calabi-Yau fourfold $\hat Y$
from the F-theory construction.
At the same time, the general structure of the intersection form on
middle cohomology indicates that
we should minimize the
intersection numbers on the divisor $\Sigma$
that supports the original nonabelian factor, such that the
tadpole caused by a given flux configuration is
minimal. As a specific example of the  exotic $\U(1)$
charges arise from flux breaking, we construct an explicit 4D F-theory
model that  combines   the two optimizations described above,
leading to a surprisingly large value of $q$:
\begin{equation}
    q_\mathrm{max}= 657\,,
\end{equation}
for light vector-like charged matter fields.  A similar construction
can give truly massless  chiral matter fields with charges of 465 or
greater.

This chapter is organized as follows: In \S\ref{sec:u1:non-ade}, 
we first extend the flux breaking formalism developed in Chapter~\ref{chap:fluxbreaking}
to include chiral matter from breaking non-ADE gauge groups.
In \S\ref{sec:u1:U1breaking}, we go through the general
framework of vertical flux breaking from a geometric nonabelian group
to an isolated $\U(1)$ gauge factor, and illustrate with a specific
class of simple examples from the breaking $\SU(3) \rightarrow\U(1)$.
In Section
\ref{sec:u1:G2model}, we present the  explicit 4D F-theory model with
$q_\mathrm{max}=657$ for vector-like matter, and related models with
comparably large charges for massless chiral fields. The $\U(1)$ model comes from
a $G_2\rightarrow\U(1)$ breaking on the \cy
fourfold with the fifth highest $h^{3,1}$ in the
Kreuzer-Skarke (KS) database of toric hypersurface
constructions
\cite{Kreuzer:1997zg,Scholler:2018apc}. We describe this model in some
detail, and give qualitative arguments  for
why this model may give, or at least be close to, the upper
bound on decoupled $\U(1)$ charges in the 4D
F-theory landscape. In \S\ref{sec:u1:coupling},
we extend our discussion to the case of $\U(1)$
coupled to other gauge groups, with an example of
even slightly larger $q_\mathrm{max}$ when the
$\U(1)$ is coupled to an $E_6$. We finally
conclude in Section
\ref{sec:u1:conclusion}, and give more geometric properties
of our $\U(1)$ model in Appendix \ref{sec:u1:equivalence}.

\section{Vertical flux breaking for non-ADE gauge groups}
\label{sec:u1:non-ade}

The vertical flux breaking formalism, which is the central tool for 
constructing the models in this chapter, has been developed in 
Chapter~\ref{chap:fluxbreaking}. Nevertheless, the chiral index
formula \eqref{fullchiR'} assumes ADE groups for the unbroken
gauge group $G$, while this chapter uses $G_2$ as the unbroken gauge group.

Let $G$ be a simple nonabelian gauge group, which may be ADE or non-ADE 
but does not support chiral matter.
Then the chiral indices are given by the
following: for a weight $\beta=-b_{i}\alpha_{i}$ in a representation
$R$ of $G$ that is localized on the matter curve $C_R=\Sigma\cdot D_R$,
by analysis following
\cite{Braun_2012,Marsano_2011,KRAUSE20121,Grimm:2011fx} the chiral
index is
\begin{equation} \label{eq:u1:chi}
    \chi_\beta=\sum_{i}b_{i}\frac{\left<\alpha_{i},\alpha_{i}\right>}{\left<\alpha_{\mathrm{max}},\alpha_{\mathrm{max}}\right>}\Theta_{i D_R}\,,
\end{equation}
where $\alpha_{\mathrm{max}}$ is the longest $\alpha_{i}$.
For ADE groups, the normalization factor is 
$\langle \alpha_i, \alpha_i \rangle / \langle \alpha_{\mathrm{max}}, \alpha_{\mathrm{max}} \rangle = 1$ 
for all $i$, agreeing with \eqref{fullchiR'} in Chapter~\ref{chap:fluxbreaking}. 
For $G_2$, however, there are two possible root lengths and the short roots satisfy
$\langle \alpha_i, \alpha_i \rangle / \langle \alpha_{\rm max}, \alpha_{\rm max} \rangle = 1/3$. 
This factor enters the chiral indices computed in \S\ref{subsec:u1:fluxbackground},
and is crucial to ensure 4D anomaly cancellation.

\section{Flux breaking to $\U(1)$}
\label{sec:u1:U1breaking}

We now turn to abelian $\U(1)$ factors in $G'$,
which are a key feature of vertical flux breaking.
We start by giving the general framework for flux breaking  of a
simple nonabelian factor to $\U(1)$ and then give a simple
illustrative example of breaking $\SU(3) \rightarrow\U(1)$ using
vertical fluxes.

\subsection{$\U(1)$ factors from flux breaking}

Although every root of a simple Lie algebra corresponds to a linear
combination of Cartan generators, the reverse is
seldom true. In fact, we can write down arbitrary
linear combinations of Cartan generators, while
there is only a finite number of roots. Following
the logic of vertical flux breaking described in
\S\ref{sec:u1:non-ade},
suppose that we have an F-theory model over a threefold base $B$ that
contains a single nonabelian gauge factor $G$.
We then turn on vertical flux parameters $\phi_{j\beta}$ giving
some non-vanishing fluxes $\Theta_{i\alpha}$ through \eqref{eq:review:Mred-simple}.
If we impose the condition (summing as above by convention over
doubled indices $i$)
\begin{equation} \label{eq:u1:exoticu1}
    p_{i}\Theta_{i\alpha}=0\,,
\end{equation}
for all $\alpha$, while
\begin{equation}
    \sum_{i} p_{i}\left<\alpha_{i},\alpha_{i}\right>^{-1}\alpha_{i}\,,
\end{equation} is not along any roots, then the Cartan generator
\begin{equation}
    p_{i}T_{i}\,,
\end{equation}
is preserved but does not belong to any nonabelian subgroup of
$G$. 
Such generators thus form
the abelian part of the preserved gauge group $G'$. More generally, there may be $\U(1)$'s
that are combinations of Cartan generators from
multiple gauge factors. These $\U(1)$'s, however, are
not relevant to our analysis below, and we focus on
$\U(1)$'s coming from  $G$ with a single simple nonabelian factor as above.
We focus attention in particular on cases
involving vertical flux breaking of
such a gauge group
$G$,  where no nonabelian gauge factor
remains and we have a single residual $\U(1)$ gauge
factor on the gauge divisor.
 As
long as Eq.\ (\ref{eq:u1:exoticu1})
is satisfied, the
coefficients $p_{i}$ look arbitrary and the
resulting $\U(1)$ can naively be arbitrarily complicated, which leads
to arbitrarily exotic matter, although as we shall discuss there are
upper bounds from other flux constraints. This
feature gives great power for building $\U(1)$
models from vertical flux breaking, as one can flexibly tune
suitable $p_{i}$ to get a desired $\U(1)$, with specific matter content. In
contrast, in field theory for example, the $\U(1)$ realized
after breaking through a Higgsing process is determined by the representation
and vev of the Higgs field, which substantially constrains the
resulting possible $\U(1)$ factors and associated charges.  While these
kinds of Higgsed $\U(1)$ fields are transparent from the low-energy physics
point of view, they are
much harder
to study in general in F-theory as they involve deformations of the
Weierstrass model that are in some cases unknown
\cite{Raghuram:2018hjn}. In contrast, vertical flux
breaking seems to rely much on the UV
physics of string theory, and while we have a clear way of analyzing these
systems from the geometry of fluxes,
so far we do not see any
clear approach to attaining a low-energy description of the
breaking.

The condition in Eq.\ (\ref{rlowerbound}) also holds for $\U(1)$ factors.
Now $\mathrm{rank}(G')$ also counts the number of $\U(1)$'s that
descend from $G$. In particular, to break a high-rank nonabelian
factor to a single $\U(1)$, the
gauge factor must arise on a divisor with $h^{1, 1} (\Sigma)\geq
\mathrm{rank} (G)$.

\subsection{A simple example: breaking $\SU(3) \rightarrow\U(1)$}
\label{subsec:u1:simpleexample}

It is useful to demonstrate the above techniques with a simple example of
$\U(1)$ models before discussing the maximization of $\U(1)$ charges.
Let us consider the base $B$ as a $\mathbb P^1$-bundle over $\mathbb F_0=\mathbb P^1\times\mathbb P^1$, with an $\SU(3)$ supported on
$\mathbb F_0$.\footnote{The analysis here is independent of how the
$\SU(3)$ is realized on $\mathbb F_0$.} Notice that $\SU(3)$ has rank
$2$ and $h^{1,1}(\mathbb F_0)=2$. Therefore by Eq.\ (\ref{rlowerbound}), $B$
is the simplest base that supports the breaking to $\U(1)$
described in the last subsection. Since the models in the coming sections
have the same divisor geometry, this subsection also serves as a
warm-up exercise for those constructions.

First, we describe the geometry of $B$. Let the two
$\mathbb P^1$'s on $\mathbb F_0$ be $s,f$. Then $B$ has three
independent divisors: $\Sigma$ as the section $\mathbb F_0$,
and $S,F$ as the $\mathbb P^1$ bundles on $s,f$ respectively. $\Sigma$
is also the gauge divisor. The only nontrivial intersection number is $\Sigma\cdot S\cdot F=1$. Generically
there are (anti-)fundamentals $\mathbf 3$ and $\bar{\mathbf 3}$, as
well as the adjoint $\mathbf 8$ as the matter content of the model.

Eq.\ (\ref{rlowerbound}) tells us that we can at most reduce the rank of
the gauge group by one when satisfying primitivity. In other words,
to preserve at least a $\U(1)$,
the nonzero vertical flux should always satisfy the constraint
\begin{equation}
    a\Theta_{1\alpha}+b\Theta_{2\alpha}=0\,,
\end{equation}
for all $\alpha=S,F$, where $1,2$ are the Cartan indices for $\SU(3)$,
and $a,b$ are some coprime integers. There are two possible cases: for
generic $a,b$ we obtain the breaking $\SU(3)\rightarrow\U(1)$, but if
$a=0$, $b=0$, or $a=b$, these coefficients align with some roots of
$\SU(3)$ and we get $\SU(3)\rightarrow\SU(2)$ instead. We focus on
the former case with the $\U(1)$ generator $T=aT_1+bT_2$. The flux
constraint is then solved by
\begin{equation}
    \left(\phi_{1S},\phi_{2S},\phi_{1F},\phi_{2F}\right)=\frac{1}{3}\left(\left(a-2b\right)n_{S},\left(2a-b\right)n_{S},\left(a-2b\right)n_{F},\left(2a-b\right)n_{F}\right)\,,
\end{equation}
where $n_S,n_F$ are flux parameters to be chosen, such that all
$\phi$'s are integers to satisfy flux quantization.

Now we turn to the condition of primitivity. In the
F-theory limit where the elliptic and exceptional
fibers shrink to zero volume, only the K\"ahler
form of $\Sigma$ contributes in Eq.\ 
(\ref{eq:fluxbreaking:intPrimitivity}). Let the
K\"ahler form of $\Sigma$ be
\begin{equation}
    [J_\Sigma]=t_1 \Sigma\cdot F+t_2\Sigma\cdot S\,,
\end{equation}
where $t_1,t_2$ are K\"ahler moduli. Eq.\ (\ref{eq:fluxbreaking:intPrimitivity}) then implies
\begin{equation}
    t_1 n_S+t_2 n_F=0\,.
\end{equation}
To ensure stabilization within the K\"ahler cone
where $t_1,t_2>0$, we require $n_S,n_F$ to be both
nonzero and have opposite signs.

Assuming the tadpole constraint %Eq.\ 
(\ref{eq:review:tadpole}) is satisfied,
now we are free to choose the parameters $a,b,n_S,n_F$ and
calculate the resulting $\U(1)$ charges. First notice that under
the breaking, the (anti-)fundamentals give charges $a,b,a-b$ and
their conjugates, and the adjoint gives charges $a+b,a-2b,2a-b$ and
their conjugates. As an example with small flux parameters, let us
choose $(a,b,n_S,n_F)=(-2,5,2,-1)$. Then we obtain the following spectrum:
\begin{equation}
    q=2,\,3,\,5,\,7,\,9,\,12\,.
\end{equation}
To be more precise, we can also calculate the chiral spectrum of
these charges. Using Eq.\ (\ref{eq:u1:chi}), we see that the chiral
spectrum induced from the adjoint is
\begin{equation}
    14\times\mathbf 1_3+4\times\mathbf 1_{12}+10\times\mathbf 1_{-9}\,.
\end{equation}
It is easy to check that this chiral spectrum is free of both pure
gauge and gauge-gravity anomalies
since $\sum q_i =\sum q_i^3 =  0$. More generally, for any
such $a,b$, we have the following chiral spectrum from the adjoint:
\begin{equation}
    2(b-a)\times\mathbf 1_{a+b}+2a\times\mathbf
    1_{a-2b}+2b\times\mathbf 1_{2a-b}\,,
\label{eq:u1:infinite-family}
\end{equation}
which is remarkably always anomaly-free.
One can perform a similar analysis for the (anti-)fundamentals,
although it depends more on the geometry of $B$.

Notice that the charges are coprime. Therefore through such a
simple construction, we already obtain some relatively large
$\U(1)$ charges. In the next Section, we will optimize this
procedure subject to the tadpole constraint, to obtain our
%main
extremal result $q_\mathrm{max}=657$.

Some relevant comments can be made
here regarding the connection of these spectra with related 6D models.
The family of models described in Eq.\ (\ref{eq:u1:infinite-family}) is
very similar in structure to an infinite family of 6D U(1) models with
arbitrarily large charges encountered in \cite{Taylor:2018khc}.  In
the 6D case, charges arise from complicated Weierstrass models (see,
e.g., \cite{Raghuram:2021wvx}), and the infinite family is apparently
rendered unphysical by the automatic enhancement mechanism
\cite{Raghuram:2020vxm,Cvetic:2021vsw}, which guarantees the
appearance of an additional U(1) factor.  In the 4D case, the charges
arise from the distinct physical mechanism of flux breaking, so the
infinite family of anomaly-consistent models is bounded by the
tadpole, and automatic enhancement does not seem to occur.  It would
be interesting to better understand how the automatic enhancement
story differs in this context.  It is also interesting to observe that
because $a, b$ are coprime, this family of models can
 contain massless
or light matter fields that generate the full charge lattice, in
accord with the massless charge sufficiency conjecture formulated for
6D F-theory in \cite{Morrison:2021wuv}.  In this case, however, the
nonzero multiplicities of massless or light matter depend upon the
choice of flux.  It would be interesting to look further into the
question of whether the light fields always
generate the full charge lattice
for arbitrary choices of flux.

\section{A $\U(1)$ model with $q_\mathrm{max}=657$}
\label{sec:u1:G2model}
In this section, we construct a $\U(1)$ model with
$q_\mathrm{max}=657$ using vertical flux breaking.
We describe the geometry of the fourfold $\hat Y$
and the base $B$, as well as the vertical flux
background in detail. Then we give qualitative and
heuristic arguments towards $q_\mathrm{max}=657$
being (close to) an upper bound in the 4D F-theory landscape.
Notice that there are other nonabelian gauge
factors in this model, but they are completely
decoupled from the $\U(1)$ we construct, hence we
still call it a $\U(1)$ model, and the analysis of
\S\ref{sec:u1:U1breaking} applies essentially unchanged.

It is useful to first recap our strategy. From Eq.\ 
(\ref{eq:u1:exoticu1}), we see that the more exotic
the $\U(1)$ or the coefficient $p_{i}$ is, the
larger integer $\phi_{i\alpha}$ we need to turn
on. From Eq.\ (\ref{eq:review:tadpole}), the size of
$\phi_{i\alpha}$ is bounded from above by the
Euler characteristic $\chi(\hat Y)$ and the
intersection numbers on $\Sigma$ that arise in
$[G_4]\cdot[G_4]$. Therefore to obtain the largest
$q_\mathrm{max}$, we shall maximize $\chi(\hat Y)$
while minimizing the intersection numbers on
$\Sigma$. Although the list of elliptic \cy
fourfolds is far from complete, the KS database
provides a set of good toric representatives especially
at large $\chi$. Scanning through the KS database
leads us to consider the \cy fourfold with the
fifth largest $h^{3,1}$ and $\chi$. We now describe its geometry in detail.

\subsection{Geometry}
\label{subsec:u1:geometry}
The fourfold $\hat Y$ has the following Hodge numbers:
\begin{gather} \label{u1:eq:hodge}
    h^{1,1}=256\,,\quad h^{2,1}=0\,,\quad h^{3,1}=289384\,,\nonumber\\
    h^{2,2}=44+4h^{1,1}-2h^{2,1}+4h^{3,1}=1158604\,,\nonumber\\
    \chi=6(8+h^{1,1}-h^{2,1}+h^{3,1})=1737888\,.
\end{gather}
Notice that there are many more fourfolds with the
same $\chi$, but they all have much larger
$h^{1,1}$ and are harder to analyze, while they
very probably do not give larger $q_\mathrm{max}$, as discussed in
\S\ref{subsec:u1:bound}.

$\hat Y$ is a \cy hypersurface in an ambient
toric fivefold, a (singular) weighted projective
space $\mathbb P^{1,80,492,1148,1722}$ \cite{Scholler:2018apc}. It can also
be understood as a generic elliptic fibration over
a toric base $B$ to be specified below. The
equivalence of the two descriptions is shown in
Appendix \ref{sec:u1:equivalence}. Now, $B$ can be
described as a $B_2$-bundle over $\mathbb P^1$,
where $B_2$ is a toric surface characterized by a
closed cycle of divisors (or rays in the 2D toric
fan) with self-intersection numbers
$0,6,-12//-11//-12//-12//-12//-12//-12//-12//-12$,
where $//$ represents the chain
$-1,-2,-2,-3,-1,-5,-1,-3,-2,-2,-1$ \cite{MorrisonTaylorToric,TaylorWangVacua}. Its toric rays
$v_\alpha\in \mathbb Z^2$ can be taken to be\footnote{Note that in
  \cite{TaylorWangVacua}, the indices on $v_\alpha, w_\alpha$ etc. are
taken to be  roman indices $i$; here to avoid confusion we use the
appropriate base divisor index notation $\alpha$, although when there
is possible ambiguity with integer indices $i$ indexing Cartan
divisors
as in $D_i$, we put the index as a superscript or use alternative
explicit non-integer notation.}
\begin{equation}
    v_{1}=\left(-1,-12\right)\,,\quad v_{2}=\left(0,1\right)\,,\quad...\,,\quad v_{99}=\left(0,-1\right)\,,
\end{equation}
where the intermediate rays are determined by
$v_{\alpha-1}+v_{\alpha+1}+C_{\alpha}^{2}v_{\alpha}=0$ and $C_{\alpha}^{2}$
is the self-intersection number of the divisor
corresponding to $v_\alpha$, starting at
$C_1^2=0,C_2^2=6$. Then the 3D toric fan of $B$ is
given by the rays $w_\alpha$:
\begin{equation}
    w_{0}=\left(0,0,1\right)\,,\quad w_{1\leq \alpha\leq99}=\left(v_{\alpha},0\right)\,,\quad w_{100}=\left(80,468,-1\right)\,,
\end{equation}
where $(80,468)=4v_{19}$ is the twist of the
$B_2$-bundle. We denote the corresponding divisor
classes to be $D^\alpha$, where the superscript integer
indexing the base divisors is
distinguished from the subscript for exceptional
divisors as mentioned above; when we use $\alpha$ as a subscript where
Cartan indices $i$ are also possible, as in,
e.g., $S_{i \alpha}$ we use non-integer notation for the $\alpha$'s. The cones of the fan are given by
$\left(w_{0},w_{\alpha},w_{\alpha+1}\right),\left(w_{100},w_{\alpha},w_{\alpha+1}\right)$ for $1\leq \alpha\leq98$,
as well as $\left(w_{0},w_{99},w_{1}\right),\left(w_{100},w_{99},w_{1}\right)$.

The local geometry on divisor $D^0=D^{100}$ is
clearly $B_2$, while that on divisors $D^{1\leq \alpha\leq 99}$ are all Hirzebruch surfaces
$\mathbb F_n$. In particular, we have
$h^{1,1}(D^{1\leq \alpha\leq 99})=2$. The intersection numbers on
$D^{1\leq \alpha\leq 99}$ are then determined by
$n$ only. Note that since the twist is along $v_{19}$,
the local geometry on $D^{19}$ is $\mathbb F_0$,
which has the smallest intersection numbers among all $\mathbb F_n$.

Some of
the divisors $D^{1\leq \alpha\leq 99}$ have sufficiently
negative normal bundles in $B$ that the
elliptic fibration is forced to be singular to certain
degrees, and nonabelian gauge factors
automatically arise on these divisors. Such  \emph{rigid} or \emph{geometrically
non-Higgsable} gauge groups are
present throughout the whole set of moduli space
branches associated with  elliptic \cy's over such a base \cite{MorrisonTaylorClusters,MorrisonTaylor4DClusters}. As
a result, these gauge groups cannot be broken by
any geometric deformation (corresponding to
Higgsing from the low-energy perspective), while
they can still be broken by fluxes. The method for
determining the rigid gauge groups in 4D F-theory
models has been described in \cite{MorrisonTaylor4DClusters}, and here we
summarize the result applied in this type of case where the base $B$
is a $B_2$ bundle over $\P^1$. The divisor $D^{1\leq \alpha\leq 99}$ supports $E_8$ if $C_{\alpha}^{2}=-11,-12$, $F_{4}$
if $C_{\alpha}^{2}=-5$, $G_{2}$ if $C_{\alpha}^{2}=-3$, and
$\SU\left(2\right)$ if $C_{\alpha}^{2}=-2$ and
intersects with a $G_{2}$ gauge divisor. Therefore,
the full gauge group is
\begin{equation} \label{eq:u1:totalgaugegroup}
    E_{8}^{9}\times F_{4}^{8}\times\left(G_{2}\times\mathrm{SU}\left(2\right)\right)^{16}\,.
\end{equation}
In particular, there is a $G_2$ factor supported on $D^{19}$.

Note that there may be codimension-2 $(4,6)$ singularities
localized on divisors supporting $E_8$ factors. By computing
the normal bundles on divisors, one can check that there are
four irreducible components of codimension-2 $(4,6)$ loci on
$D^3$ (with $C_3^2=-12$) and one on $D^{15}$ (with $C_{15}^2=-11$).
To remove these singularities, non-toric blowups must be performed,
contributing $5$ to the $h^{1,1}$ in Eq.\ (\ref{u1:eq:hodge}).
These singularities are associated with extra strongly coupled
(probably conformal)
sectors that have not been well understood \cite{HeckmanMorrisonVafa,DelZotto:2014hpa,Apruzzi:2018oge}.
Nevertheless, these sectors are decoupled from the gauge sectors
we are studying and should not affect our analysis.

\subsection{Flux background}
\label{subsec:u1:fluxbackground}
Now we would like to break some gauge factors in
Eq.\ (\ref{eq:u1:totalgaugegroup}) to get an exotic
$\U(1)$ using vertical flux breaking. Since
vertical flux breaking must decrease the rank of
the gauge group, we cannot have breaking like
$\SU(2)\rightarrow\U(1)$. By Eq.\ (\ref{rlowerbound}) with
$r=2$ for all gauge factors, we then see that the
only available breaking is $G_2\rightarrow\U(1)$.
One may naively consider a breaking like
$G_2\times\SU(2)\rightarrow\U(1)$ where the $\U(1)$
is a combination of Cartan generators from both
gauge factors, since a $G_2$ gauge divisor always
intersects with an $\SU(2)$ gauge divisor. It can
be shown that, however, such breaking violates an
analogous version of Eq.\ (\ref{rlowerbound}).

Here we reach one of the main points in this section:
we can minimize the intersection numbers on the
gauge divisor by performing the flux breaking on
$D^{19}$, which is locally $\mathbb F_0$ and
supports a $G_2$. This crucial feature is why we
study the fifth largest $h^{3,1}$ and $\chi$ in the
KS database but not one of the \cy's with even larger $\chi$.

Let us specify more details on $D^{19}$. The only
$D^i$'s that intersect with $D^{19}$ are
$D^0=D^{100},D^{18},D^{20}$. The curves on $D^{19}$
are then spanned by $D^0\cdot D^{19}=D^{100}\cdot D^{19}$ and $D^{18}\cdot D^{19}=D^{20}\cdot D^{19}$. Following the notation
in \S\ref{subsec:u1:simpleexample}, we denote
$\Sigma=D^{19}, S=D^0, F=D^{18}$. The only
nontrivial intersection number on $\Sigma$ is
$\Sigma\cdot S\cdot F=1$. Now to break $G_2\rightarrow\U(1)$, we turn on nonzero $\phi_{i\alpha}$ such that
\begin{equation}
    a\Theta_{1\alpha}+b\Theta_{2\alpha}=0\,,
\end{equation}
for all $\alpha=S,F$, where the index $i$ is the
Cartan index for the $G_2$, and integers $a,b$ are
coprime. The labels $1,2$ correspond to
\begin{equation}
    \kappa^{ij}=\left(
    \begin{array}{cc}
        6 & -3\\
        -3 & 2
    \end{array}\right)\,.
\end{equation} 
If $(a,b)$ is not along any root of $G_2$,
all the roots of $G_2$ are broken and the remaining
gauge group is $\U(1)$ with generator
$T=aT_{1}+bT_{2}$. The flux constraint is solved by
\begin{equation}
    \left(\phi_{1S},\phi_{2S},\phi_{1F},\phi_{2F}\right)=\left(\left(a-2b/3\right)n_{S},\left(2a-b\right)n_{S},\left(a-2b/3\right)n_{F},\left(2b-a\right)n_{F}\right)\,,
\end{equation}
where $n_S,n_F$ are flux parameters to be chosen.
To satisfy flux quantization, we see that $n_S,n_F$
must be multiples of 3 unless $b$ is a multiple of 3.
When $b$ is a multiple of 3, one can show that
$(a-2b/3)$ and $(2a-b)$ must be coprime and
$n_S,n_F$ must be integer. Since the size of $\phi$
has been bounded, to ensure the most exotic choice
of $(a,b)$ we should assume $b$ as a multiple of 3
and integer $n_S,n_F$. Now we turn to primitivity;
as in \S\ref{subsec:u1:simpleexample},
only the K\"ahler
form of $D^{19}$ contributes in Eq.\ 
(\ref{eq:fluxbreaking:intPrimitivity}) with $D_{i}$ being the
exceptional divisors from the $G_2$. Therefore, we require
$n_S,n_F$ to be both
nonzero and have opposite signs.

\begin{comment}
Now we turn to the condition of primitivity. In the
F-theory limit where the elliptic and exceptional
fibers shrink to zero volume, only the K\"ahler
form of $D^{19}$ contributes in Eq.\ 
(\ref{eq:u1:intprimitivity}) with $D_{i}$ being the
exceptional divisors from the $G_2$. Let the
K\"ahler form of $D^{19}$ be
\begin{equation}
    [J_{19}]=t_1 \Sigma\cdot F+t_2\Sigma\cdot S\,,
\end{equation}
where $t_1,t_2$ are K\"ahler moduli. Eq.\ (\ref{eq:u1:intprimitivity}) then implies
\begin{equation}
    t_1 n_S+t_2 n_F=0\,.
\end{equation}
To ensure stabilization within the K\"ahler cone
where $t_1,t_2>0$, we require $n_S,n_F$ to be both
nonzero and have opposite signs.
\end{comment}

With the above information, we can easily calculate the tadpole from this flux:
\begin{equation}
    \frac{1}{2}\int_{\hat Y} G_{4}\wedge G_{4}=-2\left(a^{2}-ab+\frac{b^{2}}{3}\right)n_{S}n_{F}\,.
\end{equation}
We see that to minimize the tadpole and satisfy
primitivity, we should choose e.g.
$(n_S,n_F)=(1,-1)$. This ensures the tadpole
to be positive. Then Eq.\ (\ref{eq:review:tadpole}) becomes
\begin{equation} \label{u1:eq:g2tou1tadpole}
    a^{2}-ab+\frac{b^{2}}{3}\leq36206\,.
\end{equation}
To maximize the $\U(1)$ charges, we should choose
$(a,b)$ such that the above is the closest to
saturation. The ratio between $a$ and $b$ is now
determined by the matter spectrum charged under the
$G_2$. Let us first focus on the adjoint
$\mathbf{14}$ of $G_2$. After the breaking, the
W-bosons become charged singlets with the $\U(1)$
charges
\begin{equation}
    a,\,b,\,3a-b,\,2a-b,\,a-b,\,3a-2b\,,
\label{eq:u1:general-g2-spectrum}
\end{equation}
and their conjugates. There are also two uncharged
singlets. To find out the largest possible
$q_\mathrm{max}$, we then maximize one of the
charges in the above, subject to the tadpole constraint
and the assumption of $b$ being multiple of 3. It
turns out that there are multiple choices giving
the same largest $q_\mathrm{max}$. For example,
$(a,b)=(329,657)$ (or $(\phi_{1S},\phi_{2S})=(-109,1)$)
gives the largest $b=657$,
with the full set of $\U(1)$ charges from the adjoint being
\begin{equation}
    q=1,\, 327,\, 328,\, 329,\, 330,\, 657\,.
\end{equation}
Therefore, we have reached one of
the main results of this chapter, a $\U(1)$ 4D F-theory model with
\begin{equation}
    q_{\max}=657\,.
\end{equation}

To complete the discussion, we still need to look
at other representations. There is also
bifundamental matter $(\mathbf{7},\mathbf{2})$
charged under $G_2\times\SU(2)$ before breaking \cite{MorrisonTaylorClusters}. It
breaks into representations of $\SU(2)\times\U(1)$
after the breaking, so the $\U(1)$ is still coupled
to other gauge factors. One can, however, turn on
one more unit of vertical flux to break the
adjacent $\SU(2)$ completely. Then the
bifundamental also breaks into $\U(1)$ charged
singlets and the $\U(1)$ we constructed is fully
decoupled. The same calculation as above shows that
the bifundamental only gives a subset of $\U(1)$
charges coming from the adjoint, with the maximum being $q=329$.

It is informative to study the chiral spectrum of these large
charges. Interestingly, Eq.\ (\ref{eq:u1:chi}) implies that the chiral
indices from the adjoint are proportional to $(n_S+n_F)$, hence
vanish in the above example. In particular, it means that the
charge $q=657$ must belong to vector-like matter, which is not exactly
massless if including interactions in the superpotential. A careful
calculation using the approach of \cite{BeasleyHeckmanVafaI}
shows that the multiplicity of vector-like $\mathbf 1_{657}$ in this model is indeed nonzero.
On the
other hand, there is chiral matter from the bifundamental. Since the
adjacent $\SU(2)$ is completely broken, we can effectively consider
two copies of $\mathbf 7$ localized on $C_{\mathbf 7}=D^{18}\cdot D^{19}=\Sigma\cdot F$. Eq.\ (\ref{eq:u1:chi}) then gives the following chiral spectrum:
\begin{equation}
    438\times\mathbf 1_1+220\times\mathbf 1_{-328}+218\times\mathbf 1_{329}\,,
\end{equation}
which is again anomaly-free as expected. Therefore, if we restrict
to the truly massless chiral fields only, $q_\mathrm{max}$ is not as
large as 657. There are still ways to go beyond $q=329$ for chiral
fields. For example, in the same model as above, we can choose
$(n_S,n_F)=(2,-1)$ instead. Then there are chiral fields from the
adjoint, and the same calculation of $q_\mathrm{max}$ from the adjoint
gives $q_\mathrm{max}=465\simeq 657/\sqrt{2}$.

\begin{comment}
An interesting implication of such massless chiral fields is that
from the low-energy perspective, we can apparently
give these fields a vacuum expectation value to Higgs the symmetry to
a discrete abelian group
$\U(1)\rightarrow\mathbb Z_k$,
where as suggested above $k$ can be as large as 465.\footnote{We thank
Paul Oehlmann for raising this point.} Similar to
the argument in \cite{Raghuram:2018hjn}, this Higgsing indirectly
implies that in the 4D F-theory landscape, there exist discrete gauge
symmetries as large as $\mathbb Z_{465}$.
This is much larger than
the highest record $\mathbb{Z}_6$ \cite{Anderson:2019kmx} currently known for discrete gauge
symmetries from
%torsion in the Mordell-Weil
Tate-Shafarevich/Weil-Ch\^atelet groups of smooth elliptic
\cy threefolds, as is realized from Higgsing U(1)
factors in the Mordell-Weil group
\cite{Braun:2014oya,Morrison:2014era}. The
explicit construction of such large discrete
symmetries in F-theory, however, is far from clear, in contrast to our
explicit construction of large $\U(1)$ charges. We leave this interesting
direction to future work.
\end{comment}

One may naively expect, from the low-energy perspective,
that we can
give the above massless chiral fields a vacuum expectation value to Higgs the symmetry to
a discrete abelian group
$\U(1)\rightarrow\mathbb Z_k$.
The above example then suggests that $k$  could be as large as 465 for
such discrete symmetries in 4D.\footnote{We thank
Paul Oehlmann for raising this point.}
This is much larger than
the largest size  $\mathbb Z_6$ currently known 
(\cite{Anderson:2019kmx} and references therein)
for discrete gauge
symmetries from 
Tate-Shafarevich/Weil-Ch\^atelet groups of smooth elliptic
\cy threefolds or fourfolds \cite{Braun:2014oya,Morrison:2014era}.
Nevertheless in 4D, there are various Yukawa couplings involving these
chiral fields, which can induce a potential and stabilize these
vacuum expectation values. As a result, although here we do not demonstrate
it explicitly, we expect that such Higgsing to discrete gauge symmetries is
not possible in our setup.

One should be reminded that this kind of $\U(1)$ model
is very rare in the 4D F-theory landscape, as we
have almost saturated the tadpole bound, and
arranged all fluxes to be along several specific
directions. In particular, with these constructions there is almost no room
to turn on horizontal flux for moduli
stabilization.
A generic $\U(1)$ model is expected to have
fluxes spreading over many directions, with only
a small amount of flux along each direction, hence giving small $\U(1)$ charges.

\subsection{Towards an upper bound}
\label{subsec:u1:bound}
One important question regarding $\U(1)$ charges of
massless fields is
whether an upper bound on such $q$ exists, and if so what that upper
bound is. With our
current technologies, it seems impossible to precisely determine
the value of the upper
bound with certainty, since the lists of elliptic \cy fourfolds
$\hat Y$ and bases $B$, as well as tools for
building $\U(1)$ models, are rather limited.  One can
certainly attempt to seek models that
exceed our result $q_\mathrm{max}=657$. Here, however, we
provide some heuristic reasons for why we expect our
result may give, or at least be close to, an actual upper bound
on $q$ within the 4D
F-theory landscape.
\begin{itemize}
    \item The most straightforward way to find
    other large $\U(1)$ charges is to generalize
    the method of vertical flux breaking to other known
    geometries. There are four known \cy
    fourfolds with
    $h^{3,1}$ and $\chi$ larger than those in our
    model  (these are all in the KS database; Euler characters of, e.g.,
     CICY fourfolds are much smaller, with $\chi \leq 2610$
\cite{Gray:2014fla}). The bases from these fourfolds are also $B_2$-bundles
    over $\mathbb P^1$, with the same $B_2$
but with different twists \cite{TaylorWangVacua}.
    Note that none of these twists are along a
    $G_2$ gauge divisor, so the local geometries of
    the $G_2$ gauge divisors are never as simple as
    $\mathbb F_0$. In fact, the same construction
    as in our model needs to be done on $\mathbb F_3, \mathbb F_6, \mathbb F_9, \mathbb F_{12}$
    respectively when $h^{3,1}$ and $\chi$
    increase. Therefore, the increase of
    intersection numbers on $G_2$ gauge divisors
    surpasses the slight increase of $\chi$, and
    leads to lower $q_\mathrm{max}$. The geometries
    with the same $\chi$ but lower $h^{3,1}$ in the
    KS database have much larger $h^{1,1}$ and are
    harder to analyze. Although we do not have any
    quantitative statements, the general
    expectation is that these geometries contain many more rigid
    gauge groups, and the divisor geometries are generically
    more complicated with higher $h^{1,1}$.  Due to
    such complexity, we may not
    expect there to be a $G_2$ gauge divisor as simple as
    $\mathbb F_0$. Even if there is such a gauge divisor, by
    the same construction the resulting charge should not be
    significantly larger than 657.
    
    \item In principle, there may be elliptic \cy fourfolds with much larger $\chi$ than that in
    our model, thus potentially giving much larger
    $q_\mathrm{max}$. 
From what is known of the structure of elliptic threefolds and
fourfolds, however, it seems unlikely that $\chi$ of any elliptic
fourfold
can exceed those that are known and mentioned above.  While this
cannot be proven rigorously, we summarize some arguments for this here.
    The situation for elliptic threefolds is fairly clear: there are
a finite number of
    elliptic \cy threefolds \cite{GrossFinite}
and all of the allowed bases have
    been classified by the minimal model program \cite{Grassi1991}.
The elliptic \cy threefold with the largest $h^{2, 1}$ is known to
be the generic
elliptic fibration over $\mathbb F_{12}$ \cite{Taylor:2012dr}, which has the largest known
(absolute value of) Euler character $|\chi| = 960$.  The distinctive ``shield'' shape of the Hodge
numbers for all toric hypersurface \cy threefolds has 3 peaks with maximum
$h^{1, 1}+h^{2, 1}$, which are all
realized by elliptic fibrations over toric bases.
(Because of the alternating signs in the Euler character,  $h^{1, 1}
+h^{2, 1}$ may be a better proxy for the Euler character of fourfolds
than the threefold Euler character $2 (h^{1, 1}-h^{2, 1})$.)
A systematic classification of the allowed bases, including all toric
bases
\cite{MorrisonTaylorToric} and non-toric bases  giving \cy threefolds with
$h^{2, 1}\geq 150$ \cite{TaylorWangNon-toric} shows that the toric hypersurface \cy threefolds
in the KS database \cite{Kreuzer:2000xy} accurately
capture the boundary of the set of possible Hodge numbers. 
In particular, there is known to be
no \cy threefold with larger (absolute value
of) Euler character or $h^{1, 1}+h^{2,1}$  among generic elliptic
fibrations with  $h^{2, 1}\geq 150$ over any base surface.
This gives extremely strong (but not airtight) evidence that the
largest values of the Euler character and $h^{1, 1}+h^{2,1}$ for
elliptic \cy threefolds are realized by elliptic fibrations
over toric bases and are found at the boundary points of the KS database.

While    it is far from clear whether the analogous
    statement is true for fourfolds, it seems very plausible that this
    should be true. 
The
 shape of the Hodge shield (in $h^{1, 1},h^{3,1}$) for CY fourfolds
 takes a very similar, although more spiky, form to that for
 threefolds, with again 3 prominent cases with maximum $h^{1, 1}+h^{3,1}$
\cite{Kreuzer:1997zg,Scholler:2018apc}, corresponding again to the
largest known CY fourfold Euler characters.
From the
    perspective of the analogous minimal model program (the Mori program)
    for threefold bases, it is expected that the largest $h^{3,1}$
    will come from a minimal threefold base that is either Fano, a 
$\mathbb P^1$ bundle over a surface $B_2$, or a
    $B_2$-bundle over $\mathbb P^1$.  The last of these classes seem to
    give the largest possible values for $h^{3,1}$ and $\chi$
    \cite{Klemm_1998,Halverson:2015jua,TaylorWangVacua}, and  as we
    have discussed here the bases we have used with large $h^{3,1}$
    are all $B_2$ bundles over $\mathbb P^1$.  If the fourfold case
    follows the better understood pattern of geometries for
    threefolds, these are indeed the elliptic CY fourfolds with
    largest $\chi$.
As for threefolds,
currently most known elliptic
    \cy fourfolds come from hypersurfaces in toric
    ambient spaces or elliptic fibrations on toric
    threefold bases \cite{TaylorWangMC,HalversonLongSungAlg,TaylorWangLandscape}, so elliptic \cy fourfolds with
    larger $\chi$, if they exist, would very probably
    involve non-toric constructions. 
It has recently been
    noticed that, unlike in 6D, non-toric
    constructions of elliptic \cy fourfolds and threefold
    bases seem to give an additional large class of
    4D F-theory models \cite{Braun:2014pva,Li:2022aek} with
    qualitatively novel features.
The extent
    of such geometries is certainly an open
    question, although, as for elliptic \cy threefolds, it is known that
    the number of topological types of
elliptic \cy fourfolds is finite up to
birational equivalence \cite{di2021birational}.
From analogy with CICY fourfolds, however, where the Euler characters
as mentioned above are generally much smaller than those of toric
hypersurfaces, and from experience with non-toric bases for elliptic
threefolds
\cite{TaylorWangNon-toric}, it seems natural to expect that  non-toric
bases will not give larger Hodge numbers or Euler characters than the
examples already known.  Thus,
we think that it is not unreasonable to
    believe that there may be no fourfolds
    with $\chi$ significantly larger than that in our model.
Rigorous results in these directions, however, are clearly an important
direction for further work.

    \item
It is natural to consider $\U(1)$ factors from
    breaking of gauge groups other than $G_2$, but
such $\U(1)$'s are unlikely to give larger
    $q_\mathrm{max}$. First, consider $\U(1)$ factors arising
from
    gauge groups with higher rank. Eq.\     (\ref{rlowerbound}) then requires $h^{1,1}(\Sigma)>2$, so we cannot use bases
    as simple as a $B_2$-bundle over $\mathbb P^1$ in the same
    way,
    and we are forced to consider  more
    complicated divisor geometries, which may lead to tighter tadpole
    constraints, as discussed previously.  Moreover, from the
    calculation in our model, it seems that the
    optimization of $q_\mathrm{max}$ can be done by
    localizing almost all the
    flux $\phi_{i\alpha}$ on one of the exceptional
    divisors.
    Therefore, the presence of additional Cartan
    directions should not significantly change the
    optimization process. The third reason arises when considering gauge groups with any rank: $G_2$ has the
    most exotic root vectors due to the presence of
    $3$ in the components, so generically the
    resulting $\U(1)$ charges from $G_2$ are larger
    than those from other gauge groups. All these
    reasons lead us to expect that the $G_2$ breaking is
    likely to give the largest $q_\mathrm{max}$.
    
    \item Finally, the possibilities of $\U(1)$
    models provided by vertical flux breaking
    clearly greatly exceed other available methods in
    literature. In contrast to the construction analyzed
    here,
    realizing $\U(1)$ with additional rational
    sections and nontrivial Mordell-Weil group
    relies heavily on the global geometry. Given the
    difficulty of building such models even with charges up to $q =
    6$, as
    summarized in \S\ref{sec:u1:intro}, it is
    reasonable to expect that such a construction
    can never exceed, or even approach, the result found here. Moreover, as
    discussed in Chapter~\ref{chap:fluxbreaking},
    vertical flux breaking provides even more
    flexibility than Higgsing in field theory,
    which also corresponds to geometric
    deformation in F-theory. Therefore, we expect
    our result to exceed any charges obtained from
    Higgsing arguments.
\end{itemize}

Readers should be warned that all the above are
only heuristic arguments. It is possible that
some of these arguments are not true in general,
and that larger, or even much larger, $q_\mathrm{max}$ can be found in
F-theory. Although
F-theory is so far the most promising approach to
exploring global aspects of the nonperturbative string landscape, we also cannot exclude
the possibility that there are compactifications
in other corners of string theory that give rise
to even larger $q_\mathrm{max}$. We have not
explored non-geometric or non-supersymmetric
constructions in this regard at all. Clearly much more work needs
to be done to rigorously construct an upper bound on for $q$, but
hopefully the work and arguments presented here provide a starting
point for further analysis.

\section{Coupling to other gauge groups}
\label{sec:u1:coupling}

So far we have  focused on a single $\U(1)$
gauge factor, decoupled from other gauge
groups; it is also interesting to study a $\U(1)$ factor
coupled to other gauge groups. In particular, this
is phenomenologically interesting since the
hypercharge $\U(1)$ and various $\U(1)$ extensions
to the Standard Model have this structure. 
Although the hypercharge $\U(1)$ in the
Standard Model cannot be obtained from purely vertical
flux breaking \cite{Li:2022aek}, one may still apply vertical
flux breaking to build various $\U(1)$ extensions. On
the theoretical side, we expect that
$q_\mathrm{max}$ may exceed the value of 657 found above when the
$\U(1)$ is coupled to other gauge groups, since there are many more
possible breaking scenarios with other gauge factors and more parameters.
A full analysis of this problem is
beyond the scope of this chapter.  Here we
simply demonstrate  a single example with a slightly larger $q_\mathrm{max}$ than that in
\S\ref{subsec:u1:fluxbackground}.
While there may be larger values possible, for similar reasons to
those discussed in \S\ref{subsec:u1:bound}, we do not expect
enormously larger values for $\U(1)$ charges even when other gauge
factors are included.
Note that in many cases when the $\U(1)$ factor couples to one or more
nonabelian factors, the global structure of the group may have a quotient
by a discrete component of the center, such as in the Standard Model
group where the global structure seems likely to be $(SU(3) \times
SU(2) \times U(1))/\Z_6$ (see, e.g., \cite{LawrieEtAlRational,GrimmKapferKleversArithmetic,CveticLinU1,TaylorTurnerGeneric,TaylorTurner321}).  In such cases it is often conventional to
use fractional values for $\U(1)$ charges, as is often used in the
(unbroken) Standard Model.  
In our discussion here, as mentioned in
a footnote in \S\ref{sec:u1:intro}, we always treat
$\U(1)$ charges as integers, with
the minimal $\U(1)$ charge being $q = 1$.  With this normalization,
while the approach of flux breaking provides $\U(1)$ factors with much
larger charges than those available  directly from F-theory Weierstrass
models (as analyzed in, e.g., \cite{LawrieEtAlRational}), we
expect bounds of similar magnitude on $q_\mathrm{max}$ in the presence of
nonabelian factors to those found above for pure $\U(1)$ factors.

Following the construction in Section
\ref{sec:u1:G2model}, here we build a similar $\U(1)$
model with $q_\mathrm{max}=672>657$, but where the
$\U(1)$ is coupled to an $E_6$. This time, we use
the known \cy fourfold with the largest $h^{3,1}$
and $\chi$. It has been argued that this geometry
plausibly supports the most flux vacua in the 4D
F-theory landscape \cite{TaylorWangVacua}. The geometry is the same as
that in \S\ref{subsec:u1:geometry} except the
twist of the $B_2$-bundle. To be precise, we
replace the toric ray $w_{100}$ of $B$ in Section
\ref{subsec:u1:geometry} by
\begin{equation}
    w_{100}=(84,492,-1)\,,
\end{equation}
where $(84,492)=12v_{15}$ is the twist. The
fourfold has $\chi=1820448$. The rigid
gauge groups are the same as before. We notice
that the divisor $D^{15}$ now has local geometry
$\mathbb F_0$ and supports a rigid $E_8$. Although
we cannot construct a single $\U(1)$ gauge group,
Eq.\ (\ref{rlowerbound}) allows us to perform the
breaking $E_8\rightarrow E_6\times \U(1)$ on this
divisor.\footnote{As noted above, there are
codimension-2 $(4,6)$ singularities on $D^{15}$ associated
with extra strongly coupled sectors. Unless
these sectors have direct conflict with vertical
flux breaking (which we do not see immediately),
they should be irrelevant to the matter coming
from the adjoint of $E_8$, which is localized on
the bulk of $D^{15}$ instead of on matter curves.}

\begin{figure}[t]
\centering
\includegraphics[width=0.5\columnwidth]{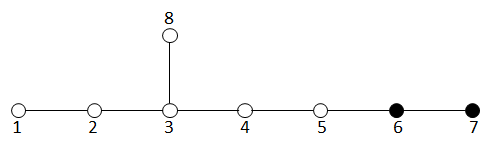}
\caption{The Dynkin diagram of $E_8$. The Dynkin node labelled
$i$ corresponds to the exceptional divisor $D_i$. The solid
nodes are the ones we break by vertical flux,
while we preserve a linear combination of their
corresponding Cartan generators. The unbroken
nodes form the Dynkin diagram of $E_6$.}
\label{fig:u1:dynkine8}
\end{figure}

Now the calculation is similar to that in Section
\ref{subsec:u1:fluxbackground}. The breaking can be done by imposing (see Figure \ref{fig:u1:dynkine8})

\begin{equation}
    \Theta_{1\alpha}=\Theta_{2\alpha}=\Theta_{3\alpha}=\Theta_{4\alpha}=\Theta_{5\alpha}=\Theta_{8\alpha}=a\Theta_{6\alpha}+b\Theta_{7\alpha}=0\,,
\end{equation}
where $\alpha$ stands for $S=D^0$ and $F=D^{14}$,
and $a,b$ are coprime integers.
We further require that there is no $E_8$ root
with the sixth and seventh components along
$(a,b)$. Now the above flux constraints are solved by
\begin{gather}
    (\phi_{1S},\phi_{2S},\phi_{3S},\phi_{4S},\phi_{5S},\phi_{6S},\phi_{8S})=(2,4,6,5,4,3,3)(a-2b)n_S\,,\nonumber\\
    \phi_{7S}=(2a-3b)n_S\,,
\end{gather}
and similarly for $\phi_{iF}$. Since $(a-2b)$ and
$(2a-3b)$ are also coprime, $n_S,n_F$ must be
integers. Primitivity still requires $n_S,n_F$ to
have opposite signs, and we choose $(n_S,n_F)=(\pm 1,\mp 1)$ to minimize the tadpole as before. Then
Eq.\ (\ref{eq:review:tadpole}) becomes
\begin{equation} \label{u1:eq:e8tofe6tadpole}
    a^2-3ab+3b^2\leq37926\,.
\end{equation}
Note the similarity to Eq.\ (\ref{u1:eq:g2tou1tadpole}).

We now look at the matter spectrum. The adjoint
$\mathbf{248}$ of $E_8$ breaks into $E_6$
fundamentals $\mathbf{27}$ (and their conjugate)
and singlets that are charged under the $\U(1)$.
To be precise, the charged representations are
\begin{equation}
    \mathbf{27}_{a},\,\mathbf{27}_{a-3b},\,\mathbf{27}_{-2a+3b},\,\mathbf{1}_{3b},\,\mathbf{1}_{3a-3b},\,\mathbf{1}_{3a-6b}\,,
\end{equation}
and their conjugates. 
Similar to the example in \S\ref{subsec:u1:fluxbackground},
the flux induces no chiral spectrum and the above representations
belong to vector-like matter only.
The maximum possible charge can be
obtained by maximizing the charge of one of the
singlets while satisfying Eq.\ 
(\ref{u1:eq:e8tofe6tadpole}). For example, $(a,b)=(2,-111)$ gives
maximum $3a-6b=672$ and the charged representations are
\begin{equation}
    \mathbf{27}_{2},\,\mathbf{27}_{335},\,\mathbf{27}_{-337},\,\mathbf{1}_{-333},\,\mathbf{1}_{339},\,\mathbf{1}_{672}\,,
\end{equation}
and their conjugates. Therefore, we have reached
the result $q_\mathrm{max}=672$ for $\U(1)$
coupled to $E_6$. Note that, as discussed above, the charge is 672 by
the definition we are using here where all $\U(1)$ charges are integers, while
the spectrum is invariant under the $\mathbb Z_3$
center of the gauge group. Therefore, it is also
natural to normalize the charges with units of
$1/3$, in which units the maximum charge would become
$q_\mathrm{max}=224$. 
One should thus be careful when
interpreting the result in this Section, while the
same ambiguity does not occur in \S\ref{sec:u1:G2model}.

It is worth emphasizing again that these models
with large $\U(1)$ charges require very
non-generic flux configurations, and small $\U(1)$
charges are exponentially preferred in the
landscape (assuming a measure where all flux configurations satisfying
the tadpole constraint are equally weighted). Heuristically, if one collects all the
$\U(1)$ charges of massless states across  the
landscape, one may expect a distribution that
peaks near $q=0$ and decays as $q$ grows \cite{Andriolo:2019gcb}. This
matches the expectation from phenomenology where
the $\U(1)$ charges are always small.

We see that no matter how large the
charges are, some Yukawa couplings such as
$\mathbf{27}^3$ in the above models are allowed to  be possible
from the structure of the unbroken gauge group,
while some other couplings are indeed forbidden by
the inclusion of exotic $\U(1)$ charges. The above
breaking can be straightforwardly generalized to
conventional grand unified theories such as
$E_6\rightarrow \SU(5)\times\U(1)$. Therefore,
vertical flux breaking can be useful in
phenomenological model building, in which the
$\U(1)$ extension may naturally explain issues
like proton decay \cite{Marsano:2009wr,Grimm:2010ez}. In particular, such
constructions can be easily incorporated into the
recently proposed natural construction of Standard
Model-like structure in F-theory \cite{Li:2021eyn,Li:2022aek}, as vertical flux
breaking is used in both contexts.

\section{Conclusion}
\label{sec:u1:conclusion}

In this chapter, we have studied $\U(1)$ charged
massless fields in string theory. 
While the completeness hypothesis
\cite{Polchinski:1998rr,Banks:2010zn,Harlow:2018tng} suggests that
states should exist with all possible charges under a U(1) gauge
field, in general massless or light fields have small charges.
Even in 6D, 
the upper bound
on  possible $\U(1)$ charges $q$ for  massless fields
is not completely understood, with very few
explicit string theory constructions of $\U(1)$
models available, and in 4D the question is addressed very little in
the literature. Using the formalism of vertical
flux breaking, we have efficiently constructed 4D F-theory models containing
$\U(1)$ charges for light fields that can be
as large as $q=657$, and massless chiral fields with charges that can
be at least as large as $q = 465$.
%This large
%charge greatly exceeds the previous results in literature. 
The string theory construction is
fully explicit, using fourfolds with large Euler
characteristics in the KS database and exotic flux
background. While a rigorous proof is far from
complete, we have some reasons to believe that our
result gives a plausible upper bound for $\U(1)$
charges in the 4D F-theory landscape, when the
$\U(1)$ is decoupled. We have found that the
$\U(1)$ charges can become slightly larger when the
$\U(1)$ factor is coupled to other gauge groups,  although we expect a similar
upper bound when for properly normalized charges.

It is worth emphasizing that the St\"uckelberg mechanism is also
available in, for example, type IIB string theory (see e.g.
\cite{Cvetic:2012xn} and the references therein), and one can construct
similar $\U(1)$ models in such a perturbative setup. The nonperturbative
nature of F-theory, however, allows us to explore a much wider range of
compactifications, in particular those containing exceptional gauge
groups as in our examples. As a result, F-theory opens up new corners in
the string landscape by raising the $\U(1)$ charges that can be explicitly constructed
to a much higher value. It remains interesting to compare the upper
limits from both F-theory and perturbative string theories.

These results lead in a number of interesting directions. First, as
described in \S\ref{sec:u1:intro}, this class of constructions
may give new insights in the context of the
Swampland program, as  it greatly expands the
view on which $\U(1)$ charged massless fields in
the low-energy theory can be consistently coupled
to gravity. 
It would be interesting to understand better how these 4D
constructions of theories with light fields having large U(1) charges
fit with the related 6D analyses of
automatic enhancement and of massless charge sufficiency and
the completeness hypothesis
\cite{Raghuram:2020vxm,Cvetic:2021vsw,Morrison:2021wuv}.
%As discussed in \S\ref{subsec:u1:fluxbackground}, these results
%also suggest interesting structure for large discrete gauge groups
%in 4D F-theory models
%that may be realized by Higgsing on massless chiral fields with large
%U(1) charges.
%We also learnt more on how the charge
%completeness becomes invalid when restricted to
%massless fields. 
More generally, we have found that the
formalism of vertical flux breaking %opens up large
%new  in the 4D F-theory landscape, 
leads
to large classes of new F-theory models ranging from exotic
$\U(1)$ charges to natural Standard Model-like
constructions. 
At the same time, the large charges we have found here are expected to
be exponentially rare in any natural measure on the landscape, and
this work motivates a more careful study of the impact of the
distribution of fluxes on the structure of the gauge group and matter
content. While this approach using flux breaking is certainly not completely new
in the literature, it has not been fully utilized until
now. 
We hope that this formalism will lead to many more
exciting results in  F-theory  constructions of 4D supergravity
models.
 Finally, this perspective offers new
ways of thinking about $\U(1)$ extensions in the
Standard Model or grand unified theories. As shown
in \S\ref{sec:u1:coupling}, the $\U(1)$ charges
we get, albeit exotic, are still potentially
relevant to particle phenomenology. 
It is interesting that the possible
appearance of such charges is supported from the UV perspective by explicit
string theory constructions.

We hope that future work will lead to a solidification of the
arguments for the upper bound on $q$, and the application of
this construction of exotic $\U(1)$ charges to
broader setups as mentioned in the previous
paragraphs.

\chapter{Conclusion and outlook}
\label{chap:conclusion}

String theory is a top-down framework that potentially unifies all the fundamental forces 
and gives answers to the deepest fundamental questions in the Standard Model of particle 
physics. Despite decades of research, our current knowledge in string theory is still
too limited for understanding how our Universe may be realized in this framework. One
of the approaches to study string theory is to explore a large set of string vacua,
known as the string landscape, and look for their statistical properties. F-theory, as
a nonperturbative version of type IIB string theory, is particularly suitable for this
approach. In particular, F-theory provides the currently largest set of string vacua, in
which the low-energy gauge dynamics are encoded in pure geometry and fluxes. Therefore,
there have been multiple proposals for constructing Standard-Model-like solutions in
F-theory, but it is not clear which method gives rise to the Standard Model in the most
``natural'' way. In this thesis, we have proposed to utilize flux breaking of rigid gauge 
groups in F-theory, since rigid gauge groups arise ubiquitously in F-theory without
fine-tuning any moduli, and at the same time contain the Standard Model gauge group as
a subgroup.

In Chapter~\ref{chap:fluxbreaking}, we have developed a systematic formalism of gauge 
symmetry breaking with fluxes. Vertical and remainder fluxes are used to break rigid
gauge groups, and the former induces chiral matter at the same time. Interestingly,
there can be nontrivial chiral spectra after flux breaking even when the unbroken gauge
group does not have complex representations. The resulting chiral index is controlled by
a large number of flux parameters with a linear Diophantine structure, hence it is fairly
likely to have small number of generations. We have written down streamlined procedures
for studying flux breaking with intersection theory on the compactified manifold.
Applying our formalism, we have obtained the Standard Model gauge group and chiral matter
by breaking rigid $E_6,E_7$ gauge groups with an intermediate $\SU(5)$ unification group.
We have further constructed a global example for such $E_6$ models.

Then, in Chapter~\ref{chap:e7}, we have further studied the phenomenological aspects of
the rigid $E_7$ models. We have found that $E_7$ unification models, while not possible
in the conventional field theory approach, can be realized in F-theory and lead to many
features compatible with the Standard Model but not available in other unification 
groups. More specifically, breaking the $E_7$ group leads to a set of approximate 
$\U(1)$ global symmetries descending from $E_7$, which exponentially suppress the 
undesired couplings. For example, dimension-4 and -5 proton decay is ubiquitously 
suppressed, and it is easy to induce doublet-triplet splitting. Under these symmetries,
the Higgs and the chiral matter have different string theory origins, which help ensure
the right Yukawa couplings and remove exotic couplings. We have found that these features
come from both the group-theoretic structure of $E_7$, and the geometric structure of 
$E_7$ realized in F-theory as a rigid gauge group, thus these features do not require 
fine-tuning of moduli. We have constructed explicit global examples of such $E_7$ models,
and compared our models to the existing Standard-Model like solutions in the literature.

Finally, in Chapter~\ref{chap:u1}, we have applied the flux breaking formalism to 
construct large $\U(1)$ charges in 4D F-theory models. Vertical flux breaking can produce
exotic $\U(1)$ gauge groups through the St\"uckelberg mechanism. We have applied vertical
flux on a compactified manifold with a $G_2$ gauge group and a very large Euler 
characteristic, which is found in the Kreuzer-Skarke database. The resulting charges in
the light matter spectrum are as high as $q=657$, and $q=465$ for massless chiral matter.
This result is much higher than the previous bounds set by 6D F-theory models, as flux
breaking in 4D has played a crucial role in constructing our models. We have argued
heuristically that our results may set an upper bound for decoupled $\U(1)$ charges in
4D F-theory.

Although these results allow us to understand many aspects of Standard-Model-like 
solutions in the F-theory landscape, these are far from the full story and many
important open questions remain. First, while we can reliably extract the discrete data such as
gauge groups and matter spectra from our models, our current technologies are not
sufficient to compute continuous data such as masses and couplings in our models. These
data depend on many factors that are still not fully understood, including moduli
dependence and corrections from the K\"ahler potential. Computing these quantities are
particularly challenging given the nonperturbative nature of F-theory. In addition, so 
far there is still no systematic way to stabilize moduli with horizontal flux in 
F-theory, although it is unlikely that moduli stabilization would heavily alter the 
discrete data due to the rigidity of the gauge groups. In any case, without explicit 
values of the masses and couplings, we can never ensure that
the exotic Higgs and other vector-like exotics are sufficiently heavy to be 
phenomenologically safe, and issues such as doublet-triplet splitting, the $\mu$-problem,
and the Yukawa hierarchy, remain not quantitatively understood.

At the same time, even at the discrete level, it is not yet fully clear whether the $E_7$
models are the right venue for natural Standard-Model-like solutions. Codimension-3
$(4,6)$ singularities in these models are not yet understood. It seems that these
singularities are resolved by non-flat fibers instead of fibral curves, so it is not clear if 
they can induce exotic matter to alter the phenomenology. Similar issues arise in $E_8$
models, which generically have codimension-2 $(4,6)$ singularities supporting strongly
coupled matter. In principle, it may be possible to break the $E_8$ gauge group with 
fluxes and turn the strongly coupled matter into ordinary gauge theory matter, in 
particular the Standard Model matter, but the formulation of $E_8$ flux breaking is not
clear at this moment. Since $E_8$ factors are much common in the F-theory landscape than
$E_7$, it is important to understand if $E_8$ models can also give rise to 
Standard-Model-like solutions like $E_7$ models. If so, we will be able to unlock a large 
portion of the string landscape for Standard Model constructions. Utilizing the duality between F-theory
and heterotic strings, in which constructing $E_8,E_7$ models are relatively easy, may help clarify some of these issues.
If it is not possible to utilize $E_8$ models, it is still
important to see how exactly $E_7$ factors are ubiquitous in the F-theory landscape.
For this reason, we also need to extend our techniques in, for example, computing the
vector-like spectrum on general types of geometry, such that a large-scale scan of 
$E_7$ unification models in the landscape is feasible.

More broadly speaking, since at this moment we do not observe any supersymmetry in our
Universe, we must look for ways to incorporate supersymmetry breaking (hence electroweak
symmetry breaking in the context of the minimally supersymmetric Standard Model) into
our models, and understand how many of our results crucially rely on supersymmetry.
In addition, our models only focus on the particle physics aspect, hence for realistic 
unification they must be merged with string theory models with de Sitter cosmology, 
constructing which is still a huge challenge in string theory; see recent progress in 
\cite{McAllister:2024lnt,McAllister:2025qwq}. Departing from the superstring framework and exploring solutions in 
nonsupersymmetric string theories may also be an important and promising route.

We also need a precise notion of naturalness when comparing different string theory 
models, or in other words, a reasonable and physical measure defined on the string
landscape. So far, we have been naively using the measure of counting Calabi-Yau 
topologies for naturalness arguments. While this measure allows us to extract some
nontrivial statistical properties of the string landscape, it is far from guaranteed that
this is even remotely close to the correct measure, which may depend on more consistency
conditions and/or the transition dynamics between different string vacua. Even within
the counting measure, there are subtleties of overcounting or undercounting the flux
configurations, the triangulations of polytopes in toric geometry, and so on. Without the
correct measure, the physical interpretation of the string landscape remains very vague.

Despite all the above limitations and open questions, we hope that the results in this
thesis have provided some important steps on the long-term goal of applying string theory
to our Universe, probably in a \emph{natural} way. We also hope that the formalism and
techniques we have developed can serve as foundations for future research in string
phenomenology.

%%% Appendices of thesis  %%%%%%%%%%%%%%%%%%%%%%%%%%%%%%%%%%%%%%%%%%%%%%%%%%%%%%%%%%%%%%%%%%%%%%%%

\appendix
\chapter{Appendix for Chapter~\ref{chap:fluxbreaking}}
\label{chap:appendix3}

\section{Flux-induced St\"uckelberg mechanism}
\label{sec:Stuckelberg}

In this appendix, we review \cite{Donagi:2008ca,Grimm:2010ks} how vertical gauge-breaking flux
induces St\"uckelberg masses for the gauge bosons in broken
$\U(1)$ directions. As we will see, the masses are indeed
given by nonzero $\Theta_{i\alpha}$. For simplicity, here we
ignore all numerical factors and signs, which are not
important to the results.

This effect is perhaps most easily understood in the dual M-theory
picture. Consider M-theory compactified on $\hat Y$. We need
the parts of the supergravity action $S_{11D}$ involving
$G_4$:
\begin{equation}
    S_{11D}\supset\int_{\mathbb R^{2,1}\times\hat Y}\left(G_4\wedge *G_4+C_3\wedge G_4\wedge G_4\right)\,.
\end{equation}
Here the first term is the kinetic term for $G_4$ and the
second term is the Chern-Simons coupling. We now expand $C_3$
and $G_4$ with the following relevant terms:
\begin{equation}
    C_{3}\supset A^{\alpha}\wedge\left[D_{\alpha}\right]+A^{i}\wedge\left[D_{i}\right]\,,\quad G_{4}\supset F^{\alpha}\wedge\left[D_{\alpha}\right]+F^{i}\wedge\left[D_{i}\right]+G_{\rm int}\,,
\end{equation}
where $A$ are the $\U(1)$ gauge fields in 3D, $F=dA$, and
$G_{\rm int}$ is the flux in compactified or internal directions
i.e. the $G_4$ in the main text. We can then integrate over
$\hat Y$ and get the 3D effective action. Recall that
in the F-theory limit, $A^{\alpha}$ lives in chiral
multiplets and can be dualized into axions, while $A^{i}$
lives in vector multiplets giving the gauge bosons
in 4D. In particular, $A^{\alpha}$ and $A^i$
decouple in the kinetic term. The terms involving
$A^{\alpha}$ and its derivative are therefore
\begin{equation}
    S_{3D}\supset\int_{\mathbb{R}^{2,1}}\left(K_{\alpha
      \beta}F^{\alpha}\wedge*F^{\beta}+\Theta_{i\alpha}A^{i}\wedge F^{\alpha}\right)\,,
\end{equation}
where
\begin{equation}
    K_{IJ}=\int_{\hat Y}\left[D_{I}\right]\wedge*\left[D_{J}\right]\,,
\end{equation}
is the metric. There are also terms proportional to
$A^{\alpha}\wedge F^{i}$, but they are the same as
$A^{i}\wedge F^{\alpha}$ by integration by parts. 

We then construct the axion dual. First notice that since $dF^{\alpha}=0$, we can add a Lagrange multiplier $a_{\alpha}$ to the action i.e. a term $a_{\alpha}dF^{\alpha}$. Performing integration by parts, we have
\begin{equation}
    S_{3D}\supset\int_{\mathbb{R}^{2,1}}\left(K_{\alpha \beta}F^{\alpha}\wedge*F^{\beta}+\left(da_{\alpha}+\Theta_{i\alpha}A^{i}\right)\wedge F^{\alpha}\right)\,.
\end{equation}
The equation of motion gives
\begin{equation}
    *F^{\beta}=K^{\alpha \beta}\left(da_{\alpha}+\Theta_{i\alpha}A^{i}\right)\,.
\end{equation}
We finally integrate out $F^\alpha$ and get
\begin{equation}
    S_{3D}\supset\int_{\mathbb{R}^{2,1}}\left(K^{\alpha\beta}\left(da_{\alpha}+\Theta_{i\alpha}A^{i}\right)\wedge*\left(da_{\beta}+\Theta_{j\beta}A^{j}\right)\right)\,.
\end{equation}
By gauge transformations, the gauge fields $A^i$ can ``eat''
the axions $a_\alpha$ and become massive as long as there are
enough axion fields. The masses are determined by the
eigenvalues of the mass matrix
$K^{\alpha\beta}\Theta_{i\alpha}\Theta_{j\beta}$. Its
null space, hence massless $\U(1)$ directions, corresponds to
linear relations between $\Theta_{i\alpha}$ i.e.
$c_i\Theta_{i\alpha}=0$ for all $\alpha$. This justifies the flux constraints
imposed in the main text for gauge breaking.

\section{Embeddings of Standard Model gauge group into $E_7$}
\label{sec:embeddingcount}
In this Appendix, we count different embeddings of $\gsm$
into $E_7$ giving SM matter representations. It is stated in
\S\ref{ssubsec:fluxbreaking:SMchiral} that the root embedding of
$\SU(3)\times\SU(2)$ is unique up to automorphisms, and there
are 4 distinct choices of hypercharge $\U(1)$. Here we prove
these claims.

To prove the uniqueness of the root embedding  up to automorphisms
we proceed in a somewhat
explicit constructive fashion.
We can describe $E_8$ explicitly as a lattice consisting of all points
\begin{equation}
  (x_1, x_2, \ldots, x_8) \in \{(\Z)^8\cup (\Z +1/2)^8: \sum_i x_i
  \equiv 0 \;({\rm mod}\ 2)\}\,.  
\label{eq:}
\end{equation}
The roots of $E_8$ are the 240 elements of this lattice satisfying $r
\cdot r = 2$.
$E_7$ can be realized as the
orthogonal complement of any root $r_8$ of $E_8$, so without loss of
generality we pick $r_8 =(1, -1, 1, -1, 1, -1, 1, -1)/2$.
There are 126 roots $r$ of $E_8$ satisfying $r \cdot r_8 = 0$,
corresponding to the roots of $E_7$.  To embed $SU(3) \subset E_7$ as
a root embedding, we wish to choose roots $r_1, r_2 \in E_7$ such that
$r_1 \cdot r_2= -1$.  Choosing arbitrarily $r_1 = (1, 1, 0, 0, 0, 0,
0, 0)$ from the 126 equivalent roots of $E_7$, there are 32 roots
satisfying the condition on $r_2$, from which we pick arbitrarily $r_2
= (0, -1, -1, 0, 0, 0, 0, 0)$.  There are 30 roots of $E_7$ that are
perpendicular to $r_1, r_2$, so we embed the $SU(2)$ with the
arbitrary choice $r_7 = (0, 0, 0, 1, 1, 0, 0, 0)$.  Assuming
momentarily that all 32 choices of $r_2$ give 30 choices of $r_7$
(which we will prove below shortly) this gives $126*32*30 =120,960$
root embeddings of $SU(3) \times SU(2)$ into $E_7$.

We now show that
our given choice is equivalent to the one illustrated in
Figure~\ref{dynkine7}.  To identify the root associated with node 3 in
that diagram, we need a root $r_3 \in E_7$ such that $r_3 \cdot  r_1 =
0, r_3 \cdot r_2 = -1, r_3 \cdot r_7 = -1$.  There are 4 such roots,
among them we pick $r_3 =(0, 0, 1, -1, 0, 0, 0, 0)$.  Continuing in
this fashion, there are 3 choices for $r_4$, and 2 choices for $r_5$,
after which $r_6$ is uniquely determined.  Multiplying out
$120,960*24*240 = 696,729,600$, which is exactly the size of the
automorphism group of $E_8$.  Given one choice of embedding given by
the sequence of roots described above, each automorphism of $E_8$ will
give a distinct embedding.  Thus, we must have at least this many
independent sequences of choices based on the equivalent embeddings.
If there were any inequivalent root embeddings, they would have given
rise to a larger number of choices at some step in the process.  This
proves that indeed all of the root embeddings are  equivalent at each stage.
Note that there are also exotic non-root embeddings of $SU(3) \times
SU(2)$ into $E_7$\footnote{We would like to thank Andrew Turner for
  discussions on this point.}, but these cannot be realized by flux breaking and
are not relevant to the discussion here.

We can now relate this analysis to the choices of hypercharge.
Without loss of generality, we can now put the non-abelian
part of $\gsm$ in nodes $1,2,7$ in Figure \ref{dynkine7}. Let
the hypercharge of $R'$ with given $b_{i'}$ be $q_Y=a_{i'} b_{i'}$, where $i'=3,4,5,6$ and $a_{i'}$ are numbers to be
solved. We then break the $\mathbf{56}$ into representations of
$\gsm$ and require them to be the SM representations. There
is only one $(\mathbf 3,\mathbf 2)$, which has
$b_{i'}=(2,3/2,1,1/2)$. Therefore,
\begin{equation}
    2a_3+3a_4/2+a_5+a_6/2=1/6\,.
\end{equation}
Similarly, by looking at $(\bar{\mathbf 3},\mathbf 1)$ and $(\mathbf 1,\mathbf 2)$, we get
\begin{align}
    a_3+3a_4/2+a_5+a_6/2&=-2/3\;\mathrm{or}\;1/3\,,\\
    a_3+a_4/2+a_5+a_6/2&=-2/3\;\mathrm{or}\;1/3\,,\\
    a_4/2+a_5+a_6/2&=\pm 1/2\,,\\
    a_4/2+a_6/2&=\pm 1/2\,,\\
    a_4/2-a_6/2&=\pm 1/2\,.
\end{align}
It is then straightforward to deduce that the only
possibilities of $a_{i'}$ are
\begin{equation}
    a_{i'}=(5/6,-1,0,0),(-1/6,1,-1,0),(-1/6,0,1,-1),(-1/6,0,0,1)\,.
\end{equation}
These four $a_{i'}$ correspond to the four choices in Eq.\ (\ref{hyperchargeChoice}), and the four roots that enhance
$\SU(3)\times\SU(2)$ to $\SU(5)$, which are equivalent under
automorphisms as described above. This finishes our proof.

Note that we have assumed here that the U(1) hypercharge assignment
for the states coming from the $\mathbf{56}$
gives the SM values, with no exotics.  This does not rule
out a choice of U(1) where the states coming from the $\mathbf{56}$
include some exotics and omit some SM states, while the
states coming from the $\mathbf{133}$ can complete the SM
states and contain other exotics, as we found explicitly for the
breaking pattern described in \S\ref{ssubsec:fluxbreaking:exotic}.  While in that
case, there was no flux choice that gives just the SM
chiral matter content with no exotics, we have not ruled out the
possibility that some other U(1) hypercharge assignment may allow in
principle for a similar situation, where fine tuning of the fluxes may
reduce to only SM chiral matter.  We leave a more detailed
investigation of this question for further work.

\section{Resolution of $E_6$ model}
\label{sec:resolution}

In this appendix, we describe more about the resolution of
$E_6$ model used in the main text. As a starting point, a
generic $E_6$ model can be described by a Tate model \cite{BershadskyEtAlSingularities,KatzEtAlTate}, where
$Y$ is given by the locus of
\begin{equation}
    y^2+a_{1,1}sxyz+a_{3,2}s^2yz^3=x^3+a_{2,2}s^2x^2z^2+a_{4,3}s^3xz^4+a_{6,5}s^5z^6\,,
\end{equation}
where the class of $a_{i,j}$ is $-(iK_B+j\Sigma)$. We now
resolve this model by performing blowups. We denote
\begin{equation}
    Y_1\stackrel{(x,y,s|e_{1})}{\longrightarrow}Y\,,
\end{equation}
as the blowup from $Y$ to $Y_1$ by the redefinition
\begin{equation}
    x\rightarrow xe_1\,,\quad y\rightarrow ye_1\,,\quad s\rightarrow se_1\,.
\end{equation}
The resulting locus $e_1=0$ is a divisor in the ambient
space, denoted by $E_1$. Using the same notation, we can then
write down the resolution as the following steps \cite{Esole:2017kyr,Bhardwaj_2019,Jefferson:2021bid}:
\begin{equation}
    \hat Y\stackrel{(y,e_4|e_6)}{\longrightarrow}Y_5\stackrel{(y,e_3|e_5)}{\longrightarrow}Y_4\stackrel{(e_2,e_3|e_4)}{\longrightarrow}Y_3\stackrel{(x,e_2|e_3)}{\longrightarrow}Y_2\stackrel{(y,e_1|e_2)}{\longrightarrow}Y_1\stackrel{(x,y,s|e_1)}{\longrightarrow}Y\,.
\end{equation}
This resolution smooths out all singularities on $Y$ up to codimension-3. The exceptional divisors on $\hat Y$ are given by
\begin{align}
    D_1 &= E_5\cap \hat Y\,, \nonumber \\
    D_2 &= E_6\cap \hat Y\,, \nonumber \\
    D_3 &= (-E_1+2E_2-E_3-E_4)\cap \hat Y\,, \nonumber \\
    D_4 &= (E_1-2E_2+E_3+2E_4-E_6)\cap \hat Y\,, \nonumber \\
    D_5 &= (E_3-E_4-E_5)\cap \hat Y\,, \nonumber \\
    D_6 &= (E_1-E_2)\cap \hat Y\,.
\end{align}
Using the above information, the intersection numbers between
divisors on $\hat Y$ can then be computed using the
techniques in \cite{Esole:2017kyr}.

\chapter{Appendix for Chapter~\ref{chap:e7}}
\label{appendix:e7}

\section{Toric hypersurfaces}
\label{appendix:Hypersurfaces}

In this Appendix, we explain how to count $h^{1,1}$, or the number of
divisors, of a threefold hypersurface in an ambient toric 
fourfold, 
following the general
approach of Danilov and Khovanskii \cite{Danilov_1987}.\footnote{We thank Manki Kim for teaching us these techniques.} This
technique is useful in \S\ref{sec:ExplicitConstruction}.
To simplify the discussion, we focus on simple cases where there is a
triangulation such that both the ambient space and the hypersurface
are smooth. We also assume that the hypersurface does
not have any base locus.

The geometry of the hypersurface can be understood from its 
stratification. First we look at the stratification of the ambient space.
A $d$-dimensional toric variety is given by a 
disjoint union of algebraic tori $(\mathbb C^*)^k$, where $0\leq k\leq d$.
These algebraic tori, called strata, are associated with the cones of a
toric (polyhedral)
fan. To be more precise,
for a toric fan $\Sigma$ with $n$-dimensional cones
$\sigma^{(n)}\in\Sigma(n)$ (where $0\leq n\leq d$), the toric variety
$\mathbb P_\Sigma$ is given by
\begin{equation}
    \mathbb P_\Sigma=\coprod_n \coprod_{\sigma^{(n)}\in\Sigma(n)}T_{\sigma^{(n)}}\,,\quad T_{\sigma^{(n)}}\cong (\mathbb C^*)^{d-n}\,.
\end{equation}
Notice that the unique $\sigma^{(0)}$ corresponds to the prime stratum
$(C^*)^d$, which is the defining feature of toric varieties. The 
one-dimensional cones $\sigma^{(1)}$ are also given by the toric rays 
$\vec v$, associated with prime toric divisors $D_{\vec v}$.

Now consider a hypersurface $Z$ as a divisor in $\mathbb P_\Sigma$. We
abuse notation and use $Z$ to also denote its divisor class:
\begin{equation}
    Z=\sum_{\vec v} a_{\vec v} D_{\vec v}\,.
\end{equation}
Note that the prime toric divisors are not all independent and there
are multiple choices of the coefficients $a_{\vec v}$ for the same $Z$;
our final results are independent of such a choice. 
We assume
 that all
the strata of $\mathbb P_\Sigma$ intersect $Z$ transversely. Then $Z$
admits the following stratification
\begin{equation}
\label{eq:Zstratasigma}
    Z=\coprod_n \coprod_{\sigma^{(n)}\in\Sigma(n)}Z_{\sigma^{(n)}}\,,\quad Z_{\sigma^{(n)}}=Z\cap T_{\sigma^{(n)}}\,.
\end{equation}
Note that the dimension of the strata $Z_{\sigma^{(n)}}$ is $d-n-1$. To 
understand the geometry of $Z_{\sigma^{(n)}}$, it is useful to 
construct the Newton polytope $\Delta$ of $Z$
\begin{equation}
    \Delta=\left\{\vec m \mid \vec m\cdot\vec v\geq -a_{\vec v},\forall \vec v\in\Sigma(1)\right\}\,.
\end{equation}
The Newton polytope encodes information about the holomorphic sections
of the line bundle $\mathcal O_{\mathbb P_\Sigma}(Z)$. Below we restrict
to the case where this line bundle is big, i.e. $\Delta$ is
also $d$-dimensional.

The faces of $\Delta$ %also
 encode
 the geometry of $Z$ in the following
way. From $\Delta$ we can construct the so-called normal fan 
$\Sigma(\Delta)$, where each $k$-dimensional face $\Theta^{(k)}$ is
associated with a $(d-k)$-dimensional cone in $\Sigma(\Delta)$.\footnote{The
explicit construction of $\Sigma(\Delta)$ is more complicated but is
not important for our purpose.} The
resulting toric variety $\mathbb P_{\Sigma(\Delta)}$ is a blowdown of
$\mathbb P_\Sigma$,
 which is singular in general. The corresponding
blowdown of $Z$ is denoted by $Z(\Delta)$. An important fact is that 
$Z(\Delta)$ is an ample divisor in $\mathbb P_{\Sigma(\Delta)}$. Now
given the one-to-one correspondence between faces of $\Delta$ and cones
of $\Sigma(\Delta)$, we can write the stratification of $Z(\Delta)$ as
(again, assuming all strata of $\mathbb P_{\Sigma(\Delta)}$ intersect
$Z(\Delta)$ transversely)
\begin{equation}
    Z(\Delta)=\coprod_k\coprod_{\Theta^{(k)}}Z_{\Theta^{(k)}}\,.
\end{equation}
Note that the dimension of the strata $Z_{\Theta^{(k)}}$ is $k-1$. Now
including the blowups from $Z(\Delta)$ back to $Z$, the stratification of
$Z$ can be written as
\begin{equation}
\label{eq:ZstrataTheta}
    Z=Z_{\Theta^{(d)}}\coprod_{\Theta^{(d-1)}}Z_{\Theta^{(d-1)}}\coprod_{k\geq 2}\coprod_{\Theta^{(d-k)}}E_{\Theta^{(d-k)}}\times Z_{\Theta^{(d-k)}}\,,
\end{equation}
where
\begin{equation}
    E_{\Theta^{(d-k)}}=\coprod^{k-1}_{i=0}\left(\coprod(C^*)^i\right)\,,
\end{equation}
is the exceptional set associated with $\Theta^{(d-k)}$ resulting from 
the blowups. The geometry of $Z_{\sigma^{(n)}}$ can then be seen by
comparing Eqs. (\ref{eq:Zstratasigma}) and (\ref{eq:ZstrataTheta}).

A great advantage of studying the stratification of $Z$ is that the Hodge
numbers of $Z$ can be computed using the Hodge-Deligne numbers together with
the stratification \cite{Danilov_1987,Braun:2014xka} (see also \cite{Jefferson:2022ssj} for more recent review and 
applications). In our case where $Z$ is smooth, the Hodge-Deligne numbers
are just certain signed combinations of the Hodge numbers, but they behave
nicely under disjoint unions and products. One can then compute the
Hodge-Deligne numbers of $Z$ by combining those of its strata, which are
easy to get. Although the formulas for general Hodge numbers are more
complicated, it can be shown that for $d\geq 4$, $h^{1,1}(Z)$ is simply
given by counting the irreducible components of $Z_{\sigma^{(1)}}$. In
terms of $\Delta$, we should look at the faces
\begin{equation}
    \Theta_i=\left\{\vec m \mid \vec m\cdot\vec v\geq -a_{\vec v},\forall \vec v\neq \vec v_i;\vec m\cdot\vec v_i=-a_{\vec v_i}\right\}\,.
\end{equation}
All $\Theta_i$'s are nontrivial when $Z$ does not have any base locus,
but they can have different dimensions and contribute differently to
$h^{1,1}(Z)$:
\begin{itemize}
    \item $\mathrm{dim}(\Theta_i)=0$: $Z_{\sigma^{(1)}}$ is given by
    the components in $E_{\Sigma^{(0)}}\times Z_{\Sigma^{(0)}}$. For generic
    moduli, however, $Z_{\Sigma^{(0)}}$ is an empty set and such $\Theta_i$
    does not contribute to $h^{1,1}(Z)$.
    \item $\mathrm{dim}(\Theta_i)=1$: $Z_{\sigma^{(1)}}$ is given by
    the components in $E_{\Sigma^{(1)}}\times Z_{\Sigma^{(1)}}$. Notice
    that $Z_{\Theta_i}$ is a degree $n=l^*(\Theta_i)+1$ hypersurface
    in $\mathbb C^*$, where $l^*$ denotes the number of interior points. For
    generic moduli, this hypersurface is a collection of $n$
    points, so there are $n$ copies of an irreducible
    component in $Z_{\sigma^{(1)}}$, contributing $n$ to $h^{1,1}(Z)$.
    \item $\mathrm{dim}(\Theta_i)=k\geq 2$: $Z_{\sigma^{(1)}}$ is given by
    the components in $E_{\Sigma^{(k)}}\times Z_{\Sigma^{(k)}}$. We see that $Z_{\sigma^{(1)}}$ is irreducible, contributing 1 to $h^{1,1}(Z)$.
\end{itemize}
Finally, the above procedure overcounts $h^{1,1}(Z)$ by $d$, since there
are $d$ linear relations between prime toric divisors in $\mathbb P_\Sigma$. One can check that
the above procedure reproduces the famous Batyrev formula for toric
hypersurface Calabi-Yau manifolds \cite{batyrev1993dual}.

For applications in \S\ref{sec:ExplicitConstruction}, it is now clear that to obtain $h^{1,1}(Z)=h^{1,1}(\mathbb P_\Sigma)$, we can pick $Z$ such that $\mathrm{dim}(\Theta_i)\geq 2$ for all rays $\vec v_i$. It is straightforward to check
this condition for the example in \S\ref{sec:ExplicitConstruction}.

\section{Flux quantization}
\label{appendix:Quantization}

In this Appendix, we compute $c_2(\hat Y)$ and determine the impact of
flux quantization on the vertical flux parameters in \S\ref{sec:ExplicitConstruction}. The 
computation of $c_2(\hat Y)$ involves an explicit choice of resolution. 
We consider the singular Weierstrass model in Eq. \eqref{eq:e7}, and
resolve it by performing blowups. We denote
\begin{equation}
    Y_1\stackrel{(x,y,s|e_{1})}{\longrightarrow}Y\,,
\end{equation}
as the blowup from $Y$ to $Y_1$ by the redefinition
\begin{equation}
    x\rightarrow xe_1\,,\quad y\rightarrow ye_1\,,\quad s\rightarrow se_1\,.
\end{equation}
The resulting locus $e_1=0$ is a divisor in the ambient
space, denoted by $E_1$. Using the same notation, we can then
write down the resolution as the following steps \cite{Esole:2017kyr,Bhardwaj_2019,Jefferson:2021bid}:
\begin{equation}
    \hat Y\stackrel{(e_4,e_5|e_7)}{\longrightarrow}Y_6\stackrel{(e_2,e_4|e_6)}{\longrightarrow}Y_5\stackrel{(e_2,e_3|e_5)}{\longrightarrow}Y_4\stackrel{(y,e_3|e_4)}{\longrightarrow}Y_3\stackrel{(x,e_2|e_3)}{\longrightarrow}Y_2\stackrel{(y,e_1|e_2)}{\longrightarrow}Y_1\stackrel{(x,y,s|e_1)}{\longrightarrow}Y\,.
\end{equation}
This resolution smooths out all singularities on $Y$ up to codimension
3. The exceptional divisors on $\hat Y$ are given by
\begin{align}
    D_1 &= (E_1-E_2)\cap \hat Y\,, \nonumber \\
    D_2 &= (-E_1+2E_2-E_3-E_5-E_6)\cap \hat Y\,, \nonumber \\
    D_3 &= (E_1-2E_2+E_3+2E_5+E_6-E_7)\cap \hat Y\,, \nonumber \\
    D_4 &= E_7\cap \hat Y\,, \nonumber \\
    D_5 &= (E_3-E_4-E_5)\cap \hat Y\,, \nonumber \\
    D_6 &= (-E_3+2E_4+E_5-E_6-E_7)\cap \hat Y\,, \nonumber \\
    D_7 &= (-E_1+2E_2-E_3-2E_5+E_7)\cap \hat Y\,.
\end{align}
Using the above information, we can then compute $c_2(\hat Y)$ using the
techniques in \cite{Esole:2017kyr,Jefferson:2021bid}. The computation
involves a pushforward formula from $\hat Y$ to $Y$ for the total Chern
class, and homology relations to relate all $D_i\cdot D_j$ to $D_i\cdot D_\alpha$. The result is
\begin{align}
[c_{2}(\hat Y)] & =  [c_{2}(B)]+11\pi^{*}K_{B}^{2}\nonumber \\
& +\left(-12D_{0}+14D_{1}+30D_{2}+48D_{3}+41D_{4}+28D_{5}+17D_{6}+27D_{7}\right)\cdot\pi^{*}K_{B}\nonumber \\
 & +\left(2D_{1}+6D_{2}+12D_{3}+12D_{4}+8D_{5}+6D_{6}+8D_{7}\right)\cdot\pi^{*}\Sigma\,.
\end{align}
It is known that the first row of the above is even \cite{Collinucci:2010gz}. Therefore, the potentially odd terms are
$D_i\cdot\pi^* K_B$ for $i=4,6,7$. To determine the parity of these terms, it is
more convenient to work with their pushforward
$\pi_*(D_i\cdot\pi^*K_B)=\Sigma\cdot K_B$.
For the model in \S\ref{sec:ExplicitConstruction}, we calculate
\begin{equation}
\Sigma\cdot K_{B}=K_{\Sigma}+N_{\Sigma}=-\sigma_{A}\cdot\left(9F_{\sigma}\cdot F_{E_{1}}+6F_{\sigma}\cdot F_{E_{2}}+11F_{\sigma}\cdot F_{F}+6F_{E_{1}}\cdot F_{F}\right)\,,
\end{equation}
which has odd coefficients. Notice that
\begin{equation}
\Sigma\cdot\left.(F_{E_1}+F_F)\right|_B=B\cdot\sigma_{A}\cdot\left(F_{E_{1}}+F_{F}\right)=\sigma_{A}\cdot\left(F_{\sigma}\cdot F_{E_{1}}+F_{\sigma}\cdot F_{F}+2F_{E_{1}}\cdot F_{F}\right)\,,
\end{equation}
has the same parity as $\Sigma\cdot K_B$, so the pullback
\begin{equation}
    (D_4+D_6+D_7)\cdot\pi^*\left.(F_{E_1}+F_F)\right|_B\,,
\end{equation}
has the same parity as $[c_2(\hat Y)]$. We see that flux quantization as
in Eq. \eqref{eq:review:fluxquant} can be satisfied by turning on half-integer $\phi_{iE_1}$ and $\phi_{iF}$ for $i=4,6,7$. From Eq. (\ref{eq:phi-n}),
this is the same as half-integer flux parameters $n_{E_1},n_F,\phi_{6E_1},\phi_{6F}$.

\chapter{Appendix for Chapter~\ref{chap:u1}}
\label{chap:appendix5}

\section{Various descriptions of the geometry}
\label{sec:u1:equivalence}

In this section, we compare two descriptions of
the geometry in Section \ref{subsec:u1:geometry}, and
show that they are equivalent. One description is
a generic elliptic fibration over a given
threefold base, and another one is the
anticanonical hypersurface in a 5D (singular)
weighted projective space.

Let us provide more details on the elliptic fibration. Since the elliptic curve is the \cy hypersurface in weighted projective space $\mathbb P^{2,3,1}$, the elliptic fibration on a toric base can also be written as the \cy hypersurface in a 5D toric ambient space. The toric ambient space is a $\mathbb P^{2,3,1}$ fibration over the base, given by the following toric rays:
\begin{equation}
    (w_i,-2,-3)\,,\quad(0,0,0,1,0)\,,\quad(0,0,0,0,1)\,.
\end{equation}
The resulting fivefold is singular due to the
presence of rigid gauge groups. Those
singularities on gauge divisors can be resolved by
adding ``tops'' into the toric fan \cite{Candelas:1996su}. We do not
describe the details here, but one can show that
after adding all the tops from Eq.\
(\ref{eq:u1:totalgaugegroup}), the convex hull of the
toric fan is a reflexive polytope with vertices
\begin{gather}
    \left(0,0,1,-2,-3\right)\,,\quad\left(-1,-12,0,-2,-3\right)\,,\quad\left(0,1,0,-2,-3\right)\,,\nonumber\\\left(0,0,0,1,0\right)\,,\quad\left(0,0,0,0,1\right)\,,\quad\left(80,468,-1,-2,-3\right)\,,\quad\left(42,246,0,-2,-3\right)\,.
\end{gather}
Notice again that $(80,468)=4v_{19}$ and $(42,246)=6v_{15}$.

Now we perform the following $\mathrm{SL}(5)$ transformation on the vertices:
\begin{equation}
    \left(\begin{array}{ccccc}
0 & 0 & 1 & 0 & 0\\
-1 & 0 & 0 & 0 & 0\\
-12 & 1 & 0 & 0 & 0\\
-26 & 2 & 2 & 1 & 0\\
-39 & 3 & 3 & 0 & 1
\end{array}\right)\cdot\left(\begin{array}{ccccccc}
0 & -1 & 0 & 0 & 0 & 80 & 42\\
0 & -12 & 1 & 0 & 0 & 468 & 246\\
1 & 0 & 0 & 0 & 0 & -1 & 0\\
-2 & -2 & -2 & 1 & 0 & -2 & -2\\
-3 & -3 & -3 & 0 & 1 & -3 & -3
\end{array}\right)=\left(\begin{array}{ccccccc}
1 & 0 & 0 & 0 & 0 & -1 & 0\\
0 & 1 & 0 & 0 & 0 & -80 & -42\\
0 & 0 & 1 & 0 & 0 & -492 & -258\\
0 & 0 & 0 & 1 & 0 & -1148 & -602\\
0 & 0 & 0 & 0 & 1 & -1722 & -903
\end{array}\right)\,.
\end{equation}
The geometry described by the polytope should be
$\mathrm{SL}(5)$ invariant. On the right hand
side, the first six columns are precisely the
toric rays of $\mathbb P^{1,80,492,1148,1722}$
mentioned in Section \ref{subsec:u1:geometry}, and
the last column represents an exceptional divisor
resulting from resolving the singularity in this
weighted projective space. These data match those
in the KS database. Hence we have proved the
equivalence between these two descriptions of the geometry.

%%% Bibliography (biblatex)  %%%%%%%%%%%%%%%%%%%%%%%%%%%%%%%%%%%%%%%%%%%%%%%%%%%%%%%%%%%%%%%%%%%%%%

\defbibheading{bibintoc}{\chapter*{#1}\addcontentsline{toc}{backmatter}{\refname}} 
% this sets the title of contents name for bibliography to \refname (= References)
% change "backmatter" to "chapter" if you prefer a bold face entry in the table of contents

\printbibliography[title={\refname},heading=bibintoc]

% biblatex also supports chapter-by-chapter bibliography, https://tex.stackexchange.com/a/296502/119566
% see the biblatex manual, section 3.14.3

\end{document}